%% file: paper_jhugen.tex
\documentclass[showpacs,aps,superscriptaddress,letterpaper,nofootinbib]{revtex4}

\usepackage{xspace}
\usepackage{graphicx}
\usepackage{epsfig}
\usepackage{float}
\usepackage[nointegrals]{wasysym}
\usepackage{hyperref}
\usepackage{url}
\usepackage{subfig}
\usepackage{amsmath}
\usepackage{xcolor}
\usepackage{amssymb,array}

\newcommand{\Offshell}{Off-shell}
\newcommand{\offshell}{off-shell}

\newcommand{\onshell}{on-shell}

\newcommand{\be}{\begin{equation}}
\newcommand{\ee}{\end{equation}}
\newcommand{\ba}{\begin{eqnarray}}
\newcommand{\ea}{\end{eqnarray}}

\newcommand{\Z}{Z}

\newcommand{\PH}{H}

\newcommand{\ttH}{t\bar{t}H}

\newcommand{\Hboson}{$H$~boson\xspace}
\newcommand{\Hff}{H\!f\!\bar{f}}

\def\sss{\scriptscriptstyle}

\begin{document}

\begin{flushright}
\vbox{
\begin{tabular}{l}
{ HU-EP-20/02}
\end{tabular}
}
\end{flushright}

\vspace{0.6cm}

\title{New features in the JHU generator framework: Constraining Higgs boson properties from on-shell and off-shell production}
\author{Andrei V. Gritsan \thanks{e-mail: gritsan@jhu.edu}}
\affiliation{Department of Physics and  Astronomy, Johns Hopkins University, Baltimore, MD 21218, USA}
\author{Jeffrey Roskes  \thanks{e-mail: hroskes@jhu.edu}}
\affiliation{Department of Physics and Astronomy, Johns Hopkins University, Baltimore, MD 21218, USA}
\author{Ulascan Sarica  \thanks{e-mail: ulascan.sarica@cern.ch}}
\affiliation{Department of Physics and Astronomy, Johns Hopkins University, Baltimore, MD 21218, USA}
\affiliation{Department of Physics, University of California, Santa Barbara, CA 93106, USA}
\author{Markus Schulze \thanks{e-mail:  markus.schulze@physik.hu-berlin.de}}
\affiliation{Institut f\"ur Physik, Humboldt-Universit\"at zu Berlin, D-12489 Berlin, Germany}
\author{Meng Xiao  \thanks{e-mail: meng.xiao@cern.ch}}
\affiliation{Department of Physics and Astronomy, Johns Hopkins University, Baltimore, MD 21218, USA}
\affiliation{Zhejiang Institute of Modern Physics, Department of Physics, Zhejiang University, Hangzhou, 310027, P. R. China}
\author{Yaofu Zhou  \thanks{e-mail: yzhou49@jhu.edu}}
\affiliation{Department of Physics and Astronomy, Johns Hopkins University, Baltimore, MD 21218, USA}
\affiliation{Department of Physics, Missouri University of Science and Technology, Rolla, MO 65409, USA}
%

\date{February 21, 2020}

\begin{abstract}
\vspace{2mm}
We present an extension of the JHUGen and MELA framework, which includes an event generator and 
library for the matrix element analysis. It enables simulation, optimal discrimination, reweighting techniques, and 
analysis of a bosonic resonance and the triple and quartic gauge boson interactions with the most general anomalous couplings.
The new features, which become especially relevant at the current stage of LHC data taking, are the simulation of gluon fusion 
and vector boson fusion in the off-shell region, associated $ZH$ production at NLO QCD including the $gg$ initial state, 
and the simulation of a second spin-zero resonance. We also quote translations of the anomalous coupling
measurements into constraints on dimension-six operators of an effective field theory. 
Some of the new features are illustrated with projections for experimental measurements 
with the full LHC and HL-LHC datasets. 
\end{abstract}


\maketitle

\thispagestyle{empty}


\section{Introduction}
\label{sect:cp_intro}

We present a coherent framework for the measurement of couplings of the Higgs ($H$) boson and a possible second spin-zero resonance. 
Our framework includes a Monte Carlo generator and matrix element techniques for optimal analysis of the data. We build upon the earlier
developed framework of the JHU generator and MELA analysis package~\cite{Gao:2010qx,Bolognesi:2012mm,Anderson:2013afp,Gritsan:2016hjl}
and extensively use matrix elements provided by MCFM~\cite{Campbell:2010ff,Campbell:2011bn,Campbell:2013una,Campbell:2015vwa,Campbell:2015qma}.
Thanks to the transparent implementation of standard model (SM) processes in MCFM, we extend them to add the most general scalar and 
gauge couplings and possible additional states. This allows us to build on the previously studied 
topics~\cite{Nelson:1986ki,Soni:1993jc,Plehn:2001nj,Choi:2002jk,Buszello:2002uu,Hankele:2006ma,Accomando:2006ga,
Godbole:2007cn,Hagiwara:2009wt,Gao:2010qx,DeRujula:2010ys,Christensen:2010pf,Bolognesi:2012mm,Ellis:2012xd,Chen:2012jy,
Artoisenet:2013puc,Anderson:2013afp,Chen:2013waa,Maltoni:2013sma,
Azatov:2014jga,Cacciapaglia:2014rla,Denner:2014cla,Dolan:2014upa,Englert:2014ffa,Gonzalez-Alonso:2014eva,
Ballestrero:2015jca,Greljo:2015sla,Hespel:2015zea,Kauer:2015dma,Kauer:2015hia,Kilian:2015opv,Mimasu:2015nqa,
Degrande:2016dqg,Dwivedi:2016xwm,Gritsan:2016hjl,deFlorian:2016spz,Azatov:2016xik,
Denner:2017vms,Deutschmann:2017qum,Greljo:2017spw,Goncalves:2017gzy,Jager:2017owh,
Brass:2018hfw,Gomez-Ambrosio:2018pnl,Goncalves:2018pkt,Harlander:2018yns,Harlander:2018yio,Lee:2018fxj,Kalinowski:2018oxd,Perez:2018kav,
Jaquier:2019bfs,Denner:2019fcr,Banerjee:2019twi} and present phenomenological results in a unified approach.
This framework includes many options for production and decay of the $H$ boson.
Here we consider gluon fusion (ggH), vector boson fusion (VBF), and associated production with 
a vector boson ($VH$) in both \onshell\ $H$ and \offshell\ $H^*$ production, with decays to two vector bosons.
In the \offshell\ case, interference with background processes is included. 
Additional heavy particles in the gluon fusion loop and a second resonance interfering with the SM processes are also considered. 
In the $VH$ process, we include next-to-leading order QCD corrections, as well as the gluon fusion process for $ZH$.
The processes with direct sensitivity to fermion $\Hff$ couplings, such as $t\bar{t}H$, $b\bar{b}H$, $tqH$, or $H\to\tau^+\tau^-$,
are discussed in Ref.~\cite{Gritsan:2016hjl}.

In an earlier version of our framework, we focused mostly on the Run-I targets and their possible extensions as
documented in Refs.~\cite{Gao:2010qx,Bolognesi:2012mm,Anderson:2013afp}. It was adopted in Run-I 
analyses using Large Hadron Collider (LHC) data~\cite{Chatrchyan:2012xdj,Chatrchyan:2012jja,Chatrchyan:2013mxa,Chatrchyan:2013iaa,
Aad:2013xqa,Khachatryan:2014iha,Khachatryan:2014ira,Khachatryan:2014kca,Khachatryan:2015mma,Khachatryan:2015cwa,
Aad:2015mxa,Khachatryan:2016tnr}.
Some new features in this framework have been reported earlier~\cite{deFlorian:2016spz} and have been used
for LHC experimental analyses. Most notably, this framework was employed in recent Run-II measurements of the 
$HVV$ anomalous couplings from the first joint analysis of \onshell\ production and decay~\cite{Sirunyan:2017tqd,Sirunyan:2019nbs}, 
from the first joint analysis of \onshell\ and \offshell\ \Hboson\ production~\cite{Sirunyan:2019twz}, 
for the first measurement of the CP structure of the Yukawa interaction between the \Hboson and top quark~\cite{Sirunyan:2020sum},
in the search for a second resonance in interference with the continuum background~\cite{Sirunyan:2018qlb,Sirunyan:2019pqw},
and in projections to future  \onshell\ and \offshell\ \Hboson\ measurements at the High Luminosity (HL) LHC~\cite{Cepeda:2019klc}. 
In this paper, we document, review, and highlight the new features critical for exploring the full Run-II dataset at the LHC
and preparing for Run-III and the HL-LHC. 
We also broaden the theoretical underpinning, allowing interpretation in terms of either anomalous couplings or 
an effective field theory (EFT) framework.

Both Run-I and Run-II of the LHC have provided a large amount of data on \Hboson properties 
and its interactions with other SM particles, as analyzed by the ATLAS and CMS experiments. 
The \Hboson has been observed in all accessible production channels, 
gluon fusion, weak vector boson fusion, $VH$ associated production, and top-quark associated 
production~\cite{Khachatryan:2016vau,Sirunyan:2018koj,Aad:2019mbh,
Sirunyan:2018kst,Aaboud:2018zhk,Sirunyan:2018hoz,Aaboud:2018urx},
and its production strength is consistent with the SM prediction within the uncertainties~\cite{deFlorian:2016spz}.
Also its decay channels into gauge bosons ($ZZ, WW, \gamma\gamma)$ have been observed and 
do not show significant deviations within the uncertainties
~\cite{Khachatryan:2016vau, 
Sirunyan:2018koj, 
Aad:2019mbh}. 
The fermionic interactions have been established for the third generation quarks ($t,b$) and the $\tau$ 
lepton~\cite{
Sirunyan:2018kst, 
Aaboud:2018zhk, 
Sirunyan:2018hoz, 
Aaboud:2018urx, 
Sirunyan:2017khh, 
Aaboud:2018pen}, 
and so far, they are consistent with the SM within the uncertainties. 

While this picture shows that Nature does not radically deviate from the SM dynamics, 
it should be noted that many generic extensions of the SM predict deviations  
only below the current precision.
Open questions remain, for example about CP-odd mixtures, the Yukawa coupling hierarchy, and other 
states involved in electroweak symmetry breaking. 
These questions can be addressed in the years to come by fully utilizing the existing and upcoming LHC data sets. 
In particular, the study of kinematic tails of distributions involving the \Hboson is becoming accessible for the first time. 
These signals involve off-shell \Hboson production and strong interference effects with irreducible backgrounds 
that are subject to the electroweak unitarization mechanism in the SM.
This feature turns the kinematic tails into particularly sensitive probes of the mechanism of electroweak symmetry breaking 
and possible extensions beyond the SM. 
Moreover, the study of electroweak production of the \Hboson  (VBF and $VH$)
is probing $HVV$ interactions over a large range of momentum transfer, 
which can expose possible new particles that couple through loops. 
Even the direct production of new resonances will first show up as deviations from the expected high-energy tail of kinematic distributions. 
Hence, analyzing these newly accessible features in off-shell \Hboson production is 
of paramount importance to understand electroweak symmetry breaking in the SM and 
possible extensions involving new particles. 
In the following, we review the framework and demonstrate its capabilities through examples of possible analyses.
The technical details of the framework are described in the manual, which
can be downloaded at~\cite{jhugen}, together with the source code.

\section{Parameterization of anomalous interactions}
\label{sect:cp_couplings}

\subsection{\texorpdfstring{$H$}{H} boson interactions}
\label{subsec:hvv}

We present our parameterization of anomalous couplings relevant for \onshell\ and \offshell\ \Hboson production and decay.
\begin{figure}[t]
\centering
\subfloat[$HVV$]{\includegraphics[width=0.16\linewidth]{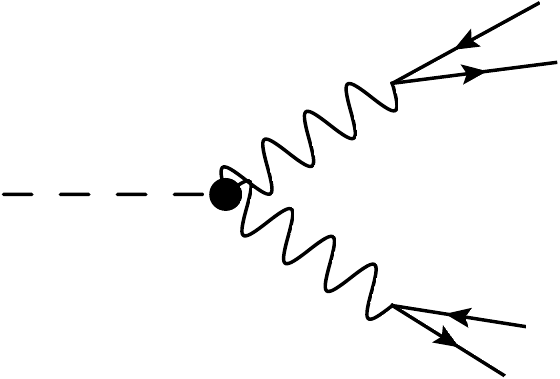}}
\quad\quad\quad\quad\quad
\subfloat[$Hf\bar{f}$]{\includegraphics[width=0.13\linewidth]{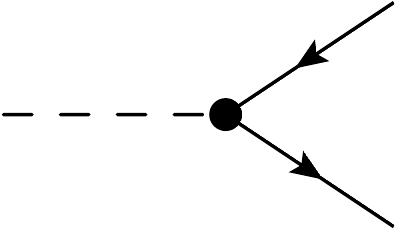}}
\quad\quad\quad\quad\quad
\subfloat[$HVf\bar{f}$]{\includegraphics[width=0.15\linewidth]{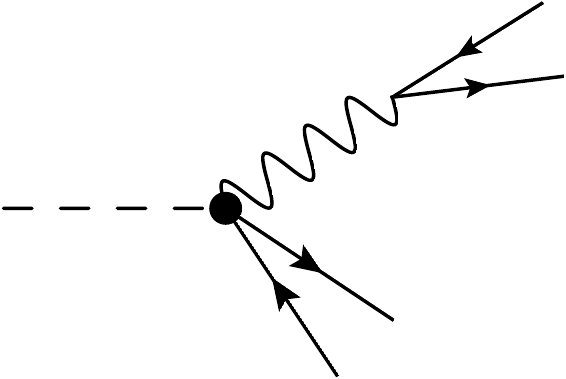}}
\quad\quad\quad\quad\quad
\subfloat[$Hf\bar{f}f\bar{f}$]{\includegraphics[width=0.13\linewidth]{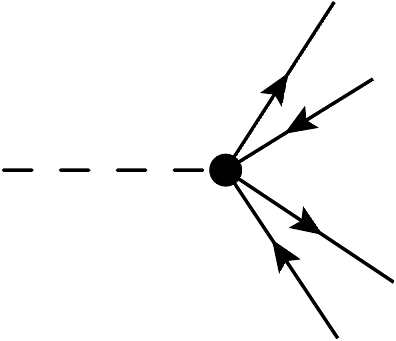}}
\caption{Vertices relevant for $HVV$ and $\Hff$ interactions.}
\label{fig:feynman1}
\end{figure}
%
Following the notation of Refs.~\cite{Gao:2010qx,Bolognesi:2012mm,Anderson:2013afp},
the $HVV$ scattering amplitude of a spin-zero boson $H$ and two vector bosons $VV$ with polarization vectors and momenta
$\varepsilon_{1}^\mu$, $q_1$ and $\varepsilon_{2}^\mu$, $q_2$, as illustrated in Fig.~\ref{fig:feynman1}(a),
is parameterized by
\begin{eqnarray}
  A({H V_1 V_2}) \!=\! \frac{1}{v} &\bigg\{  &
                          M_{V_1}^2 \bigg(g_{1}^{VV} 
+ \frac{\kappa_1^{VV}q_{1}^2 + \kappa_2^{VV} q_{2}^{2}}{\left(\Lambda_{1}^{VV} \right)^{2}} + \frac{\kappa_3^{VV} (q_1+q_2)^2}{\left(\Lambda_{Q}^{VV} \right)^{2}} 
+ \frac{2 q_1\cdot q_2}{M_{V_1}^2} g_2^{VV}\bigg) (\varepsilon_1 \cdot \varepsilon_2) 
\nonumber \\ &&
                        -2 g_2^{VV} \, {(\varepsilon_1 \cdot q_2)(\varepsilon_2 \cdot q_1)}
                        -2  g_4^{VV} \, {\varepsilon_{\varepsilon_1\,\varepsilon_2\,q_1\,q_2}}
                               \bigg\}\,,
\label{eq:HVV}
\end{eqnarray}
where $M_{V_1}$ is the vector boson's pole mass, $v$ is the SM Higgs field vacuum expectation value,
and $g_{1,2,4}^{VV}$, $\kappa_{1,2}^{VV}/(\Lambda_{1}^{VV})^2$, and $\kappa_3^{VV}/(\Lambda_{Q}^{VV})^2$
are coupling constants to be measured from data.
This parametrization represents the most general Lorentz-invariant form.

At tree level in the SM, only the CP-even $HZZ$ and $HWW$ interactions contribute via  $g_1^{ZZ}=g_1^{WW}=2$. 
The loop-induced interactions of $HZ\gamma$, $H\gamma\gamma$, and $Hgg$ contribute effectively via the CP-even $g_{2}^{VV}$ terms 
and are parameterically suppressed by $\alpha$ or $\alpha_s$. 
The CP-violating couplings $g_{4}^{VV}$ are generated only at three-loop level in the SM and are therefore tiny.
Beyond the SM, all of these couplings can receive additional contributions, which do not necessarily have to be small.
For example, the $Hgg$ interaction can be parameterized through a fermion loop, as discussed later in application to Eq.~(\ref{eq:g2gg}).  
The fermions in the loop interact with the $H$ boson as illustrated in Fig.~\ref{fig:feynman1}(b), with the couplings $\kappa_f$ and $\tilde\kappa_f$ and the amplitude 
\begin{eqnarray}
&& { A}(\Hff) = - \frac{m_f}{v}
\bar{\psi}_{f} \left ( \kappa_f  + \mathrm{i} \, \tilde\kappa_f  \gamma_5 \right ) {\psi}_{f}\,,
\label{eq:ampl-spin0-qq} \label{eq:Hffcoupl}
\end{eqnarray}
where $\bar{\psi}_{f}$ and ${\psi}_{f}$ are the Dirac spinors and $m_f$ is the fermion mass.
One may equivalently choose to express the couplings through a Lagrangian (up to an unphysical global phase)
\begin{eqnarray} 
&& {\cal L}_{hff} = - \frac{m_f}{v} \bar{\psi}_{f} \left ( \kappa_f + \mathrm{i} \, \tilde\kappa_f \gamma_5 \right ) \psi_{f} \, h \,,
  \label{eq:lagrang-spin0-qq}
\end{eqnarray}
which allows a connection to be made between the couplings $\kappa_f$ and $\tilde \kappa_f$ and anomalous operators in an effective field theory.
In the SM, the dominant contribution to gluon fusion comes from a top quark loop with $(\kappa_t, \tilde\kappa_t)= (1,0)$.

The couplings $\kappa_i^{VV}/(\Lambda_i^{VV})^2$ in Eq.~(\ref{eq:HVV}) are introduced to allow for additional momentum dependence.
Below, we also show that these terms can be reinterpreted as the contact interactions shown in Figs.~\ref{fig:feynman1}(c) and \ref{fig:feynman1}(d).
By symmetry we have $\kappa_1^{ZZ}=\kappa_2^{ZZ}$, but we do not enforce 
$\kappa_1^{WW}=\kappa_2^{WW}$ for $W^\pm$ bosons.
Note that 
$\kappa_1^{\gamma\gamma}=\kappa_2^{\gamma\gamma}=\kappa_1^{gg}=\kappa_2^{gg}=\kappa_1^{Z\gamma}=0$, 
while $\kappa_1^{\gamma Z}=\kappa_2^{Z\gamma}$ may contribute~\cite{Khachatryan:2014kca}.
The coupling $\kappa_3^{VV}/(\Lambda_Q^{VV})^2$ allows for scenarios which violate the gauge symmetries of the SM. 

For the $Hgg$ couplings entering the gluon fusion process we also consider the full one-loop dependence instead of the effective $g_{2,4}^{gg}$ couplings in Eq.~(\ref{eq:HVV}).
This feature is important for correctly describing off-shell Higgs production and additional broad, heavy resonances, where the $q^2$-dependence of the interaction cannot 
be approximated as a constant coupling. 
In addition to the closed quark loop with explicit dependence on the bottom and top quark masses, we allow for the insertion of fourth generation $b^\prime$ and $t^\prime$ quarks into the loop. 
\\

If a gauge boson in Eq.~(\ref{eq:HVV}) is coupled to a light fermion current, we replace its polarization vectors by 
\begin{eqnarray}
\label{eq:Vff}
      \varepsilon_{i}^\mu(q_i) \to j^\mu_i =  e \,  \frac{ \bar{\psi}_{f'}  \gamma^\mu \left( g^{Vf'f}_\mathrm{L} \, \omega_\mathrm{L}+g^{Vf'f}_\mathrm{R} \, \omega_\mathrm{R} \right) \psi_{f}  }
                                                 {q_i^2 - M_V^2 + \mathrm{i} M_V \Gamma_V}\,,
\end{eqnarray}
where $e$ is the electron electric charge, $\Gamma_V$ is the gauge boson's width, $\omega_\mathrm{L,R}$ are the left- and right-handed chirality projectors, and the $g^{Vf'f}_\mathrm{L,R}$ are the corresponding couplings of the gauge boson $V$ to fermions.
We also allow for exchanges of additional spin-1 bosons $V'$ between the \Hboson and the fermions.
Hence, we add 
\begin{eqnarray}
     \varepsilon_{i}^\mu(q_i) \to j^\mu_i  
     -    \, \frac{ \bar{\psi}_{f'}  \gamma^\mu \left( e_\mathrm{L}^{V'f'f}  \omega_\mathrm{L}  + e_\mathrm{R}^{V'f'f}  \omega_\mathrm{R} \right) \psi_{f}  }
                                                 {q_i^2 - M_{V'}^2 + \mathrm{i} M_{V'} \Gamma_{V'}}\,,
\label{eq:Zprff}                                                 
\end{eqnarray}
with the chirality and flavor dependent couplings $e_{\mathrm{L,R}}^{V'f'f}$.
In this approach, we allow for flavor changing interactions ($f' \neq f$) in both the neutral and charged $V'$ interactions. 
In the case where the $V'$ boson is very heavy, the limit $M_{V'}^2/q_i^2 \to \infty$  yields the contact interaction
\begin{eqnarray}
     \varepsilon_{i}^\mu(q_i) \to j^\mu_i  
     + \frac{1}{M_{V'}^2} \; { \bar{\psi}_{f'}  \gamma^\mu \left( e_\mathrm{L}^{V'f'f}  \omega_\mathrm{L}  + e_\mathrm{R}^{V'f'f}  \omega_\mathrm{R} \right) \psi_{f}  }
\label{eq:HVff}
\end{eqnarray}
in Fig.~\ref{fig:feynman1}(c-d).
We note that these contact terms and new $V^\prime$ states are not the primary interest in this study because their 
existence would become evident in resonance searches and in electroweak measurements, without 
the need for $H$ boson production.
Moreover, the $HZf{\bar f}$ contact terms are equivalent to the already constrained $\kappa_{1,2}^{ZZ}$ 
and $\kappa_2^{Z\gamma}$ terms~\cite{Gonzalez-Alonso:2014eva,Greljo:2015sla} 
if coupling flavor universality is assumed.
Under the approximation that the $Z$ boson has a narrow width,
this correspondence, given in Eq.~(\ref{eq:POtranslation}), only involves real couplings.
For example, in the limit where $\Gamma_Z \ll M_Z$,
a nonzero $\kappa_1^{ZZ}/(\Lambda_1^{ZZ})^2$ in Eq.~(\ref{eq:HVV})
is equivalent to shifting $g_1^{ZZ} \to g_1^{ZZ} + 2\kappa_1^{ZZ} ( M_Z \big/ \Lambda_1^{ZZ})^2$
and activating a contact interaction $g_1^{ZZ'}=\kappa_1^{ZZ} ( M_{Z'} \big/ \Lambda_1^{ZZ})^2$, $e^{Z'f'f}_\lambda= e \, g_\lambda^{Zf'f}$.
\\

The parameterization of the amplitude in Eq.~(\ref{eq:HVV}) can be related to a fundamental Lagrange density function.
Here, we closely follow the so-called Higgs basis of Ref.~\cite{deFlorian:2016spz}, which is based on an effective field theory expansion 
up to dimension six.
The relevant $\mathrm{SU(3)\times SU(2)\times U(1)}$ invariant Lagrangian for \Hboson interactions with gauge bosons (in the mass eigenstate parameterization) reads
\begin{eqnarray}
\label{eq:EFT_hvv}
 {\cal L}_{\rm hvv} =& & {h \over v} \left [ 
  \left (1 +  \delta c_z \right )  {(g^2+g^{\prime 2}) v^2 \over 4} Z_\mu Z_\mu
+ c_{zz} {g^2 + g'{}^2 \over  4} Z_{\mu \nu} Z_{\mu\nu}  
+c_{z \Box} g^2 Z_\mu \partial_\nu Z_{\mu \nu}
+ \tilde c_{zz}  {g^2 + g'{}^2  \over  4} Z_{\mu \nu} \tilde Z_{\mu\nu}
\right . \nonumber \\ & & \left . 
 +\left (1 +  \delta c_w \right )  {g^2 v^2 \over 2} W_\mu^+ W_\mu^- 
+ c_{ww}  {g^2 \over  2} W_{\mu \nu}^+  W_{\mu\nu}^- 
+ c_{w \Box} g^2 \left (W_\mu^- \partial_\nu W_{\mu \nu}^+ + {\mathrm h.c.} \right )  
 + \tilde c_{ww}  {g^2 \over  2} W_{\mu \nu}^+   \tilde W_{\mu\nu}^- 
 \right . \nonumber \\ & & \left . 
+ c_{z \gamma} {e \sqrt{g^2 + g'{}^2}  \over  2} Z_{\mu \nu} A_{\mu\nu} 
+ \tilde c_{z \gamma} {e \sqrt{g^2 + g'{}^2} \over  2} Z_{\mu \nu} \tilde A_{\mu\nu}
+ c_{\gamma \Box} g g' Z_\mu \partial_\nu A_{\mu \nu}
\right . \nonumber \\ & & \left . 
+ c_{\gamma \gamma} {e^2 \over 4} A_{\mu \nu} A_{\mu \nu} 
+ \tilde c_{\gamma \gamma} {e^2 \over 4} A_{\mu \nu} \tilde A_{\mu \nu} 
+  c_{gg} {g_s^2 \over 4 } G_{\mu \nu}^a G_{\mu \nu}^a   
+  \tilde c_{gg} {g_s^2 \over 4} G_{\mu \nu}^a \tilde G_{\mu \nu}^a  
\right ]\,,
\end{eqnarray} 
in accordance with Eq.~(II.2.20) in Ref.~\cite{deFlorian:2016spz}%
\footnote{We note that the so-called Higgs basis is based on a set of Lagrangians 
for Higgs physics that do not contain the whole SM. Hence, it is not a complete 
operator basis in the strict mathematical sense. In this work, however, all contributions have  
direct relations to the Warsaw basis, which fulfills the requirements of a complete basis.}. 
The fields and real-valued couplings, as well as the corresponding dimension-six operators, are defined in Ref.~\cite{deFlorian:2016spz};
for example, $g^2+g'^2=e^2/(s_w c_w)^2$, $e^2= 4 \pi \alpha$ and $g_s^2=4 \pi \alpha_s$.
We note that when restricting the discussion to the dimension-six effective field theory (see Eq.~(II.2.38) in Ref.~\cite{deFlorian:2016spz}),
Eq.~(\ref{eq:EFT_hvv}) is parameterized by ten real degrees of freedom, so not all of the coefficients are independent. 
For example, the coefficients $\delta c_w$, $c_{ww}$, $\tilde c_{ww}$, $c_{w \Box}$, and $c_{\gamma \Box}$ can be expressed 
through linear combinations of the other couplings.
The redundancy was introduced intentionally in Ref.~\cite{deFlorian:2016spz} for easier connections to observable quantities in Higgs physics. 

The generality of our amplitude parameterization allows us to uniquely represent each EFT coefficient in Eq.~(\ref{eq:EFT_hvv}) by an anomalous coupling in Eq.~(\ref{eq:HVV}). 
Limiting our couplings to real-valued numbers, we find 
\begin{eqnarray}  &&
\label{eq:EFT_ci}
    \delta c_z  = \frac12 g_1^{ZZ} - 1\,,
    \quad\quad
    c_{zz} = -\frac{2 s_w^2 c_w^2}{e^2} g_2^{ZZ}\,,
    \quad\quad
    c_{z \Box} = \frac{M_Z^2 s_w^2}{e^2} \, \frac{\kappa_1^{ZZ}}{(\Lambda_1^{ZZ})^2}\,,
    \quad\quad
    \tilde c_{zz} = -\frac{2 s_w^2 c_w^2}{e^2} g_4^{ZZ}\,,
    \nonumber \\ &&
     \delta c_w = \frac12 g_1^{WW} - 1\,,
    \quad\quad
    c_{ww} = -\frac{2 s_w^2 }{e^2} g_2^{WW}\,,
    \quad\quad
    c_{w \Box} = \frac{M_W^2 s_w^2}{e^2} \, \frac{\kappa_1^{WW}}{(\Lambda_1^{WW})^2}\,,
    \quad\quad
    \tilde c_{ww} = -\frac{2 s_w^2}{e^2} g_4^{WW}\,,
    \nonumber\\ &&
     c_{z \gamma} = -\frac{2 s_w c_w}{e^2} g_2^{Z\gamma}\,,
    \quad\quad
    \tilde c_{z \gamma} = -\frac{2 s_w c_w}{e^2} g_4^{Z\gamma}\,,
    \quad\quad
    c_{\gamma \Box} = \frac{s_w c_w}{e^2} \, \frac{M_Z^2}{(\Lambda_1^{Z\gamma})^2} \kappa_2^{Z\gamma}\,,
   \nonumber\\ &&
     c_{\gamma \gamma} = -\frac{2}{e^2} g_2^{\gamma\gamma}\,,   
    \quad\quad
   \tilde c_{\gamma \gamma} = -\frac{2}{e^2} g_4^{\gamma\gamma}\,,
    \quad\quad
     c_{gg} = -\frac{2}{g_s^2} g_2^{gg}\,,
   \quad\quad
   \tilde c_{gg} = -\frac{2}{g_s^2} g_4^{gg}\,.
\end{eqnarray} 
The Lagrangian for SM $HVV$ interactions is retained by setting $\delta c_z = \delta c_w=0 $ and all other $c_i=0$. 
Hence, only the CP-even $HZZ$ and $HWW$ interactions remain at tree level.

Not every anomalous coupling in Eq.~(\ref{eq:HVV}) has a corresponding term in the EFT Lagrangian of Eq.~(\ref{eq:EFT_hvv}). 
For example, the gauge invariance violating term $\kappa_3^{VV}/(\Lambda_Q^{VV})^2$ has no correspondence because 
${\cal L}_{\rm hvv}$ is gauge invariant by construction. 
Similarly, charge symmetry in ${\cal L}_\mathrm{hvv}$ enforces $\kappa_1^{WW}=\kappa_2^{WW}$, which 
does not necessarily have to be true in our amplitude setting. 
For a unique comparison at the level of dimension-six interactions, the above mentioned dependencies amongst EFT coefficients 
also have to be enforced in the amplitude parameterization of Eq.~(\ref{eq:EFT_hvv}).
We quote these relations later in Section~\ref{subsec:symmrel}.
Correspondences to other EFT bases are obviously possible. 
As an illustration, we quote relationships of the CP violating couplings to the Warsaw basis~\cite{Grzadkowski:2010es}
in Appendix~\ref{sect:appendix_Warsaw}.
\\

The dimension-six Lagrangian for $HVf\bar{f}$ contact interactions  
(cfg. Eq.~(II.2.24) in Ref.~\cite{deFlorian:2016spz}) reads
\begin{eqnarray}
\label{eq:EFT_hvff}
{\cal L}_{hvff} = 2 e \, {h \over v}  
\Bigg\{ &&
\frac{W^+_\mu}{\sqrt{2} s_w}  \left( 
 \bar u_\mathrm{L} \gamma^\mu \delta g_\mathrm{L}^{hWq} d_\mathrm{L} 
+\bar u_\mathrm{R} \gamma^\mu \delta g_\mathrm{R}^{hWq} d_\mathrm{R} 
+\bar \nu_\mathrm{L} \gamma^\mu \delta g_\mathrm{L}^{hW\ell} e_\mathrm{L} 
\right) 
\\
+&&
\frac{W^-_\mu}{\sqrt{2} s_w} \left( 
 \bar d_\mathrm{L} \gamma^\mu \delta g_\mathrm{L}^{hWq} u_\mathrm{L} 
+\bar d_\mathrm{R} \gamma^\mu \delta g_\mathrm{R}^{hWq} u_\mathrm{R} 
+\bar e_\mathrm{L} \gamma^\mu \delta g_\mathrm{L}^{hW\ell} \nu_\mathrm{L} 
\right)
\nonumber \\ \nonumber
+ &&
\frac{Z_\mu}{s_w c_w}
\bigg( 
\sum_{f=u,d,e,\nu}
\bar f_\mathrm{L} \gamma^\mu  \delta g^{hZf}_\mathrm{L} f_\mathrm{L}
+
\sum_{f=u,d,e}
\bar f_\mathrm{R} \gamma^\mu  \delta g^{hZf}_\mathrm{R} f_\mathrm{R}
\bigg)
 \Bigg\}\,.
\end{eqnarray}
It contributes to the amplitude shown in Fig.~\ref{fig:feynman1}(c). 
A relationship to our framework with anomalous couplings can be obtained in the 
limit $M_{V'}^2 \big/ q_i^2 \to \infty$. 
It is given by 
\begin{eqnarray}
\delta g^{hWf}_\lambda = \frac{M_W^2}{M_{W'}^2} \frac{\sqrt{2} s_w}{e} g_1^{WW} e_\lambda^{W'f'f}\,,
\quad\quad
\delta g^{hZf}_\lambda = \frac{M_Z^2}{M_{Z'}^2} \frac{s_w c_w}{2 e} g_1^{ZZ} e_\lambda^{Z'ff}\,,
\end{eqnarray}
where $\lambda=\mathrm{L, R}$.
Similar to the above, the coefficients $\delta g^{hVf}_\lambda$ are not independent couplings and can be 
expressed through other coefficients of the dimension-six effective field theory. 

\begin{figure}[t]
\centering
\subfloat[$VVV$]{\includegraphics[width=0.11\linewidth]{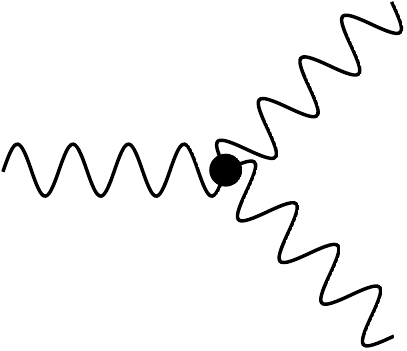}}
\quad\quad\quad\quad\quad
\subfloat[$VVVV$]{\includegraphics[width=0.10\linewidth]{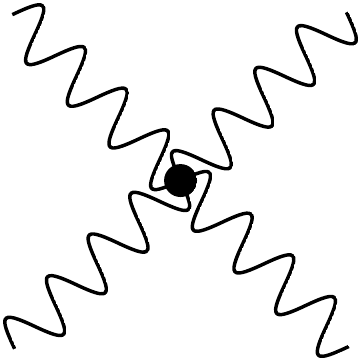}}
\caption{Gauge boson self interactions related to the $HVV$ vertices.}
\label{fig:feynman2}
\end{figure}

\subsection{Gauge boson self-interactions}
\label{subsec:vvvv}

In studying off-shell \Hboson production, some Feynman diagrams not involving an \Hboson also contribute.
In particular, the processes $gg \to WW$ and $q q' \to q q' + WW/ZZ/Z\gamma^*/\gamma^*\gamma^*(\to4f)$
involve diagrams with triple and quartic gauge boson self couplings, shown in Fig.~\ref{fig:feynman2}, instead of an \Hboson vertex. 
Since there is an intricate interplay between gauge boson self couplings and \Hboson gauge couplings (which 
guarantees unitarity of the cross section at high energies), we also consider gauge boson self couplings in our study. 
Their parameterization reads
\begin{eqnarray}
 & A({VW^+W^-}) = (-e) \,  d^{VWW} \, \bigg\{  
                          (\varepsilon_V \cdot \varepsilon_+)(q_{12}^V \cdot \varepsilon_-)
                         +(\varepsilon_+ \cdot \varepsilon_-)(q_{23}^V \cdot \varepsilon_V)
                         +(\varepsilon_V \cdot \varepsilon_-)(q_{31}^V \cdot \varepsilon_+)
                         + d^V_4 \varepsilon_{\varepsilon_V\,\varepsilon_+\,\varepsilon_-\,p_1}
                               \bigg\} \,, \;\;
\label{eq:VVV}                               
\\
 & A({V_1 V_2 W^+W^-}) = (+e^2) \,  d^{VVWW} \, \bigg\{  
                               (\varepsilon_{1} \cdot \varepsilon_{+})  (\varepsilon_{2} \cdot \varepsilon_{-})
                              +(\varepsilon_{1} \cdot \varepsilon_{-})  (\varepsilon_{2} \cdot \varepsilon_{+})
                             -2(\varepsilon_{1} \cdot \varepsilon_{2})  (\varepsilon_{+} \cdot \varepsilon_{-})
                               \bigg\} \,,
\label{eq:VVVV}
\end{eqnarray}
where $q_{ij}^V = d_i^V p_i - d_j^V p_j$ is the relative momentum transfer. 
We fix $d^{\gamma WW}  = 1$ and $d^{Z WW} = {c_w}/{s_w}$ per convention and allow all other couplings to vary. 
In the SM, their values are
\begin{eqnarray}
\label{eq:VVVVSM}
&& d_1^V =d_2^V =d_3^V =1\,,
\quad
d_4^V=0\,, 
\\
&& d^{\gamma\gamma WW} = 1\,,
\quad
d^{Z\gamma WW} = \frac{c_w}{s_w}\,,
\quad
d^{ZZWW} = \frac{c_w^2}{s_w^2}\,,
\quad
d^{WWWW} = \frac{1}{2 s_w^2} \,. 
\nonumber
\end{eqnarray}
Extensions of the gauge sector of the SM lead to modifications of the above couplings. 
For example, the CP-violating term $d_4^V$ in Eq.~(\ref{eq:VVVVSM}) can be non-zero.
The relevant contributions of the dimension-six Lagrangian for the triple and quartic gauge boson self-interactions are 
(see Eqs.~(3.12, 3.14, 3.15)  in Ref.~\cite{Falkowski:2001958})
\begin{eqnarray} 
\label{eq:EFT_tgc}
 {\cal L}_{\rm tgc}  = &  &
 \mathrm{i}  e    \left ( W_{\mu \nu}^+ W_\mu^-  -  W_{\mu \nu}^- W_\mu^+ \right ) A_\nu   
 + \mathrm{i}  e  \left [  (1 + \delta \kappa_\gamma )  A_{\mu\nu}\,W_\mu^+W_\nu^-   
+ \tilde \kappa_\gamma  \tilde A_{\mu\nu}\,W_\mu^+W_\nu^-  \right ]
\nonumber \\  &  + & \mathrm{i} e \frac{c_w}{s_w}  \left [  (1 + \delta g_{1,z})   \left ( W_{\mu \nu}^+ W_\mu^-  -  W_{\mu \nu}^- W_\mu^+ \right ) Z_\nu 
 + (1 +  \delta \kappa_z) \, Z_{\mu\nu}\,W_\mu^+W_\nu^-   
 +  \tilde  \kappa_z \,  \tilde Z_{\mu\nu}\,W_\mu^+W_\nu^-   \right ] \,,
\\
\label{eq:EFT_qgc}
 {\cal L}_{\rm qgc}  = &  &
 e^2 ( W^+_\mu A_\mu W^-_\nu A_\nu - W^+_\mu W^-_\mu A_\nu A_\nu  ) 
 + \frac{e^2}{2 s_w^2} (1+2 c_w^2 \delta g_{1,z}) ( W^+_\mu W^+_\mu W^-_\nu W^-_\nu -  W^+_\mu W^-_\mu W^+_\nu W^-_\nu )
 \nonumber \\ &+ & 
 e^2 \frac{c_w^2}{s_w^2}  (1+2\delta g_{1,z} ) ( W^+_\mu Z_\mu W^-_\nu Z_\nu -  W^+_\mu W^-_\mu Z_\nu Z_\nu )
 \nonumber \\ &+ &  
 e^2 \frac{c_w}{s_w} (1+\delta g_{1,z} ) ( W^+_\mu Z_\mu W^-_\nu A_\nu + W^+_\mu A_\mu W^-_\nu Z_\nu  - 2  W^+_\mu W^-_\mu Z_\nu A_\nu  )
 \,.
\end{eqnarray} 
The anomalous coefficients in Eqs.~(\ref{eq:EFT_tgc},\ref{eq:EFT_qgc}) are related to couplings in Eqs.~(\ref{eq:VVV},\ref{eq:VVVV}) by 
\begin{eqnarray}  
  \delta \kappa_\gamma = \frac12 \left( d_1^\gamma - 1 \right) \,,
  \quad\quad 
  \tilde \kappa_\gamma =  \frac12 d_4^\gamma \,,
  \quad\quad 
  \delta \kappa_z = \frac12 \left( d_1^Z - 1 \right) \,,
  \quad\quad 
  \tilde  \kappa_z =  \frac12 d_4^Z \,,
  \nonumber \\ 
  \delta g_{1,z} = d_2^Z - 1= d_3^Z - 1= \frac12 \left( \frac{s_w^2}{c_w^2} d^{ZZWW}-1 \right) = \frac{s_w}{c_w} d^{Z \gamma WW} -1
  \,.
\end{eqnarray}
Similar to the case of $\mathcal{L}_\mathrm{hvv}$, not all coefficients in Eqs.~(\ref{eq:EFT_tgc},\ref{eq:EFT_qgc}) 
are independent in the effective field theory framework, and we discuss their dependence in the next subsection. 
Moreover, additional anomalous triple and quartic contributions, the $\lambda_{\gamma,Z}, \tilde \lambda_{\gamma,Z}$ 
terms in Ref.~\cite{Falkowski:2001958,deFlorian:2016spz}, can arise. 
These additional terms are unrelated to any of the \Hboson contributions, and therefore, we do not consider them here.

\subsection{Coupling relations}
\label{subsec:symmrel}

In the previous subsections we related our anomalous couplings to the effective field theory coefficients of the so-called Higgs basis \cite{deFlorian:2016spz}. 
As mentioned above, not all of the EFT coefficients are independent when limiting the discussion to dimension-six interactions\footnote{
It should be noted that contributions of dimension-eight can invalidate the relations.  See the comments in Section~II.2.1.d of Ref.~\cite{deFlorian:2016spz}.}. 
The linear relations for the dependent coefficients can be found in Ref.~\cite{deFlorian:2016spz} and they translate into 
relations amongst our anomalous couplings. 
Enforcing these relations allows a unique comparison between the two frameworks, based on a minimal set of degrees of freedom.
We find for the $HVV$ interactions 
\begin{eqnarray}
  g_1^{WW} &=& g_1^{ZZ} + \frac{\Delta M_W}{M_W} \,,    
  \label{eq:deltaMW}
  \\
  g_2^{WW} &=& c_w^2 g_2^{ZZ} + s_w^2 g_2^{\gamma\gamma} + 2 s_w c_w g_2^{Z\gamma}\,,
  \label{eq:g2WW}
  \\
  g_4^{WW} &=& c_w^2 g_4^{ZZ} + s_w^2 g_4^{\gamma\gamma} + 2 s_w c_w g_4^{Z\gamma}\,,
  \label{eq:g4WW}
  \\
  \frac{\kappa_1^{WW}}{(\Lambda_1^{WW})^2} (c_w^2-s_w^2) &=& \frac{\kappa_1^{ZZ}}{(\Lambda_1^{ZZ})^2}
                                                           +2 s_w^2 \frac{g_2^{\gamma\gamma}-g_2^{ZZ}}{M_Z^2} 
                                                          +2 \frac{s_w}{c_w} (c_w^2-s_w^2) \frac{g_2^{Z\gamma}}{M_Z^2}\,,
 \label{eq:kappa1WW}
  \\
  \frac{\kappa_2^{Z\gamma}}{(\Lambda_1^{Z\gamma})^2} (c_w^2-s_w^2) &=&  2 s_w c_w \left( \frac{\kappa_1^{ZZ}}{(\Lambda_1^{ZZ})^2} 
                                                                              + \frac{ g_2^{\gamma\gamma} - g_2^{ZZ}}{M_Z^2}  \right)
                                                                               +2 (c_w^2-s_w^2) \frac{g_2^{Z\gamma}}{M_Z^2}\,.
 \label{eq:kappa2Zgamma}
 \end{eqnarray}
The term $\Delta M_W$ in Eq.~(\ref{eq:deltaMW}) induces a shift in the $W$ boson mass. Given that $M_W$ is experimentally measured to high precision 
one can assume $\Delta M_W \approx 0$. 
The couplings $e_\lambda^{V'f'f}$ for $HVf\bar{f}$ contact interactions in Eq.~(\ref{eq:HVff}) are equal to 
the corresponding $V\bar{f}f$ couplings $g_\lambda^{Vf'f}$ in the SM. 
Therefore, one can often neglect them as they are strongly constrained by electroweak precision measurements. 
The  gauge boson self couplings in Eqs.~(\ref{eq:VVV}-\ref{eq:VVVVSM}) are determined by $HVV$ couplings in Eq.~(\ref{eq:HVV}) through  
\begin{eqnarray}
  d_1^\gamma &=& 1 + (g_2^{\gamma\gamma}-g_2^{ZZ}) c_w^2  + g_2^{Z\gamma} \left(\frac{c_w}{s_w}-2s_w c_w \right)\,,
   \label{eq:d1}
  \\
  d_4^{\gamma} &=& (g_4^{\gamma\gamma}-g_4^{ZZ}) c_w^2 + g_4^{Z\gamma} \left( \frac{c_w}{s_w}-2 s_w c_w \right)\,,
   \label{eq:d2}
  \\
  d_1^Z &=& 1 -2 \frac{s_w^2 c_w^2}{c_w^2-s_w^2} \left(g_2^{\gamma\gamma}  -g_2^{ZZ} \right) -2 s_w c_w g_2^{Z\gamma} - \frac{M_Z^2}{2 (c_w^2-s_w^2) }  \frac{\kappa_1^{ZZ}}{(\Lambda_1^{ZZ})^2}\,,
   \label{eq:d3}
  \\
  d_2^Z &=& d_3^Z = 1 - \frac{s_w^2}{c_w^2-s_w^2} \left(g_2^{\gamma\gamma}-g_2^{ZZ} \right) - \frac{s_w}{c_w} g_2^{Z\gamma} - \frac{M_Z^2}{2(c_w^2-s_w^2)} \frac{\kappa_1^{ZZ}}{(\Lambda_1^{ZZ})^2}\,,
   \label{eq:d4}
  \\
  d_4^Z &=& -\frac{s_w^2}{c_w^2} d_4^{\gamma}\,,
   \label{eq:d5}
  \\
  d^{ZZWW} &=& \frac{c_w^2}{s_w^2} \left( 2 d_2^Z -1 \right)\,,
  \quad\quad
  d^{Z\gamma WW} = \frac{c_w}{s_w} d_2^Z\,.
   \label{eq:d6}
\end{eqnarray}

\subsection{Correspondence to a Pseudo Observable framework}

Here we briefly quote relations between our parameterization and the so-called Pseudo Observable framework \cite{Gonzalez-Alonso:2014eva}.
Similar to our work, the Pseudo Observables are derived from on-shell amplitudes. 
For the $H \to ZZ/Z\gamma^*/\gamma^*\gamma^* \to 4\ell$ amplitude we find the relations 
\begin{eqnarray}  &&
\kappa_{ZZ} = \frac{1}{2} g_1^{ZZ}
+ \frac{M_Z^2-\mathrm{i}M_Z\Gamma_Z}{(\Lambda_1^{ZZ})^2} \kappa_1^{ZZ}\,,
\quad \quad
\varepsilon_{ZZ} = g_2^{ZZ}\,,
\quad \quad 
\varepsilon_{ZZ}^\mathrm{CP} = g_4^{ZZ}\,,
\nonumber \\ &&
\quad \quad
\varepsilon_{\gamma\gamma} = g_2^{\gamma\gamma}\,,
\quad \quad 
\varepsilon_{\gamma\gamma}^\mathrm{CP} = g_4^{\gamma\gamma}\,,
\quad \quad
\varepsilon_{Z\gamma} = - g_2^{Z\gamma}\,,
\quad \quad 
\varepsilon_{Z\gamma}^\mathrm{CP} = - g_4^{Z\gamma}\,,
\nonumber \\ &&
\varepsilon_{Z f_\lambda} = 
\frac{M_Z^2-\mathrm{i}M_Z\Gamma_Z}{2 (\Lambda_1^{ZZ})^2} \kappa_1^{ZZ} e g_{\lambda}^{Zff} - \frac{M_Z^2-\mathrm{i}M_Z\Gamma_Z}{2 (\Lambda_1^{Z \gamma})^2} \kappa_2^{Z\gamma} e Q_f \,,
\label{eq:POtranslation}
\end{eqnarray}
for the couplings given in Eqs.~(9--11) and Eqs.~(20--21) of Ref.~\cite{Gonzalez-Alonso:2014eva}. 
Similarly, the relations for the  $H\to W^+ W^- \to 2\ell \, 2\nu$ amplitude read
\begin{eqnarray} &&
\kappa_{WW} = \frac{1}{2} g_1^{WW}
+ \frac{M_W^2-\mathrm{i}M_W\Gamma_W}{2 (\Lambda_1^{WW})^2} ( \kappa_1^{WW} + \kappa_2^{WW})\,,
\quad \quad
\varepsilon_{WW} = g_2^{WW}\,,
\quad \quad 
\varepsilon_{WW}^\mathrm{CP} = g_4^{WW}\,,
\quad\quad
\nonumber \\ &&
\varepsilon_{W \ell_\lambda}^* =  
\frac{M_W^2-\mathrm{i}M_W\Gamma_W}{2 (\Lambda_1^{WW})^2} \kappa_1^{WW} e g_{\lambda}^{W\ell\nu} \,,
\quad\quad
\varepsilon_{W \ell'_\lambda} =  \frac{M_W^2-\mathrm{i}M_W\Gamma_W}{2 (\Lambda_1^{WW})^2} \kappa_2^{WW} e g_{\lambda}^{W\ell'\nu'}\,.
\end{eqnarray}
Note that the imaginary terms in these relations are proportional to $\Gamma_V/M_V$,
so that in the limit $\Gamma_V \ll M_V$, real couplings in one framework translate to real couplings in the other.
The $g^{Vf'f}_\lambda$ are the chiral couplings of fermions to gauge bosons in Eq.~(\ref{eq:Vff}).
Similar to the effective field theory framework, the $\kappa_3^{VV}/(\Lambda_Q^{VV})^2$ term in Eq.~(\ref{eq:HVV})
does not have a counter piece in the Pseudo Observable framework. 
For all other couplings, there is a unique correspondence to our parameterization in Eq.~(\ref{eq:Vff}). 
Gauge boson self couplings can also be incorporated in the Pseudo Observable framework 
(see Refs.~\cite{Gonzalez-Alonso:2015bha,Greljo:2015sla}),
but we do not explicitly quote the relations to our framework here. 

\subsection{Unitarization}
\label{subsec:unitarization}

The above interactions describe all possible dynamics involving the $H$ boson as appearing in gluon fusion $gg\to H$, vector boson fusion $VV\to H$, associated production $V\to VH$, 
and its decays to bosons and fermions. 
For \emph{\onshell} $H$ boson production and decay, the typical range of invariant masses is $\mathcal{O}(100\,\mathrm{GeV})$.
However, in associated and off-shell production of the $H$ boson, there is no kinematic
limit on $q^2_{\sss Vi}$ or $q_H^2$ other than the energy of the colliding beams. 
When anomalous couplings with $q^2$-dependence are involved, this sometimes leads to cross sections growing with energy, 
which leads to unphysical growth at high energies. 
Obviously, these violations are unphysical and an artifact of the lacking knowledge of a UV-complete theory. 
Therefore, one should dismiss regions of phase space where a violation of unitarity happens. 
To mend this issue, we allow the option of specifying smooth cut-off scales $\Lambda_{V1,i}, \Lambda_{V2,i}, \Lambda_{H,i}$
for anomalous contributions with the form factor scaling 
\begin{eqnarray}
\frac{\Lambda_{V1,i}^2 \Lambda_{V2,i}^2 \Lambda_{H,i}^2}
{(\Lambda_{V1,i}^2+|q^2_{\sss V1}|)(\Lambda_{V2,i}^2+|q^2_{\sss V2}|)(\Lambda_{H,i}^2+|(q_{\sss V1}+q_{\sss V2})^2|)} 
\,.
\label{eq:formfact-spin0}
\end{eqnarray}
Studies of experimental data should include tests of different form-factor scales when there is
no direct bound on the $q^2$-ranges. 
An alternative approach is to limit the $q^2$-range in experimental analysis by restricting the data sample,
using, for example, a requirement on the transverse momentum $p_T$ of the reconstructed particles. 
The experimental sensitivity of both approaches is equivalent and no additional tools are required for the latter approach. 
However, such restrictions of the data sample lead to statistical fluctuations and therefore noisy results. 
They are also difficult experimentally since each new restriction requires re-analysis of the data, rather than simply
a change in the signal model. Moreover, while $p_T$ of the particles and $q^2$ of the intermediate vector bosons are correlated, 
this correlation is not 100\%. Therefore, it is not possible to have a fully consistent analysis in all channels using this approach. 
Finally, we note that other unitarization prescriptions have been presented in Refs.~\cite{Alboteanu:2008my,Perez:2018kav}.

\section{Parameterization of cross sections}
\label{sect:appendix_XS}

In this Section, we discuss the relationship between the coupling constants
and the cross section of a process involving the \Hboson.  We denote the coupling
constants as $a_n$, which could stand for $g_n$, $c_n$, or $\kappa_n$
as used in Section~\ref{sect:cp_couplings}.
The cross section of a process $i\to H\to f$ can be expressed as
\begin{eqnarray}
\frac{\mathrm{d}\sigma(i\to H\to f)}{\mathrm{d}s}\propto
\frac{\left(\sum  \alpha_{jk}^{(i)}a_ja_k\right)\left(\sum \alpha_{lm}^{(f)}a_la_m\right)}{(s-M_H^2)^2 + M_H^2\Gamma_{\rm tot}^2} \,.
\label{eq:diff-cross-section1}
\end{eqnarray}
where $\left(\sum  \alpha_{jk}^{(i)}a_ja_k\right)$ describes the production for a particular initial state $i$
and $\left(\sum  \alpha_{lm}^{(f)}a_la_m\right)$ describes the decay for a particular final state $f$.
Here we assume real coupling constants $a_n$, though these formulas can also be extended to complex couplings. 
The coefficients $\alpha_{jk}^{(i)}$ and $\alpha_{lm}^{(f)}$ evolve with $s$ and may be functions of kinematic observables.
These coefficients can be obtained from simulation, as we discuss in Section~\ref{sect:cp_mc}.
In this Section, we discuss integrated cross sections, and for this reason we deal with 
$\alpha_{jk}^{(i)}$ and $\alpha_{lm}^{(f)}$ as constants that have already been integrated over the kinematics. 
We will come back to the kinematic dependence in Section~\ref{sect:exp_kinematics}.

In the narrow-width approximation for \onshell\ production, we integrate Eq.~(\ref{eq:diff-cross-section1}) over $s$ in the relevant 
range, $\sim M_H\Gamma_{\rm tot}$ around the central value of $M_H^2$, to obtain the cross section for the process of interest
\begin{eqnarray}
\sigma(i\to H\to f)\propto
\frac{\left(\sum  \alpha_{jk}^{(i)}a_ja_k\right)\left(\sum \alpha_{lm}^{(f)}a_la_m\right)}{\Gamma_{\rm tot}} \,.
\label{eq:diff-cross-section2}
\end{eqnarray}
One can express the total width as
\begin{eqnarray}
\Gamma_{\rm tot}= \Gamma_{\rm known} + \Gamma_{\rm other} \,,
\label{eq:width_other}
\end{eqnarray}
where $\Gamma_{\rm known}$ represents decays to known particles and $\Gamma_{\rm other}$ represents
other unknown final states, either invisible or undetected in experiment. 

Without direct constraints on $\Gamma_{\rm other}$, if results are to be interpreted in terms of couplings via
the narrow-width approximation in Eq.~(\ref{eq:diff-cross-section2}), assumptions must be made on $\Gamma_\text{other}$.
However, in the case of the $ZZ$ and $WW$ final states, there is an interplay between the massive vector boson 
or the \Hboson\ going \offshell,  resulting in a sizable \offshell\ $H^*$ production~\cite{Kauer:2012hd}
with $(s-M_H^2)\gg M_H\Gamma_{\rm tot}$ in Eq.~(\ref{eq:diff-cross-section1}).
The resulting cross section in this region $s>(2M_W)^2$ is independent of the width. 
It should be noted that Eq.~(\ref{eq:diff-cross-section1}) represents only the signal part of the \offshell\ process 
with the \Hboson\ propagator. 
The full process involves background and its interference with the signal~\cite{Kauer:2012hd,Caola:2013yja}, 
as we illustrate in Section~\ref{sect:exp_offshell}.
Nonetheless, the lack of width dependence in the \offshell\ region is the basis for the measurement of the \Hboson's
total width $\Gamma_{\rm tot}$~\cite{Caola:2013yja}, provided that the evolution of Eq.~(\ref{eq:diff-cross-section1}) with $s$ is known. 
Therefore, a joint analysis of the \onshell\ and \offshell\ regions provides a simultaneous measurement 
of $\Gamma_{\rm tot}$ and of the cross sections corresponding to each coupling $a_n$ in a process $i\to H^{(*)}\to f$, 
as illustrated in Refs.~\cite{Khachatryan:2015mma,Sirunyan:2019twz}.
In a combination of multiple processes, the measurement can be further interpreted
as constraints on $\Gamma_{\rm other}$ and the couplings, following Eqs.~(\ref{eq:diff-cross-section2}) and (\ref{eq:width_other}),
and with the help of the identity
\begin{eqnarray}
{\Gamma_{\rm known}} = \sum_{f} \Gamma_f 
  = {\Gamma_{\rm tot}^{\rm SM}} \times
\sum_{f} \left( \frac{\Gamma_f^{\rm SM}}{\Gamma_{\rm tot}^{\rm SM}} \times \frac{\Gamma_f }{ \Gamma_f^{\rm SM}} \right)
= \sum_{f}\Gamma_f^{\rm SM}\sum_{lm}\alpha_{lm}^{(f)}a_la_m 
\,.
\label{eq:width}
\end{eqnarray}
The coefficients $\alpha_{lm}^{(f)}$ describe couplings to the known states and are normalized 
in such a way that $R_{f}(a_n)=\left(\sum  \alpha_{lm}^{(f)}a_la_m\right)=1$ in the SM,
and otherwise  $R_{f}(a_n)=\Gamma_f/\Gamma_f^{\rm SM}$.

In the following, we proceed to discuss the \onshell\ part of the measurements using the narrow-width approximation. 
In Table~\ref{tab:cformalism}, we summarize all the coefficients and functions $R_{f}$ needed to calculate 
${\Gamma_{\rm known}}$ in Eq.~(\ref{eq:width}).
These expressions with explicit coefficients $\alpha_{lm}^{(f)}$ help us to illustrate the relationship between the coupling 
constants introduced in Section~\ref{sect:cp_couplings} and experimental cross-section measurements. 
We will also use these expressions in Section~\ref{sect:exp_onshell} in application to particular measurements. 
For almost all calculations, we use the JHUGen framework implementation discussed in Section~\ref{sect:cp_mc}.
The only exceptions are $R_{\gamma\gamma}$ and $R_{Z\gamma}$, which are calculated using HDECAY~\cite{Djouadi:2018xqq,Fontes:2017zfn}.
The calculations are performed at LO in QCD and EW, with the $\overline{\mathrm{MS}}$-mass for the 
top quark $m_{t} =162.7$\,GeV and the \onshell\ mass for the bottom quark $m_{b} =4.18$\,GeV, and QCD scale $\mu = M_H/2$.

For all fermion final states $H\to q\bar{q}$, where we generically use $q=b,c,\tau,\mu$ to denote either quarks or leptons,
in the limit $m_q\ll M_H$ we obtain
\begin{eqnarray}
\label{eq:ratio-1}
R_{qq} = \kappa_q^2+\tilde\kappa_q^2
\,.
\end{eqnarray}

\begin{table}[t]
\begin{center}
\captionsetup{justification=centerlast}
\caption{
Partial widths $\Gamma_f$ of the dominant $H\to f$ decay modes in the SM in the narrow-width approximation~\cite{deFlorian:2016spz} 
and their modifications with anomalous couplings at $M_H=125$\,GeV, where $\Gamma_{\rm tot}^{\rm SM}=4.088\times 10^{-3}$\,GeV. 
Final states with $\Gamma_f^{\rm SM}<\Gamma_{\mu\mu}^{\rm SM}$  are neglected. 
\label{tab:cformalism}
}
\begin{tabular}{cccc}
\hline\hline
\vspace{-0.3cm} \\
$H\to f$ channel &    $\Gamma_f^{\rm SM}/\Gamma_{\rm tot}^{\rm SM}$        &      $\Gamma_f / \Gamma_f^{\rm SM}$   & Eq. \\
\vspace{-0.3cm} \\
\hline
\vspace{-0.3cm} \\
$H\to b\bar{b}$ &  0.5824 & $(\kappa_b^2+\tilde\kappa_b^2)$  & Eq.~(\ref{eq:ratio-1}) \\
$H\to W^+W^-$ &  0.2137 & $R_{WW}(a_n)$  & Eq.~(\ref{eq:ratio-2}) \\
$H\to gg$ &  0.08187 & $R_{gg}(a_n)$  & Eq.~(\ref{eq:ratio-4}) \\
$H\to \tau^+\tau^-$ &  0.06272 & $(\kappa_\tau^2+\tilde\kappa_\tau^2)$  & Eq.~(\ref{eq:ratio-1}) \\
$H\to c\bar{c}$ &  0.02891 & $(\kappa_c^2+\tilde\kappa_c^2)$  & Eq.~(\ref{eq:ratio-1}) \\
$H\to ZZ/Z\gamma^*/\gamma^*\gamma^*$ &  0.02619 & $R_{ZZ/Z\gamma^*/\gamma^*\gamma^*}(a_n)$  & Eq.~(\ref{eq:ratio-3}) \\
$H\to \gamma\gamma$ &  0.002270 & $R_{\gamma\gamma}(a_n)$  & Eq.~(\ref{eq:ratio-5}) \\
$H\to Z\gamma$ &  0.001533 & $R_{Z\gamma}(a_n)$  & Eq.~(\ref{eq:ratio-6}) \\
$H\to\mu^+\mu^-$ &  0.0002176 & $(\kappa_\mu^2+\tilde\kappa_\mu^2)$  & Eq.~(\ref{eq:ratio-1}) \\
\vspace{-0.3cm} \\
\hline\hline
\end{tabular}
\end{center}
\end{table}

For the gluon final state $H\to gg$, we allow for top and bottom quark contributions through the couplings from Eq.~(\ref{eq:Hffcoupl}).
In addition, we introduce a new heavy quark $Q$ with mass $m_Q\gg M_H$ and couplings to the \Hboson $\kappa_Q$ and $\tilde\kappa_Q$.
The result is
\begin{eqnarray}
\label{eq:ratio-4}
R_{gg} = &&
1.1068\, \kappa_t^2 + 0.0082\, \kappa_b^2 - 0.1150\, \kappa_t\kappa_b
	+ 2.5717\, \tilde\kappa_t^2 + 0.0091\, \tilde\kappa_b^2 - 0.1982\, \tilde\kappa_t\tilde\kappa_b \\
    &&	+\, 1.0298\, \kappa_Q^{2} + 2.1357\, \kappa_Q \kappa_t - 0.1109\, \kappa_Q \kappa_b
	+ 2.3170\, \tilde\kappa_Q^{2} + 4.8821\, \tilde\kappa_Q \tilde\kappa_t - 0.1880\, \tilde\kappa_Q \tilde\kappa_b
	\,.
	\nonumber 
\end{eqnarray}
The $\kappa_Q$ and $\tilde\kappa_Q$ couplings are connected to the $g_2^{gg}$ and $g_4^{gg}$ point-like 
interactions introduced in Eq.~(\ref{eq:HVV}) through
\begin{eqnarray}
\label{eq:g2gg}
g_{2}^{gg} =-\alpha_s \kappa_Q/(6\pi)
\,,~~~~~~~~~
g_{4}^{gg} =-\alpha_s\tilde\kappa_Q/(4\pi)
\,
\end{eqnarray}
in the limit where $m_Q\gg M_H$.
The function $R_{gg}$ also describes the scaling of the gluon fusion cross section with anomalous coupling contributions. 
 Setting $\kappa_q=\kappa_t=\kappa_b$ and $\tilde\kappa_q=\tilde\kappa_t=\tilde\kappa_b$, we find the ratio 
$\sigma(\tilde\kappa_q=1)/\sigma(\kappa_q=1)=2.38$, which differs from the ratio for a very heavy quark
$\sigma(\tilde\kappa_Q=1)/\sigma(\kappa_Q=1)=(3/2)^2=2.25$ due to finite quark mass effects. 
The latter ratio follows from the observation $\sigma(g_4^{gg}=1)=\sigma(g_2^{gg}=1)$.
In experiment, it is hard to distinguish the point-like interactions $g_2^{gg}$ and $g_4^{gg}$, or equivalently $\kappa_Q$ 
and $\tilde\kappa_Q$, from the SM-fermion loops. 
In the $H\to gg$ decay, there is no kinematic difference. In the gluon fusion production, there are effects in the tails of 
distributions, such as the transverse momentum, or in the \offshell\ region, as we discuss in Section~\ref{sect:exp_offshell}.
However, in Section~\ref{sect:exp_onshell} these effects are negligible and we do not distinguish
the $g_2^{gg}$ and $g_4^{gg}$ couplings from the SM-fermion loops. 

For the $H\to WW\rightarrow$\,four-fermion final state, we set $\Lambda_1^{WW}=100$\,GeV in Eq.~(\ref{eq:HVV}) 
in order to keep all numerical coefficients of similar order, and rely on the $\kappa_1^{WW}=\kappa_2^{WW}$ relationship to obtain 
\begin{eqnarray}
\label{eq:ratio-2}
R_{WW} = &&  \left(\frac{g_1^{WW}}{2}\right)^2 
+ 0.1320  \left(\kappa_1^{WW}\right)^2 
+ 0.1944  \left(g_2^{WW}\right)^2 
+ 0.08075  \left(g_4^{WW}\right)^2  
\\	
 && +\, 0.7204 \left(\frac{g_1^{WW}}{2}\right)  \kappa_1^{WW}
+ 0.7437 \left(\frac{g_1^{WW}}{2}\right)  g_2^{WW}
+ 0.2774\, \kappa_1^{WW} g_2^{WW} 
\,.
\nonumber 
\end{eqnarray}

For the $H\to ZZ/Z\gamma^*/\gamma^*\gamma^*\rightarrow$\,four-fermion final state, 
we set $\Lambda_1^{Z\gamma}=\Lambda_1^{ZZ}=100$\,GeV  in Eq.~(\ref{eq:HVV}) and rely on the $\kappa_2^{Z\gamma}$ and 
$\kappa_1^{ZZ}=\kappa_2^{ZZ}$ parameters to express 
\begin{eqnarray}
\label{eq:ratio-3}
R_{ZZ/Z\gamma^*/\gamma^*\gamma^*} =
&& \left(\frac{g_1^{ZZ}}{2}\right)^2 
+ 0.1695  \left(\kappa_1^{ZZ}\right)^2 
+ 0.09076  \left(g_2^{ZZ}\right)^2 
+ 0.03809  \left(g_4^{ZZ}\right)^2  
	  \\
&& +\, 0.8095 \left(\frac{g_1^{ZZ}}{2}\right)  \kappa_1^{ZZ}
+ 0.5046  \left(\frac{g_1^{ZZ}}{2}\right)  g_2^{ZZ}
+ 0.2092\,  \kappa_1^{ZZ} g_2^{ZZ} 
	\nonumber  \\
&& +\, 0.1023  \left(\kappa_2^{Z\gamma}\right)^2 
+ 0.1901 \left(\frac{g_1^{ZZ}}{2}\right)  \kappa_2^{Z\gamma}
+ 0.07429\,  \kappa_1^{ZZ} \kappa_2^{Z\gamma}
+ 0.04710\,  g_2^{ZZ} \kappa_2^{Z\gamma}
\,.
	\nonumber
\end{eqnarray}
We set $g_2^{Z\gamma}=g_4^{Z\gamma}=g_2^{\gamma\gamma}=g_4^{\gamma\gamma}=0$ in Eq.~(\ref{eq:ratio-3}).
These four couplings require a coherent treatment of the $q^2$ cutoff for the virtual photon and are left for
a dedicated analysis. 
We note that some final states in the $H\to WW$ and $ZZ/Z\gamma^*/\gamma^*\gamma^*\rightarrow$\,four-fermion 
decays may interfere, but their fraction and phase-space overlap are very small.
We therefore neglect this effect. 

For the $H\to\gamma\gamma$ and $\Z\gamma$ final states, 
we include the $W$ boson and the top and bottom quarks in the loops and obtain
\begin{eqnarray}
\label{eq:ratio-5}
R_{\gamma\gamma} = && 
1.60578\left(\frac{g_1^{WW}}{2}\right)^2 + 0.07313\,  \kappa_t^2 -0.68556\left(\frac{g_1^{WW}}{2}\right)\kappa_t
	+ 0.00002\,  \kappa_b^2 - 0.00183\, \kappa_t\kappa_b
	 \\
&&	+\, 0.00846\left(\frac{g_1^{WW}}{2}\right)\kappa_b 
	+ 0.16993\,  \tilde\kappa_t^2 + 0.00002\,  \tilde\kappa_b^2 - 0.00315\, \tilde\kappa_t\tilde\kappa_b  
	\nonumber 
	\,.
\end{eqnarray}
\begin{eqnarray}
\label{eq:ratio-6}
R_{Z\gamma} = && 
1.118600 \left(\frac{g_1^{WW}}{2}\right)^2 + 0.003489\, \kappa_t^2 - 0.125010  \left(\frac{g_1^{WW}}{2}\right)\kappa_t
	+ 0.000003\, \kappa_b^2 - 0.000183 \kappa_t\kappa_b 
	\\
&&	+\,  0.003100 \left(\frac{g_1^{WW}}{2}\right)\kappa_b 
	+ 0.012625\, \tilde\kappa_t^2 + 0.000005\, \tilde\kappa_b^2 - 0.000467\, \tilde\kappa_t\tilde\kappa_b  
	\nonumber 
	\,.
\end{eqnarray}
The point-like interactions $g_2^{\gamma\gamma}$ and $g_4^{\gamma\gamma}$ or  $g_2^{Z\gamma}$ and $g_4^{Z\gamma}$ 
could be considered in Eqs.~(\ref{eq:ratio-5}) and~(\ref{eq:ratio-6}). 
However, following the approach in Eq.~(\ref{eq:ratio-3}), these are left to a dedicated analysis. 
Within the SM EFT theory approach, a fully general study is available in Ref.~\cite{Brivio:2019myy}.
We do not consider higher-order corrections,
such as terms involving $\kappa_{1,2}^{WW}$, $g_2^{WW}$, or $g_4^{WW}$, in Eqs.~(\ref{eq:ratio-5}) and~(\ref{eq:ratio-6}).
We also neglect the $H\to\gamma^*\gamma$ contribution. 

To conclude the discussion of the cross sections,
we note that the relative contribution of an individual coupling $a_n$, either to production 
$\left(\sum  \alpha_{jk}^{(i)}a_ja_k\right)$ or to decay $\left(\sum  \alpha_{lm}^{(f)}a_la_m\right)$,
can be parameterized as an effective cross-section fraction 
\begin{eqnarray}
 f_{an}^{(i,f)} = \frac{\alpha_{nn}^{(i,f)}a_n^2}{\sum_{m}\alpha_{mm}^{(i,f)}a_m^2}\times{\rm sign}\left(\frac{a_n}{a_1}\right)
 \,,
\label{eq:fgn}
\end{eqnarray}
where the sign of the $a_n$ coupling relative to the dominant SM contribution $a_1$ is incorporated into the $f_{an}$ definition. 
In the denominator of Eq.~(\ref{eq:fgn}), the sum runs over all couplings contributing to the $i\to H$ or $H\to f$ process.
By convention, the interference contributions are not included in the effective fraction 
definition in Eq.~(\ref{eq:fgn}) so that this parameter can be more easily interpreted.

We adopt the definition of $f_{an}$ used by the LHC experiments~\cite{CMS-HIG-12-041,Khachatryan:2014kca,Aad:2015mxa}
for $HWW$, $HZZ$, $HZ\gamma$,  and $H\gamma\gamma$ anomalous couplings 
in the $H\to ZZ/Z\gamma^*/\gamma^*\gamma^*\to 2e2\mu$ process, with the $HWW$ couplings related 
through Eqs.~(\ref{eq:deltaMW})--(\ref{eq:kappa1WW}); 
$f_{\rm CP}^{\rm gg}$ in the ggH process for the effective $Hgg$ couplings~\cite{Anderson:2013afp}; 
and $f_{\rm CP}^{qq}$ for processes involving $H q\bar{q}$ fermion couplings, 
such as $H\to q\bar{q}$, with $\alpha_{mm}=1$ in Eq.~(\ref{eq:fgn}). 
The latter convention for $f_{\rm CP}^{tt}$ is extended to the $H t\bar{t}$ couplings as well, 
despite the fact that Eq.~(\ref{eq:ratio-1}) is not valid for the heavy top quark~\cite{Gritsan:2016hjl}.
It is also easy to invert Eq.~(\ref{eq:fgn}) to relate the cross section fractions to coupling ratios via
\begin{eqnarray}
\frac{a_n}{a_m}=\sqrt{\frac{\left|f_{an}\right| \alpha_{mm}}{\left|f_{am}\right| \alpha_{nn}  }}
\times{\rm sign}\left(f_{an}f_{am}\right)
 \,,
\label{eq:an}
\end{eqnarray}
where we omit the process index for either $i\to H$ or $H\to f$.
Because $\sum_n \left|f_{an}\right|=1$, only all but one of the parameters are independent.  
We choose to use the $f_{an}$ corresponding to anomalous couplings as our independent set of parameters, 
leaving for example $f_{a1}=\left(1-\sum_{n\ne1}|f_{an}|\right)$ as a dependent one. 

There are several advantages in using the $f_{an}$ parameters in Eq.~(\ref{eq:fgn}) in analyzing a given process on the LHC.
First of all, the $f_{an}$ and signal strength $\mu^{i\to f}=\sigma^{i\to f}/\sigma^{i\to f}_{\rm SM}$ form a complete 
and minimal set of measurable parameters describing the process $i\to H\to f$. 
Measuring directly in terms of couplings introduces degeneracy in Eq.~(\ref{eq:diff-cross-section2}),
because, for example, the production couplings can be scaled up and the decay couplings down without
changing the result. A similar interplay occurs between the couplings appearing in the numerator
and the denominator of Eq.~(\ref{eq:diff-cross-section2}).
Second, the $f_{an}$ parameters are independent of ${\Gamma_{\rm tot}}$, which is absorbed 
into $\mu^{i\to f}$. In contrast, the direct coupling measurement $a_n$ depends on the assumptions 
in Eq.~(\ref{eq:width_other}), including $\Gamma_{\rm other}$. 
Third, $f_{an}$ has the same meaning in all production and all decay channels of the \Hboson.
For example, the $f_{an}$ measurement in VBF production is invariant with respect to the $H\to f$ 
decay channel used. This can be seen from Eq.~(\ref{eq:diff-cross-section2}), where 
$\left(\sum  \alpha_{lm}^{(f)}a_la_m\right)/{\Gamma_{\rm tot}}$ can be absorbed into the $\mu^{i\to f}$ 
parameter.\footnote{The situation when production and decay cannot be decoupled in analysis of the data 
due to the same couplings appearing in both processes, such as in $i\to VV\to H\to VV\to f$, is discussed 
in detail in Section ~\ref{sect:exp_onshell}.}
Fourth, $f_{an}$ is a ratio of observable cross sections, and therefore it is invariant with respect to
the $a_n$ coupling scale convention. For example, the $f_{an}$ value is identical for either the $c_n$ or $g_n$ 
couplings related in Eq.~(\ref{eq:EFT_ci}). 
Fifth, in the experimental measurements of $f_{an}$ most systematic uncertainties cancel in the ratios, 
making it a clean measurement to report. 
Sixth, the $f_{an}$ are convenient parameters for presenting results as their full range is bounded 
between $-1$ and $+1$, while the couplings and their ratios are not bounded. 
Finally, the $f_{an}$ have an intuitive interpretation, as their values indicate the fractional contribution 
to the measurable cross section, while there is no convention-invariant interpretation of the coupling 
measurements. 
In the end, the measurements in individual processes can be combined, and at that point 
their interpretation in terms of couplings becomes natural. However, this becomes feasible 
only when the number of measurements is at least equal to, or preferably exceeds, the number of couplings.

\section{JHUGen/MELA framework}
\label{sect:cp_mc}

The JHUGen (or JHU generator) and MELA (or Matrix Element Likelihood Approach) framework 
is designed for the study of a generic bosonic resonance decaying into SM particles.
JHUGen is a stand-alone event generator that generates either weighted events into pre-defined histograms
or unweighted events into a Les Houches Events (LHE) file. 
A subsequent parton shower simulation as well as full detector simulation can be added using
other programs compatible with the LHE format. 
The MELA package is a library of probability distributions based on first-principle matrix elements.
It can be used for Monte Carlo re-weighting techniques and the construction of kinematic discriminants for an optimal analysis.
The packages are based on developments reported in this work and Refs.~\cite{Gao:2010qx,Bolognesi:2012mm,Anderson:2013afp,Gritsan:2016hjl}.
It can be freely downloaded at~\cite{jhugen}. 
The package has been employed in the Run-I and Run-II analyses of LHC data for the \Hboson\ property 
measurements~\cite{Sirunyan:2017tqd,Sirunyan:2019nbs,Sirunyan:2019twz,Sirunyan:2018qlb,Sirunyan:2019pqw,Sirunyan:2020sum,
Chatrchyan:2012xdj,Chatrchyan:2012jja,Chatrchyan:2013mxa,Chatrchyan:2013iaa,
Aad:2013xqa,Khachatryan:2014iha,Khachatryan:2014ira,Khachatryan:2014kca,Khachatryan:2015mma,Khachatryan:2015cwa,
Aad:2015mxa,Khachatryan:2016tnr}.

Our framework supports a wide range of production processes for spin-zero, spin-one, and spin-two resonances and their decays into SM particles. 
All interaction vertices can have the most general Lorentz-invariant structure with CP-conserving or CP-violating degrees of freedom. 
We put a special emphasis on spin-zero resonances $H$, for which we allow production through gluon fusion, associated production with one or two jets, 
associated production with a weak vector boson ($Z/\gamma^*H, WH, \gamma H$), weak vector boson fusion ($VVjj\to Hjj$), and production in association with heavy flavor quarks,
such as $t\bar{t}H$, $tH$ and $b\bar{b}H$ at the LHC.
The supported decay modes include
$H\to ZZ$ / $Z\gamma^*$ / $\gamma^*\gamma^*\to4f$,
$H\to WW \to4f$, 
$H\to Z\gamma $ / $\gamma^*\gamma \to2f\gamma$, 
$H\to \gamma\gamma$, 
$H\to \tau\tau$,
and generally $H\to f\bar{f}$,
with the most general Lorentz-invariant coupling structures.  
Spin correlations are fully included, as are interference effects from identical particles.

To extend the capabilities of our framework, JHUGen also allows interfacing 
the decay of a spin-zero particle after its production has been simulated by other MC programs (or by JHUGen itself) 
through the LHE file format. As an example, this allows production of a spin-zero $H$ boson through 
NLO QCD accuracy with POWHEG~\cite{Frixione:2007vw} and further decay with the JHUGen.
Higher-order QCD contributions are discussed in Ref.~\cite{Gritsan:2016hjl} for the $\ttH$ process
and below for the $ZH$ process.
Another interface with the MCFM Monte Carlo generator~\cite{Campbell:2010ff,Campbell:2011bn,Campbell:2013una,Campbell:2015vwa,Campbell:2015qma} allows accessing
background processes and off-shell $H^*$ boson production, including interference with the continuum.
\\

In the following we briefly outline new key features in our JHUGen/MELA framework that become available 
with this publication. 
In the subsequent Sections, we apply these new features and demonstrate how they can be used for LHC physics analyses.  
In the simulation, the values of $s^2_w$, $M_{W}$, $\Gamma_{W}$, $m_{Z}$, and $\Gamma_{Z}$ are parameters 
configurable independently, and in this paper we set $s^2_w=0.23119$, $M_{W}=80.399$\,GeV, $\Gamma_{W}=2.085$\,GeV,
$M_{Z}=91.1876$\,GeV, and $\Gamma_{Z}=2.4952$\, GeV~\cite{Heinemeyer:2013tqa, Agashe:2014kda}.

\begin{figure}[t]
\centering
\setlength{\tabcolsep}{20pt}
\begin{tabular}{m{1cm} c c c }
& {{(a) Signal}} & {{(b) Interfering background}} & {{(c) Non-interfering background}} \\[3ex]
Gluon fusion & 
\includegraphics[width=0.15\linewidth]{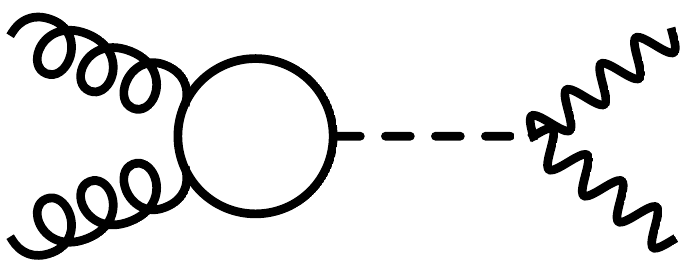} &
\includegraphics[width=0.15\linewidth]{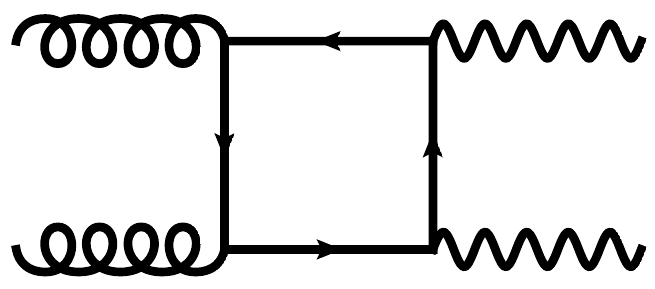} &
\includegraphics[width=0.10\linewidth]{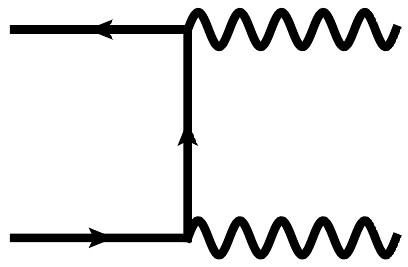} \\[3ex]
Vector boson fusion& 
\includegraphics[width=0.13\linewidth]{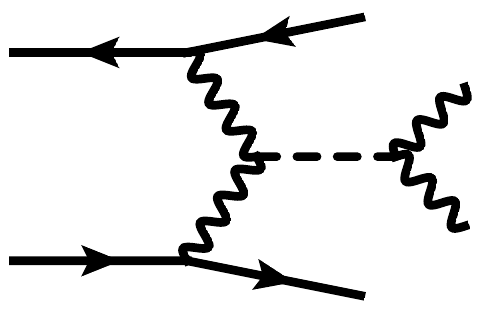} &
\includegraphics[width=0.10\linewidth]{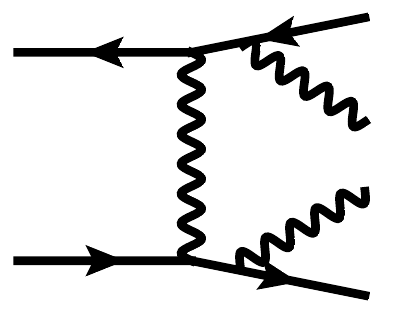} &
\includegraphics[width=0.13\linewidth]{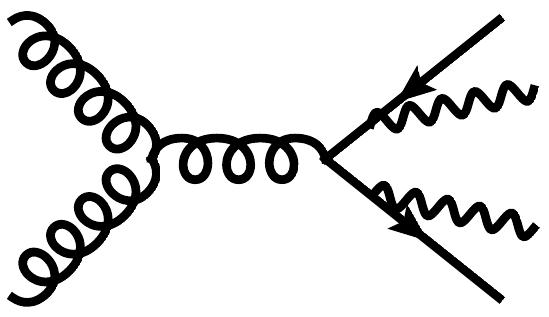} 
\end{tabular}
\caption{Sample diagrams for signal, interfering background and non-interfering background
in the processes $pp \to 4\ell  $ (gluon fusion) and $pp \to 4\ell  jj$ (weak vector boson fusion).}
\label{diag:ggvbf}
\end{figure}

\subsection{Off-shell simulation of the H boson in gluon fusion and a second scalar resonance}
\label{sect:mc_gg_offshell}

We extend our previous calculation of $gg \to H \to VV \to 4f$ by allowing $m_{4\ell}$ to be far off the $H$ resonance mass peak. 
In these regions of phase space the irredicible background from $q\bar{q}/gg \to VV \to 4f$ continuum production becomes 
significant and, in the case of the $gg$ initial state, interferes with the $H$ production amplitudes, as illustrated in Fig.~\ref{diag:ggvbf}.
The MCFM generator~\cite{Campbell:2013una} contains the SM amplitudes for this process at LO.
Our add-on extends the MCFM code and incorporates the most general anomalous couplings in the $H$ boson amplitude. 
We allow two possible parameterizations of the CP-even and CP-odd degrees of freedom: the point-like $Hgg$ couplings $g_2^{gg}, g_4^{gg}$ and
the full one-loop amplitude with heavy quark flavors, using the Yukawa-type couplings $\kappa_q, \tilde{\kappa}_q$.
Additional hypothetical fourth-generation quarks with anomalous $HQ\bar{Q}$ couplings can be included as well. 
For the study of a second $H$-like resonance $X$ with mass $m_X$ and width $\Gamma_X$, 
we allow for the same set of couplings and decay modes.

\subsection{Off-shell simulation of the H boson in electroweak production and a second scalar resonance}
\label{sect:mc_vbf_offshell}

Similar to the gluon fusion process, we extend our previous calculation of vector boson fusion $qq \to qq + H(\to VV\to 4f)$
and associated production $qq \to V + H(\to VV\to 4f)$, and allow the full kinematic range for $m_{4f}$. 
The SM implementation in MCFM~\cite{Campbell:2015vwa} includes the $s$- and $t$-channel \Hboson\ amplitudes,
the continuum background amplitudes, and their interference, as illustrated in Fig.~\ref{diag:ggvbf}.
We supplement the necessary contributions for the most general anomalous coupling structure. 
In particular, this affects the \Hboson\ amplitudes but also the triple and quartic gauge boson couplings. 
We also add amplitudes for the intermediate states $ZZ / Z\gamma^* / \gamma^*\gamma^*$ in place of $ZZ$
in both decay and production with the most general anomalous coupling structure, which are not present 
in the original MCFM implementation. It is interesting to note that the off-shell VBF process $qq \to qq + H(\to 4f)$ 
includes contributions of the $q\bar{q} \to VH(\to 4f)$ process for the case of hadronic decays of the $V$ boson.
As in the case of gluon fusion, we also allow the study of a second $H$-like resonance $X$ with mass $m_X$, width $\Gamma_X$, 
and the same set of couplings and decay modes.

\subsection{Higher-order contributions to VH production}
\label{sect:mc_vh_nlo}

\begin{figure}[t]
\centering
\setlength{\tabcolsep}{25pt}
\begin{tabular}{c c c c}
{{(a) $q\bar{q}$ LO}} & {{(b) $q\bar{q}$ NLO QCD}} & {{(c) $gg$ LO box }} & {{(d) $gg$ LO triange}} \\[3ex]
\includegraphics[width=0.12\linewidth]{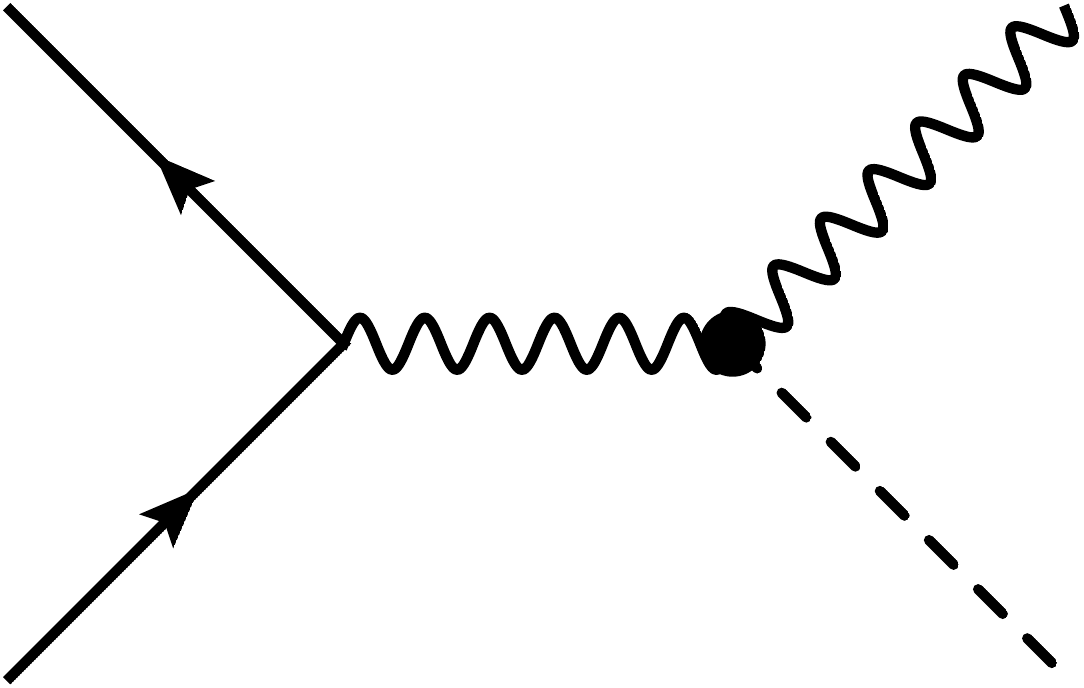} &
\includegraphics[width=0.12\linewidth]{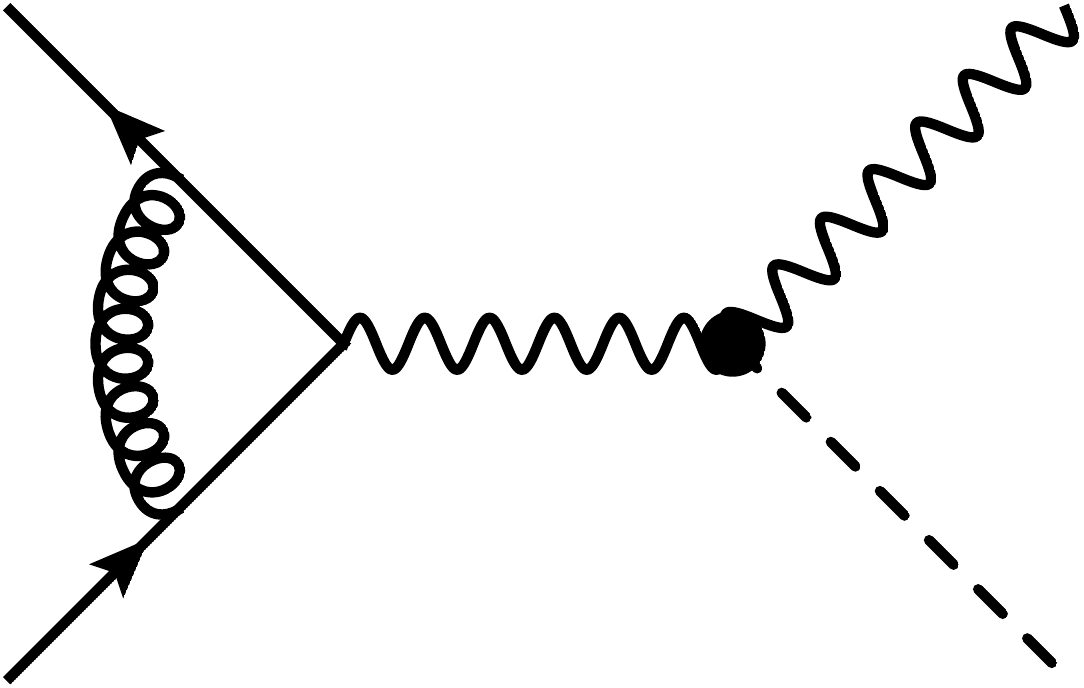} &
\includegraphics[width=0.2\linewidth]{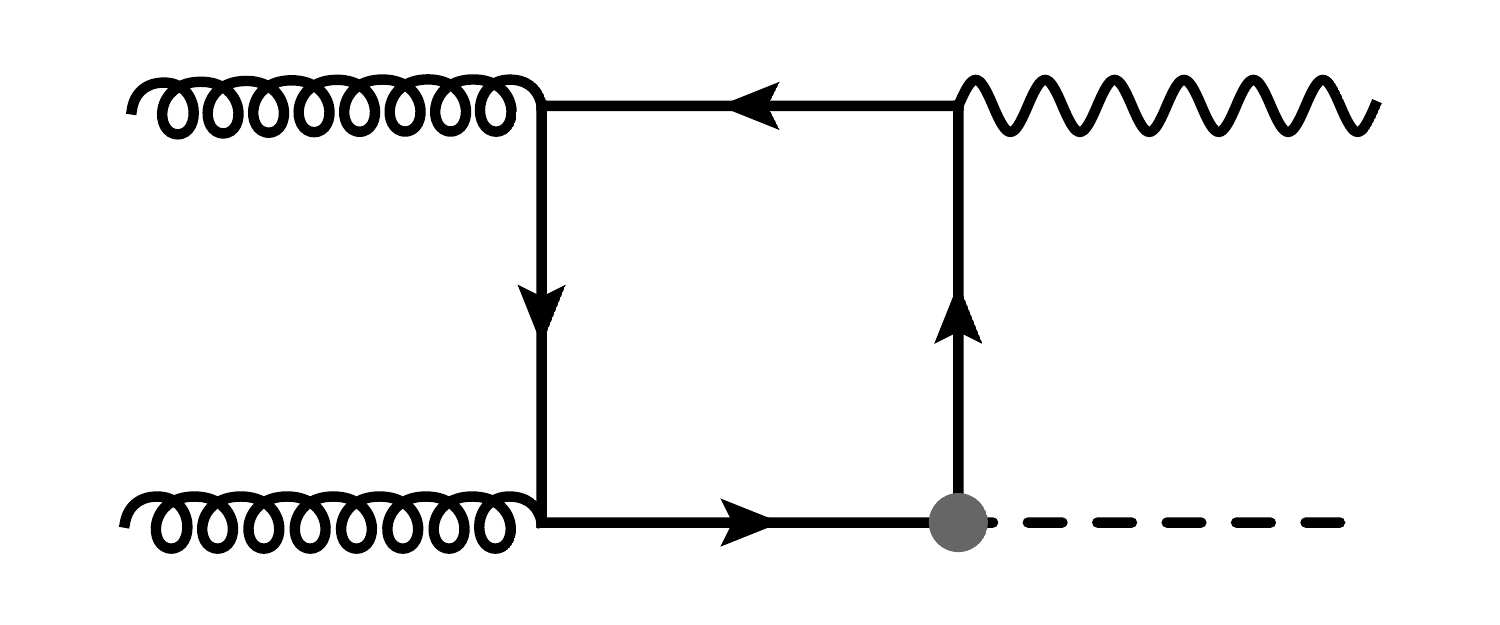} &
\includegraphics[width=0.2\linewidth]{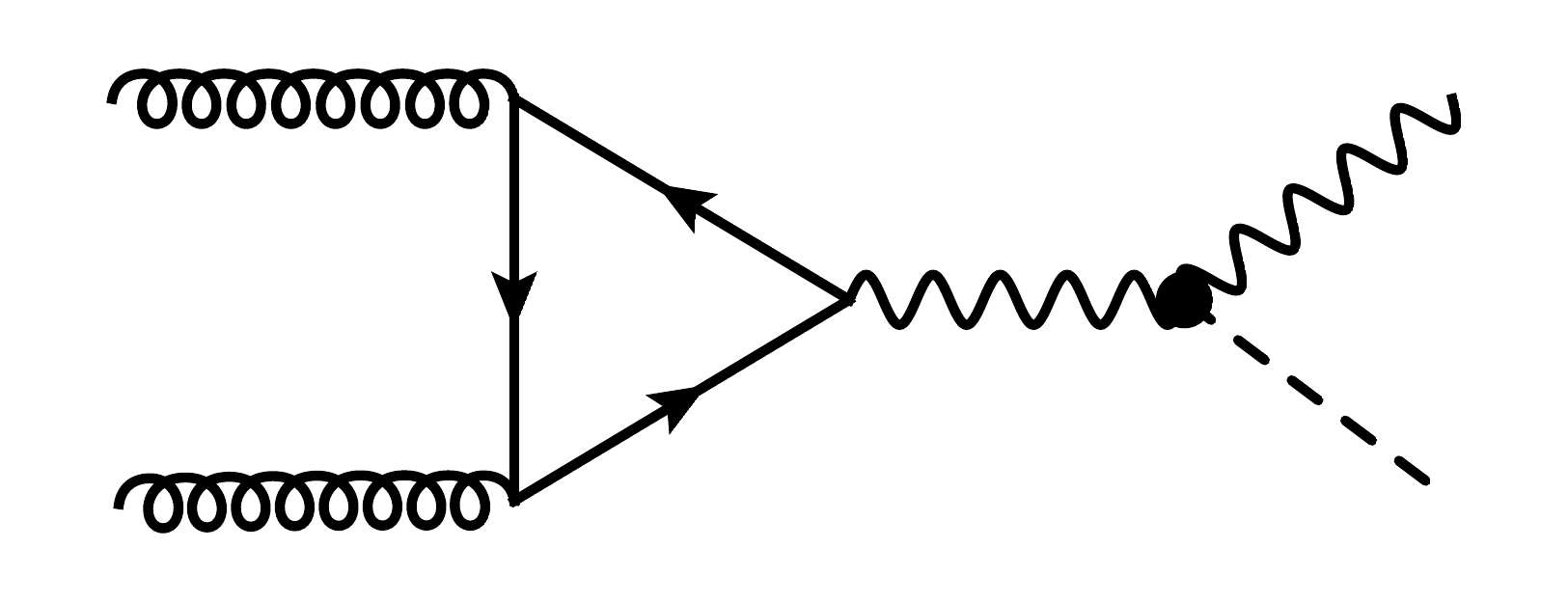} 
\end{tabular}
\caption{$ZH$ sample diagrams for leading order $q\bar{q}$ and $gg$ initial states, including higher order contributions.}
\label{diag:ppvh}
\end{figure}

We calculate the NLO QCD corrections to the associated $H$ boson production process $q\bar{q} \to V H$ where $V=\Z,W,\gamma$, 
shown in Fig.~\ref{diag:ppvh}. We use standard techniques and implement the results in JHUGen,
relying on the \textsc{COLLIER}~\citep{Denner:2016kdg} loop integral library.
This improves the physics simulation of previous studies at LO and allows demonstrating the robustness of previous matrix element method studies.
We also calculate the loop-induced gluon fusion contribution $gg \to Z H$, which is parameterically of next-to-next-to-leading 
order but receives an enhancement from the large gluon flux, making it numerically relevant for studies at NLO precision. 
In contrast to the $q\bar{q} \to VH$ process which is sensitive to $HVV$ couplings, the $gg\to ZH$ process is additionally sensitive 
to the Yukawa-type $H q\bar{q}$ couplings. In both cases we allow for the most general CP-even and CP-odd couplings. 
Strong destructive interference between triangle and box amplitudes in the SM leads to interesting physics effects that enhance 
sensitivity to anomalous $H t\bar{t}$ couplings, as we demonstrate in Section~\ref{sect:cp_vh}.

\subsection{Multidimensional likelihoods and machine learning}
\label{sect:mc_multidim_fits}

We extend the multivariate maximum likelihood fitting framework to describe the data in an optimal way 
and provide the multi-parameter results in both the EFT and the generic approaches. The main challenge in this
analysis is the fast growth of both the number of observable dimensions and the number of contributing 
components in the likelihood description of a single process with the increasing number of parameters of interest. 
We present a practical approach to accommodate both challenges, while keeping the approach generic enough
for further extensions. This approach relies on the MC simulation, reweighting tools, and optimal observables
constructed from matrix element calculations. We extend the matrix element approach by 
incorporating the machine learning procedure to account for parton shower and detector effects when 
these effects become sizable. Some of these techniques are illustrated with examples below.

\section{LHC event kinematics and the matrix element technique}
\label{sect:exp_kinematics}

\begin{figure}[t]
\centerline{
\includegraphics[width=0.45\linewidth]{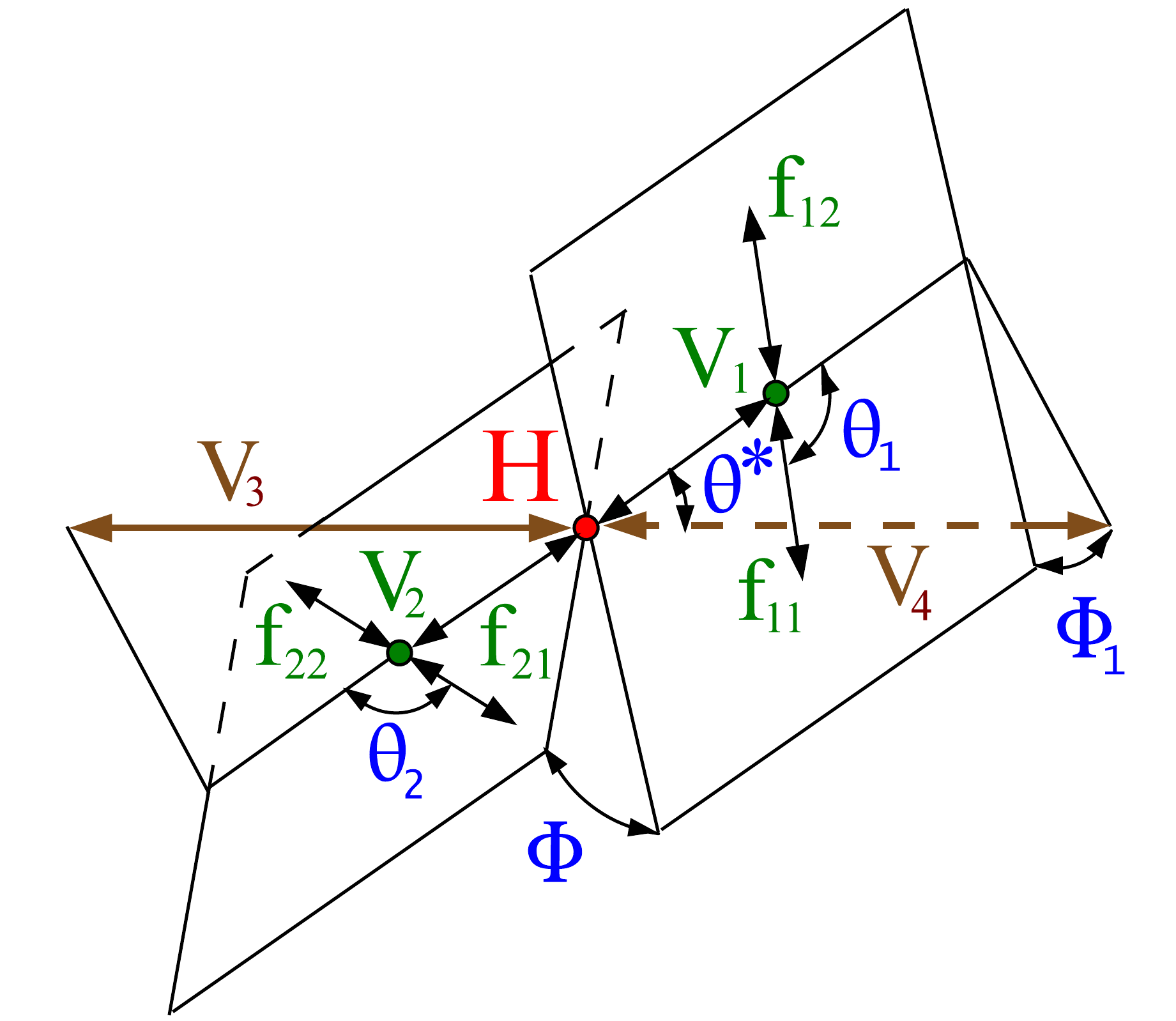}
}
\captionsetup{justification=centerlast}
\caption{
Illustrations of an \Hboson\ production and decay in three topologies:
(1) boson fusion and decay $V_3V_4\to H \to V_1V_2 \to 4f$;
(2) boson fusion with associated jets $q_{12}q_{22}\to q_{11}q_{21}(V_1V_2\to H\to V_3V_4)$;
and (3) associated production $q_{11}q_{12}\to V_1 \to V_2(H\to V_3V_4)$.
Five angles fully characterize the orientation of the production or 
decay chain and are defined in suitable frames~\cite{Gao:2010qx,Anderson:2013afp}.
}
\label{fig:kinematics}
\end{figure}

Kinematic distributions of particles produced in association with the $H$ boson or in its decay
are sensitive to the quantum numbers and anomalous couplings of the \Hboson. 
In the $1\to4$ process of the $H\to VV \to 4f$ decay, six observables
${\bf\Omega}^{\rm decay}=\{\theta_1, \theta_2, \Phi, m_1, m_2, m_{4f} \}$
fully characterize kinematics of the decay products,
while two other angles ${\bf\Omega}^{\rm prod}=\{\theta^*, \Phi_1 \}$ orient the decay frame
with respect to the production axis, as described in Ref.~\cite{Gao:2010qx} and shown in Fig.~\ref{fig:kinematics}.
The ${\bf\Omega}^{\rm prod}$ angles are random for the production of a spin-zero particle, but provide 
non-trivial information to distinguish signal from either background or alternative spin hypotheses. 
A similar set of observables can be defined in a production process.
For example, the observables
${\bf\Omega}^{\rm assoc}=\{\theta_1^{\rm assoc}, \theta_2^{\rm assoc}, \Phi^{\rm assoc}, q_1^{2,\rm assoc}, q_2^{2,\rm assoc} \}$
characterize $V\!H$ and weak or strong boson fusion (VBF or ggH) in association with two hadronic jets,
as illustrated in Fig.~\ref{fig:kinematics} and described further in Ref.~\cite{Anderson:2013afp}. 
Similar kinematic diagrams defining observables for the $t\bar{t}H$, $tqH$, and $H\to\tau\tau$ processes
are discussed elsewhere~\cite{Gritsan:2016hjl}.

In the $2\to 6$ process of associated \Hboson production and its subsequent decay to a four-fermion final state, such as VBF,
13 kinematic observables are defined, which include angles and the invariant masses of intermediate states. 
There is also the overall boost of the six-body system, which depends on QCD effects.  We decouple this boost from these considerations.
Only a reduced set of observables is available when there are no associated particles in production or when the decay chain has
less than four particles in the final state. 

\begin{figure}[ht]
\centerline{
\epsfig{figure=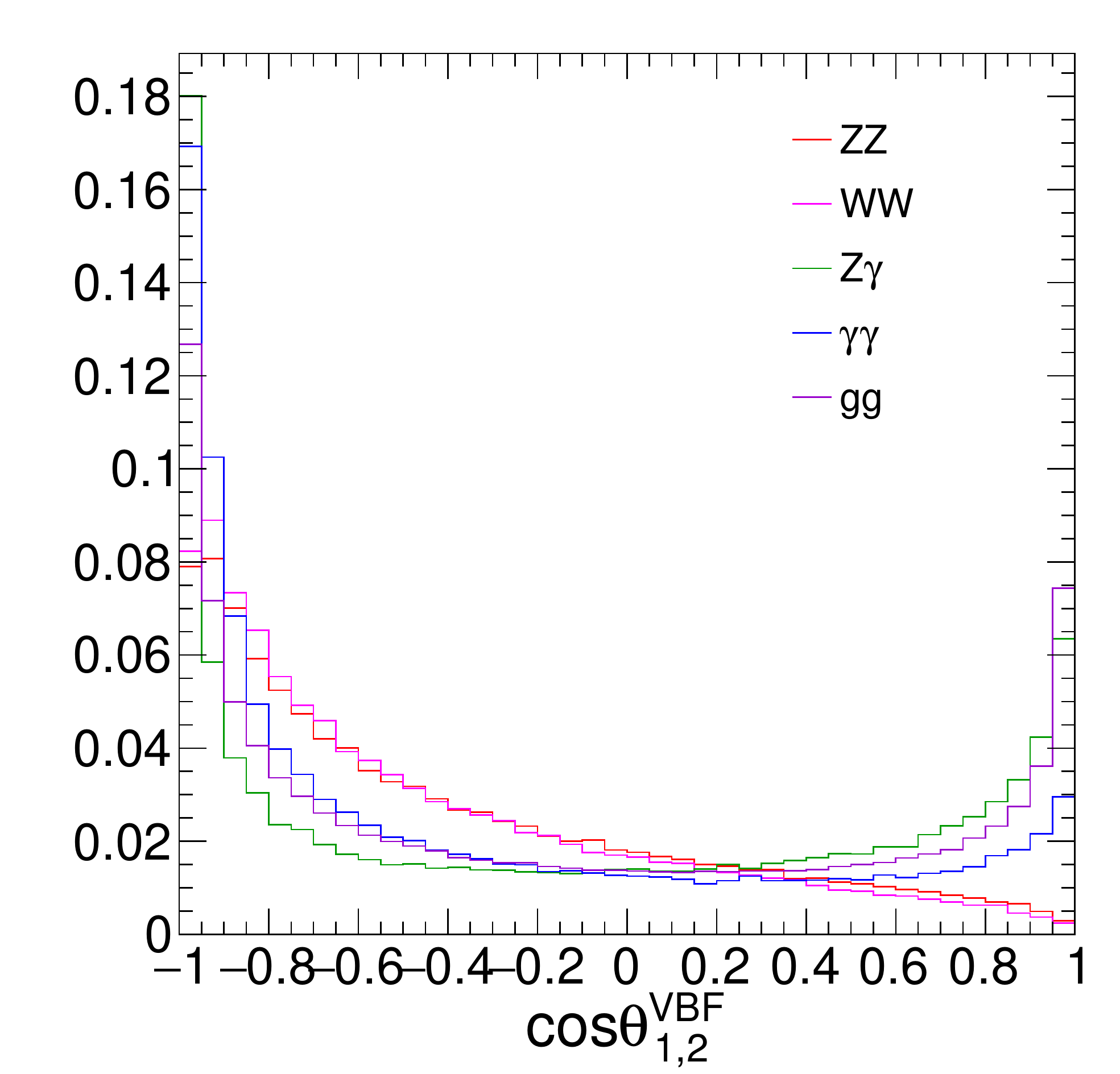,width=0.33\linewidth}
\epsfig{figure=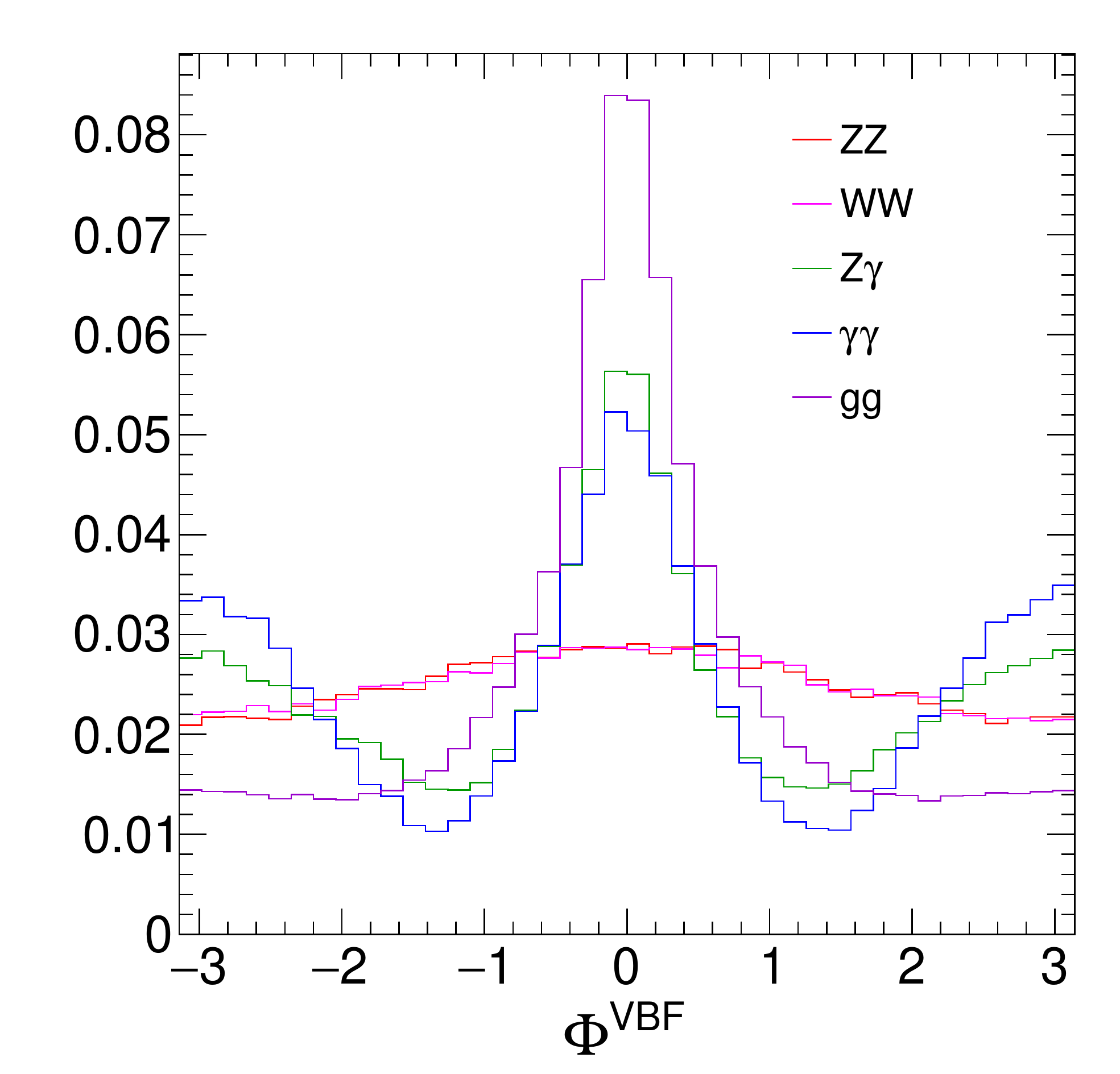,width=0.33\linewidth}
\epsfig{figure=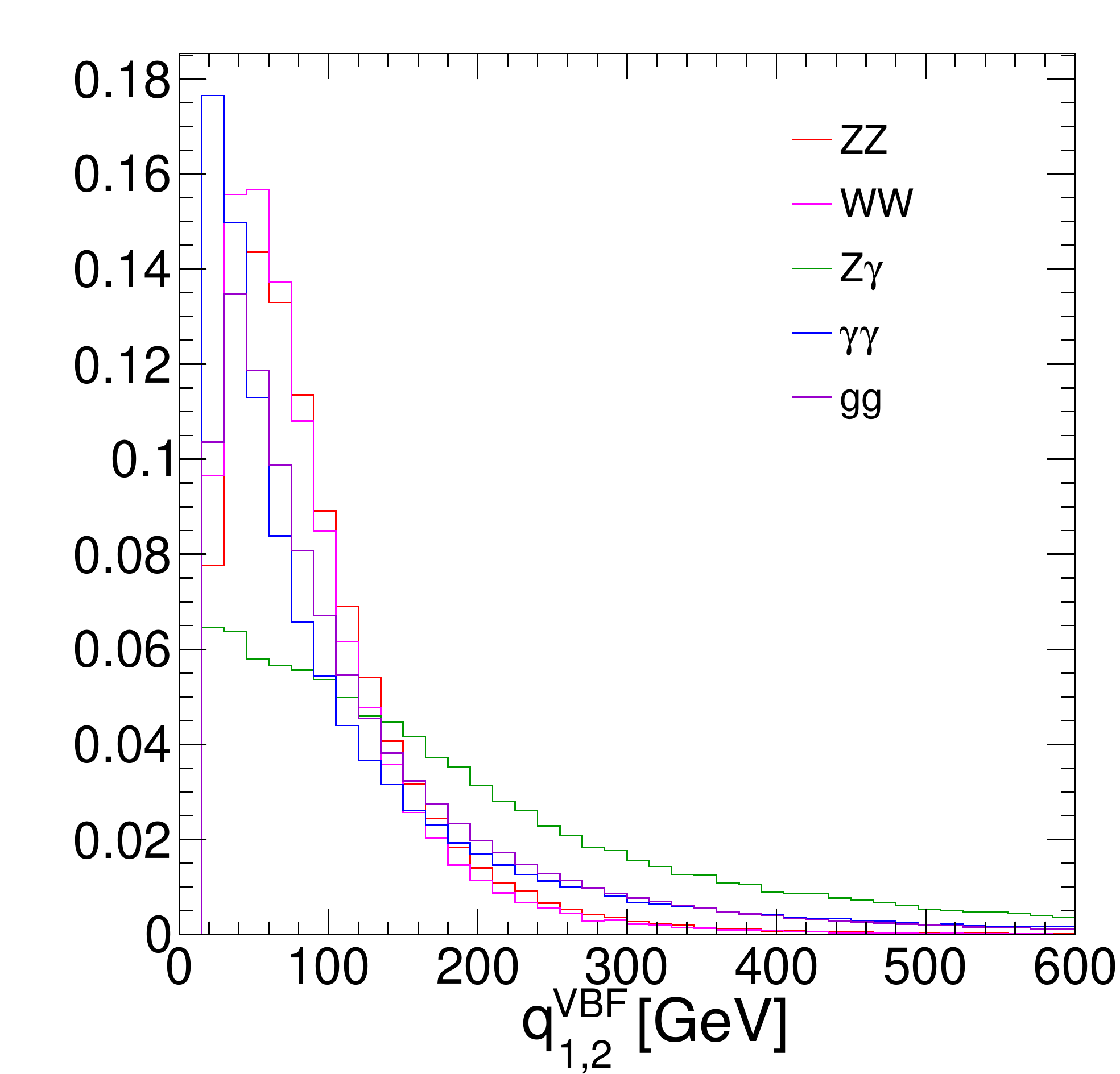,width=0.33\linewidth}
}
\centerline{
\epsfig{figure=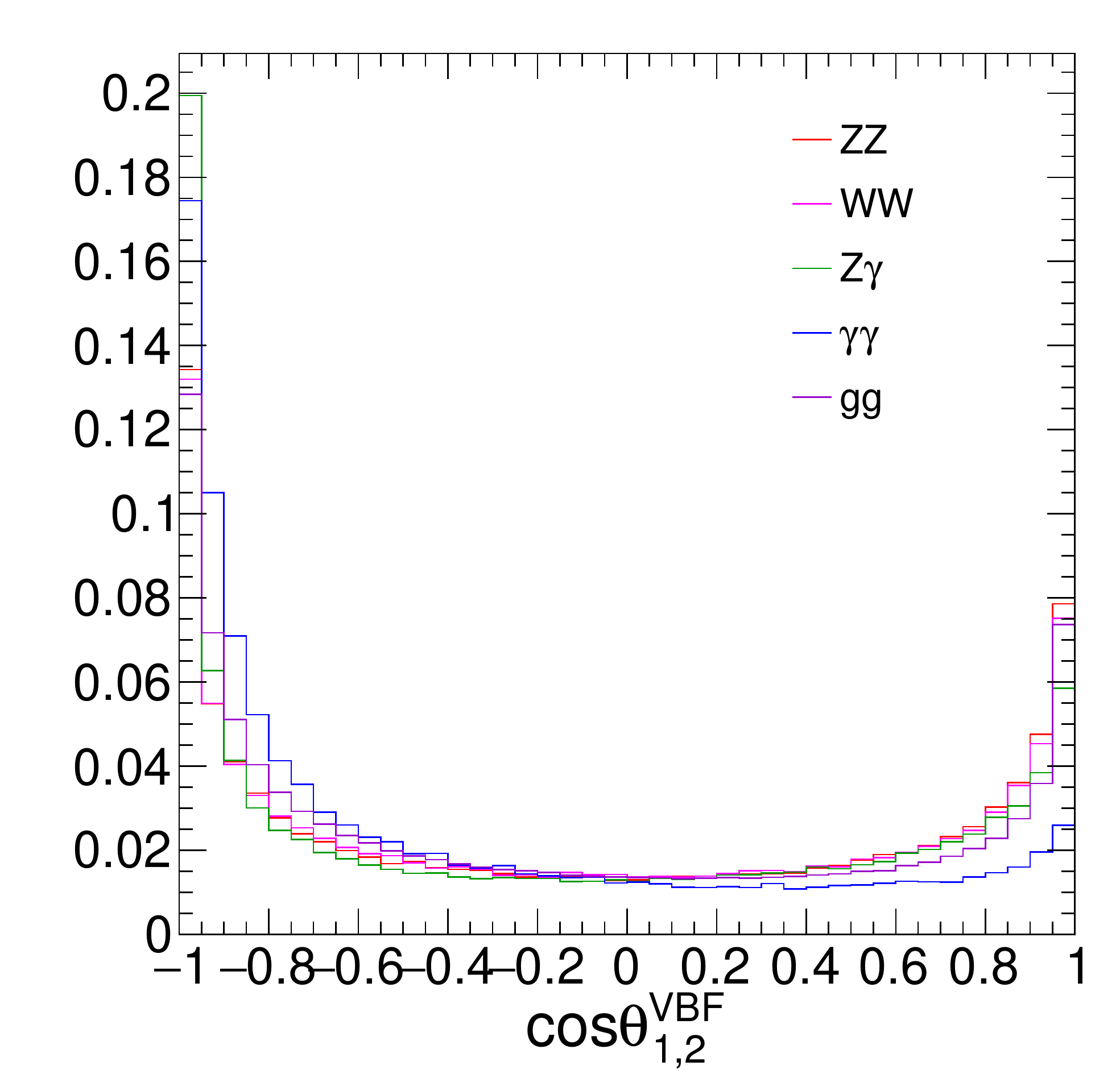,width=0.33\linewidth}
\epsfig{figure=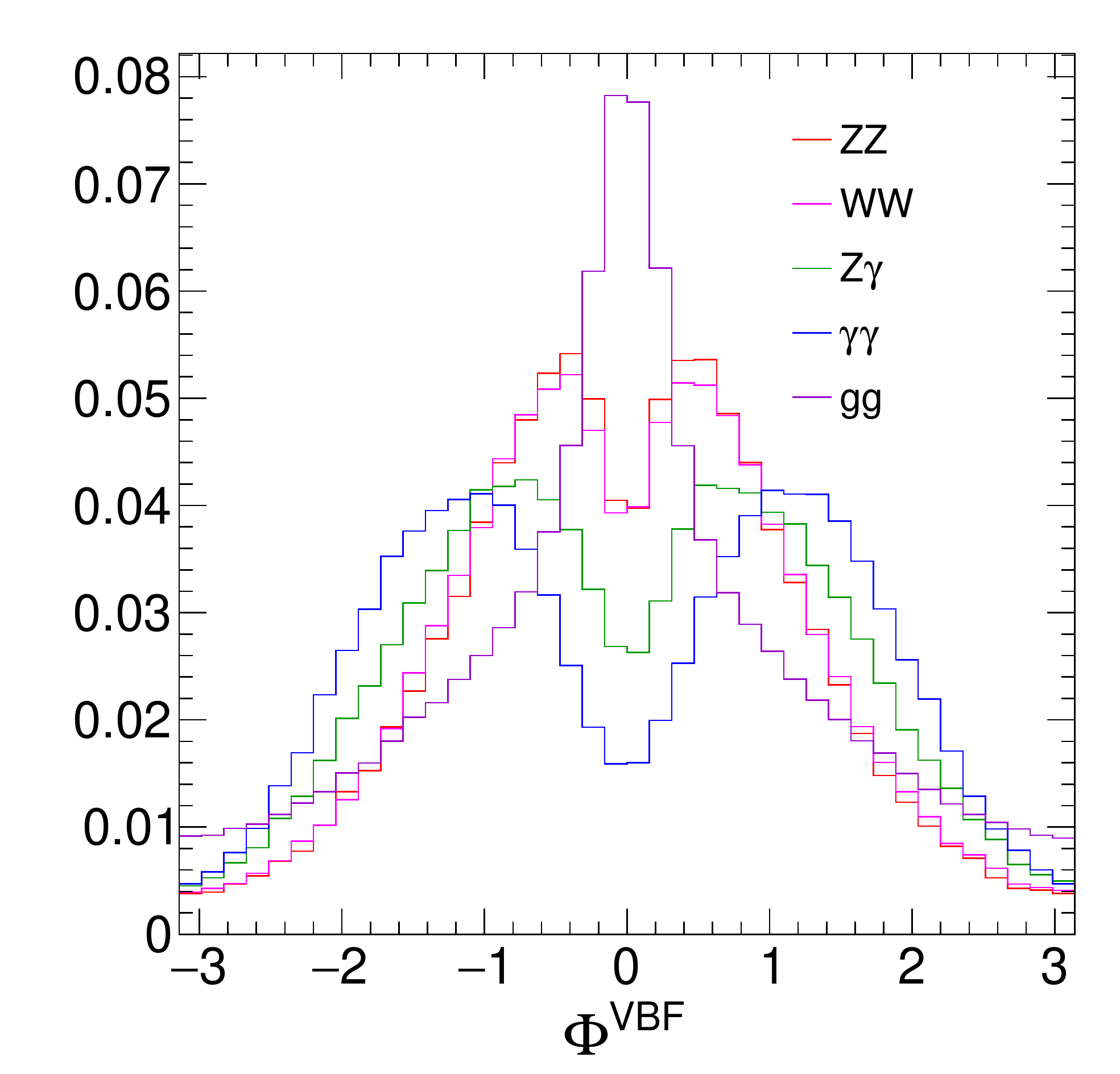,width=0.33\linewidth}
\epsfig{figure=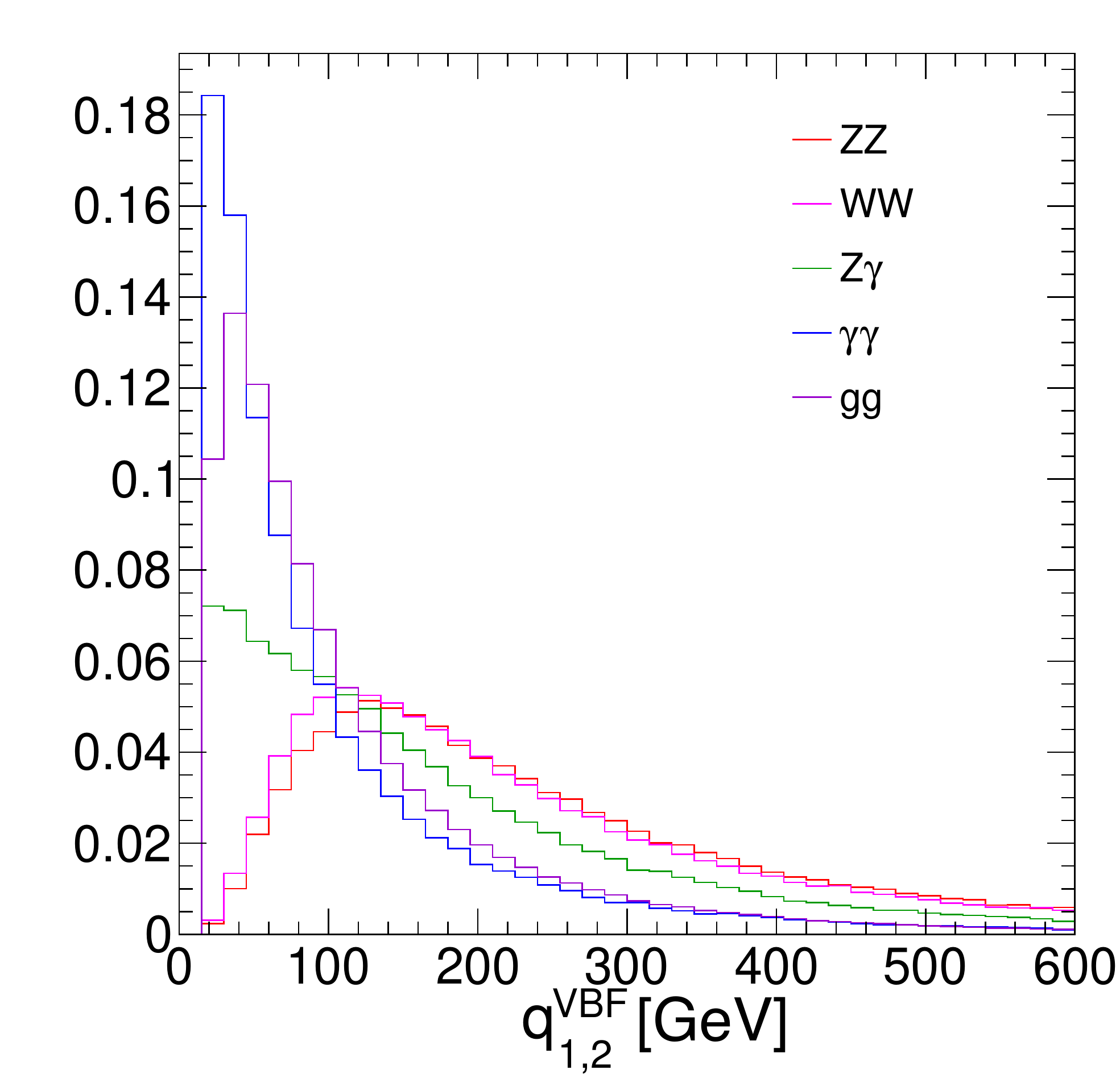,width=0.33\linewidth}
}
\captionsetup{justification=centerlast}
\caption{
Distributions of observables in vector boson fusion jet associated production:
$\{\theta_{1,2}^{\rm VBF}, \Phi^{\rm VBF}, \sqrt{-q_{1,2}^{2,\rm VBF}} \}$,
comparing $gg$, $\gamma\gamma$, $Z\gamma$, $ZZ$, and $WW$ fusion
for the SM couplings (top) and pseudoscalar couplings (bottom). 
A loose selection, $\Delta R_{JJ}>0.3$ and $p_{T}^{J}>15\,\text{GeV}$, is applied,
consistently for all processes, to avoid divergences in processes with photons and
gluons. All distributions are normalized to unit area. 
}
\label{fig:kinematics_wbf}
\end{figure}

Kinematic distributions with anomalous couplings of the \Hboson have been shown previously in 
Refs.~\cite{Gao:2010qx,Bolognesi:2012mm,Anderson:2013afp} for both decay and associated production. 
Here, we emphasize kinematics in associated production with two jets, shown in Fig.~\ref{fig:kinematics_wbf}. 
There are distinct features depending on the $gg$, $\gamma\gamma$, $Z\gamma$, $ZZ$, and $WW$ fusion,
which is reflected in the associated jet kinematics. 
Note that for the production processes we define $q_i^2$ for each vector boson, where
$q_i^2<0$ for boson fusion and we therefore plot $\sqrt{-q_i^2}$.
In this case, $\theta_i^{\rm assoc}$ angles, usually defined
in the rest frame of the vector bosons, are calculated in the $H$ frame instead. 
We would like to stress that a consistent treatment of all contributions with 
$\gamma\gamma$, $Z\gamma$, $ZZ$, and $WW$ intermediate states in weak boson fusion
is critical in a study of anomalous couplings. 

While for the SM \Hboson\ one could often neglect 
photon intermediate states when couplings to the $Z$ and $W$ bosons dominate, one generally cannot neglect
them when comparing to other contributions generated by higher-dimension operators. 
In reference to the EFT operators discussed in Section~\ref{sect:cp_couplings}, the Higgs basis 
becomes the natural one to disentangle the $H\gamma\gamma$, $HZ\gamma$, $HZZ$, and $HWW$
operators from the experimentally observed kinematics of events. This is visible, for example, in the 
$q_{1,2}^{\rm VBF}$ distributions corresponding to the pseudoscalar operators, where the 
photon intermediate states lead to a much softer spectrum compared to $W$ and $Z$. 
The advantage of the Higgs basis for experimental analysis becomes especially evident 
when considering \offshell\ effects, because there is no \offshell\ enhancement with the intermediate 
$\gamma$ states. Once experimental results are obtained in the Higgs basis, the measurements can be 
translated to any other basis. 

With up to 13 observables ${\bf\Omega}$ sensitive to the Higgs boson anomalous couplings,
it is a challenging task to perform an optimal analysis in a multidimensional space of observables,
creating the likelihood function depending on more than a dozen parameters
in Eq.~(\ref{eq:EFT_hvv}). 
Full detector simulation and data control regions in LHC data analyses may limit the number 
of available events and, as a result, the level of detail in the likelihood. Therefore, it is important 
to develop methods that are close to optimal under the practical constraints of the available 
data and simulation. 
In the rest of this Section, we discuss some of the experimental applications of the tools developed 
in our framework, which target these tasks in the study of the \Hboson kinematics.

Analysis of experimental observables typically requires the construction of a likelihood function, which is maximized
with respect to parameters of interest. The complexity of the likelihood function grows quickly both with the 
number of observables and with the number of parameters, and the two typically increase simultaneously. 
Examples of such likelihood construction will be discussed 
in Section~\ref{sect:exp_onshell}. Typically, the likelihood function will be parameterized with templates 
(histograms) of observables, using either simulated MC samples or control regions in the data. The challenge
in this approach is to keep the number of bins of observables to a practical limit, typically several bins for 
several observables, due to statistical limitations in the available data and simulation. Similar practical limitations 
appear in the number of parameters of interest, which will be discussed later. 

The information content in the kinematic observables is different, and one could pick some of the most informative 
kinematic observables of interest. The difficulty of this approach is illustrated in Fig.~\ref{fig:kinematics_wbf} where
all five observables (note that $\theta_{1,2}$ and $q_{1,2}$ each represent two independent observables)
provide important information and it is hard to pick a reduced set without substantial loss of
information.  Another approach is to create new observables optimal for the problem of interest, and in the next 
subsections we illustrate optimal observables based on both the matrix element and the machine learning techniques. 
Nonetheless, it is not possible to have a prior best set of observables universally good for all measurements
and at the same time limited in the number of dimensions for practical reasons. We note that alternative methods
may try to avoid creation of templates and parameterize the multi-dimensional likelihood function directly with 
certain approximations. We illustrated some of these methods in Refs.~\cite{Gao:2010qx,Anderson:2013afp} 
and a broader review may be found in Ref.~\cite{Brehmer:2019bvj}. However, the complexity of those methods 
also provides practical limitations on their application. 
We present some of the practical approaches in Section~\ref{sect:exp_onshell}. 

One popular example of the reduced set of bins of observables adopted for the study of the \Hboson kinematics is
the so-called Simplified Template Cross Section approach (STXS)~\cite{deFlorian:2016spz,Berger:2019wnu}.
The main focus at this stage~\cite{Berger:2019wnu} is on the three dominant \Hboson production processes, 
namely gluon fusion, VBF, and $VH$. These main production 
processes are subdivided into bins based on transverse momentum or mass of various objects, for example 
the \Hboson and associated jets. At future stages, the available information may be subdivided further. 
This approach became a strong framework for collaborative work of both theorists and experimentalists, 
as information from all LHC experiments and theoretical calculations can be combined and shared in an
efficient way. Nonetheless, as we illustrate below, this approach is still limited in its application 
for two important reasons.  First, the STXS measurements are based on the analysis of SM-like kinematics.
The measurement strategy may not be appropriate for interpretations appearing with new tensor structures or new virtual particles
(such as $\gamma^*$ in place of $Z^*$) unless a full detector simulation of such effects is performed.
Additionally, the binning of STXS may not be optimal for all the measurements of interest.

\subsection{Matrix element technique}
\label{sect:kin_oo}

The matrix element likelihood approach (MELA)~\cite{Gao:2010qx,Bolognesi:2012mm,Anderson:2013afp,Gritsan:2016hjl} 
was designed to extract all essential information from the complex kinematics of both production and decay of the \Hboson\ 
and retain it in the minimal set of observables. 
Two types of discriminants were defined for either the production or the decay process, and here we generalize 
it for any sequential process of both production and decay:
\begin{eqnarray}
\label{eq:melaAlt}
{\cal{D}_{\rm alt}}({\bf\Omega})=
\frac{{\cal P}_{\rm sig}({\bf\Omega})}{{\cal P}_{\rm sig}({\bf\Omega})+{\cal P}_{\rm alt}({\bf\Omega})}
\,, 
\end{eqnarray}
\begin{eqnarray}
{\cal{D}_{\rm int}}({\bf\Omega})=
\frac{{\cal P}_{\rm int}({\bf\Omega})}
{2\sqrt{{\cal P}_{\rm sig}({\bf\Omega})\times{\cal P}_{\rm alt}({\bf\Omega})}}
\,,
\label{eq:melaInt}
\end{eqnarray}
where ${\cal P}_{\rm sig}$, ${\cal P}_{\rm alt}$, and ${\cal P}_{\rm int}$ represent the probability distribution for 
a signal model of interest, an alternative model to be rejected (either background, a different production 
process of the \Hboson, or an alternative anomalous coupling of the \Hboson), and the interference contribution, 
which may in general be positive or negative. The probabilities are obtained from the matrix elements 
squared, calculated by the MELA library described in Section~\ref{sect:cp_mc}, and do not generally need to be normalized. 
The denominator in Eq.~(\ref{eq:melaInt}) is chosen to reduce correlation between the discriminants, 
but this choice is equivalent to that of Ref.~\cite{Anderson:2013afp}.
The above definition leads to the convenient arrangement $0\le\cal{D}_{\rm alt}\le$ 1 and $-1\le\cal{D}_{\rm int}\le$ 1. 

These discriminants retain all multidimentional correlations essential for the measurements of interest. 
For a simple discrimination of two hypotheses, the Neyman-Pearson lemma~\cite{Neyman:1933} guarantees that the ratio
of probabilities ${\cal P}$ for the two hypotheses provides optimal discrimination power. However, for a continuous set of
hypotheses with an arbitrary quantum-mechanical mixture several discriminants are required for an optimal measurement
of their relative contributions.
There are three interference discriminants when anomalous couplings appear both in production and in decay. 
Let us conside only real $g_1$ and $g_4$ couplings in Eq.~(\ref{eq:HVV}), which appear 
once in production and once in decay, as shown in Eq.~(\ref{eq:diff-cross-section2}).
The total amplitude squared would have five terms proportional to $(g_4/g_1)^m$ with $m=0,1,2,3,4$: 
%
\begin{equation}
\mathcal{P}^{\mathrm{}}\left({\bf\Omega}; g_1, g_4 \right) \\
\propto\sum_{m=0}^{4}
(g_4/g_1)^{m}  \times  \mathcal{P}^{\mathrm{}}_{m}\left({\bf\Omega} \right) \,,
\label{eq:poffshellACsimplified}
\end{equation}
where we absorb $g_1^4$ and the width into the overall normalization. 
Equation~(\ref{eq:melaAlt}) corresponds to the ratio of the $m=4$ and $m=0$ terms. Three other
ratios give rise to interference discriminants. The four discriminants may be re-arranged into two discriminants of the 
form in Eq.~(\ref{eq:melaAlt}) and two of the form in Eq.~(\ref{eq:melaInt}), in each case one observable defined 
purely for the production process and the other for the decay process. 
One could apply the Neyman-Pearson lemma to each pair of points in the parameter space of $(g_1, g_4)$,
but this would require a continuous, and therefore infinite, set of probability ratios. However, equivalent information is 
contained in a linear combination of only four probability ratios, which can be treated as four independent observables. 
Above the $2m_V$ threshold, there are also interference discriminants appearing due to interference between
the \offshell\ tail of the signal process and the background. 
A subset of equivalent optimal observables was also introduced independently in earlier work
on different topics~\cite{Atwood:1991ka,Davier:1992nw,Diehl:1993br}. 

The number of discriminants in Eqs.~(\ref{eq:melaAlt}, \ref{eq:melaInt}) is still limited if we consider just one
anomalous coupling. Nonetheless, this number grows quickly as we consider multiple anomalous couplings, 
especially the number of interference discriminants. A subset of these discriminants may contain most of the 
information, depending on the situation. For the near-term LHC measurements, the $\cal{D}_{\rm alt}$ using 
full production and decay information and $\cal{D}_{\rm int}$ using production information from correlation of
associated particles provide the most optimal information. 
In the very long term, the lowest powers of $m$ may provide most of the discriminating power 
when testing data for tiny anomalous contributions, because effects may be most visible in interference. 
Therefore, the MELA approach still allows us to select a limited set of the most optimal discriminants, 
as we illustrate with the practical applications below. 

Detector resolution effects may come into play in experimental analyses and
affect the calculations of the probabilities in Eqs.~(\ref{eq:melaAlt}, \ref{eq:melaInt}).
These can be parameterized with transfer functions. However, in most practical applications, the
``raw'' matrix element probabilities can be used, and the resulting performance degradation
is small when the distributions of the angular and mass observables are broad compared to the resolution.
One notable exception is the invariant mass
of relatively narrow resonances, such as the \Hboson mass in $H\to ZZ\to 4\ell$ or 
the $V$ boson mass in associated production $VH$. In such a case, the signal and background 
probabilities ${\cal P}_{\rm sig}$ and ${\cal P}_{\rm bkg}$ can incorporate the empirical 
invariant mass parameterization giving rise to the ${D_{\rm bkg}}$ discriminant for optimal 
background rejection. We find such an approach computationally effective without any visible
loss of performance. 
Nonetheless, below we also introduce machine learning to enhance the matrix element technique.
This approach incorporates matrix element knowledge combined with the parton shower and detector effects.

\subsection{Application of the matrix element technique to the boson fusion processes}
\label{sect:kin_vbf}

Let us illustrate the power of the matrix element technique in application to both weak boson fusion 
(VBF in the following) and strong boson fusion (ggH in the following), where at higher orders in QCD
the ggH process may include $gg, qg,$ and $qq$ initial states. 
This illustration is similar to our earlier study in Ref.~\cite{Anderson:2013afp}, but we would like to 
expand this illustration in several directions. The boson fusion process is particularly important in the 
off-shell region, which is a new feature of this work. However, most kinematic considerations apply equally 
to both the on-shell and off-shell regions. 
In the weak boson fusion, for illustration purposes we consider equal strength of $WW$ and $ZZ$ fusion, with 
$g_1^{ZZ}=g_1^{WW}$ and $g_4^{ZZ}=g_4^{WW}$ in Eq.~(\ref{eq:HVV})
and vary the relative contribution of the CP-even and CP-odd amplitudes, 
with the $f_{g4}^{\rm VBF}$ parameter representing their relative cross section fraction. 
The relative strength of $WW$ and $ZZ$ fusion is fixed in this study because the two processes 
are essentially indistinguishable in their observed kinematics, as shown in Fig.~\ref{fig:kinematics_wbf}.
In strong boson fusion, the parameter $f_{g4}^{ggH}$ represents a similar 
relative cross section fraction of the pseudoscalar coupling component. 

\begin{figure}[t]
\centering
\centerline{
\includegraphics[width=0.33\textwidth]{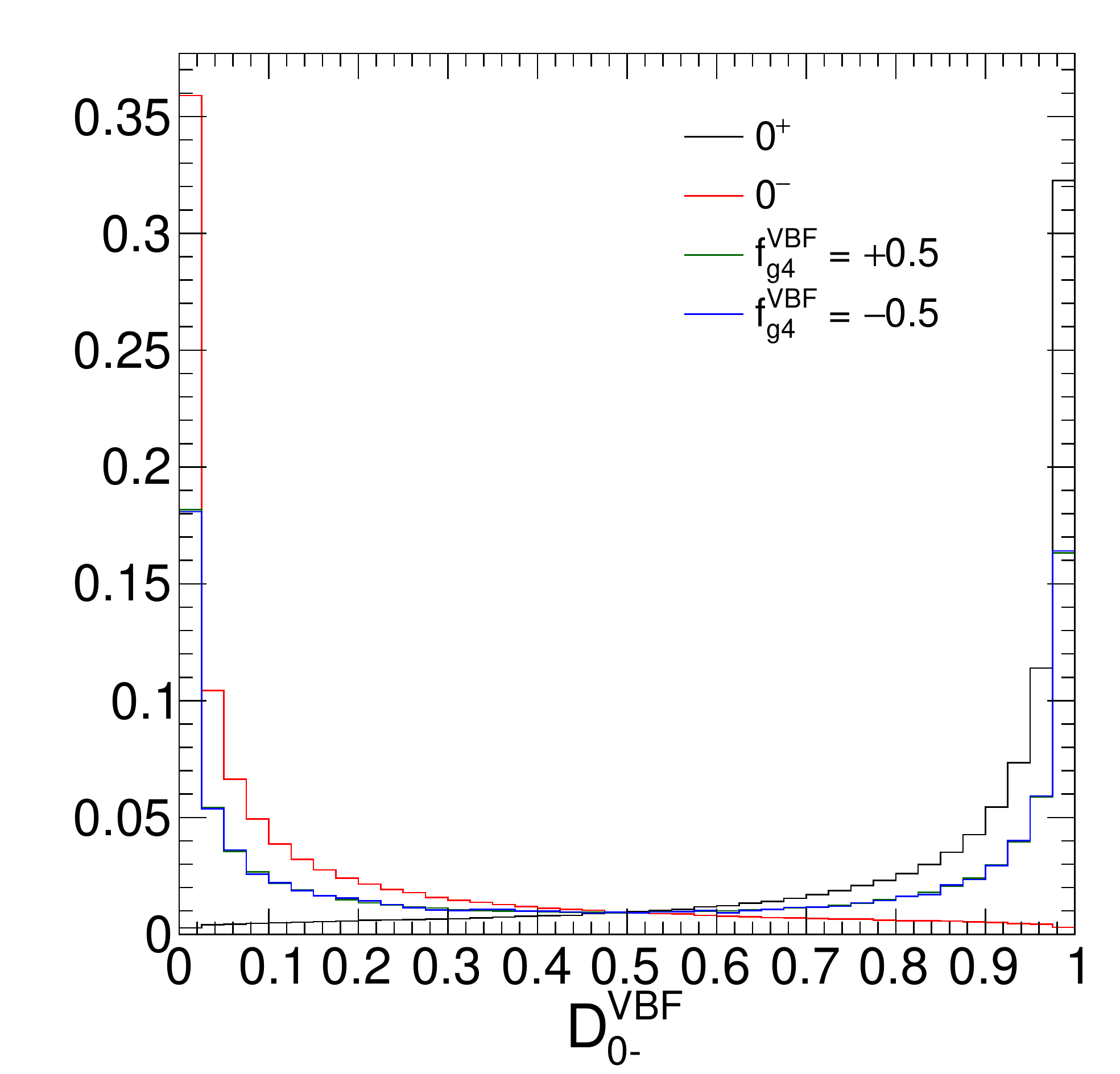}
\includegraphics[width=0.33\textwidth]{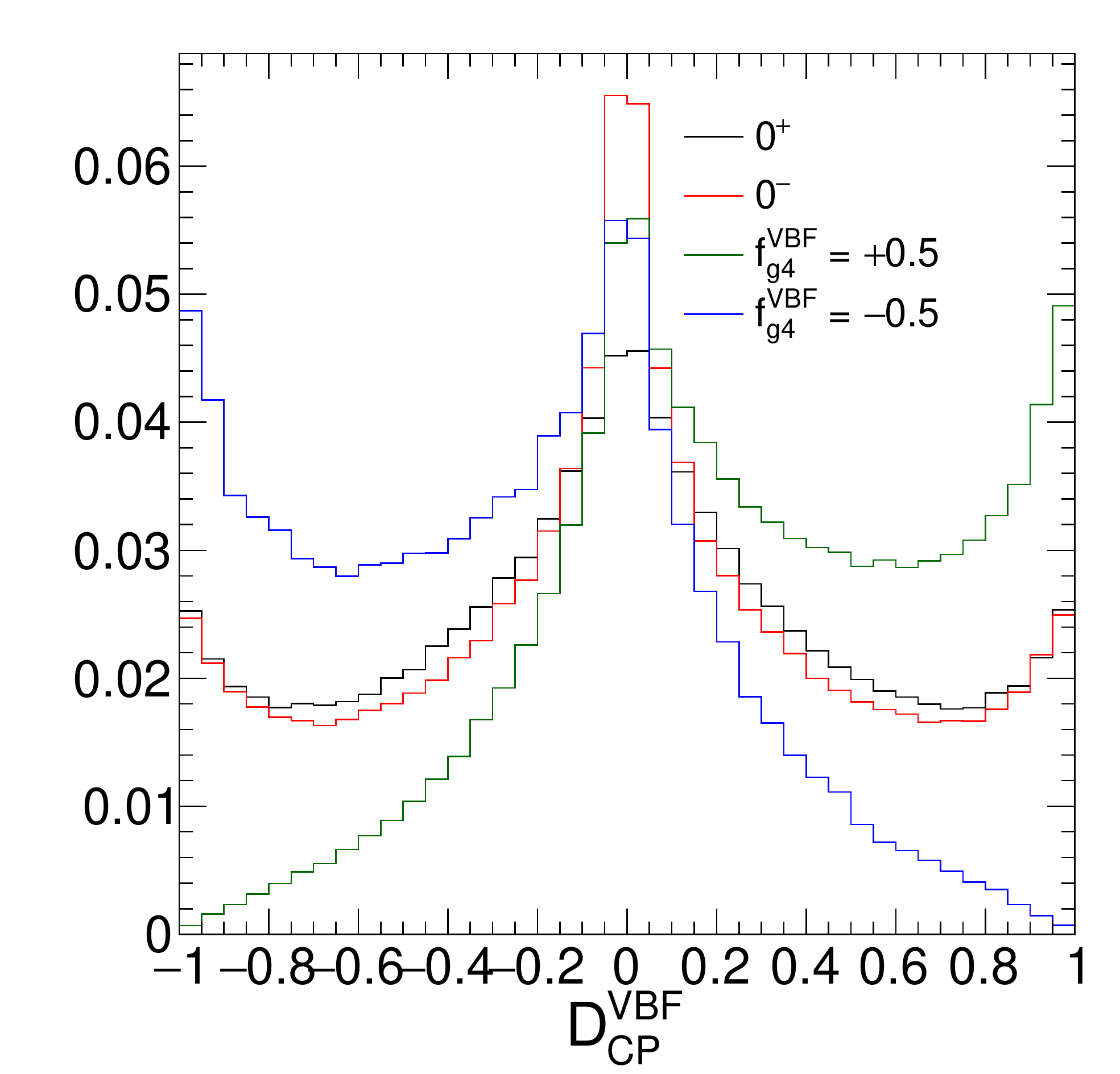}
\includegraphics[width=0.33\textwidth]{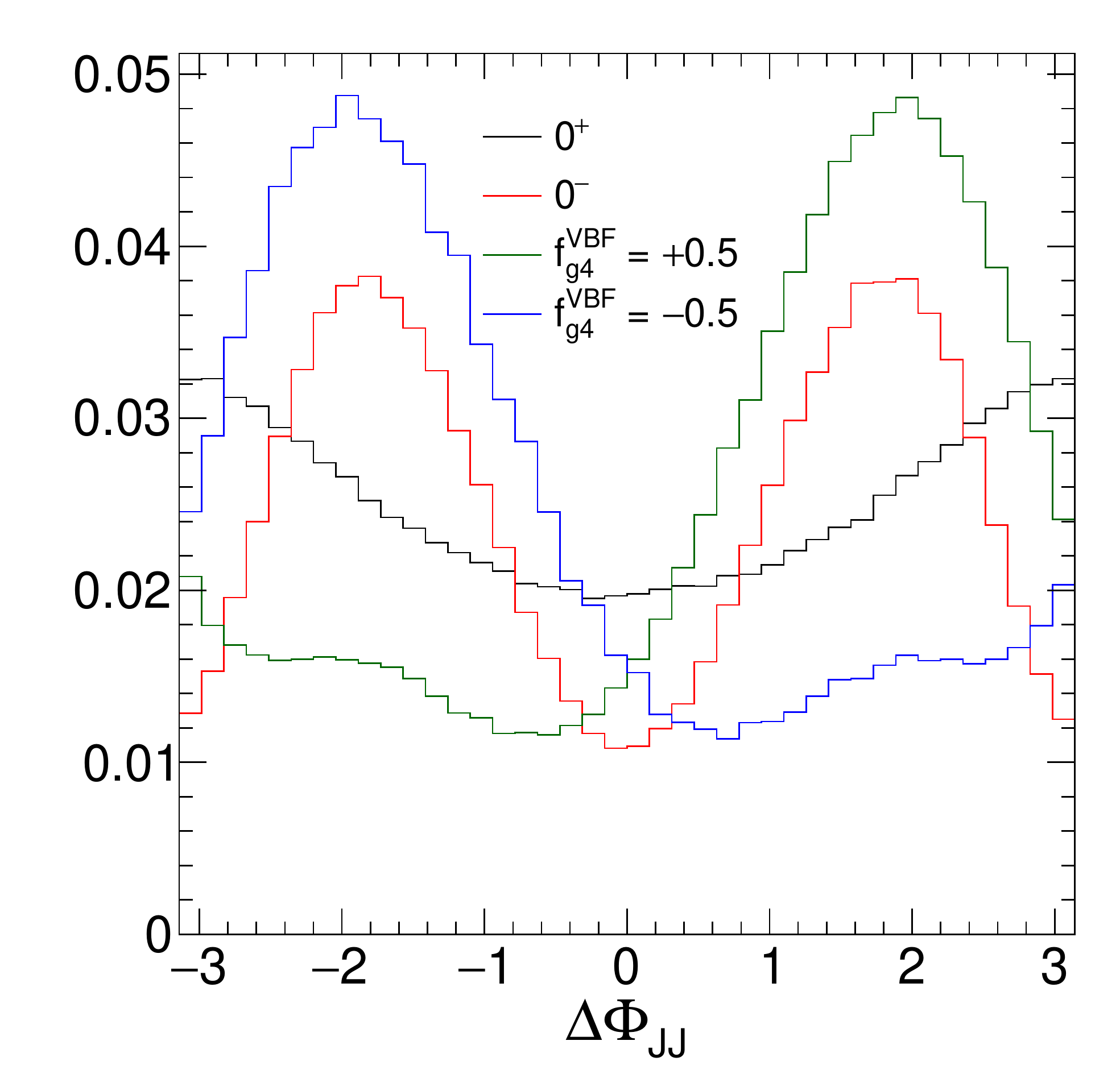}
}
\captionsetup{justification=centerlast}
\caption{
Two discriminants defined in Eq.~(\ref{eq:melaAlt}) (left) and Eq.~(\ref{eq:melaInt}) (middle) for the 
measurement of the CP-sensitive parameter $f_{g4}^{\rm VBF}$ in VBF production. 
Also shown is the $\Delta\Phi_{JJ}$ observable (right).
The values of  $f_{g4}^{\rm VBF}=\pm0.5$ correspond to 50\% mixtures of the CP-even
and CP-odd contributions.
}
\label{fig:fa3vbf}
\end{figure}

\begin{figure}[t]
\centering
\centerline{
\includegraphics[width=0.33\textwidth]{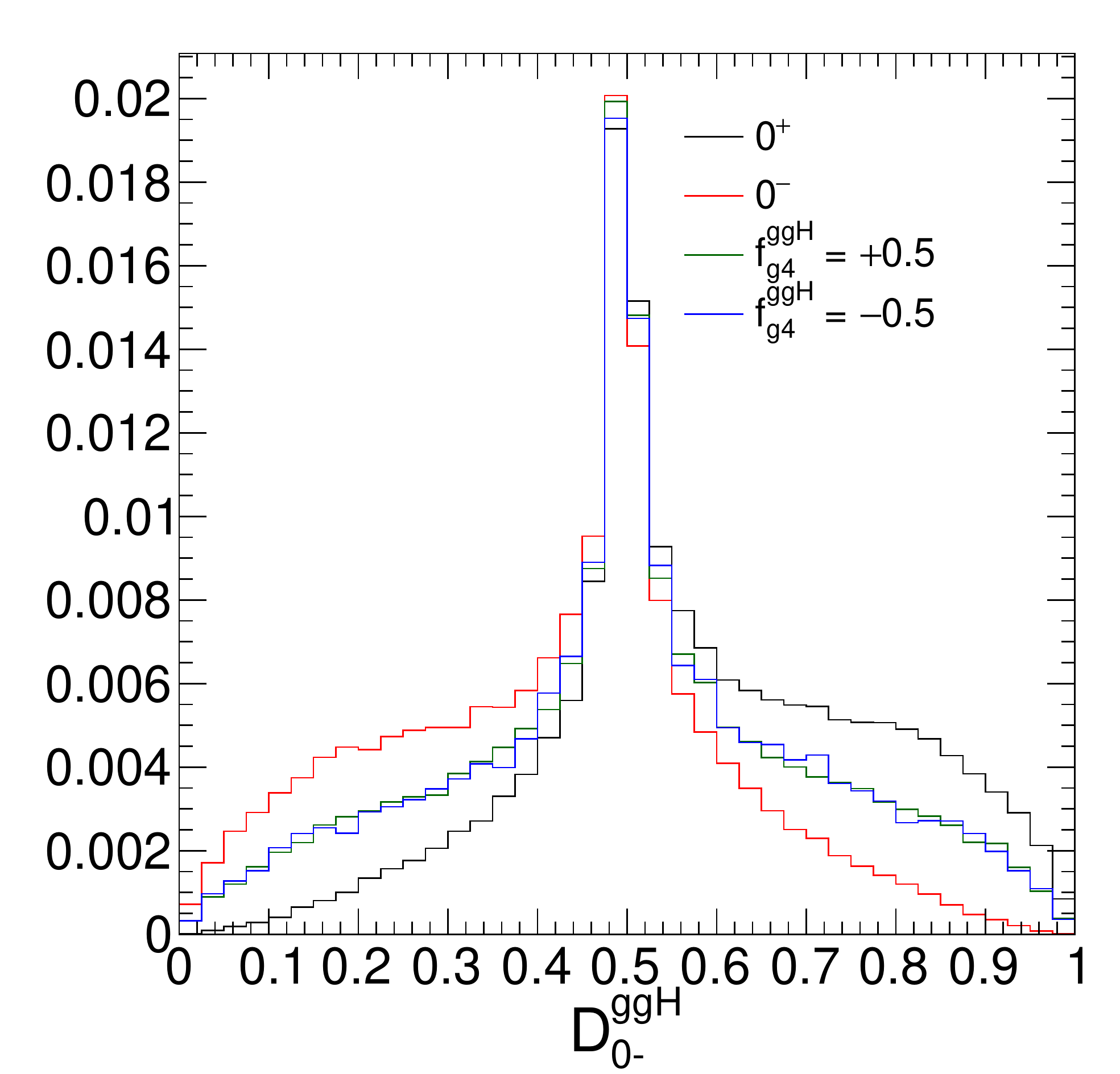}
\includegraphics[width=0.33\textwidth]{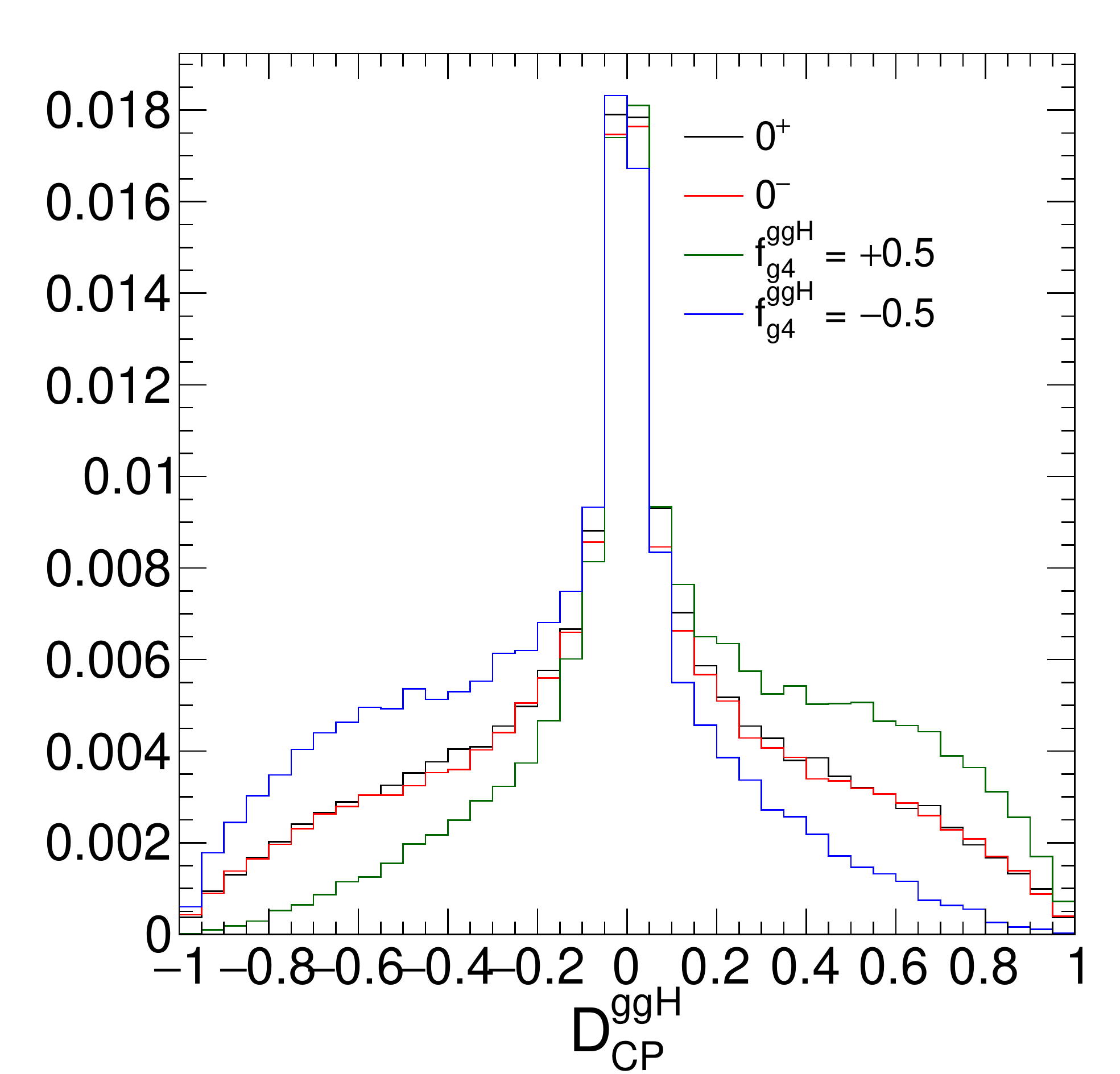}
\includegraphics[width=0.33\textwidth]{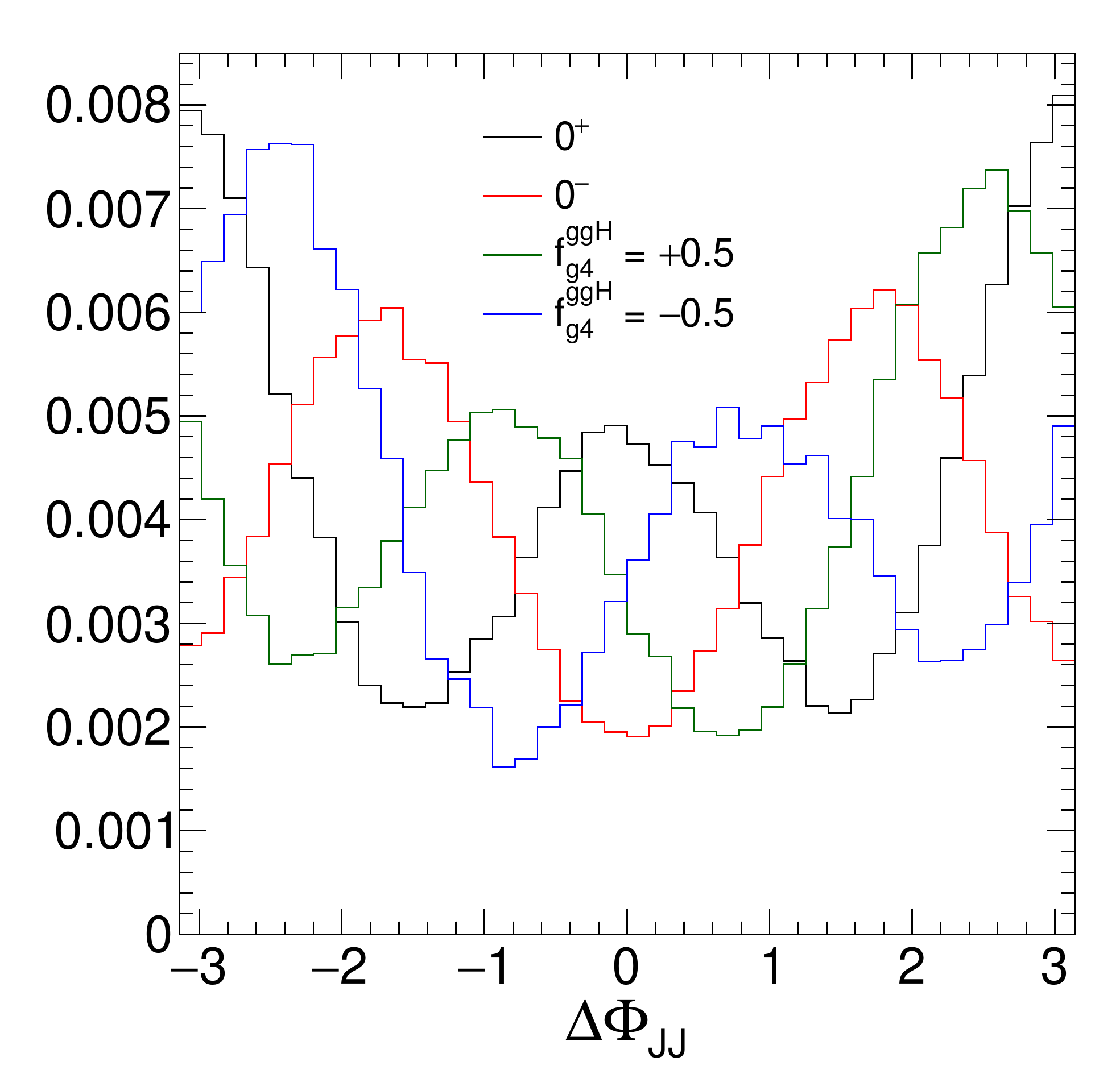}
}
\captionsetup{justification=centerlast}
\caption{
Two discriminants defined in Eq.~(\ref{eq:melaAlt}) (left) and Eq.~(\ref{eq:melaInt}) (middle) for the 
measurement of the CP-sensitive parameter $f_{g4}^{\rm ggH}$ in ggH production. 
Also shown is the $\Delta\Phi_{JJ}$ observable (right).
The values of  $f_{g4}^{\rm ggH}=\pm0.5$ correspond to 50\% mixtures of the CP-even
and CP-odd contributions.
A requirement $m_{JJ}>300$ GeV is applied to enhance the VBF-like topology of events. 
 }
\label{fig:fa3ggH}
\end{figure}

Figure~\ref{fig:fa3vbf} shows the $\mathcal{D}_{0-}$ and ${\cal D}_{\rm CP}$ discriminants,
calculated according to Eqs.~(\ref{eq:melaAlt}) and (\ref{eq:melaInt}), for the 
VBF process, to distinguish between the SM hypothesis $g_1^{ZZ}=g_1^{WW}=2$, the alternative 
hypothesis $g_4^{ZZ}=g_4^{WW}\ne0$, and the interference between these two contributions.
Figure~\ref{fig:fa3ggH} shows the same type of discriminants defined and shown for the ggH process,
enhanced with the events in the VBF-like topology using the requirement $m_{JJ}>300$ GeV
for illustration. In both cases, information from the \Hboson\ and the two associated jets, as illustrated 
in Fig.~\ref{fig:kinematics}, is used in the discriminant calculation. 
The $m_{JJ}$ requirement is based on the following observation. Among the initial states
in the ggH process, we could have $gg, qg,$ and $qq$ parton pairs. The events with the $qq$ 
initial state carry most of the information for CP measurements and have the topology most similar
to the  VBF process, which is also known to have a large di-jet invariant mass. In Section~\ref{sect:exp_onshell},
we will use this feature when developing the analysis techniques, but with the matrix-element technique
applied to isolate the VBF-like topology. 
In both the VBF and ggH cases, the azimuthal angle difference between the two jets $\Delta\Phi_{JJ}$ is also shown 
for comparison~\cite{Plehn:2001nj}. It is similar to the $\Phi^{\rm VBF}$ angle defined in Fig.~\ref{fig:kinematics} 
and shown in Fig.~\ref{fig:kinematics_wbf}, but differs somewhat because it is calculated in a different frame. 

The $\Delta\Phi_{JJ}$ angle is defined as follows. The direction of the two jets is represented by the vectors \(\vec{j}_{1,2}\) 
in the laboratory frame, and $\vec{j}_{T1,2}$ are the transverse components in the $xy$ plane. 
If we label $j_1$ as the jet going in the $-z$ direction (or less forward) and $j_2$ as the jet going in the $+z$ direction 
(or more forward), then $\Delta\Phi_{JJ}$ is the azimuthal angle difference between the first and the second jets, 
or $\phi_1-\phi_2$. In vector notation,
\begin{eqnarray}
\Delta\Phi_{JJ} =
\frac{(\hat{j}_{T1}\times\hat{j}_{T2})\cdot\hat{z}}{|(\hat{j}_{T1}\times\hat{j}_{T2})\cdot\hat{z}|}
\cdot
\frac{(\vec{j}_1-\vec{j}_2)\cdot\hat{z}}{|(\vec{j}_1-\vec{j}_2)\cdot\hat{z}|}
\cdot
\cos^{-1}\left( \hat{j}_{T1}\cdot\hat{j}_{T2}  \right) \,,
\label{eq:dPhiJJ}
\end{eqnarray}
where the angle between $\vec{j}_{T1}$ and $\vec{j}_{T2}$ defines $\Delta\Phi_{JJ}$
and the two ratios provide the sign convention.  This definition is invariant under the exchange of 
the two jets and the choice of the positive $z$ axis direction. 

The information content of the observables can be illustrated with the Receiver Operating Characteristic (ROC) curve, 
which is a graphical plot that illustrates the diagnostic ability of a binary classifier system as its discrimination threshold is varied.
Figure~\ref{fig:fa3ROC} (left) shows the ROC curves illustrating discrimination between scalar and pseudoscalar models
in the VBF process using the ${\cal D}_{0-}$ and $\Delta\Phi_{JJ}$ observables. The 
optimal observable ${\cal D}_{0-}$, which incorporates all kinematic and dynamic information, has the clear advantage.
Figure~\ref{fig:fa3ROC} (right) shows the same comparison in the ggH process. 
The gain in using the optimal observable in the ggH process is not as large as in VBF because of the smaller differences 
in dynamics of the scalar and pseudoscalar models, as both are generated by higher-dimension operators with the
same powers of $q_i^2$ in Eq.~(\ref{eq:HVV}). While the ${\cal D}_{0-}$ observable incorporates all kinematic and dynamic
information, the truly CP-sensitive observable ${\cal D}_{\rm CP}$ does not rely on dynamics.  It provides optimal
separation between the models with maximal mixing of the CP-even and CP-odd contributions and
opposite phases.  We illustrate this in Fig.~\ref{fig:fa3ROC} (middle) with a ROC curve for discrimination between 
the $f_{g4}=\pm0.5$ models in the VBF process.

\subsection{Matrix element technique with machine learning}
\label{sect:kin_ml}

The discriminants calculated with the matrix elements directly, as discussed in Section~\ref{sect:kin_oo}, 
are powerful tools in the analysis of experimental data. Most importantly, they provide scientific insight into the
problem under study. Nonetheless, there could be practical considerations limiting their 
application in certain cases. For example, events with partial reconstruction would require integration 
over unobserved degrees of freedom. Substantial detector effects or incorrect particle assignment 
in reconstructed events may lead to poor experimental resolution and would require modeling
with transfer functions. All of these effects can be taken into account, but may make calculations inefficient or impractical. 
Here we provide a practical prescription for overcoming these complications with the help
of machine learning, while still retaining the functionality of the optimal matrix-element approach. 
We achieve this by constructing the training samples and the observables used according to the
matrix-element approach.

\begin{figure}[t]
\centering
\centerline{
\includegraphics[width=0.33\textwidth]{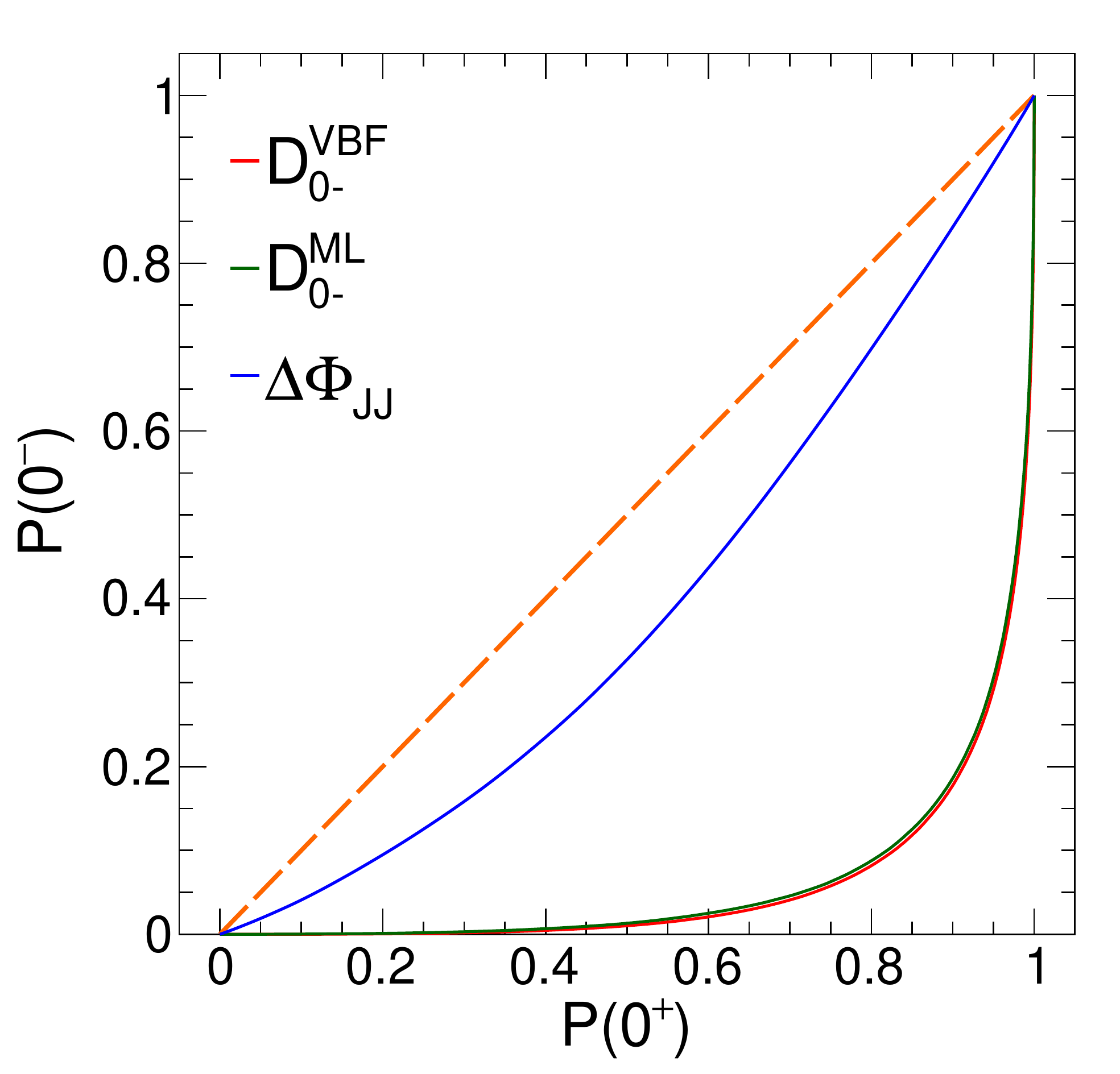}
\includegraphics[width=0.33\textwidth]{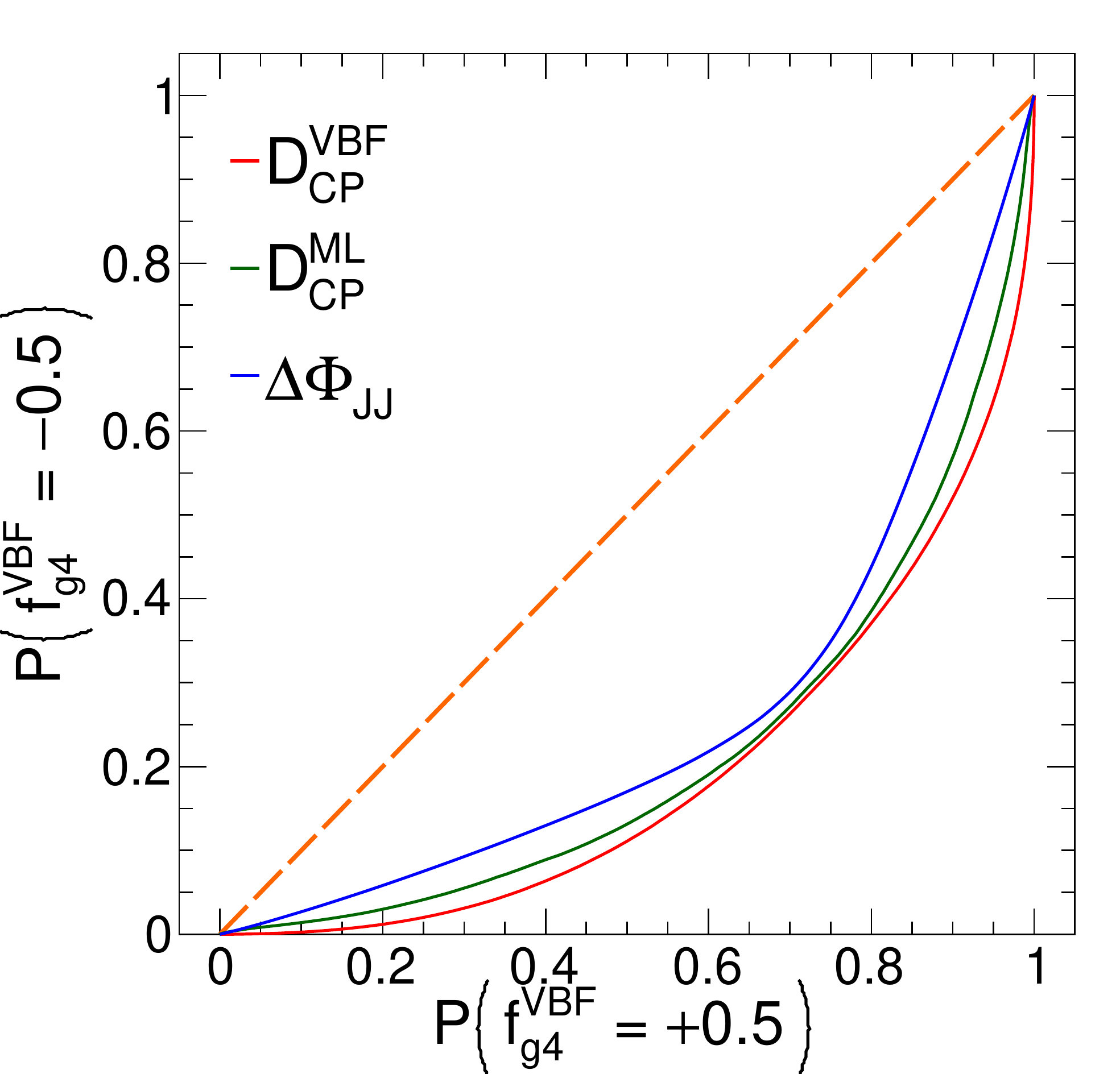}
\includegraphics[width=0.33\textwidth]{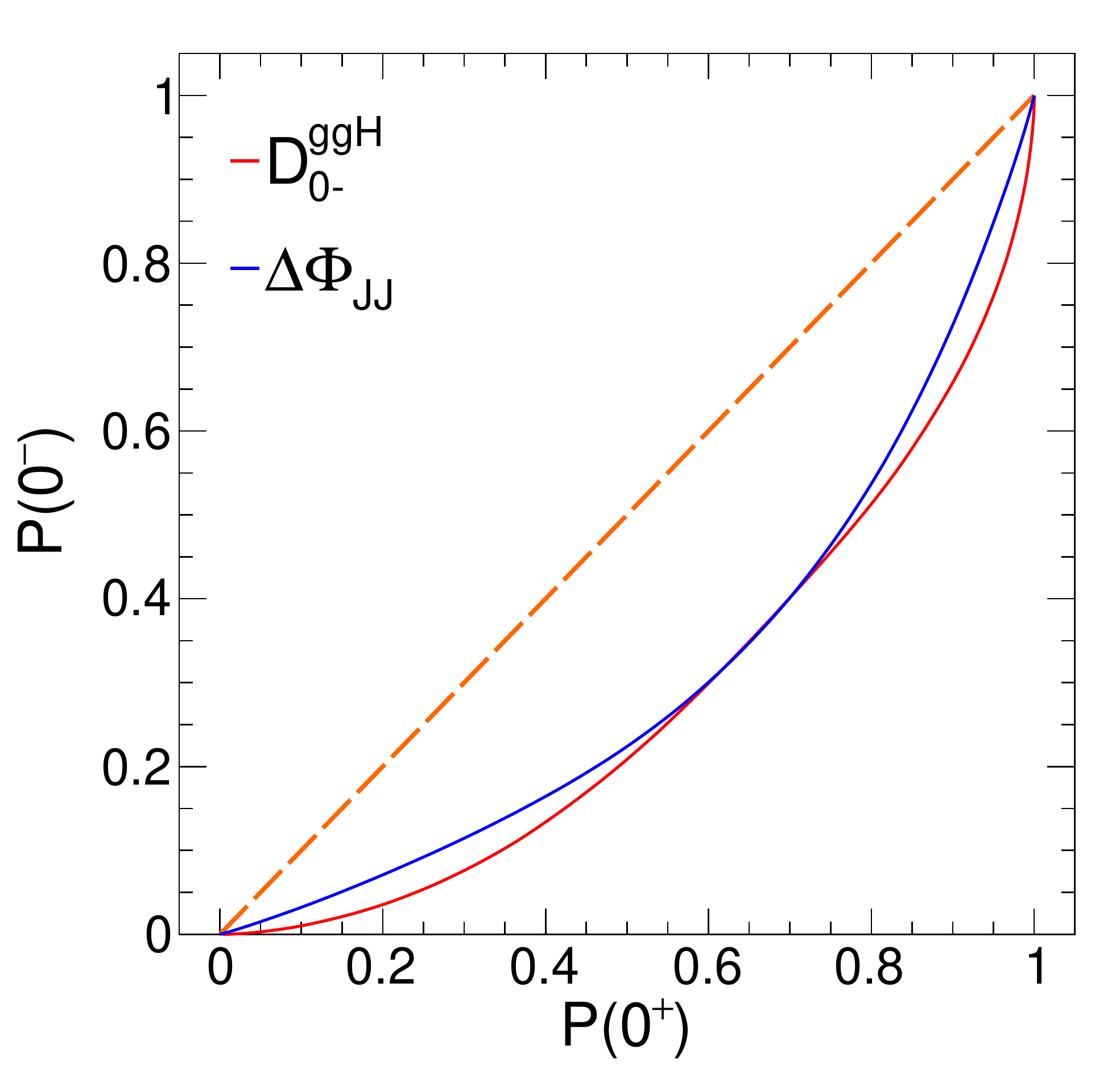}
}
\captionsetup{justification=centerlast}
\caption{
Left: a ROC curve showing the separation power between the scalar (SM-like $0^+$) and pseudoscalar ($0^-$) models 
in the VBF process using the ${\cal D}_{0-}$ and $\Delta\Phi_{JJ}$ observables. The diagonal dashed line 
shows the hypothetical no-separation scenario. The points represent the efficiency of selecting each model 
as the threshold of selection is varied. 
Right: same as the left plot, but for the ggH process, with a requirement $m_{JJ}>300$ GeV applied to enhance the VBF-like topology of events.
Middle: a ROC curve showing the separation power between the $f_{g4}^{\rm VBF}=+0.5$  and $f_{g4}^{\rm VBF}=-0.5$  models
in the VBF process using the ${\cal D}_{\rm CP}$ and $\Delta\Phi_{JJ}$ observables.
Also shown on the left and middle plots are the ROC curves representing performance of the optimal observables 
obtained with machine learning techniques. 
}
\label{fig:fa3ROC}
\end{figure}

Machine learning is a popular approach to data analysis, especially with the growing computational   
power of computers. The problem of differentiating between two models, as in Eq.~(\ref{eq:melaAlt}),
becomes a trivial task with supervised learning, where two samples of events with the signal and alternative models
are provided as input for training. One key aspect where the matrix element approach provides the insight is the
set of input observables ${\bf\Omega}$. As long as the complete set of observables, sufficient for the matrix element
calculations, is provided to the machine learning algorithm, the outcome of proper training is guaranteed to be 
a discriminant optimal for this task, equivalent to that in Eq.~(\ref{eq:melaAlt}). 
We illustrate this with such a discriminant ${\cal D}_{0-}^{\rm ML}$ in Fig.~\ref{fig:fa3ROC} (left) in application 
to the VBF process, using the Boosted Decision Tree implementation from Ref.~\cite{Hocker:2007ht}.
 
Application of the machine learning approach to the discriminant in Eq.~(\ref{eq:melaInt}) is less obvious, 
because it requires knowledge of quantum mechanics to isolate the interference component. Nonetheless, 
we provide a prescription for obtaining such a discriminant. 
A discriminant trained to differentiate the models with maximal quantum-mechanical mixing of the signal 
and alternative contributions with opposite phases becomes a machine-learning  equivalent to that 
in Eq.~(\ref{eq:melaInt}), following the discussion in Section~\ref{sect:kin_vbf}.
The complete kinematic information ${\bf\Omega}$ of the event should be provided to training. 
We illustrate this approach with such a discriminant ${\cal D}_{\rm CP}^{\rm ML}$
in Fig.~\ref{fig:fa3ROC} (middle) in application to the VBF process. 
There is a small degradation in performance of the ${\cal D}_{\rm CP}^{\rm ML}$ discriminant with respect
to the matrix element calculation, but this is attributed to the 
more challenging task of training in this case and should be recovered in the limit of perfect training. 

To summarize, the matrix element technique, expressed in Eqs.~(\ref{eq:melaAlt}) and~(\ref{eq:melaInt}), 
can be expanded with the help of machine learning with two important ingredients: (1) the complete set of 
matrix-element input observables ${\bf\Omega}$, or equivalent, has to be used, and 
(2) the machine learning process should be based 
on the carefully prepared samples according to the models discussed above. The machine learning approach 
is still based on the matrix element calculations, as the training samples are generated based on the same 
matrix elements as the discriminants in Eqs.~(\ref{eq:melaAlt}) and~(\ref{eq:melaInt}). 

\section{Application to on-shell H(125) boson production}
\label{sect:exp_onshell}

We start by investigating the on-shell production and decay of the \Hboson with its coupling to either
weak or strong vector bosons in the VBF and ggH processes. 
There has already been extensive study of the $HVV$ couplings, 
and the current challenge is in the measurement of multiple possible anomalous contributions. 
On the other hand, there have been limited studies of the anomalous $Hgg$ couplings, due to 
lower statistical precision at this time. The latter could be interpreted as both an effective coupling 
to gluons, or as a coupling to quarks in the gluon fusion loop. 
Prospects of both $HVV$ and $Hgg$ studies with either 3000 $fb^{-1}$ (HL-LHC) or 300 $fb^{-1}$ (full LHC) 
are presented below. Let us first discuss some general features in analysis of LHC data. 

For the $HVV$ studies, we will use the example of the $H\to VV\to4\ell$ decay and VBF, $VH$, or ggH production.
Equation~(\ref{eq:HVV}) defines several anomalous couplings, which we generically denote as $g_i^{VV}$.
All of these processes include the interference of several $VV$ intermediate states, 
such as ${VV}=ZZ, Z\gamma, \gamma\gamma, WW$.
In the analysis of the data (MC simulation in our case), a likelihood fit is performed~\cite{Verkerke:2003ir,Brun:1997pa}.
The probability density function for a given signal process, before proper normalization, 
is defined for the two possible numbers of couplings $N$ in the product:
\begin{eqnarray}
N=4: ~~~ &&
\mathcal{P} \left({\bf x}; \vec{f} \,\right) 
\propto\sum_{\substack{k,l,m,n=1\\k\le l\le m\le n}}^K
\mathcal{P}_{klmn}\left({\bf x}\right)
\sqrt{|f_{gk} \cdot f_{gl}\cdot f_{gm}\cdot f_{gn}|} ~\mathrm{sign}(f_{gk} \cdot f_{gl}\cdot f_{gm}\cdot f_{gn})\,,
\label{eq:psignal}
\\
N=2: ~~~ &&
\mathcal{P} \left({\bf x}; \vec{f} \,\right) 
\propto\sum_{\substack{k,l=1\\k\le l}}^K
\mathcal{P}_{kl}\left({\bf x}\right)
\sqrt{|f_{gk} \cdot f_{gl}|} ~\mathrm{sign}(f_{gk} \cdot f_{gl})\,,
\label{eq:psignalshorter}
\end{eqnarray}
where  ${\bf x}$ are the observables, but not necessarily the complete set ${\bf\Omega}$,
and $f_{gn}$ are $K$ terms corresponding to the cross-section fractions of the couplings,
defined in Eq.~(\ref{eq:fgn}).
Equations~(\ref{eq:psignal}) and~(\ref{eq:psignalshorter}) are obtained from Eq.~(\ref{eq:diff-cross-section2})
and using Eq.~(\ref{eq:an}), where the width and $f_{g1}$ are absorbed into the overall normalization. 
In the case of the electroweak process, the $HVV$ coupling appears on both the production and the decay sides.
As a result, the amplitude squared has a product of $N=4$ couplings. 
In the gluon fusion production, on the other hand, the electroweak $HVV$ couplings appear only in decay, 
and therefore $N=2$. Similarly, if one considers the $Hgg$ coupling on production, $N=2$. 

There are $(N+K-1)!/(N!(K-1)!)$ terms in either Eq.~(\ref{eq:psignal}) or Eq.~(\ref{eq:psignalshorter}).
As we explain in Section~\ref{HVV_onshell}, we consider $K=5$ in our analysis of four anomalous $HVV$ couplings.
Therefore, in the case of electroweak production ($N=4$, $K=5$), we have to deal with 70 terms. 
If we were to consider $K=13$ independent couplings in Eq.~(\ref{eq:HVV}), we would formally have to deal 
with 1820 terms describing production and decay (the actual number would be somewhat smaller because 
not all terms contribute to a given decay mode). 
While such analysis of 1820 terms is in principle feasible, at the current stage it is not practical. 
In the case of gluon fusion, there are 15 terms for $HVV$ couplings ($N=2$, $K=5$)
and 3 terms for $Hgg$ couplings ($N=2$, $K=2$). 
If both sets of anomalous couplings are considered simultaneously,
the total number of terms is the product of these, that is 45.

In the simplified analysis of LHC data, using simulation of $pp$ collisions at 13 TeV, we adopt the 
following approach. We take the analysis of the $H\to VV\to4\ell$ channel as the most interesting for illustration, 
because both production and decay information can be used.  All production modes of the \Hboson 
are included in this study and are generated with the JHU generator as discussed in Section~\ref{sect:cp_mc}. 
The JHU generator framework is also used to generate gluon fusion and electroweak background production 
of the $VV\to4\ell$ final states. The dominant $q\bar{q}\to VV\to4\ell$ background process is generated with 
POWHEG~\cite{powhegvv} and scaled to cover for other possible background contributions not modeled
otherwise~\cite{CMS-HIG-19-001}. 
All events are passed through Pythia~8~\cite{Sjostrand:2014zea} for parton shower simulation. 
The detector effects are modeled with ad-hoc acceptance selection, and the lepton and 
hadronic jet momenta are smeared to achieve realistic resolution effects. 
Going beyond the  $H\to4\ell$ channel,
inclusion of the $H\to\gamma\gamma$, $H\to\tau\tau$, and $H\to bb$ channels 
might increase the dataset by about an order of magnitude, but only for analysis of the production information. 
In addition, analysis of the $H\to WW\to2\ell2\nu$ decay may bring some information on the decay side, 
but not exceeding that from the $H\to 4\ell$ case. While we focus on the $H\to 4\ell$ channel, we comment 
on improvements which will be achieved with a combination of the above channels.

\subsection{HVV anomalous couplings}
\label{HVV_onshell}

In order to illustrate the power of the matrix element techniques and the analysis tools discussed above, 
let us consider the $HVV$ coupling of the \Hboson to two weak vector bosons 
using the $H\to 4\ell$ decay,  with vector boson fusion, associated production with the vector bosons 
$W$ and $Z$, or inclusive production, and using both \onshell\ and \offshell\ production. 
Some of these techniques have already been applied in analyses of LHC 
data~\cite{Sirunyan:2017tqd,Sirunyan:2019twz,Sirunyan:2019nbs}.
However, the rich kinematics in production and decay of the \Hboson represents particular
challenges in analysis. 

There are 13 independent $HVV$ anomalous couplings in Eq.~(\ref{eq:HVV}).  An optimal simultaneous 
measurement of all these couplings, or even a sizable subset, represents a practical challenge in data analysis 
and, as far as we know, has not been attempted experimentally yet. Here we stress that an optimal measurement 
means that the precision of any given parameter measurement is not degraded when comparing a multi-parameter 
approach with all other couplings constrained and an optimal single-parameter measurement discussed below. 
Several approaches have been adopted. 
In one approach, a small number of couplings, typically two or at most three, is considered. 
One of these is the SM-like coupling and the other could be parameterized with the cross-section fraction 
$f_{gi}$ defined above. 
While this approach is optimal for each parameter measurement, 
the problem with this approach is that correlations between measurements 
of different anomalous couplings are not considered. 

Another recently adopted approach is the STXS measurement, where cross sections of several 
\Hboson production processes are measured in several bins based on kinematics of the event. 
While this approach is attractive due to its applicability to a number of various use cases, 
the problems with this approach are that observables are not necessarily 
optimal for any given measurement, and that the kinematics of events are assumed to follow the SM when measuring
the cross section in each bin. For a correct measurement, a full detector simulation of each coupling scenario is needed, 
because the kinematics of associated particles and decay products would affect the measurement in each bin.
The STXS approach based on SM-only kinematics does not include these effects.
The latter effect is especially important because neglecting it may lead to biases in the measurements. 
In the following, we illustrate the strengths and weaknesses of each approach,
and propose a practical method based on the matrix element approach. 

First, we would like to note that it is difficult to perform an unambiguous measurement of all
13 independent $HVV$ anomalous couplings in Eq.~(\ref{eq:HVV}) in a given process. For example, 
while all these couplings contribute to the VBF production, kinematics of $WW$ and $ZZ$ fusion 
are essentially identical, as shown in Fig.~\ref{fig:kinematics_wbf}. The measurement becomes feasible
when the $WW$ and $ZZ$ couplings are related. We adopt two examples of this relationship. 
In one case, we simply set $g_i^{WW} = g_i^{ZZ}$, which could be interpreted as relationships in 
Eqs.~(\ref{eq:deltaMW}--\ref{eq:kappa1WW}) under the $c_w=1$ condition. Such results could be
re-interpreted for a different relationship of the couplings.
In the second case, we adopt the relationships in Eqs.~(\ref{eq:deltaMW}--\ref{eq:kappa2Zgamma})
without any conditions. With such a simplification, we are still left with nine parameters in the first case 
and eight parameters in the second case. To simplify the analysis further, 
we reduce the number of free parameters by setting 
$g_2^{\gamma\gamma}=g_4^{\gamma\gamma}=g_2^{Z\gamma}=g_4^{Z\gamma}=0$. 
While we do expect to observe non-zero values of $g_2^{\gamma\gamma}$ and $g_2^{Z\gamma}$
even in the SM, constraints on all four couplings are possible from decays $H\to \gamma\gamma$ 
and $Z\gamma$ with on-shell photons. We leave the exercise to include all couplings in an optimal 
analysis to future studies. In addition, we keep the $g_2^{gg}$ and $g_4^{gg}$ couplings as two 
free parameters as well. While the dedicated studies of these couplings are presented in Section~\ref{ggHJJ},
kinematics of the ggH process may affect measurements in the VBF process. 

As a reference, we take the STXS stage-1.1 binning as applied by the CMS 
experiment~\cite{CMS-HIG-19-001, deFlorian:2016spz}. In this approach, seven event categories are defined, 
which are optimal  for separating the VBF topology with two associated jets; two $VH$ categories, with leptonic and 
hadronic decay of the $V$, respectively; the VBF topology with one associated jet; two $\ttH$ categories, with 
leptonic and fully hadronic top decay, respectively; and the untagged category, which includes the rest of the events.
We call it stage-0 categorization. 
Each category of events is further split into sub-categories to match the requirements on the transverse 
momenta and invariant masses, as defined in the STXS stage-1.1 binning. In total, there are 22 categories 
defined~\cite{CMS-HIG-19-001}. 
While the above STXS stage-1.1 categorization provides fine binning for capturing some kinematic features
in production of the \Hboson, it does not keep any information from decay, it has no information sensitive 
to $CP$ violation, and more generally, it is not guaranteed to be optimal for measuring any of the 
parameters of our interest. 

Since we target the optimal analysis of four anomalous couplings expressed
through\footnote{There is an additional factor of $(-1)$ in the definition of $f_{\Lambda1}$ and $f_{\Lambda1}^{Z\gamma}$
following the convention in experimental measurements~\cite{Sirunyan:2019twz}.}
$f_{g4}$, $f_{g2}$, $f_{\Lambda1}$,  and $f_{\Lambda1}^{Z\gamma}$, we build the analysis in the following way. 
Instead of STXS stage-1.1 binning, we start from the seven categories
defined in stage-0 for isolating different event topologies. Since in this analysis we do not target fermion 
couplings\footnote{For a study of fermion couplings with this technique, see Ref.~\cite{Gritsan:2016hjl}.}, 
the two $\ttH$ categories are merged with the untagged category.
There are four discriminants relevant for this analysis, as defined by Eq.~(\ref{eq:melaAlt}): 
${\cal D}_{g4}$, ${\cal D}_{g2}$, ${\cal D}_{\Lambda1}$, and ${\cal D}_{\Lambda1}^{Z\gamma}$.
In addition, two interference discriminants, ${\cal D}_{\rm CP}$ and ${\cal D}_{\rm int}$, are defined by Eq.~(\ref{eq:melaInt}) 
for the $g_4$ and $g_2$ couplings, respectively. The two other interference discriminants are found to 
provide little additional information due to large correlations with the discriminants defined in Eq.~(\ref{eq:melaAlt}). 
The full available information is used in calculating the discriminants in the following way.
In the untagged category, the $H\to VV\to4\ell$ information is used.
In addition, the transverse momentum of the \Hboson is included, because it is sensitive to production. 
In both the VBF and $VH$ topologies with two associated jets, both production and decay information are used,
except for the two interference discriminants, where production information is chosen because it dominates. 
In the leptonic $VH$ category and the VBF topology with one associated jet, where information is in general missing,
the transverse momentum of the \Hboson is used, with finer binning than in the untagged category.
In the end, for each event in a category $j$ a set of observables ${\bf x}$ is defined. 

To parameterize the 70 terms in Eq.~(\ref{eq:psignal}) or the 15 terms in Eq.~(\ref{eq:psignalshorter}),
we rely on samples generated with JHUGen.  However, it is not necessary to generate 70 or 15 separate samples.
Instead, we generate a few samples that adequately cover the phase space and re-weight those samples using the
MELA package to parameterize the other terms.
To populate the probability distributions, we use a simulation of unweighted events with detector modeling,
and small statistical fluctuations are inevitable.
A critical step in the process is to ensure that even with these statistical fluctuations, the probability
density function $\mathcal{P}$, defined in Eqs.~(\ref{eq:psignal}) and (\ref{eq:psignalshorter}), remains positive
for all possible values of $\vec{f}$.  We detect negative probability by minimizing $\mathcal{P}$, which is a
polynomial in $\sqrt{|f_{gi}|}\cdot\mathrm{sign}(f_{gi})$. In the case of Eq.~(\ref{eq:psignal}), where the polynomial is quartic, we use
the Hom4PS program~\cite{HomotopyContinuation,Hom4PS1,Hom4PS2} to accomplish this minimization.  If negative
probability is possible, we modify $\mathcal{P}_{klmn}$ or $\mathcal{P}_{kl}$ using the cutting planes
algorithm~\cite{cuttingplanes}, using the Gurobi program~\cite{gurobi} in each iteration of the procedure,
until $\mathcal{P}$ is always positive.  We find that only small modifications to $\mathcal{P}_{klmn}$
or $\mathcal{P}_{kl}$ are needed.

\begin{figure}[t]
\centering
\includegraphics[width=0.35\textwidth]{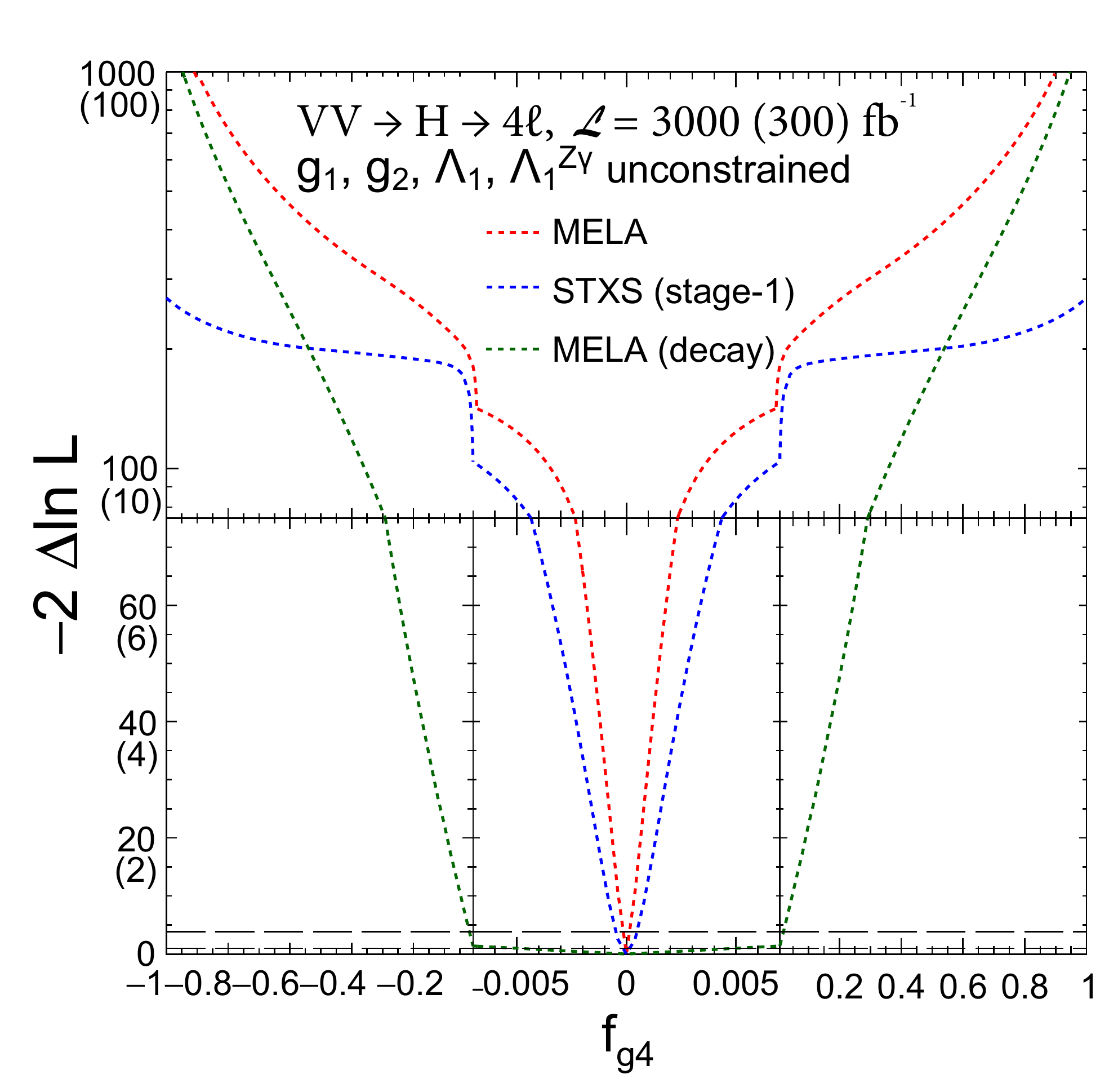}
\includegraphics[width=0.35\textwidth]{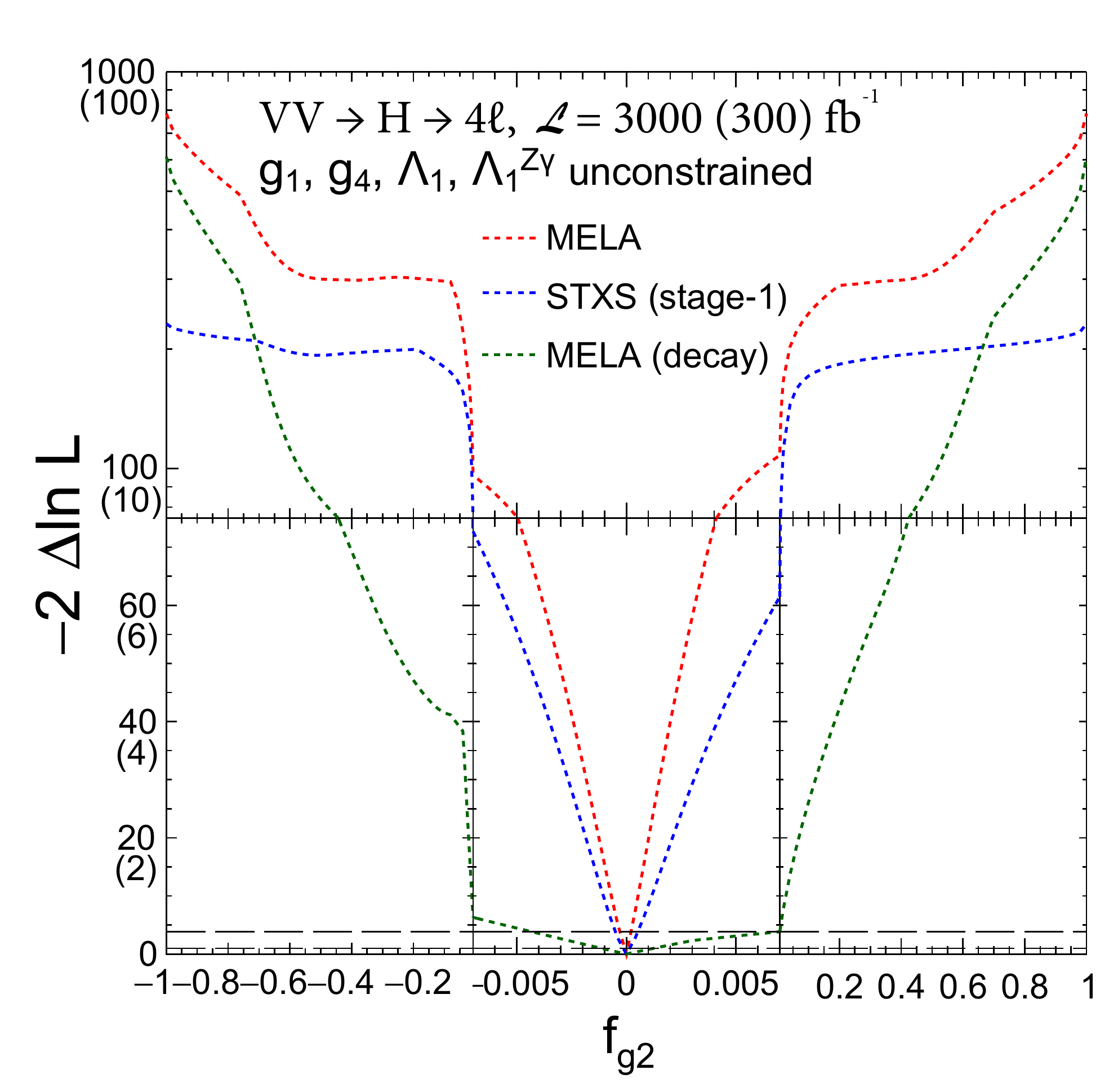}
\includegraphics[width=0.35\textwidth]{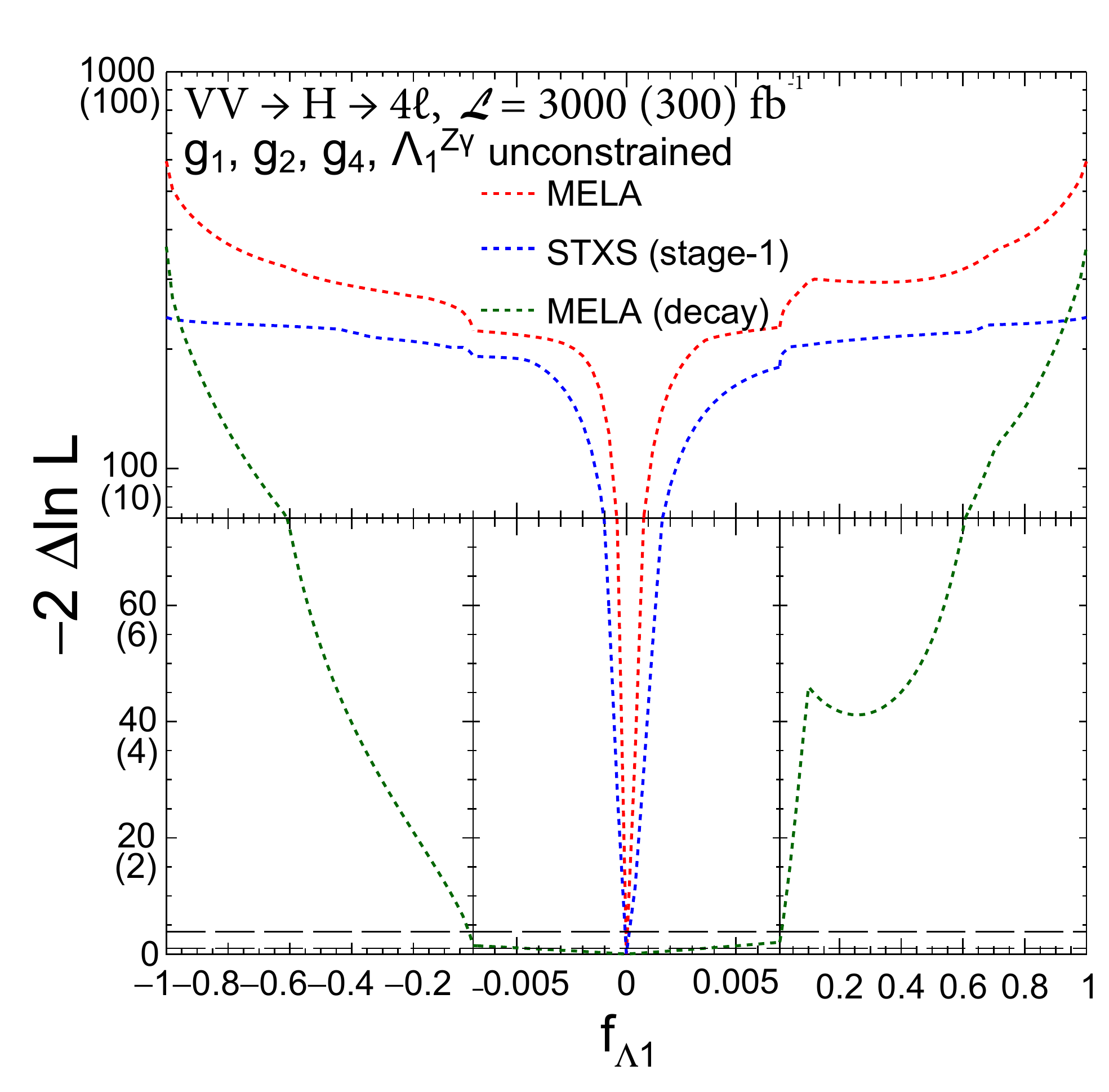}
\includegraphics[width=0.35\textwidth]{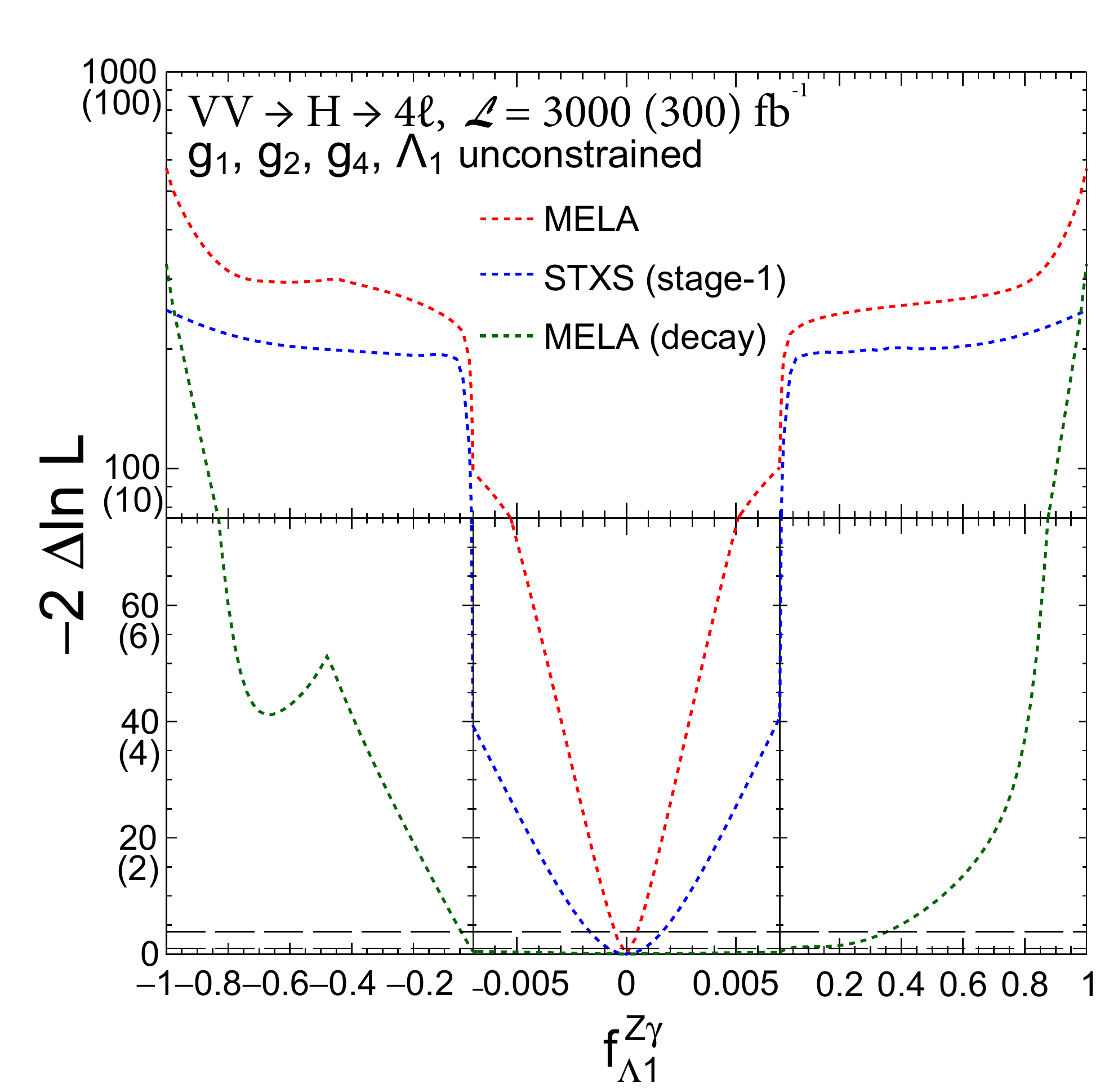}
\captionsetup{justification=centerlast}
\caption{
Expected constraints from a simultaneous fit of $f_{g4}$, $f_{g2}$, $f_{\Lambda1}$, and $f_{\Lambda1}^{Z\gamma}$
using associated production and $\PH\to4\ell$ decay with 3000 (300)\,fb$^{-1}$ data.
Three analysis scenarios are shown: using MELA observables with production and decay (or decay only) information, and using STXS binning. 
The dashed horizontal lines show the 68 and 95\% CL regions.
}
\label{fig:fg4_scan}
\end{figure}

In Fig.~\ref{fig:fg4_scan} we show the expected constraints on the four parameters of interest
$f_{g4}$, $f_{g2}$, $f_{\Lambda1}$,  and $f_{\Lambda1}^{Z\gamma}$, 
using both associated production and $\PH\to4\ell$ decay with 3000\,fb$^{-1}$ (or 300\,fb$^{-1}$) 
of data at a single LHC experiment. 
The constraints on each parameter are shown with the other parameters describing the $HVV$ and $Hgg$ couplings profiled,
including $f_{\rm CP}^{\rm gg}$ and the signal strength parameters  $\mu^{V}$ and $\mu^{f}$. The $\mu^{V}$ and $\mu^{f}$ 
parameters correspond to production strength of electroweak and other processes, respectively. 
Therefore, there are a total of seven free parameters describing $HVV$ and $Hgg$ couplings.
The MC scenario has been generated with the SM expectation.
The production information dominates in all constraints. However, as discussed in Section~\ref{subsec:unitarization},
this is due to unbounded growth of anomalous couplings with $q^2$. Since this behavior cannot 
continue forever, it is still interesting to look at the decay-only constraints, which do not rely 
on the $q^2$-dependent growth of the amplitude. Therefore, in Fig.~\ref{fig:fg4_scan} both kinds of
constraints are shown for illustration of the two limiting cases.  We point out that form factor scaling,
such as introduced in Eq.~(\ref{eq:formfact-spin0}), can be used for continuous study of this effect. 

In addition, a comparison is made to the approach where instead of the 
optimal discriminants, the STXS stage-1.1 bins are used as observables, while using full
simulation of all processes otherwise. There is a significant difference in expected precision.
The most striking effect is the lack of constraints from decay information, but there is a loss 
in precision using production information in STXS as well. 
It is interesting to point that there is still weak decay-related information in the categories
used in the STXS approach, because interference between identical leptons produces different rates of $2e2\mu$
events compared to $4e$ and $4\mu$, depending on the couplings. 

\begin{figure}[t]
	\centering
	\includegraphics[width=0.24\linewidth]{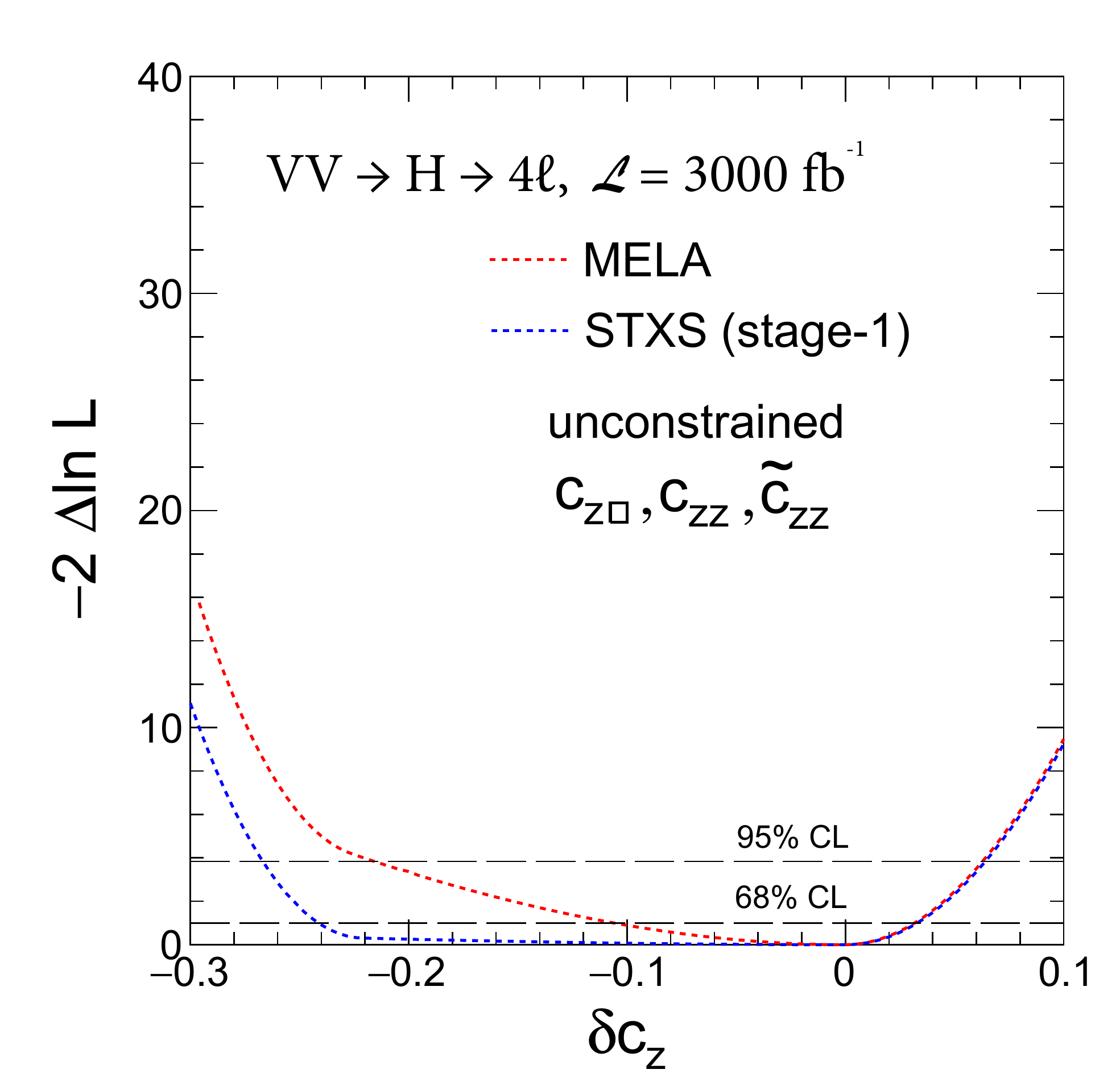}
	\includegraphics[width=0.24\linewidth]{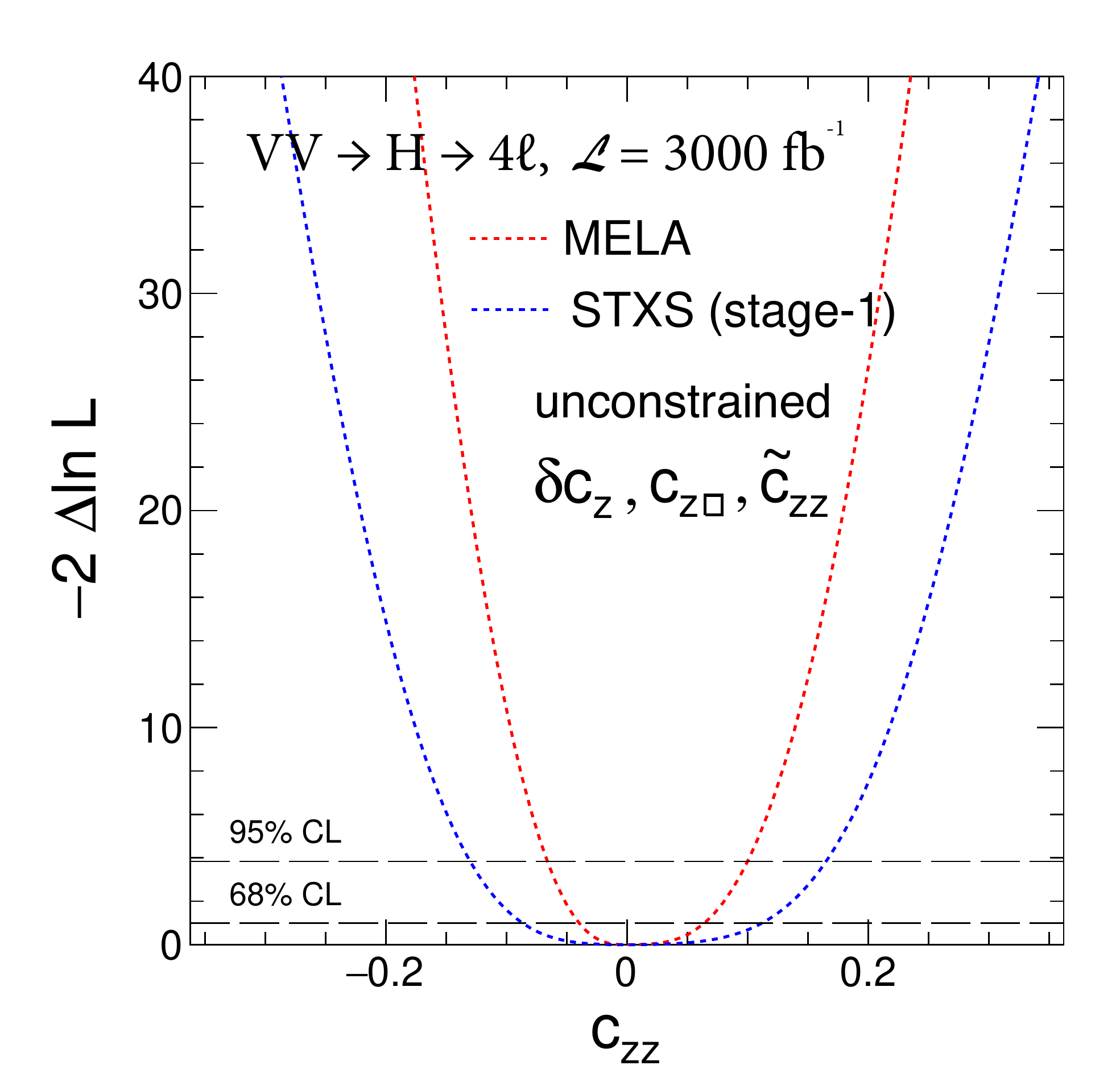}
	\includegraphics[width=0.24\linewidth]{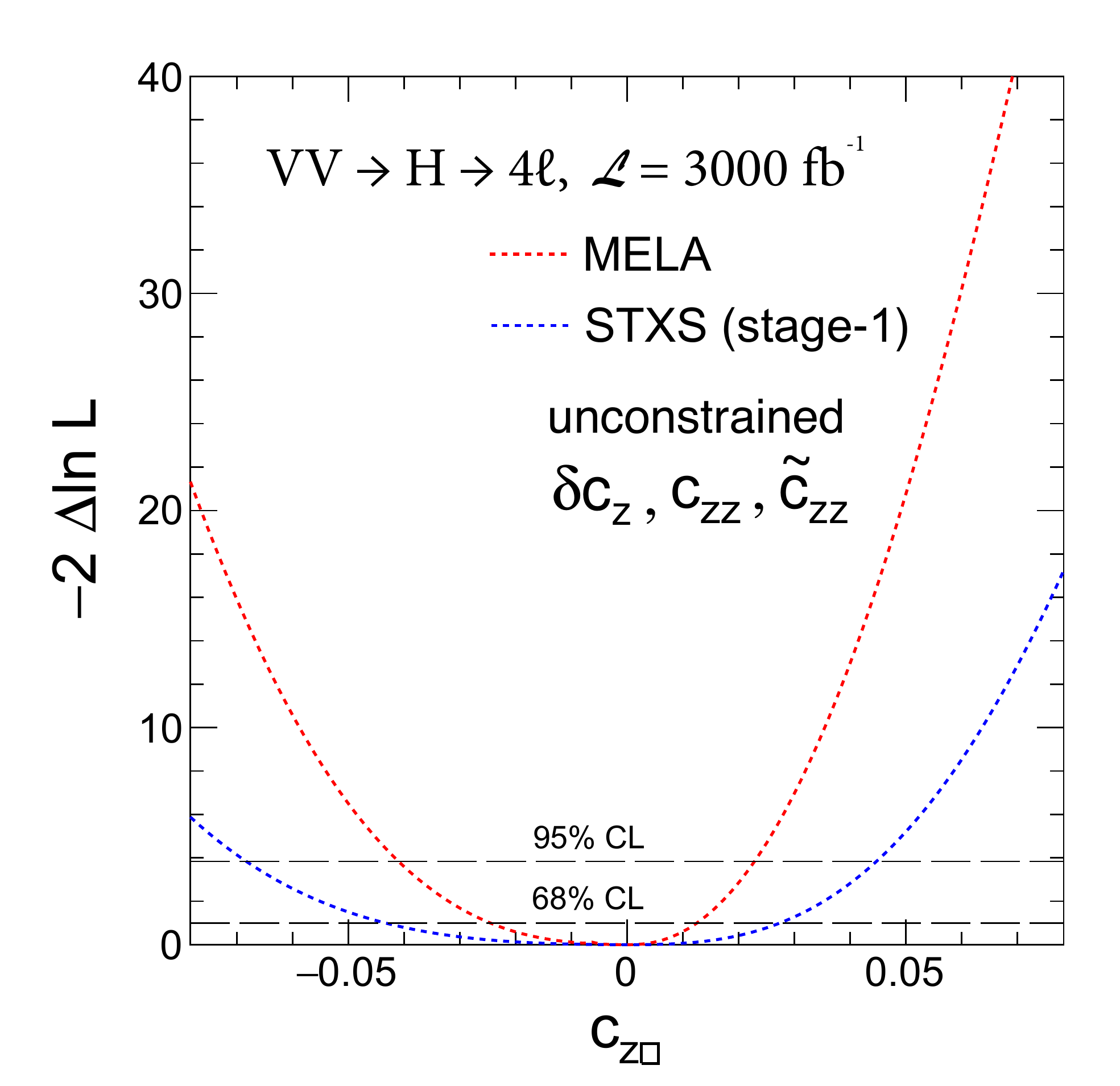}
	\includegraphics[width=0.24\linewidth]{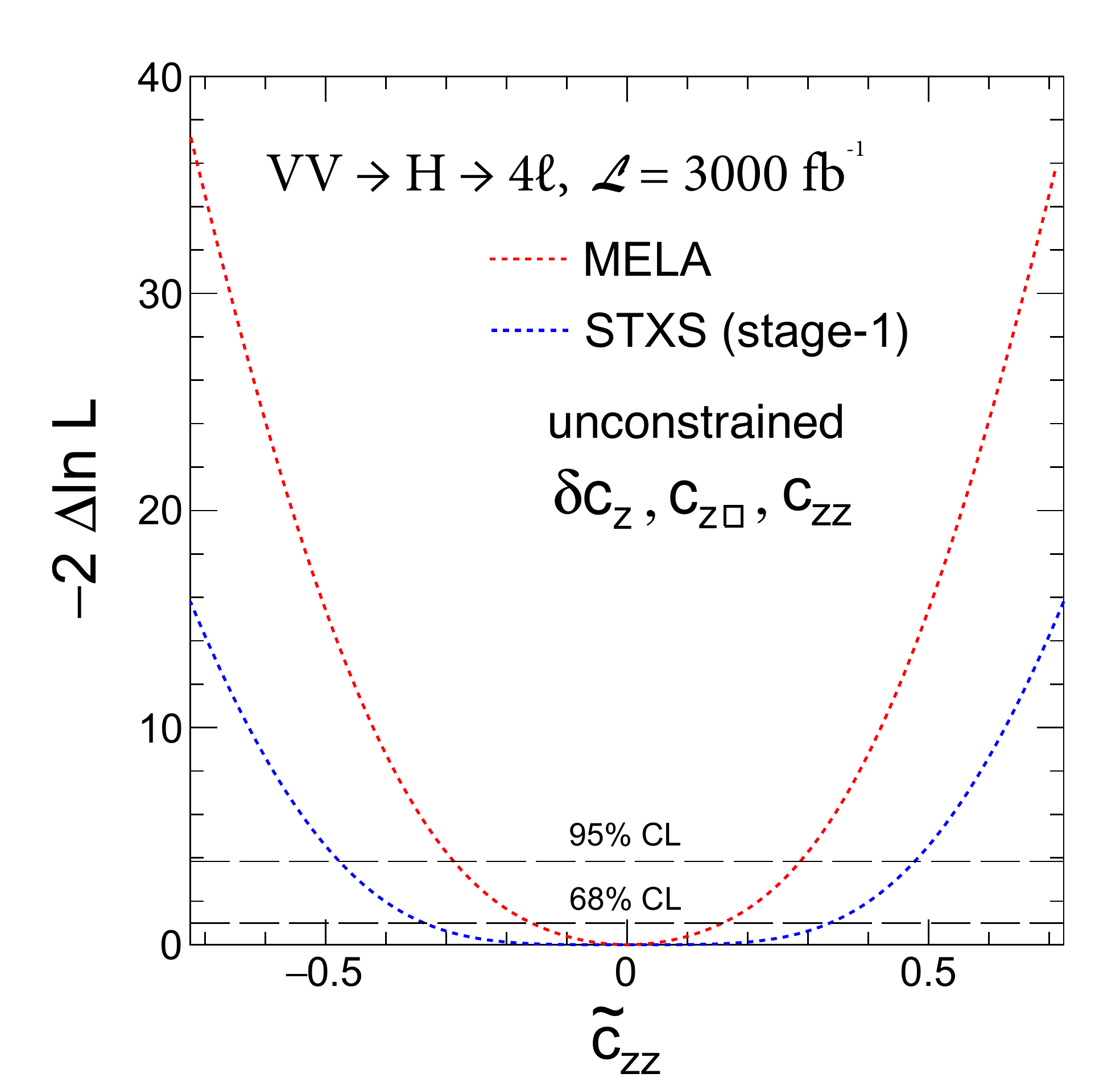}
	\captionsetup{justification=centerlast}
	\caption{
Expected constraints from a simultaneous fit of (from left to right) $\delta c_z$, $c_{zz}$, $c_{z \Box}$, and $\tilde c_{zz}$
using associated production and $\PH\to4\ell$ decay with 3000\,fb$^{-1}$ data.
The EFT coupling constraints are the result of re-interpretation from the signal strength and $f_{gi}$ 
measurements discussed in text. 
The constraints on each parameter are shown with the other parameters describing the $HVV$ and $Hgg$ couplings profiled. 
Two analysis scenarios are shown: using MELA observables and using STXS binning. 
The dashed horizontal lines show the 68 and 95\% CL regions.
	}
	\label{fig:eft_1Dscan}
\end{figure}

\begin{figure}[ht]
\centering
\includegraphics[width=0.32\textwidth]{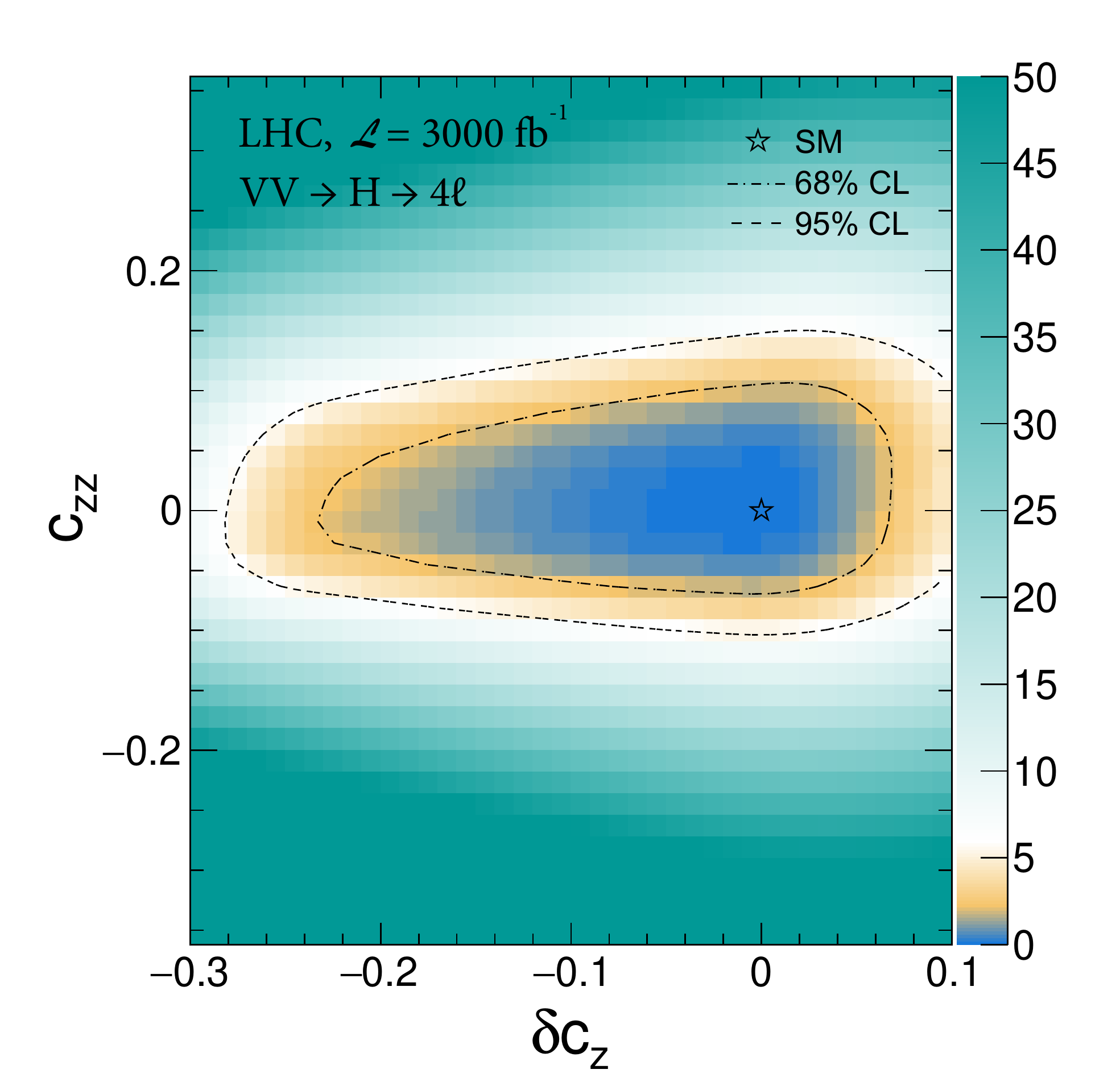}
\includegraphics[width=0.32\textwidth]{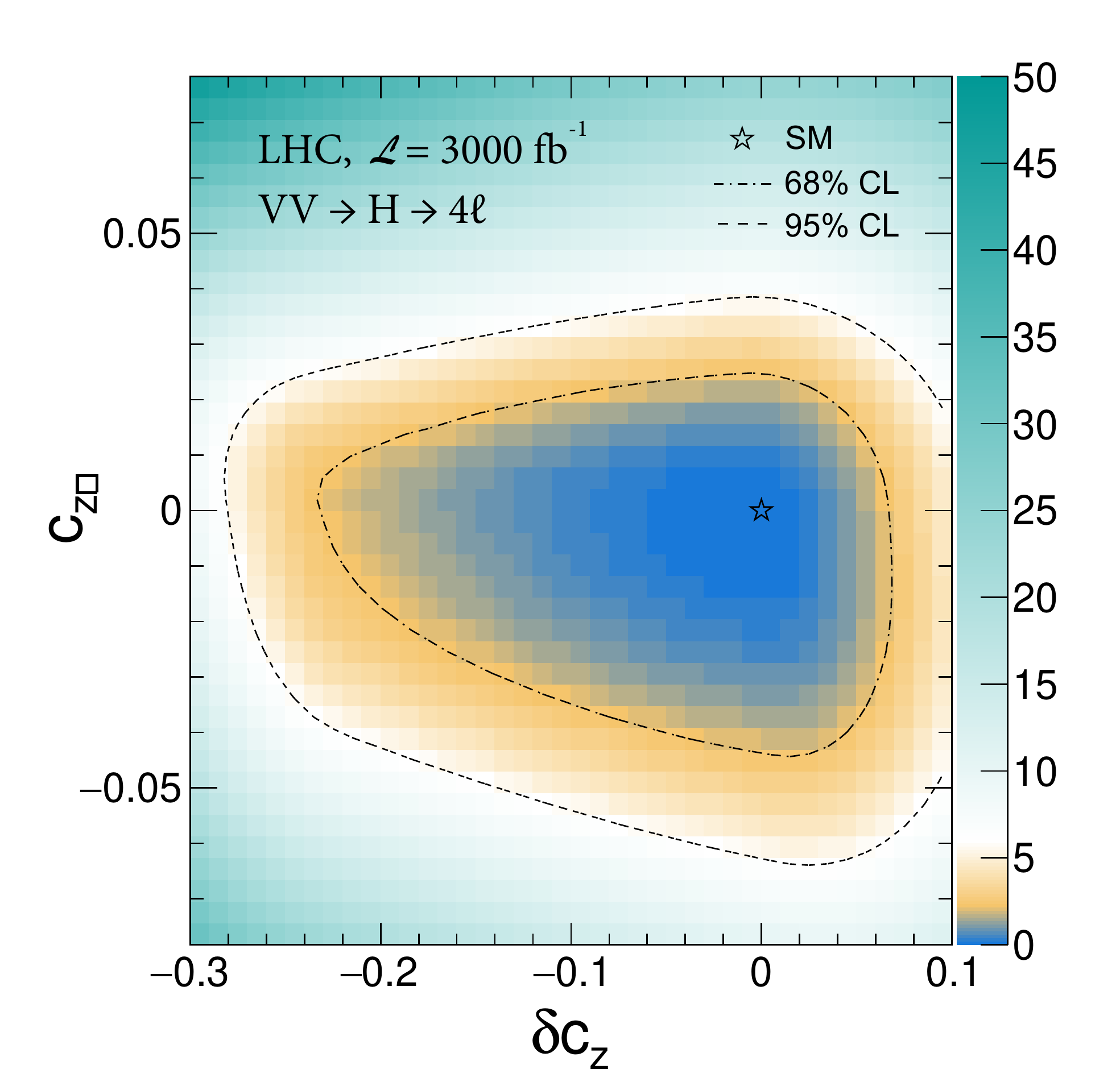}
\includegraphics[width=0.32\textwidth]{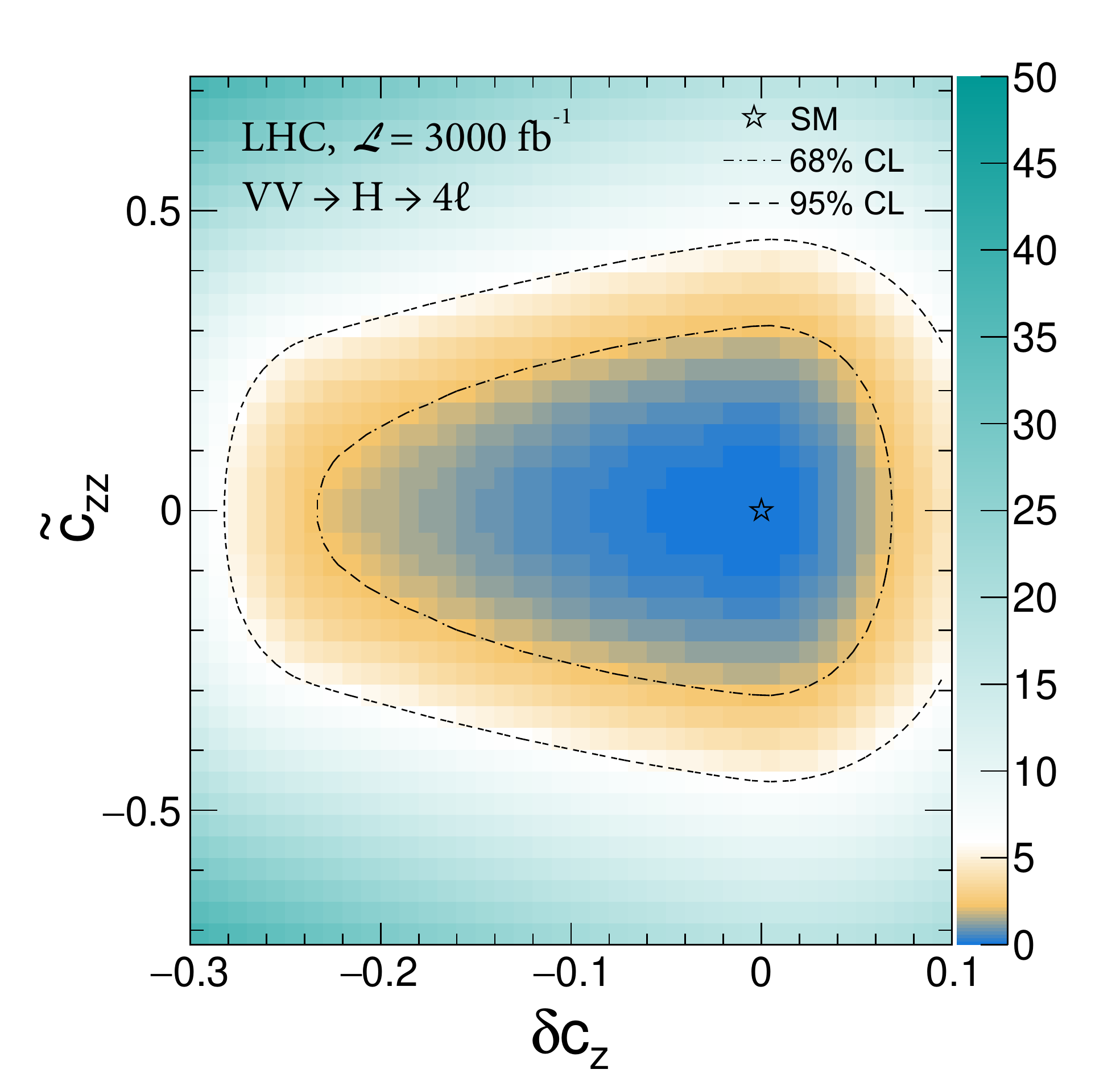}
\includegraphics[width=0.32\textwidth]{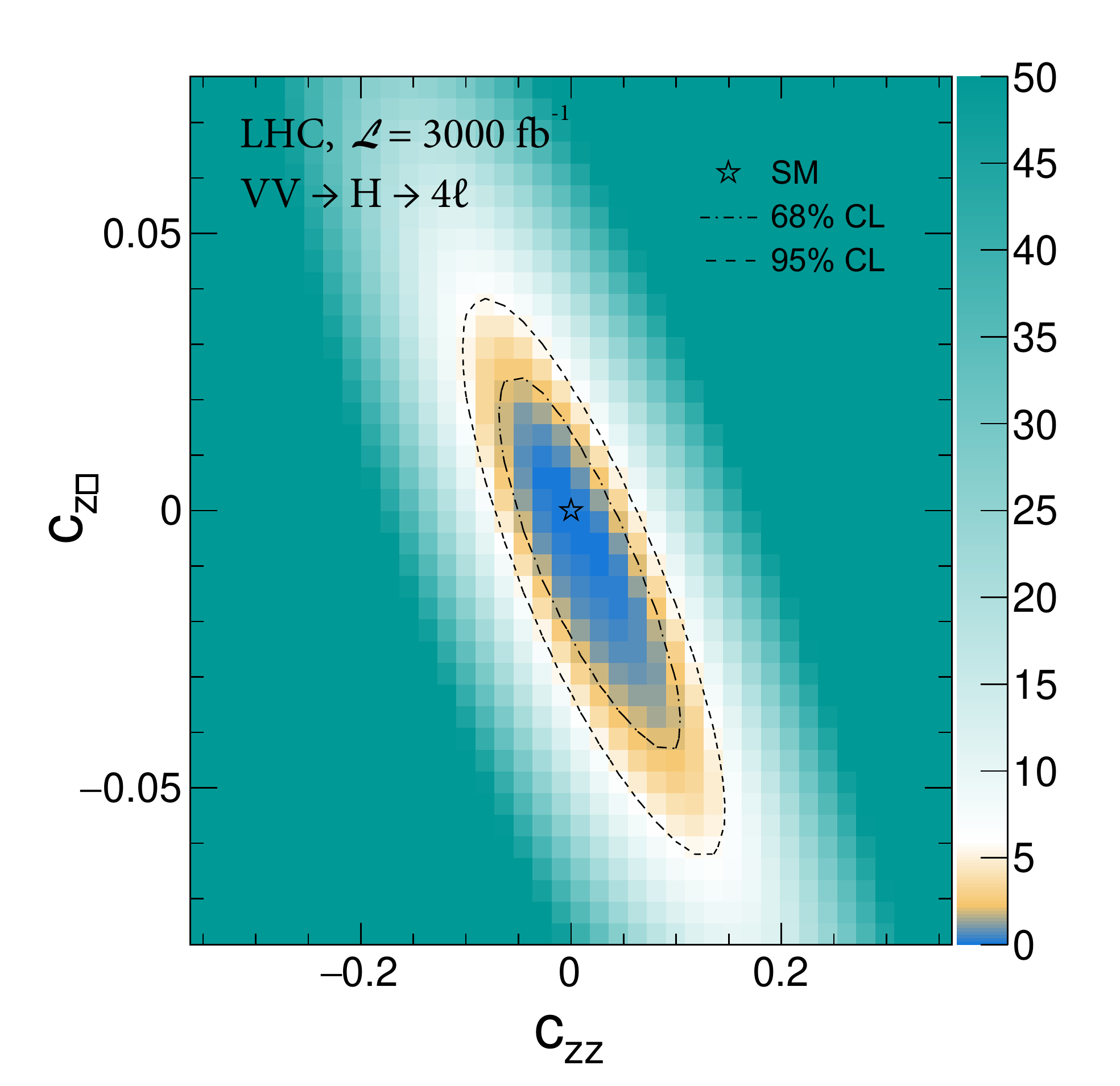}
\includegraphics[width=0.32\textwidth]{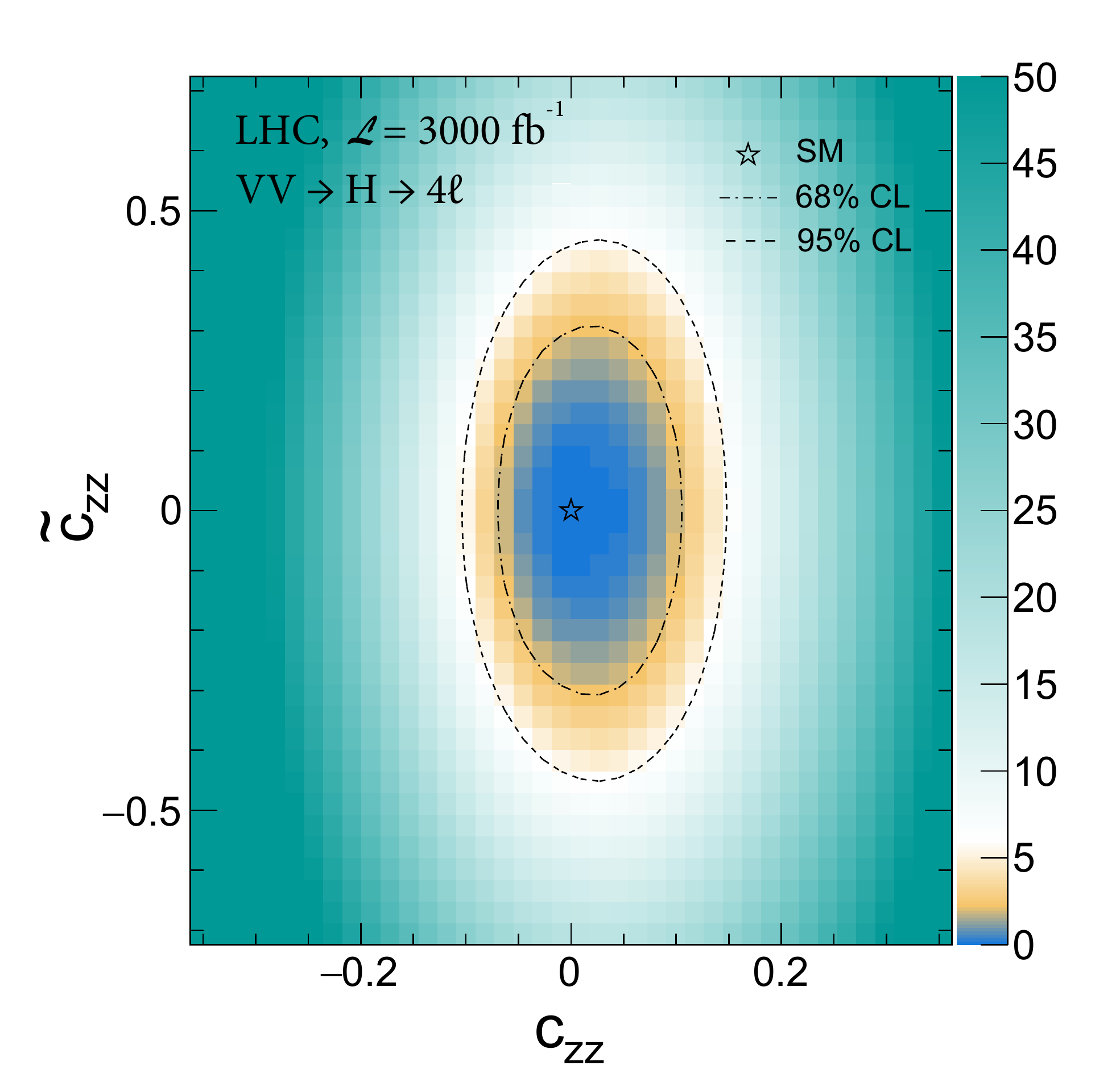}
\includegraphics[width=0.32\textwidth]{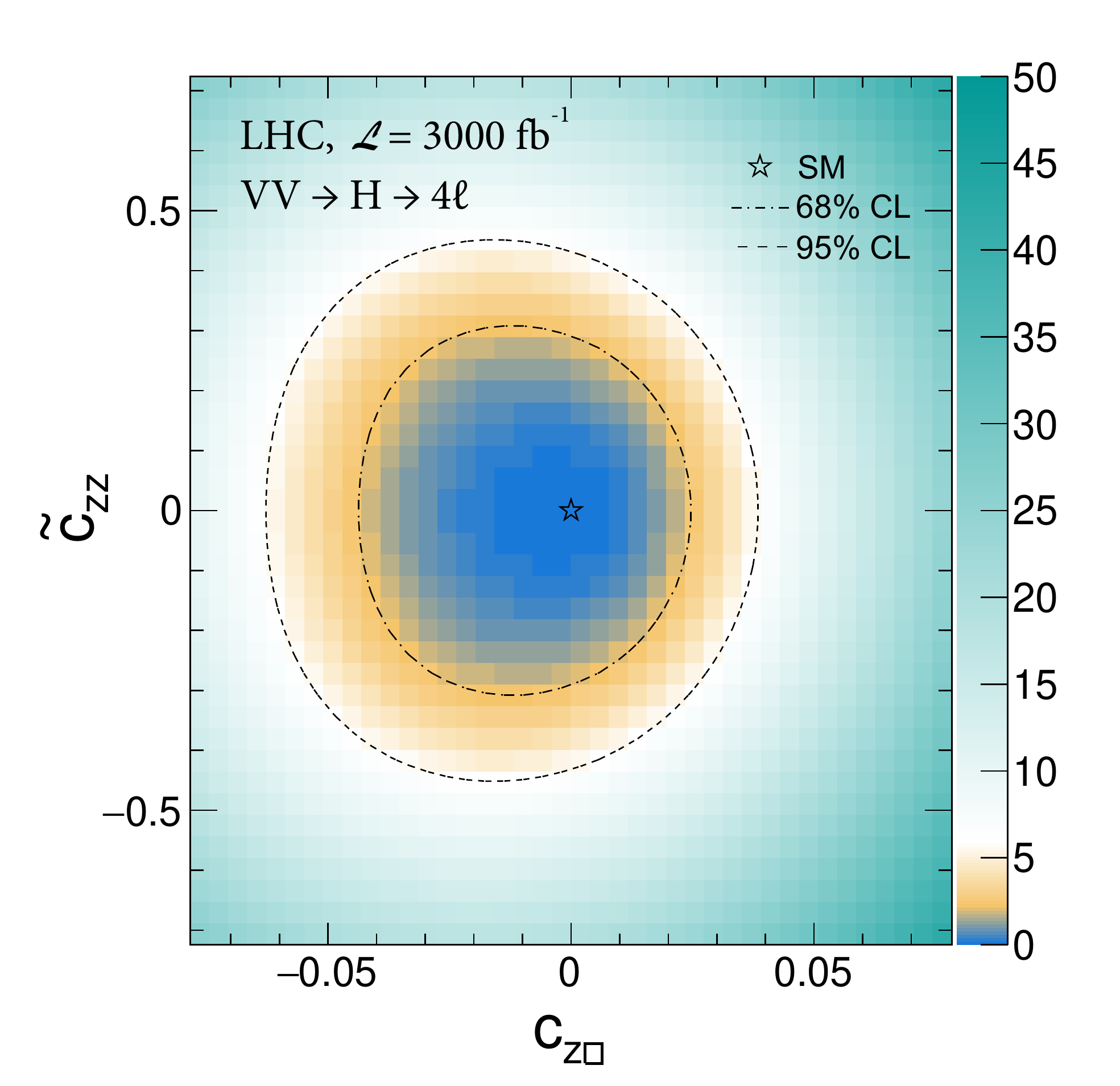}
\captionsetup{justification=centerlast}
\caption{
Expected two-dimensional constraints from a simultaneous fit of $\delta c_z$, $c_{zz}$, $c_{z \Box}$, and $\tilde c_{zz}$
as shown in Fig.~\ref{fig:eft_1Dscan} for the MELA observables. 
The constraints on each parameter are shown with the other parameters describing the $HVV$ and $Hgg$ couplings profiled. 
Top-left: $(\delta c_z, c_{zz})$; 
top-middle: $(\delta c_z, c_{z \Box})$;
top-right: $(\delta c_z, \tilde c_{zz})$; 
bottom-left: $(c_{zz}, c_{z \Box})$; 
bottom-middle: $(c_{zz}, \tilde c_{zz})$; 
bottom-right: $(c_{z \Box}, \tilde c_{zz})$.
}
\label{fig:eft_2Dscan}
\end{figure}

Since $f_{gi}$ measurements involve ratios of couplings, most systematic uncertainties that would otherwise
affect the cross section measurements cancel in the ratio. 
Therefore, the $f_{gi}$ measurements are still expected to be statistics limited with 3000\,fb$^{-1}$ of data. 
For this reason, the expected results can be easily reinterpreted for another scenario of integrated luminosity, 
as for example the expectation with 300\,fb$^{-1}$ shown in parentheses. 
However, when the $f_{gi}$ measurements are re-interpreted in terms of
couplings (as we illustrate below), both the signal strength and the $f_{gi}$ results need to be combined. This leads
to sizable systematic uncertainties affecting the couplings. In the following, we assign 5\% theoretical and
5\% experimental uncertainties on the measurements of the signal strength, which is the ratio of the measured and expected
cross sections. 

We also perform a fit with three cross-section fraction parameters $f_{g4}$, $f_{g2}$, and $f_{\Lambda1}$
with the EFT relationship among couplings following Eqs.~(\ref{eq:deltaMW}--\ref{eq:kappa2Zgamma}). 
The conclusions of this study are similar to those presented above. 
We re-interpret these results as constraints on the $\delta c_z$, $c_{zz}$, $c_{z \Box}$, and $\tilde c_{zz}$
couplings, defined in the EFT parameterization in the Higgs basis.  
This fit requires reinterpreting the process cross section and the three fractions 
in terms of couplings, and one has to take dependence of the width on the couplings into account, 
following Eq.~(\ref{eq:diff-cross-section2}).
We assume that $\Gamma_{\rm other}=0$ and express the width using Eq.~(\ref{eq:width}).
The values of $\kappa_f=\kappa_t=\kappa_b=\kappa_\tau=\kappa_\mu$ and 
$\tilde \kappa_f=\tilde\kappa_t=\tilde\kappa_b=\tilde\kappa_\tau=\tilde\kappa_\mu$ 
are left unconstrained independently for the CP-even and CP-odd fermion couplings. 
The resulting one-dimensional constraints are shown in Fig.~\ref{fig:eft_1Dscan} and 
two-dimensional contours with the other parameters profiled are shown in Fig.~\ref{fig:eft_2Dscan}.
In Fig.~\ref{fig:eft_1Dscan} it is evident again that analysis based on the optimal discriminants provides the
best constraints on the couplings of interest.

\subsection{Hgg anomalous couplings}
\label{ggHJJ}

\begin{figure}[t]
\centering
\includegraphics[width=0.42\textwidth]{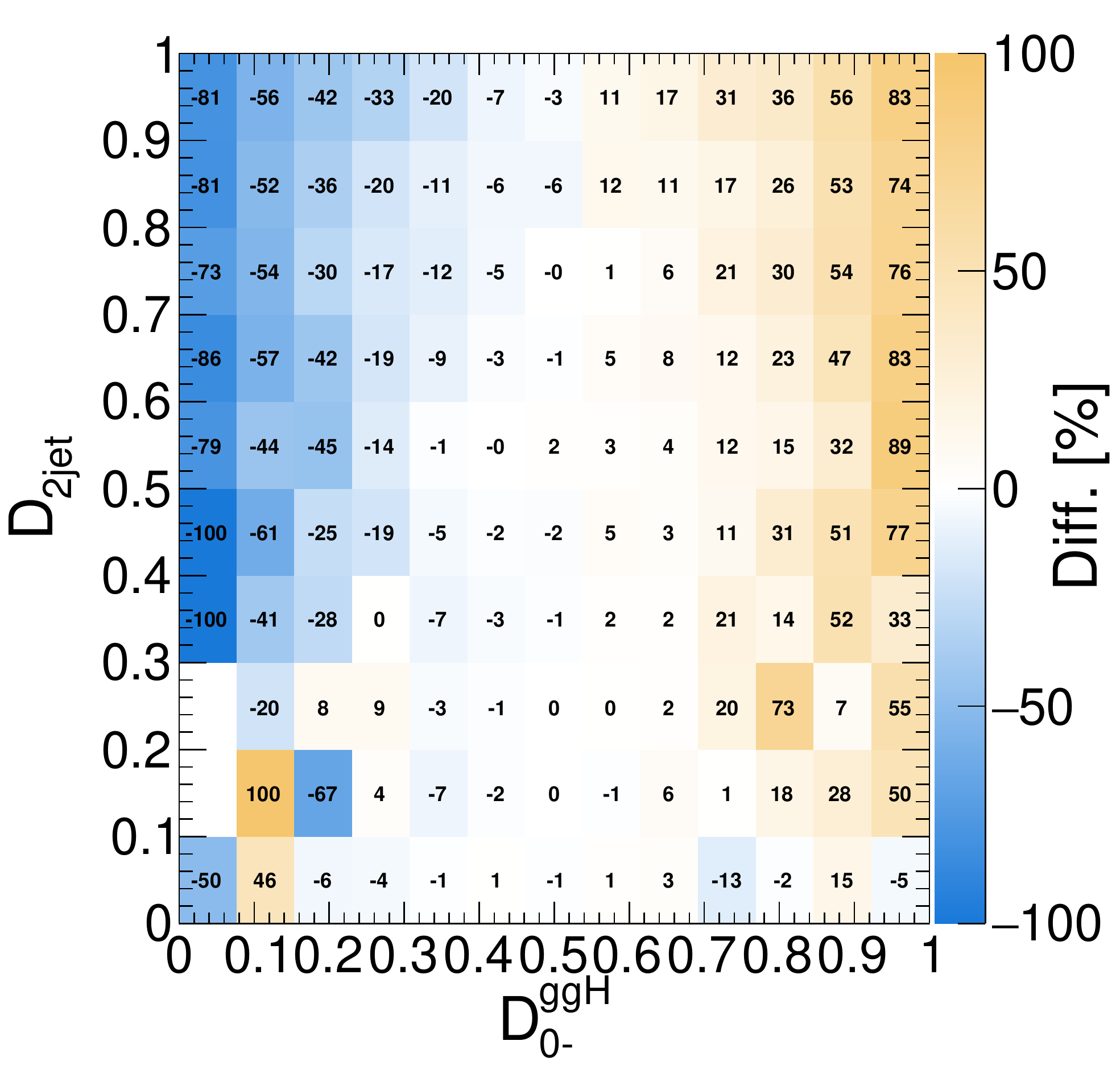}
\includegraphics[width=0.55\textwidth]{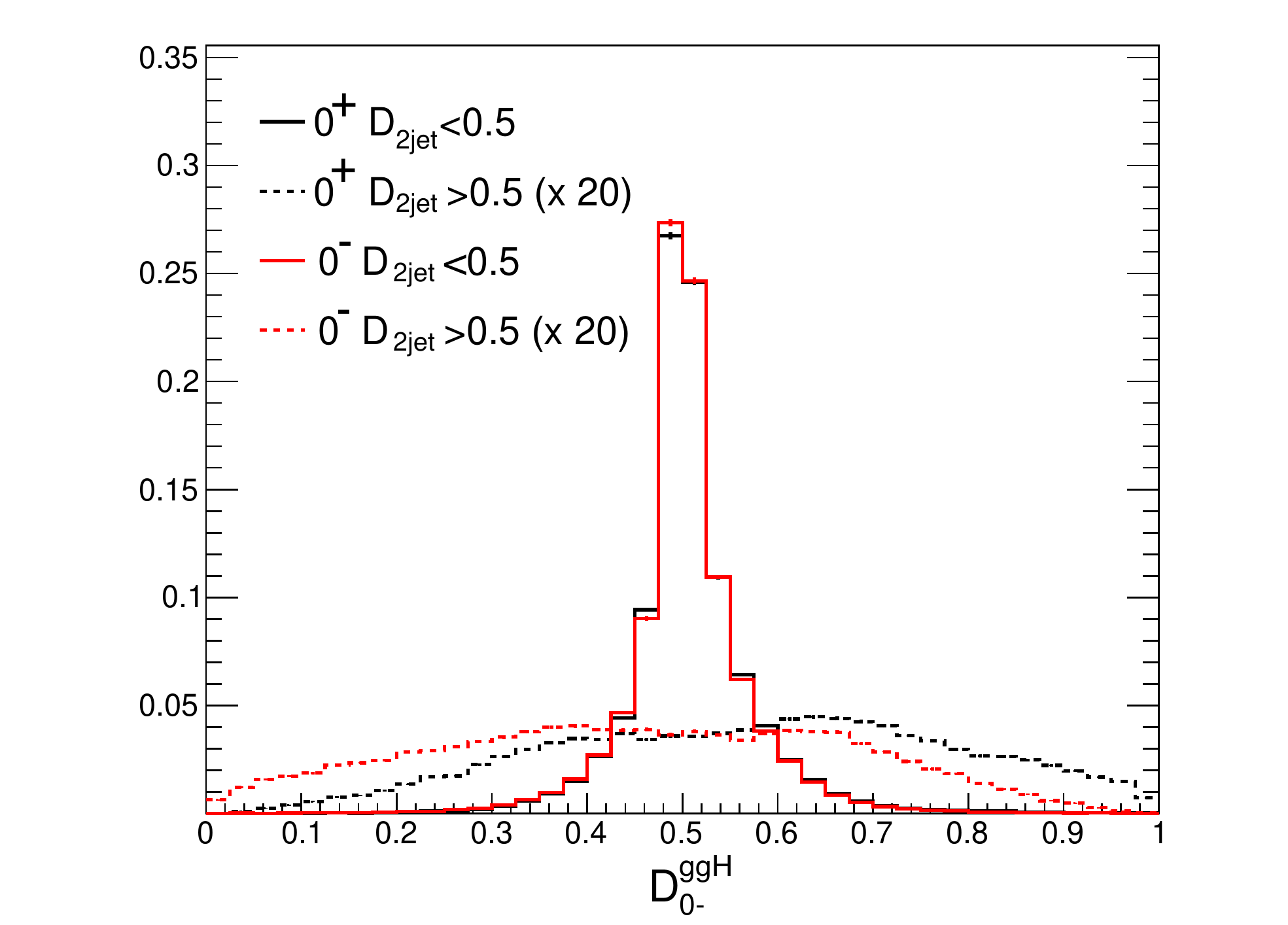}
\captionsetup{justification=centerlast}
\caption{Left: The 2D distribution of the difference between the scalar and pseudoscalar populations of events 
for the ${\cal D}_{2\rm jet}$ and ${\cal D}_{0-}^{\rm ggH}$ discriminants, both self-normalized to 1, respectively. 
Right: ${\cal D}_{0-}^{\rm ggH}$ discriminant distributions of the scalar and pseudoscalar population events 
with the requirement on the ${\cal D}_{2\rm jet}$ discriminant below or above 0.5. 
} 
\label{fig:d0m_ggH}
\end{figure}

The gluon fusion process in association with two jets allows analysis of kinematic distributions for
the measurement of potential anomalous contributions to the gluon fusion loop. Resolving the loop
effects is a separate task, which we do not attempt to perform in this work. However, we point out that 
unless the particles in the loop are light, their mass does not significantly affect the kinematics of the \Hboson 
and associated jets. The main effect is on the \Hboson's transverse momentum~\cite{deFlorian:2016spz},
where heavy particles in the loop may enhance the tail of the distribution at $p_T>200$\,GeV, 
but will not significantly affect the bulk of the distribution relevant for our study, at $p_T<200$\,GeV. 
Our analysis of the CP properties of this interaction depends primarily on the angular kinematics
of the associated jets and \Hboson, as discussed in Section~\ref{sect:kin_vbf}.
Therefore, in the rest of this work we treat the gluon fusion process without resolving the loop
contribution, allowing for any particles to contribute, either from SM or beyond. The only observable 
difference in this analysis is between the CP-even and CP-odd couplings, which can be
parameterized as the overall strength of the \Hboson's coupling to gluons and the fraction
 of the CP-odd contribution $f_{\rm CP}^{\rm gg}$ defined in Eq.~(\ref{eq:fgn}).

The analysis strategy follows the approach discussed in application to the $HVV$ measurements in the previous section, 
with the difference being the two-jet category optimized for the measurement of the gluon fusion process. In addition to
a discriminant optimal for signal over background separation, the events are described by three observables. 
The ${\cal D}_{2\rm jet}$ observables follows Eq.~(\ref{eq:melaAlt}), with the VBF and gluon fusion matrix elements 
used to isolate the VBF topology. The ${\cal D}_{0-}^{\rm ggH}$ and ${\cal D}_{\rm CP}^{\rm ggH}$ observables follow 
Eq.~(\ref{eq:melaAlt}) and Eq.~(\ref{eq:melaInt}) for separating the SM-like coupling and CP-odd coupling, but with 
one modification to the process definition. Only the quark-initiated process defines the matrix element in these two 
formulas, because only such a VBF-like topology of the gluon fusion process carries relevant CP information. 
This is illustrated in Fig.~\ref{fig:d0m_ggH}, where the left plot shows that the ${\cal D}_{0-}^{\rm ggH}$ discriminant starts
to separate the two couplings at higher values of ${\cal D}_{2\rm jet}$, which correspond to more VBF-like topology. 
The right plot shows that only at higher values of ${\cal D}_{2\rm jet}$ can one observe the separation.  Only a small 
fraction of the total gluon fusion events end up in that region. This illustrates the challenge of the CP analysis in the gluon
fusion process. The ${\cal D}_{\rm CP}^{\rm ggH}$ leads to forward-backward asymmetry in the distribution of events in the 
case of CP violation, when both CP-odd and CP-even amplitudes contribute. 

\begin{figure}[t]
\centering
\includegraphics[width=0.45\textwidth]{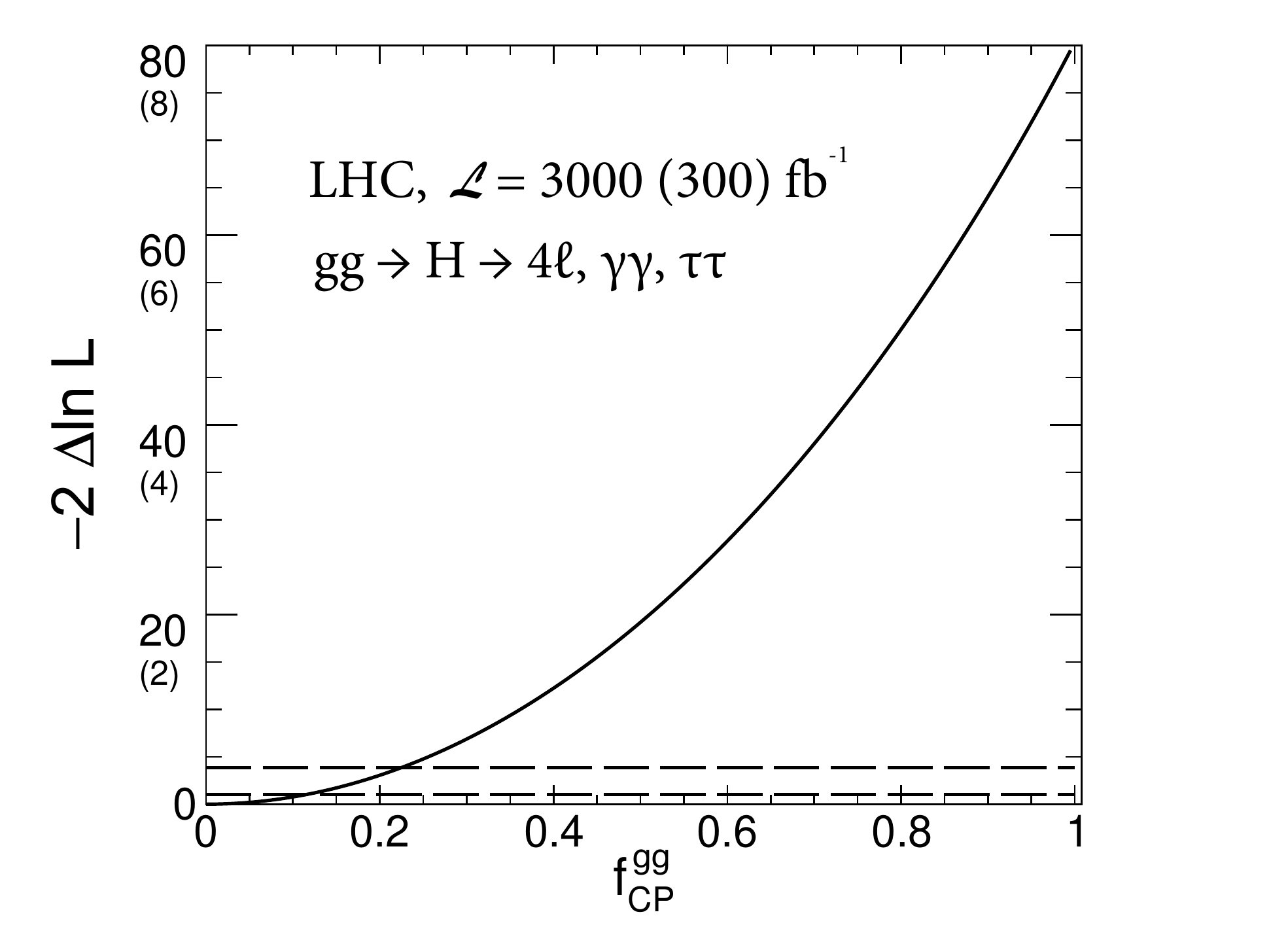}
\includegraphics[width=0.45\textwidth]{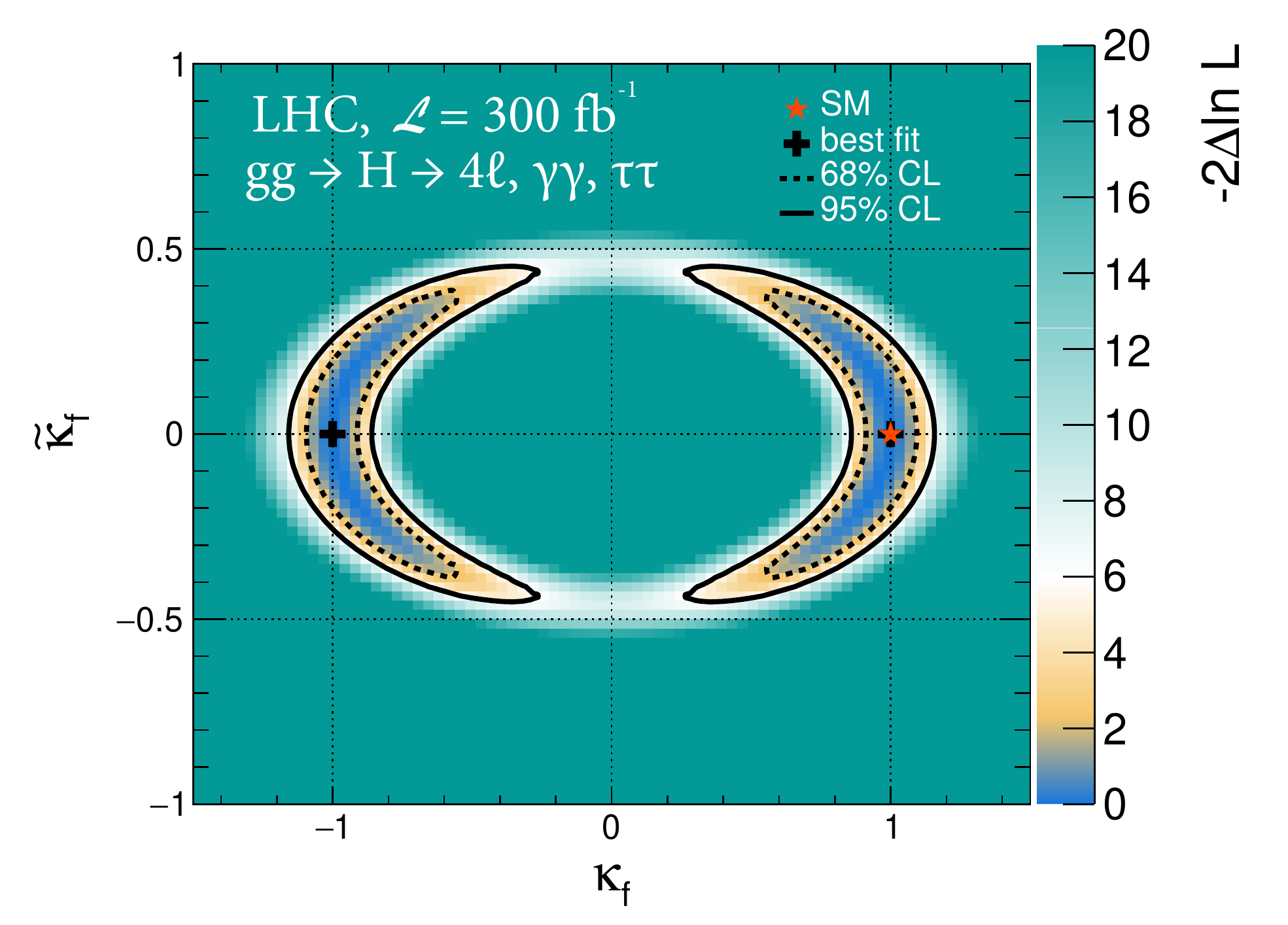}
\captionsetup{justification=centerlast}
\caption{
Expected constraints on $f_{\rm CP}^{\rm gg}$ with 3000 (300)\,fb$^{-1}$  (left) 
and $\kappa_f$ and $\tilde \kappa_f$ couplings in the gluon fusion loop with 300\,fb$^{-1}$ (right),
using the $H\to4\ell, \gamma\gamma,$ and $\tau\tau$ decays.
The dashed horizontal lines (left) and contours (right) show the 68 and 95\% CL regions.
}
\label{fig:fa3ggH_scan}
\end{figure}

A projection of $f_{\rm CP}^{\rm gg}$ sensitivity with 3000 and 300\,fb$^{-1}$ at an LHC experiment is performed. 
The overall normalization of the gluon fusion production rate in the VBF-like topology is provided by the untagged events
and events with two associated jets in a non-VBF topology. The electroweak VBF process is a background to the 
 $f_{\rm CP}^{\rm gg}$ measurement in this case, but its kinematics are still distinct enough to keep it separated in the 
fit on the statistical basis. Keeping its CP properties unconstrained has little effect on the CP analysis in the gluon fusion process. 
We use the $H\to 4\ell$ analysis to illustrate the sensitivity, but scale the expected constraints with an effective luminosity 
ratio to account for the relative sensitivity of the $H\to \gamma\gamma$ and $\tau\tau$ channels based on the typical
sensitivity in the VBF topology~\cite{Sirunyan:2017tqd,Sirunyan:2019nbs,Sirunyan:2018koj,Aad:2019mbh}. 
The expected constraints are shown in Fig.~\ref{fig:fa3ggH_scan}. 
With 3000 (300)\,fb$^{-1}$, one can separate CP-even and CP-odd $Hgg$ couplings with a confidence level
of about 9\,(3)\,$\sigma$.

We re-interpret these expected constraints on $f_{\rm CP}^{\rm gg}$ and the cross section 
as constraints on the $\kappa_f=\kappa_t=\kappa_b$ and 
$\tilde \kappa_f=\tilde\kappa_t=\tilde\kappa_b$ couplings in the gluon fusion loop, 
assuming that only the SM top and bottom quarks dominate the loop.   
There are additional considerations when re-interpreting the signal strength and $f_{\rm CP}^{\rm gg}$
in terms of couplings following Eq.~(\ref{eq:diff-cross-section2}). 
As in the $HVV$ measurements in Section~\ref{HVV_onshell}, 
we assume $\Gamma_{\rm other}=0$ and express the width $\Gamma_{\rm tot}$ using Eq.~(\ref{eq:width}).
In this approach, the overall signal strength of the VBF and VH processes, proportional to $g_1^2$, remains unconstrained. 
Using the $f_{\rm CP}^{\rm gg}$ and gluon fusion cross section constraints expected with 300\,fb$^{-1}$ of data at LHC
with the $H\to4\ell, \gamma\gamma,$ and $\tau\tau$ decays, we show the expected constraints on 
$(\kappa_f, \tilde \kappa_f)$ in Fig.~\ref{fig:fa3ggH_scan} (right). The sign ambiguity 
$(\kappa_f, \tilde \kappa_f)\leftrightarrow(-\kappa_f, -\tilde\kappa_f)$ remains unresolved with experimental data,
but the relative sign of $\kappa_f$ and $\tilde \kappa_f$ can be resolved, due to the ${\cal D}_{\rm CP}^{\rm ggH}$ observable. 
The measurement of the gluon fusion cross section alone leads to an elliptical constraint in the 2D likelihood scan, 
with the eccentricity determined by the ratio $\sigma(\tilde\kappa_f=1)/\sigma(\kappa_f=1)$ discussed earlier. 
The $f_{\rm CP}^{\rm gg}$ measurement leads to constraints within the ellipse.

Should sizable CP violation effects be hidden in the gluon fusion loop, they can be uncovered with the
HL-LHC data sample. Further improvements in the CP constraints on $(\kappa_f, \tilde \kappa_f)$ can be obtained
by measuring the $\ttH$ process, where even stronger constraints are expected~\cite{Gritsan:2016hjl}. 
However, we would like to point out that the ratio $\sigma(\tilde\kappa_f=1)/\sigma(\kappa_f=1)=0.39$ in the $\ttH$ process
is by a factor of six different from the ggH process. This large difference will lead to stronger constraints in the combination
of the two measurements under assumption of the top quark dominance in the loop
because of additional information from the ratio of cross sections. Nonetheless, the measurement 
in the gluon fusion is not limited to the top quark Yukawa coupling, but may include other BSM effects in the loop.
Therefore, it is possible for CP effects to show up in the ggH measurement, but not in $\ttH$.

\section{Application to \offshell\ H(125) boson production}
\label{sect:exp_offshell}

We continue by investigating the off-shell production and decay of the \Hboson with its coupling to either
strong or weak vector bosons. There have already been previous studies of the anomalous $HVV$ couplings 
using these tools, with the most extensive analyses from CMS~\cite{Khachatryan:2015mma,Sirunyan:2019twz}. 
Here we document and extend these studies, in particular to anomalous $Hgg$ couplings and to
anomalous couplings in background processes, which do not include the \Hboson propagator, using the EFT relationship. 
We introduced the \offshell\ effect in Section~\ref{sect:appendix_XS} and discussed simulation and analysis 
tools in Section~\ref{sect:cp_mc}. 
The special feature of \offshell\ production is the strong interference between the signal processes, which involve the \Hboson,
and background processes due to the broad invariant-mass distributions of the off-shell \Hboson. 
Analysis of the \offshell\ region is particularly important to constrain
couplings directly, without the complication of the width dependence that appears \onshell\ in Eq.~(\ref{eq:diff-cross-section2}).
Equivalently, a joint analysis of the \onshell\ and \offshell\ regions leads to constraints on
$\Gamma_{\rm other}$ in Eq.~(\ref{eq:width_other}).
Moreover, the higher $q^2$ transfer in the off-shell topology can enhance the effects of anomalous couplings.

In this Section, we set $\Gamma_{\rm tot}=4.07$\,MeV~\cite{Heinemeyer:2013tqa} and use the pole mass scheme in the gluon fusion loop 
calculations with $m_{t}=173.2$\,GeV and $m_{b}=4.75$\,GeV~\cite{Heinemeyer:2013tqa, Agashe:2014kda}.
The QCD factorization and renormalization  scales are chosen to run as $m_{4\ell}/2$.
In order to include NNLO QCD corrections in the electroweak process, a $k$~factor of 1.12~\cite{Heinemeyer:2013tqa} is applied,
see also discussion of the $k$~factor in application to the $VH$ process in Section~\ref{sect:cp_vh}.
In order to include higher-order QCD corrections in the gluon fusion process, the LO, NLO, and NNLO signal cross section 
calculations are performed using the MCFM and HNNLO~\cite{Catani:2007vq,Grazzini:2008tf,Grazzini:2013mca}  
programs for a wide range of masses using narrow width approximation. 
The ratio between the NNLO and LO, or between the NLO and LO, values is used as a weight $k$~factor.
The NNLO $k$~factors are applied to simulation as shown below. 
While this procedure is directly applicable to the signal cross section, it is approximate for background and 
for signal-background interference. However, the respective
NLO calculations are available~\cite{Melnikov:2015laa,Caola:2015psa,Grazzini:2018owa} for the mass range
$150\,\mathrm{GeV}<m_{4\ell}<2m_t$, and Ref.~\cite{Caola:2016trd} found good agreement between 
the NLO signal, background, and interference $k$~factors. 
Any differences set the scale of systematic uncertainties in our procedure. 

\begin{figure}[b]
\centerline{
\epsfig{figure=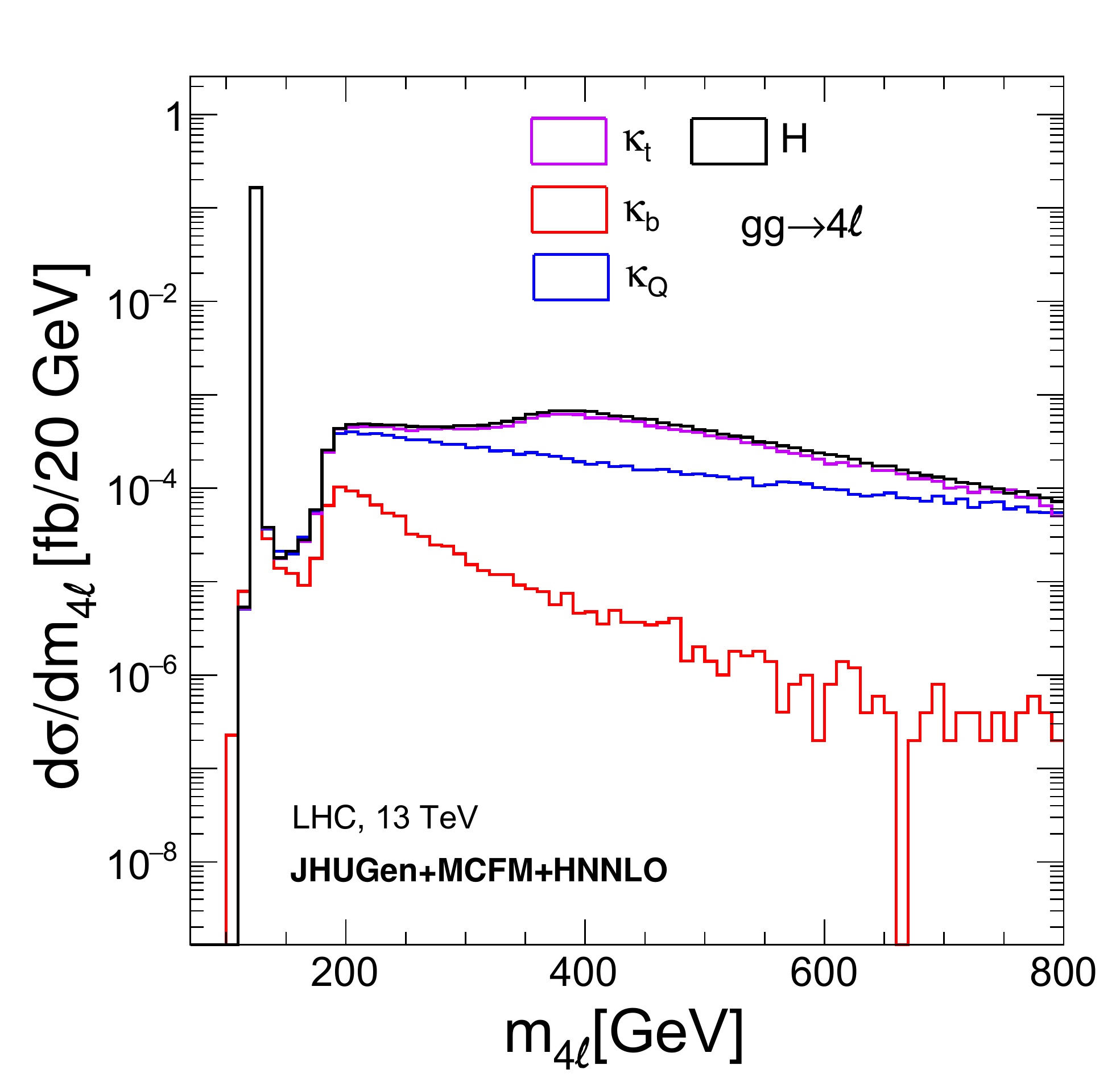,width=0.35\linewidth}
\epsfig{figure=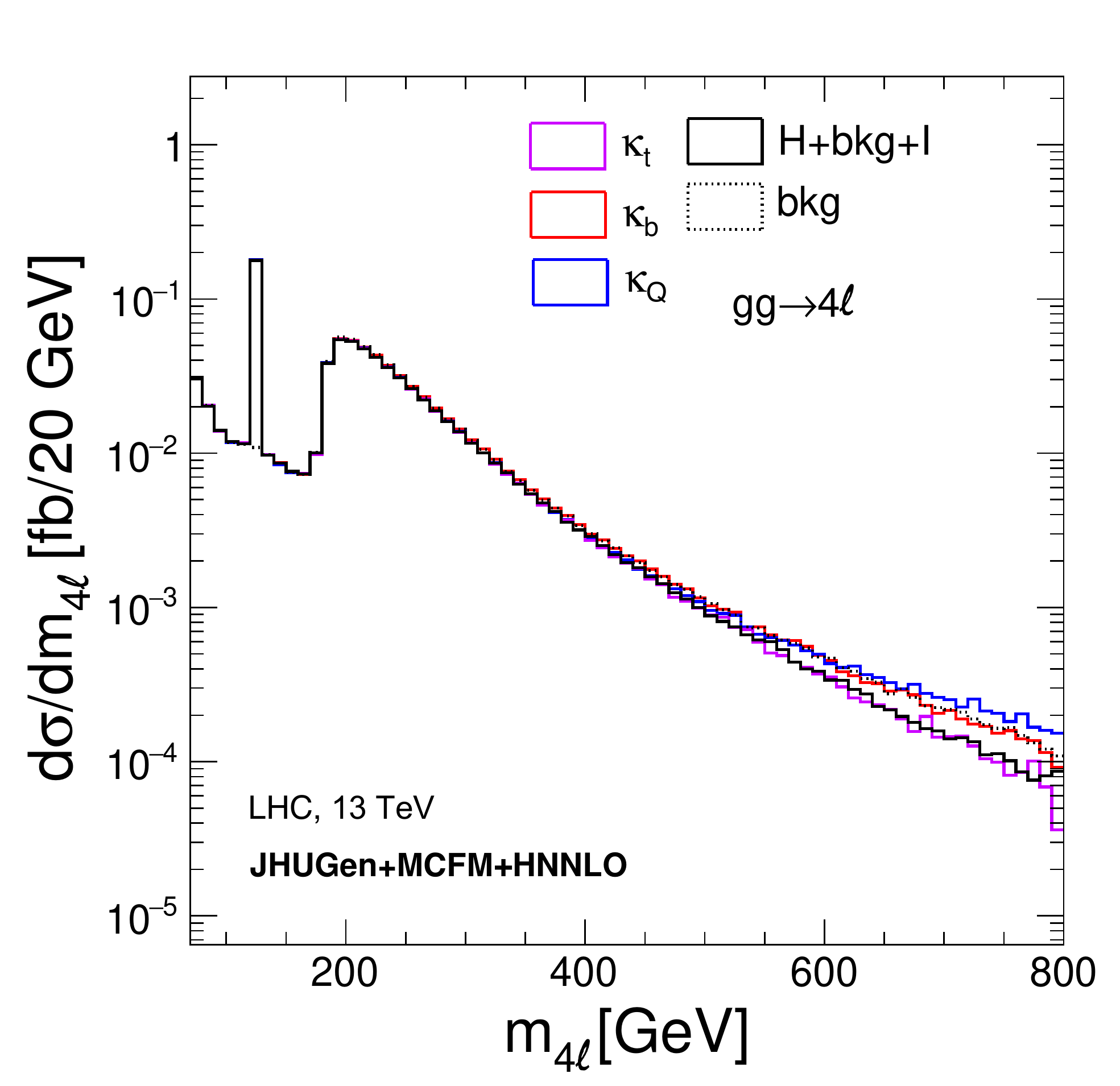,width=0.35\linewidth}
}
\captionsetup{justification=centerlast}
\caption{
The invariant mass distribution of four-lepton ($4\ell=2e2\mu$) events produced through gluon fusion at the LHC with a 13\,TeV proton collision energy. 
The different CP-even anomalous $Hgg$ couplings are simulated with JHUGen+MCFM at LO in QCD, and the NNLO $k$~factor
is calculated with the HNNLO program, assuming signal and background $k$~factors to be the same as for the SM \Hboson. 
Four off-shell scenarios are shown with the couplings chosen to match the SM on-shell $gg\to H\to 4\ell$ cross section:
SM (solid black), top-quark only (magenta), bottom-quark only (red), and a point-like effective interaction (blue)
shown in Eq.~(\ref{eq:g2gg}).
The left plot shows the $gg\to H\to 4\ell$ process with only signal, while the right plot includes interference with the 
SM $gg\to 4\ell$ background, which is also shown separately in the dotted histogram.
}
\label{fig:kinematics-ggH-offshell-1}
\end{figure}

\begin{figure}[t]
\centerline{
\epsfig{figure=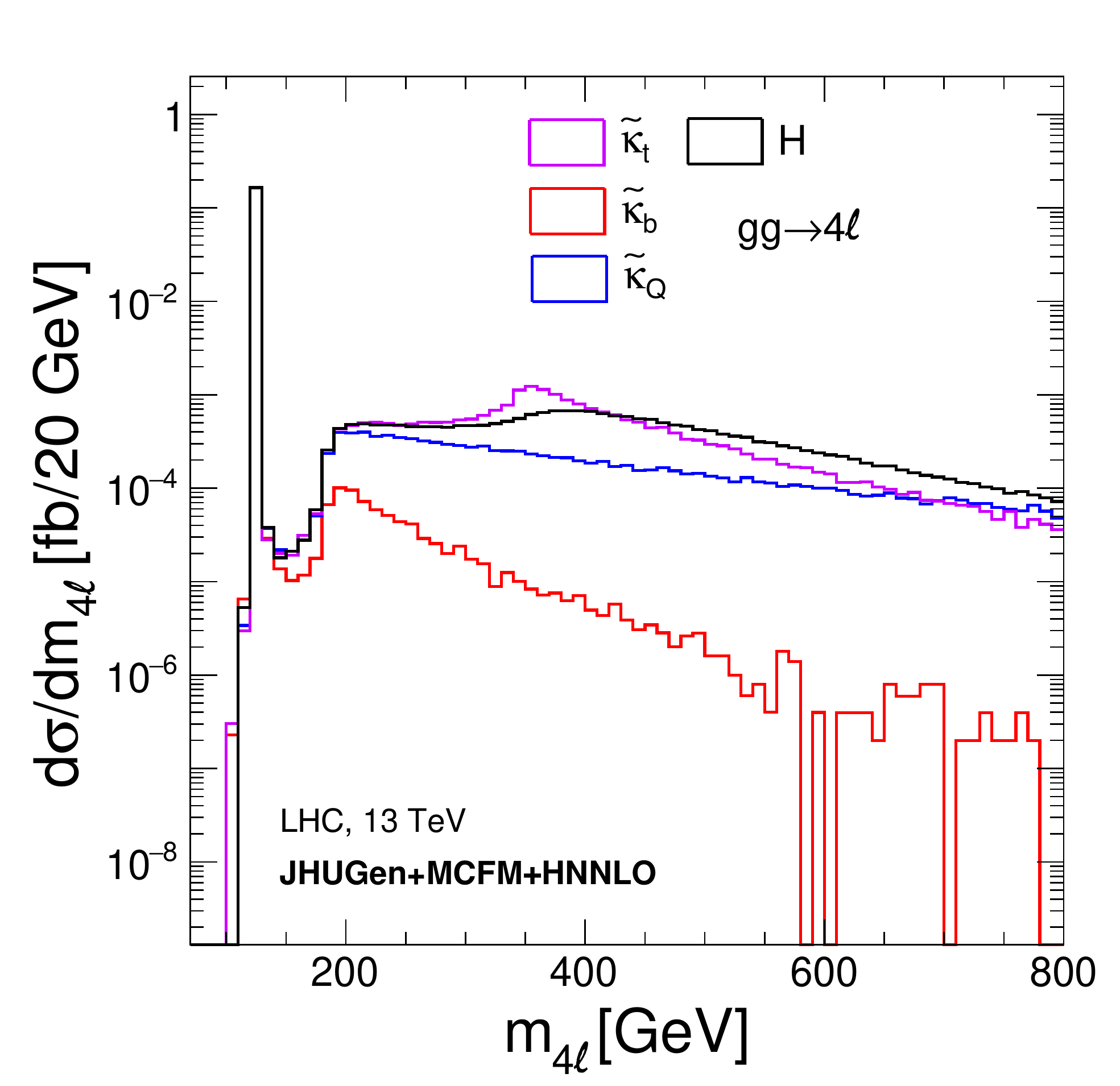,width=0.35\linewidth}
\epsfig{figure=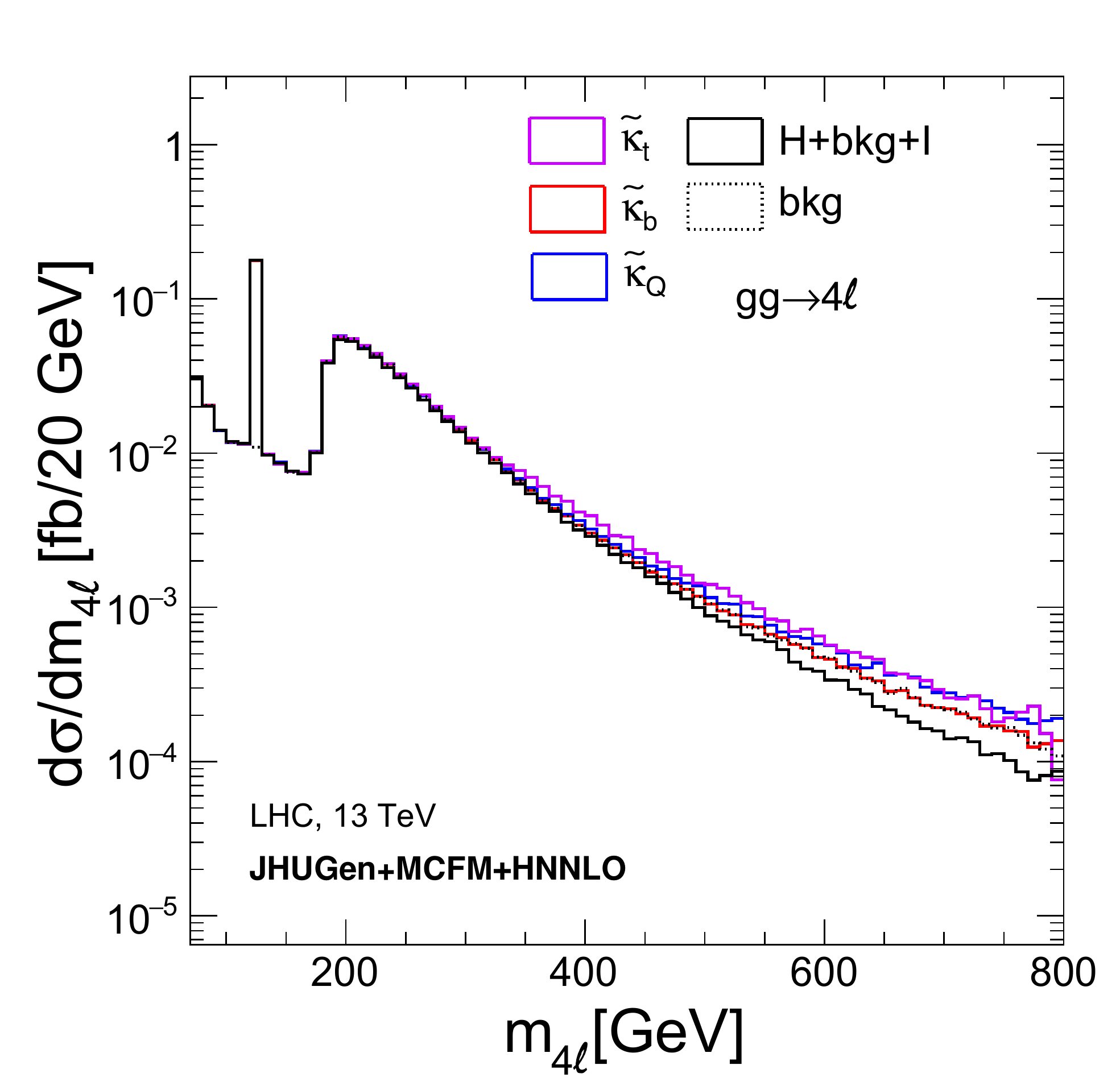,width=0.35\linewidth}
}
\captionsetup{justification=centerlast}
\caption{
The four-lepton invariant mass distributions in gluon fusion production as in Fig.~\ref{fig:kinematics-ggH-offshell-1}, 
but with the three CP-odd anomalous couplings instead, also chosen to match the SM on-shell $gg\to H\to 4\ell$ cross section.
}
\label{fig:kinematics-ggH-offshell-2}
\end{figure}

\subsection{\Offshell\ effects due to Hgg anomalous couplings}
\label{offshell_hgg}

The \offshell\ production of the \Hboson may provide a way to disentangle contributions to the gluon fusion loop
from either SM-like couplings to the top and bottom quarks, CP-odd couplings of the \Hboson, or new heavy
particles. We illustrate this in Fig.~\ref{fig:kinematics-ggH-offshell-1} for CP-even couplings and in 
Fig.~\ref{fig:kinematics-ggH-offshell-2} for CP-odd couplings. In order to illustrate the effects, we separate 
the signal contributions from the top quark, the bottom quark, and an effective point-like interaction, with both CP-even
and CP-odd couplings to the \Hboson. 
For illustration, the SM values of the $HVV$ couplings and of the \Hboson width $\Gamma_{\rm tot}$ are assumed,
but variations of these couplings are considered in Section~\ref{offshell_hvv},
and a simultaneous measurement with $\Gamma_{\rm tot}$ can be considered. 
The tools allow the modeling of the gluon fusion loop with all possible
couplings contributing simultaneously, including interference with the background $gg\to4\ell$ process. 
The effective point-like interaction is equivalent to heavy 
$t^\prime$ and $b^\prime$ quarks in the loop, and this can also be configured in the JHU generator,
with adjustable masses of the these new particles in the loop.  

While in Section~\ref{sect:exp_onshell} it was shown how the CP-even and CP-odd couplings in the gluon fusion 
loop can be separated by analyzing of \onshell\ \Hboson production in association with two jets, this approach does
not allow us to resolve different contributions to the loop. Off-shell production provides a way
to separate those contributions. As can be seen in Figs.~\ref{fig:kinematics-ggH-offshell-1} and~\ref{fig:kinematics-ggH-offshell-2},
the top quark contribution has a distinctive threshold enhancement around $2m_t$, with somewhat different 
behavior of the CP-even and CP-odd components. 
The heavy particle contribution proceeds without the $2m_t$ threshold, and
the light particle contribution is highly suppressed.
Therefore, the \offshell\ spectrum can be used to resolve the loop effects, such as to differentiate between 
the top quark and heavy BSM contributions in the loop or to set limits on the light quark Yukawa couplings
or other possible light contributions, similarly to the techniques using the \onshell\ \Hboson transverse 
momentum~\cite{deFlorian:2016spz,Bishara:2016jga}.
In all cases, the CP-odd component's interference with the background is zero
when integrated over the other observables. The actual analysis of the data will
benefit from employing the full kinematics using the matrix-element approach. 

\begin{figure}[t]
\centerline{
\epsfig{figure=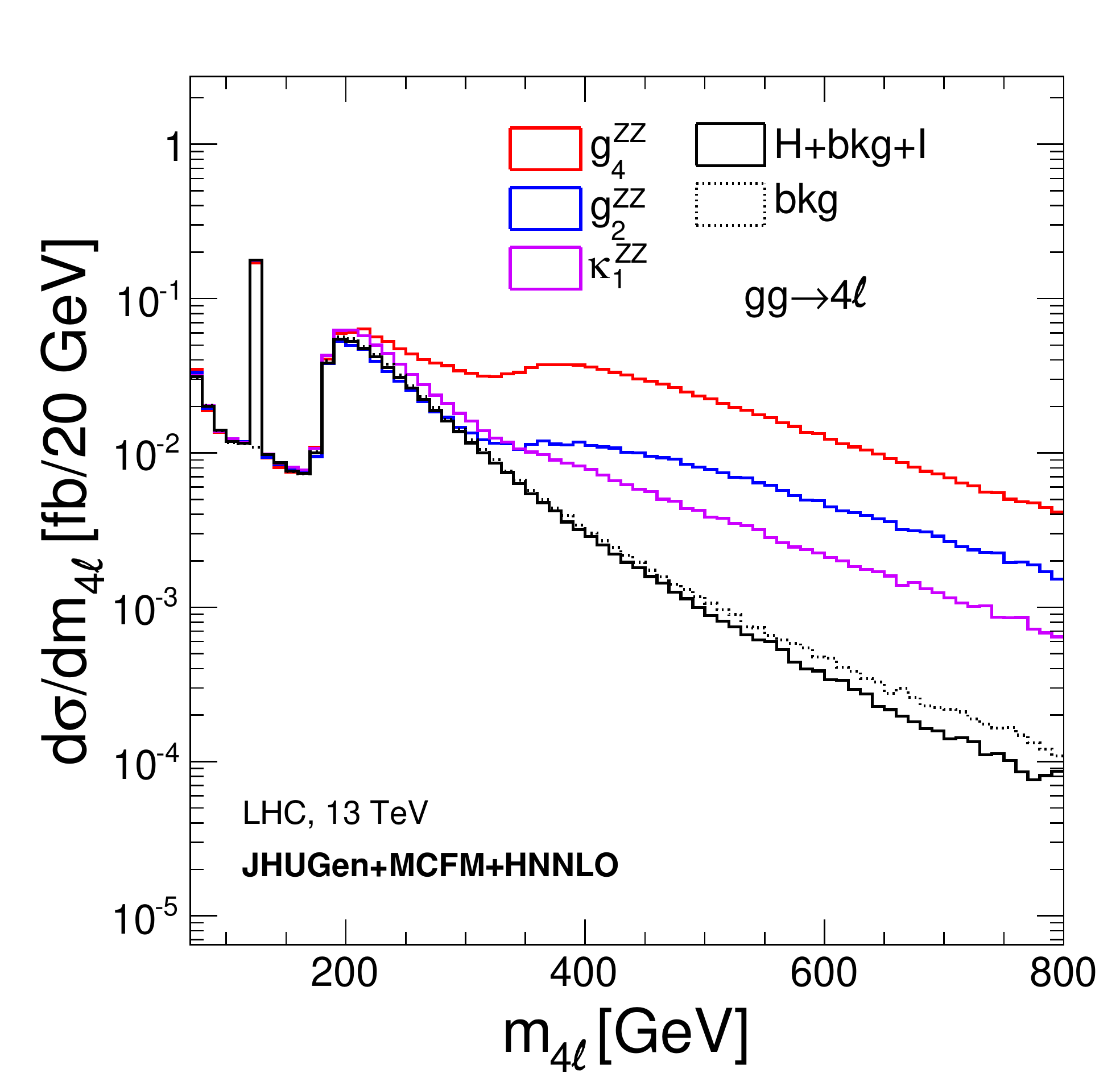,width=0.35\linewidth}
\epsfig{figure=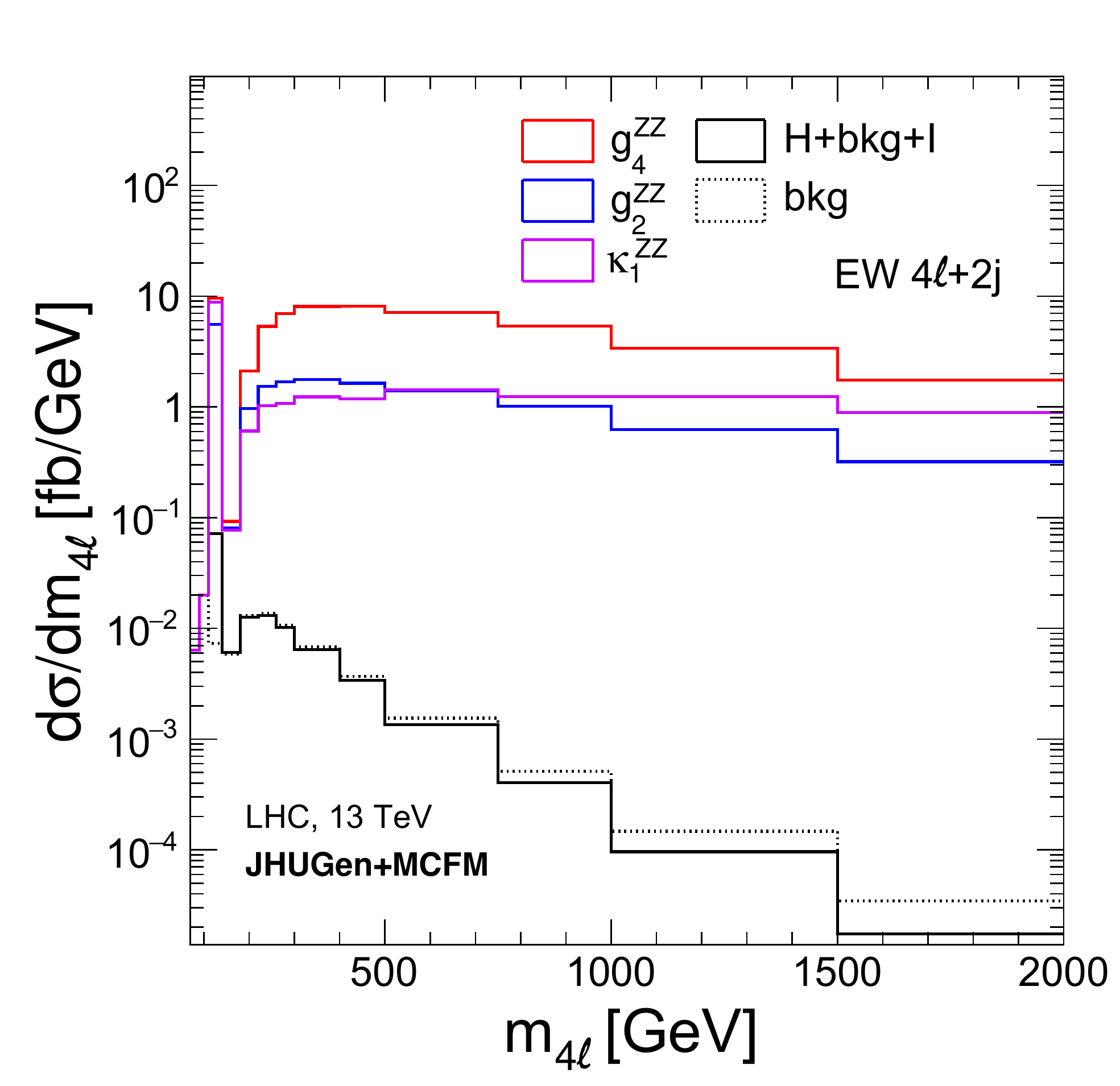,width=0.35\linewidth}
}
\captionsetup{justification=centerlast}
\caption{
The four-lepton $4\ell$ invariant mass distributions in gluon fusion (left, $4\ell=2e2\mu$) and in associated electroweak production 
with two jets (right, $\ell=e,\mu,\tau$) at the LHC with a 13\,TeV proton collision energy. The anomalous $HVV$ couplings are 
simulated with JHUGen+MCFM at LO in QCD, with the gluon fusion simulation settings and k factors matching those 
for SM in Fig.~\ref{fig:kinematics-ggH-offshell-1}. Three anomalous $HVV$ couplings are modeled 
with coupling values chosen to match the SM on-shell $gg\to H\to 2e2\mu$ cross section. 
Interference with the SM  $gg\to 4\ell$ (left) and electroweak (right) background is included. 
}
\label{fig:kinematics-offshell}
\end{figure}

\subsection{\Offshell\ effects due to HVV anomalous couplings}
\label{offshell_hvv}

The \offshell\ production of the \Hboson also allows testing the anomalous $HVV$ couplings of the \Hboson to 
two electroweak bosons, $VV=WW, ZZ, Z\gamma, \gamma\gamma$. These couplings appear in the decay $H\to 4f$
in the gluon fusion process and in both production and decay in the electroweak process. The latter includes both VBF
and $VH$ production, and in all cases interference with the gluon fusion or electroweak background is included. 
Examples of such a simulation are shown in Fig.~\ref{fig:kinematics-offshell}. Three anomalous couplings are 
shown for illustration, $g_4^{ZZ}=g_4^{WW}$, $g_2^{ZZ}=g_2^{WW}$, and $\kappa_{1,2}^{ZZ}=\kappa_{1,2}^{WW}$, 
which involve interplay of either the \Hboson\ or the $Z$ ($W$) boson going off shell. The anomalous couplings 
of the \Hboson\ to the photon are not enhanced off-shell and are not shown here, but can be considered in analysis. 
Therefore, it is important to stress here that it is natural to use the physical Higgs basis in the EFT analysis 
of the \offshell\ region, since the behavior of the couplings involving the photon is drastically different. 

Examples of applications of the tools developed here, both simulation and MELA discriminants, can be found 
in Refs.~\cite{Sirunyan:2019twz,Cepeda:2019klc} where simultaneous analysis of the \Hboson width and 
the couplings is performed both with the current LHC data and in projection to the HL-LHC. For example, 
with 3000\,fb of data, a single LHC experiment is expected to constrain 
$\Gamma_{\rm tot}=4.1^{+1.1}_{-1.1}$\,MeV, as shown in Fig.~106 of Ref.~\cite{Cepeda:2019klc}. 
With the current data sample from the LHC experiments, the \offshell\ region significantly improves
the anomalous coupling constraints, even with $\Gamma_{\rm tot}$ profiled~\cite{Sirunyan:2019twz}. 
This is evident from the enhancement observed in Fig.~\ref{fig:kinematics-offshell}. 
The expected gain is not as large at the HL-LHC, as Fig.~39 of Ref.~\cite{Cepeda:2019klc} shows,
because with access to smaller couplings, the electroweak
VBF and $VH$ production in the \onshell\ region plays a more important role. 

It has been pointed out~\cite{Azatov:2014jga,Azatov:2016xik} that the $gg \to 4\ell$ process
also provides good sensitivity for constraining the top quark electroweak couplings.
Similarly, gluon fusion in $ZH$ production is sensitive to the 
same top quark couplings and can be used to constrain them~\cite{Bylund:2016phk,Englert:2016hvy}. 
In this work, however, we separate anomalous \Hboson couplings from the rest of the electroweak 
interactions where this is possible in a consistent way. 
For the above cases, separating the effects is certainly possible,
because top quark electroweak couplings can also 
be probed in e.g. $pp \to t\bar{t}Z$, which is independent of the Higgs sector. 
Moreover, there are no EFT relations between electroweak top quark couplings and \Hboson couplings. 
However, the gauge boson self-interactions and the \Hboson couplings cannot be separated
if the EFT relations Eqs.~(\ref{eq:d1}--\ref{eq:d6})  are applied in continuum electroweak production. 
We can account for these relations as discussed in the following Section~\ref{offshell_ew}.

\begin{figure}[t]
\centerline{
\epsfig{figure=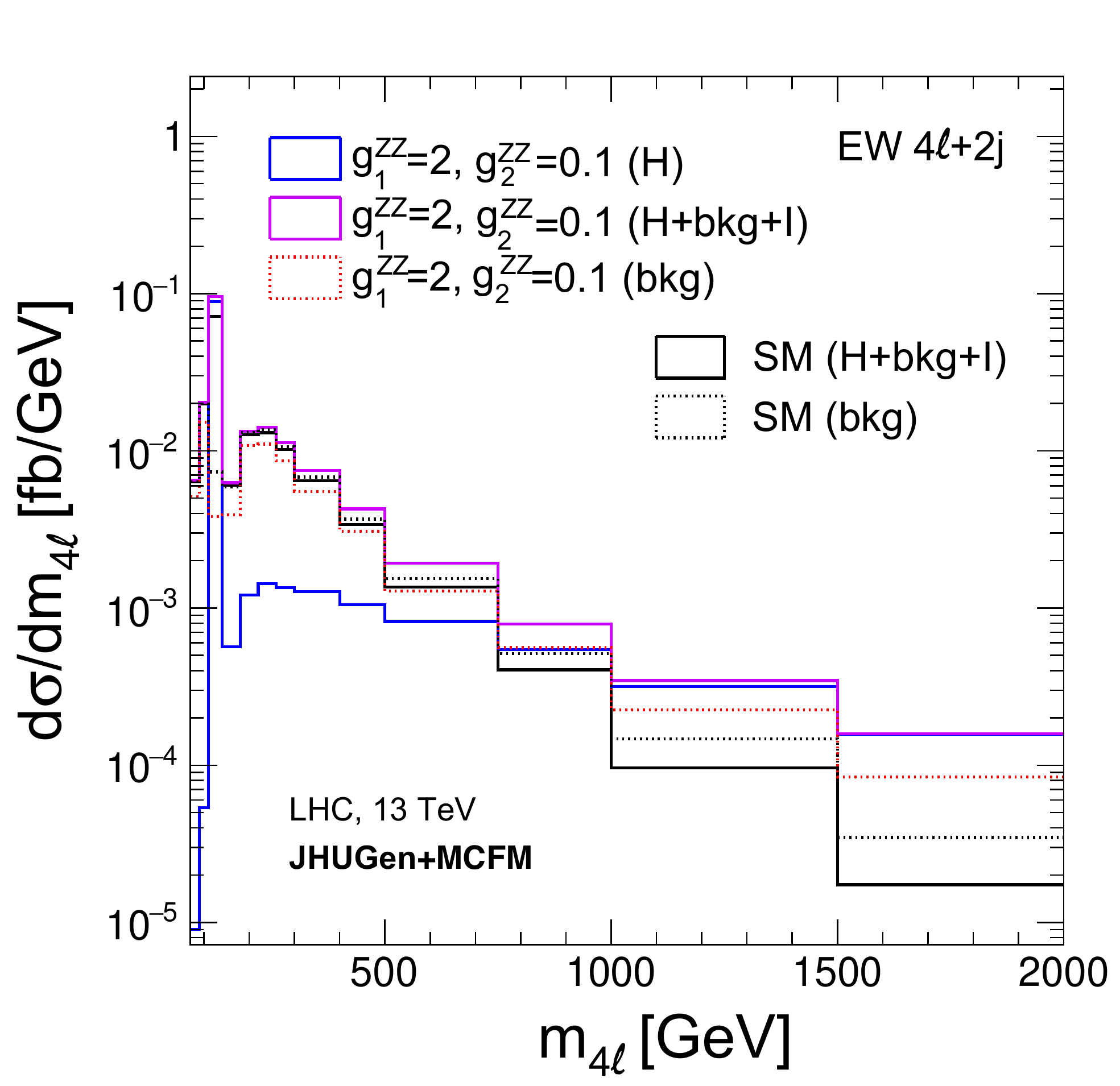,width=0.35\linewidth}
\epsfig{figure=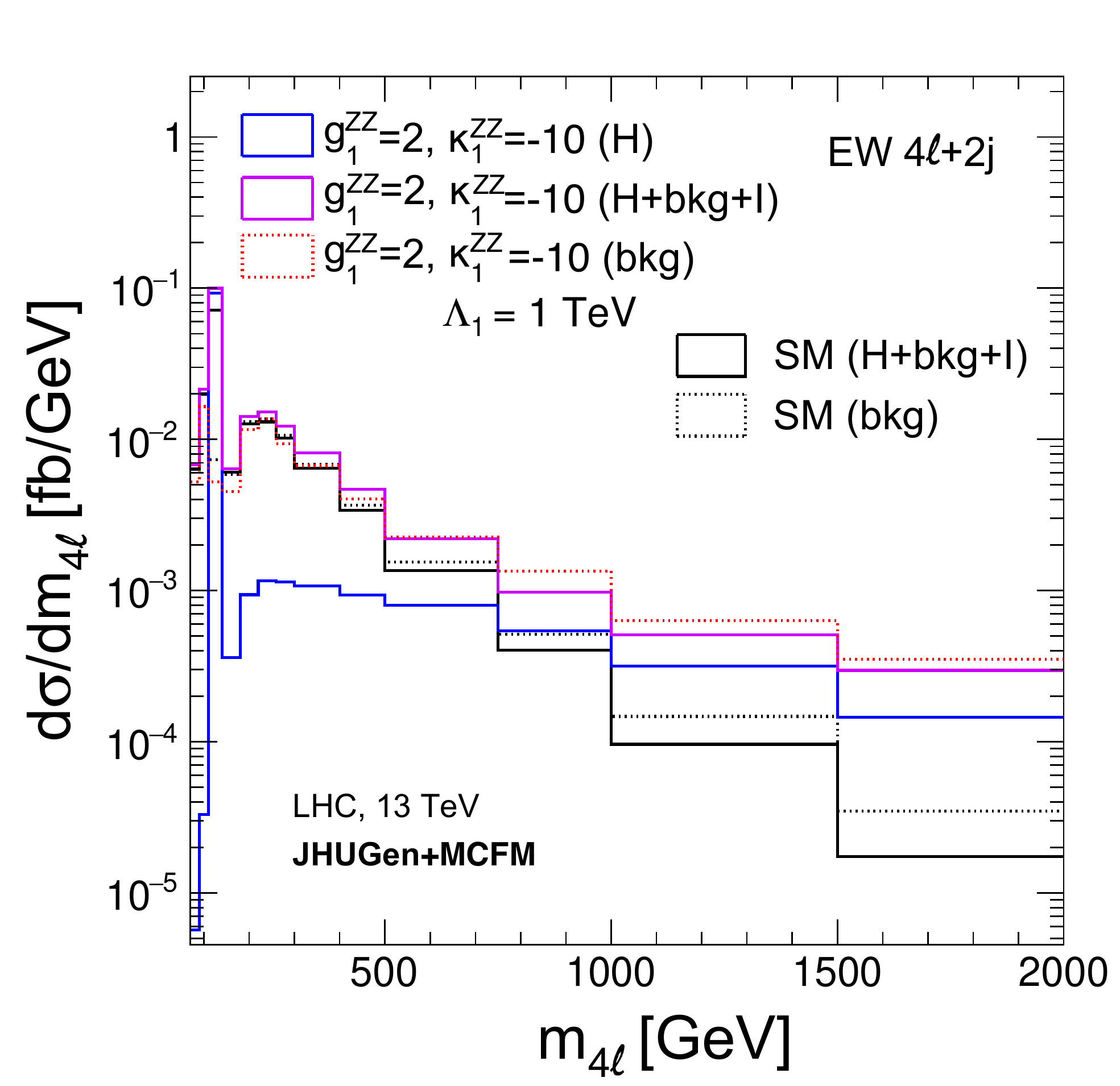,width=0.35\linewidth}
}
\captionsetup{justification=centerlast}
\caption{
The four-lepton invariant mass distributions in $4\ell$ ($\ell=e,\mu,\tau$) associated electroweak production with two jets at the LHC 
with a 13 TeV proton collision energy. The anomalous $HVV$ couplings are simulated with JHUGen+MCFM. The black distributions 
show two SM scenarios: background only (dashed) and the full contribution including the \Hboson\ (solid). 
The colored curves show an additional non-zero anomalous contribution from either $g_2^{ZZ}$ (left) or $\kappa_{1,2}^{ZZ}$ (right) 
in the \Hboson\ production component (blue solid), in the background-only component (red dashed), and including all contributions (magenta solid).
}
\label{fig:kinematics-cont-offshell}
\end{figure}

\subsection{\Offshell\ effects due to gauge boson self-interactions}
\label{offshell_ew}

In Sections~\ref{offshell_hgg} and~\ref{offshell_hvv},
only modifications of the \Hboson couplings to either strong or weak gauge bosons are considered, 
and the background contributions in either $gg\to4f$ or continuum electroweak production are assumed to be SM-like. 
However, as discussed in Section~\ref{sect:cp_couplings}, there is an intricate interplay between gauge boson self-couplings 
and \Hboson\ gauge couplings.  Therefore, under the EFT relationship, the $HVV$ anomalous contribution would affect 
the triple and quartic gauge boson self-couplings according to Eqs.~(\ref{eq:d1}--\ref{eq:d6}). 
Figure~\ref{fig:kinematics-cont-offshell} shows examples of the $m_{4\ell}$ distributions with anomalous 
gauge boson self-interactions in the EFT framework. These examples show the CP-conserving anomalous 
$g_2^{ZZ}$ and $\kappa_{1,2}^{ZZ}$ couplings, which are modified in Eqs.~(\ref{eq:deltaMW}--\ref{eq:kappa2Zgamma}) 
for the \Hboson\ couplings and in Eqs.~(\ref{eq:d1}--\ref{eq:d6}) for the gauge boson self-couplings, 
keeping all the other anomalous couplings at zero. 
The size of the anomalous contribution is taken to be similar to the current constraints
on anomalous \Hboson\ couplings~\cite{Sirunyan:2019twz} from LHC measurements. 
It is evident from Fig.~\ref{fig:kinematics-cont-offshell} that the resonant and nonresonant contributions
are of similar size in electroweak production and that there is a sizable interference between the two. 
While current analyses of LHC data typically consider the \Hboson couplings and gauge boson self-interactions
separately~\cite{Sirunyan:2019twz}, the unified framework will allow future joint constraints.

\section{Application to the ZH process at next-to-leading order}
\label{sect:cp_vh}

Production of the \Hboson in association with an electroweak gauge boson is known for its clean experimental signature and its excellent 
sensitivity to the $HVV$ couplings. During Run-II of the LHC, the experimental precision of $ZH$ 
analyses~\cite{Khachatryan:2016vau,Sirunyan:2018koj,Aad:2019mbh,Sirunyan:2018kst,Aaboud:2018zhk} 
has reached a level of accuracy that requires theory simulation beyond leading order. 
Therefore, we account for the dominant perturbative corrections at next-to-leading order QCD in this work and make them available in the JHUGen framework. 

In addition to reducing theoretical uncertainties, the simulation at higher orders also reveals sensitivity to \Hboson couplings that are invisible at the lowest order. 
In $ZH$ production, the $gg \to ZH$ sub-process enters for the first time through one-loop diagrams. 
The box diagram contribution in Fig.~\ref{diag:ppvh} yields sensitivity to the $\Hff$ couplings $\kappa_f$ and $\tilde\kappa_f$ in Eq.~(\ref{eq:Hffcoupl}), 
which are screened in the $q\bar{q}$ production process. 

We must stress that whenever we work with $ZH$ production, $\gamma H$ and $\gamma^*H$ production are 
equally important. This allows us to set constraints on anomalous $H\gamma\gamma$ and $HZ\gamma$ couplings, 
and we provide this functionality in the JHUGen framework. 

\begin{figure}[t]
\includegraphics[width=0.4\textwidth]{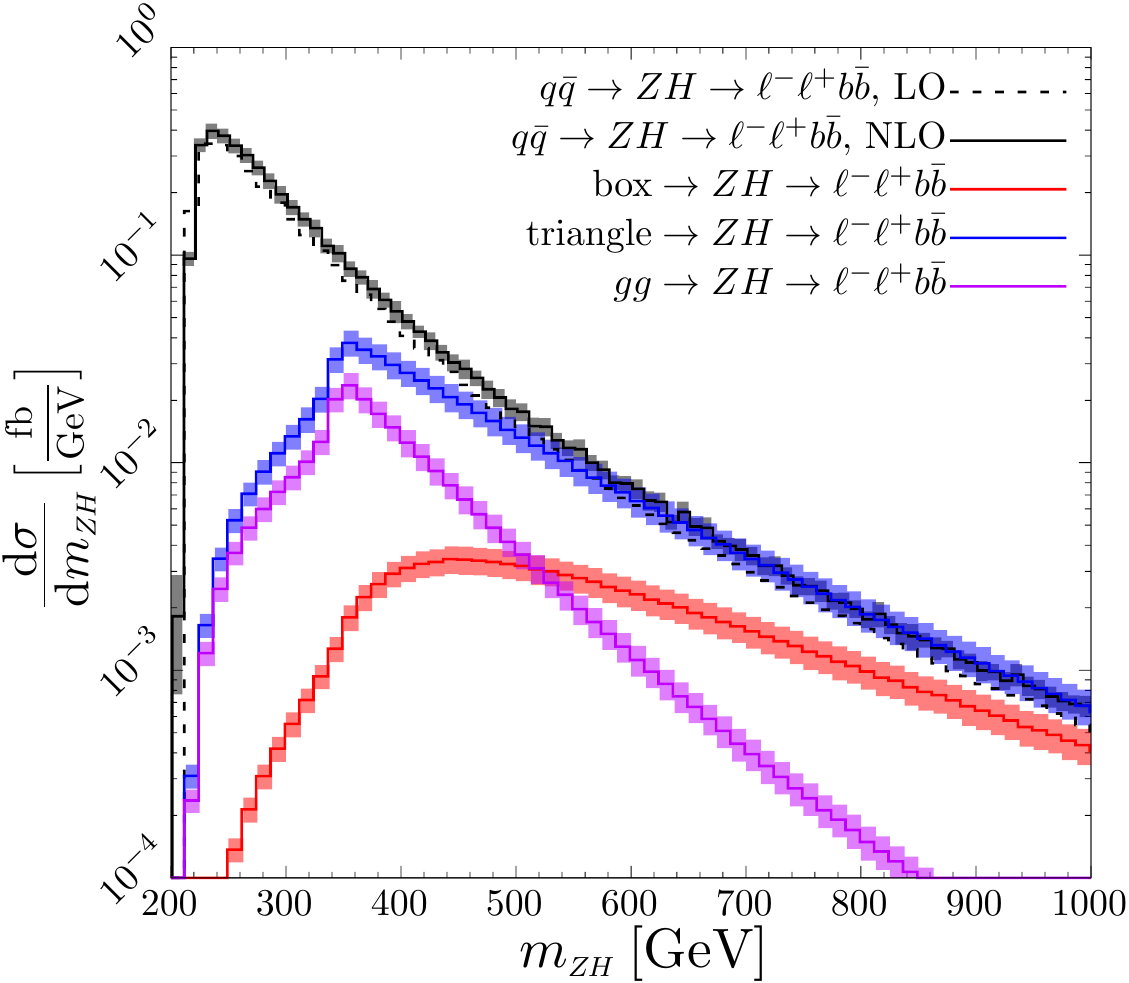}
\hspace{10mm}
\includegraphics[width=0.43\textwidth]{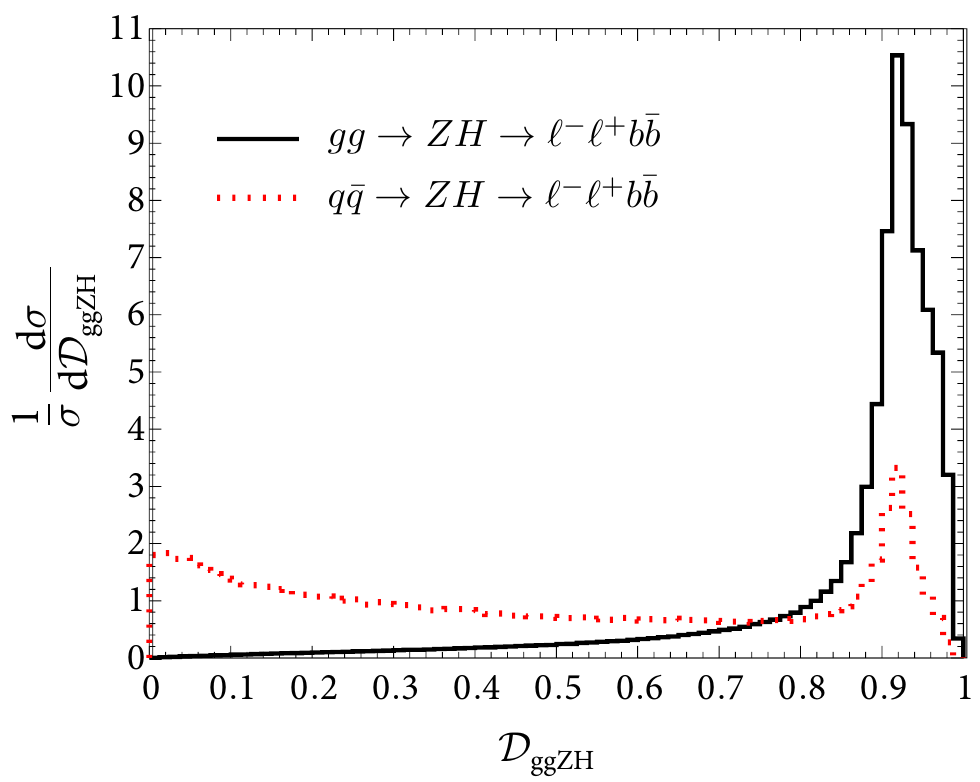}
\captionsetup{justification=raggedright}
\captionsetup{justification=centerlast}
\caption{Distribution of the $m_{ZH}$ invariant mass (left) and ${\cal D}_{\rm ggZH}$ discriminant (right)
in $gg$ and $q\bar{q}\to ZH \to \ell^-\ell^+ b \bar b$ processes under the SM hypothesis in 14 TeV LHC collisions, with several components isolated for illustration. 
The left plot shows the differential cross section for the $q\bar{q}\to ZH$ process at both LO and NLO in QCD. 
}
\label{diag:mZHSM}
\end{figure}

In Fig.~\ref{diag:mZHSM} we show the SM $m_{ZH}$ distribution assuming the decays $Z\to \ell^+\ell^- $ and $H\to b\bar{b}$  at the $14$~TeV LHC.
Contributions from $q\bar{q}$ and $gg$ initial states are shown separately, as are the triangle and box diagram parts of the $gg$ partonic process,
shown in Fig.~\ref{diag:ppvh}. 
The widths of the bands correspond to systematic uncertainties from varying the scale by a factor of two around its central value $\mu_0=m_{ZH}$. 
Close to the production threshold at $m_{ZH}= m_Z+m_H \approx 220$~GeV, the $q\bar{q}$ initial state dominates the cross section. 
Above the $2 \, m_t \approx 345$~GeV  energy, the top-quark induced $gg$ initial state becomes much more relevant. 
In particular, the triangle loop contribution, shown in blue, becomes as large as the $q\bar{q}$ contribution, shown in black, and even exceeds it at very high energies. 
However, as can be seen from the purple band in Fig.~\ref{diag:mZHSM} the quantum interference between the triangle and box diagrams is strongly destructive and 
reduces the overall $gg$ contribution significantly. 
For this reason, the $gg$ contribution plays only a marginal role in the SM description of the $ZH$ process. 

However, in anomalous coupling studies of physics beyond the SM, this strong destructive interference can be perturbed and lead to significantly larger cross sections. 
Hence, it is a sensitive probe of modifications from the SM.
The $gg\to ZH$ process has yet another interesting feature. 
A superficial inspection of the triangle loop contribution in Fig.~\ref{diag:mZHSM} suggests sensitivity to the $HVV$ couplings in Eq.~(\ref{eq:HVV}). 
Yet, an explicit calculation shows that the contributions from $g_2^{VV}$ and $g_4^{VV}$ drop out and yield a zero contribution to the complete squared one-loop amplitude.
Moreover, both the triangle and the box diagrams are only sensitive to the axial-vector coupling of the gauge boson to the fermions in the loop. 
Hence, photons do not couple to the closed fermion loop.
The process $gg \to \gamma H$ can only proceed through an intermediate $Z^*$, 
which decays into $\gamma H$ via the $\kappa_2^{Z\gamma}$ coupling in Eq.~(\ref{eq:HVV}).
Obviously, $gg\to WH$ does not exist because of charge conservation.
The absence of sensitivity to anomalous $HVV$ couplings $g_2^{VV}$ and $g_4^{VV}$ does not render the $gg\to ZH$ process completely irrelevant to their study.
If we assume, for example, that the $HZZ$ interaction involves non-zero $g_2^{ZZ}$,
which does not contribute to the triangle diagrams, the pattern of destructive interference with the box diagram would change. 
Hence, there is a strong sensitivity to the $HVV$ couplings, which is entangled with possible anomalous values of the $\Hff$ couplings. 

\begin{figure}[t]
\centering
\epsfig{figure=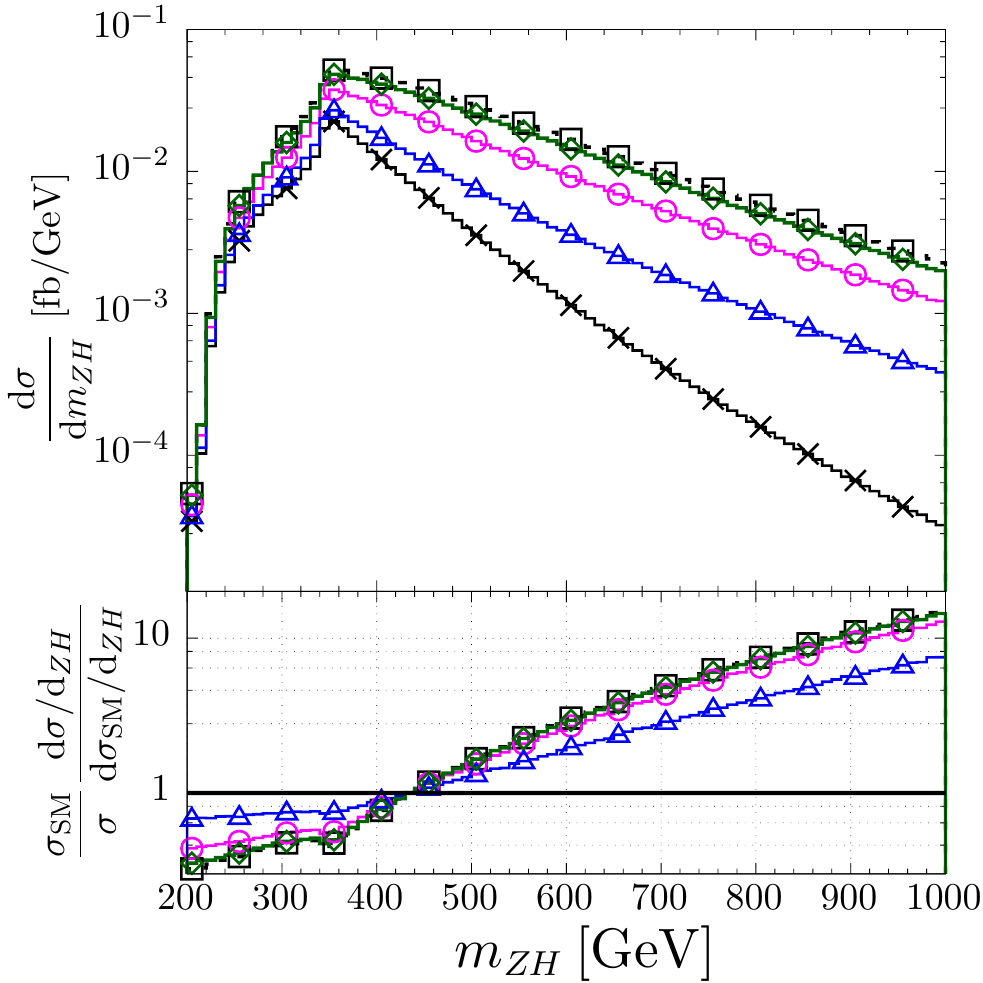,width=0.32\linewidth} \hfill
\epsfig{figure=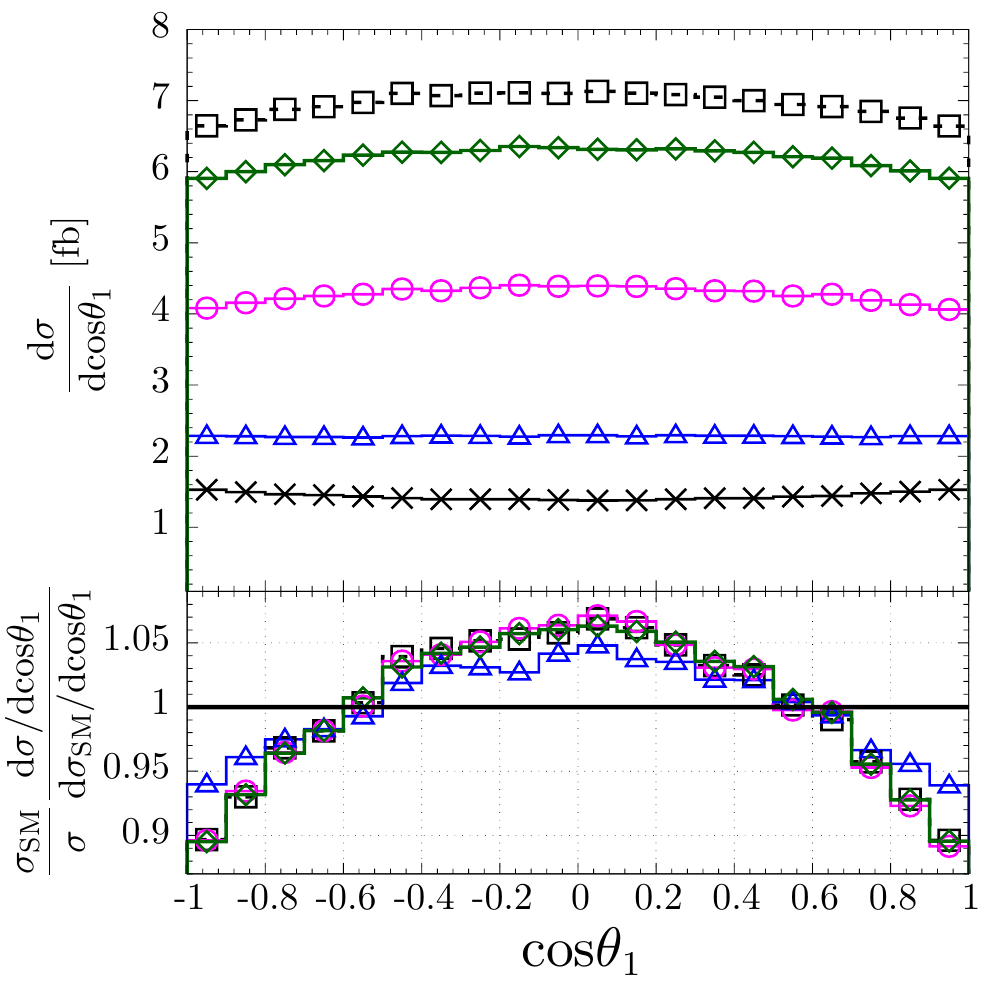,width=0.32\linewidth}\hfill
\epsfig{figure=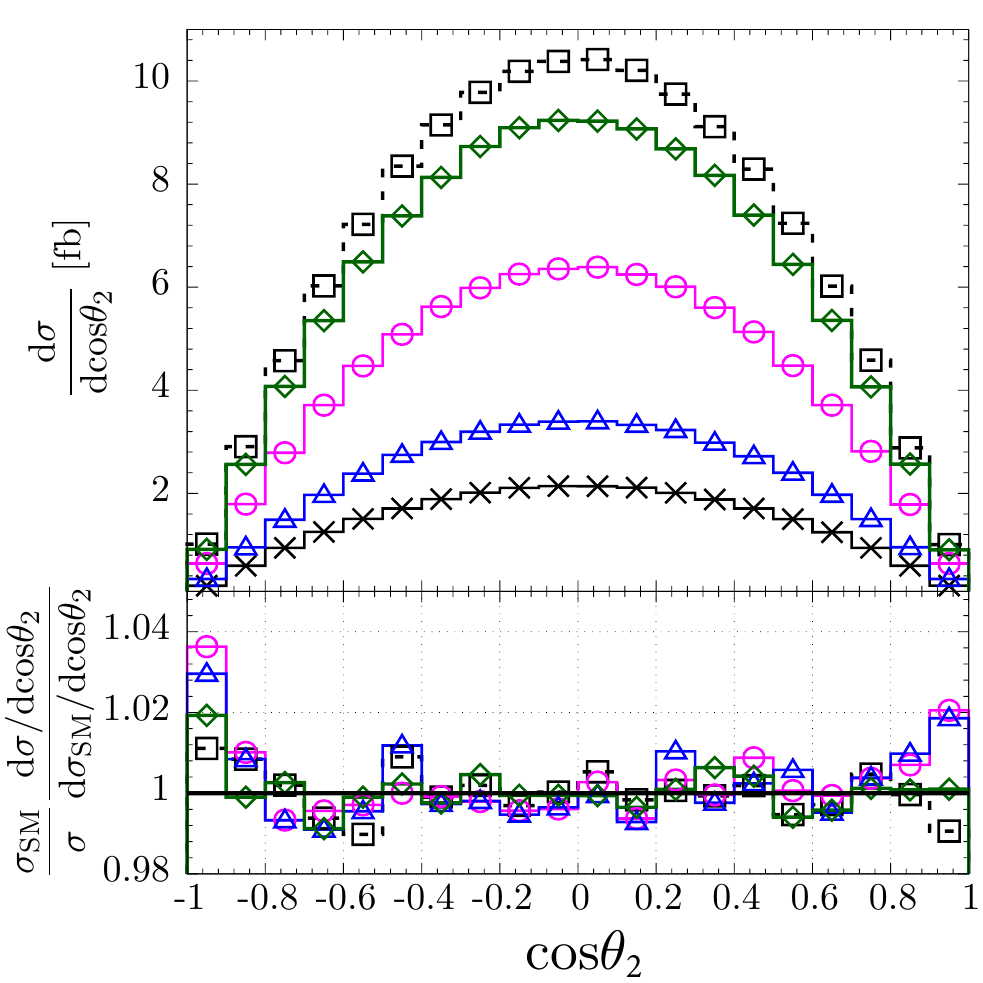,width=0.32\linewidth}
\captionsetup{justification=centerlast}
\caption{Kinematic distributions for simulated $gg \to ZH \to \ell^-\ell^+ b \bar b$ events at 14 TeV for several scenarios:
$\kappa=1,\tilde\kappa=0$ (SM, black crosses);
$\kappa=-1,\tilde\kappa=0$ (black boxes); 
$\kappa=0,\tilde\kappa=\pm 1$ (magenta circles); 
$\kappa=1/\sqrt{2},\tilde\kappa=\pm1/\sqrt{2}$ (blue triangles); 
$\kappa=-1/\sqrt{2},\tilde\kappa=\pm1/\sqrt{2}$ (green diamonds).}
\label{fig:gg_ZH}
\end{figure}

\begin{figure}[t]
\centering
\epsfig{figure=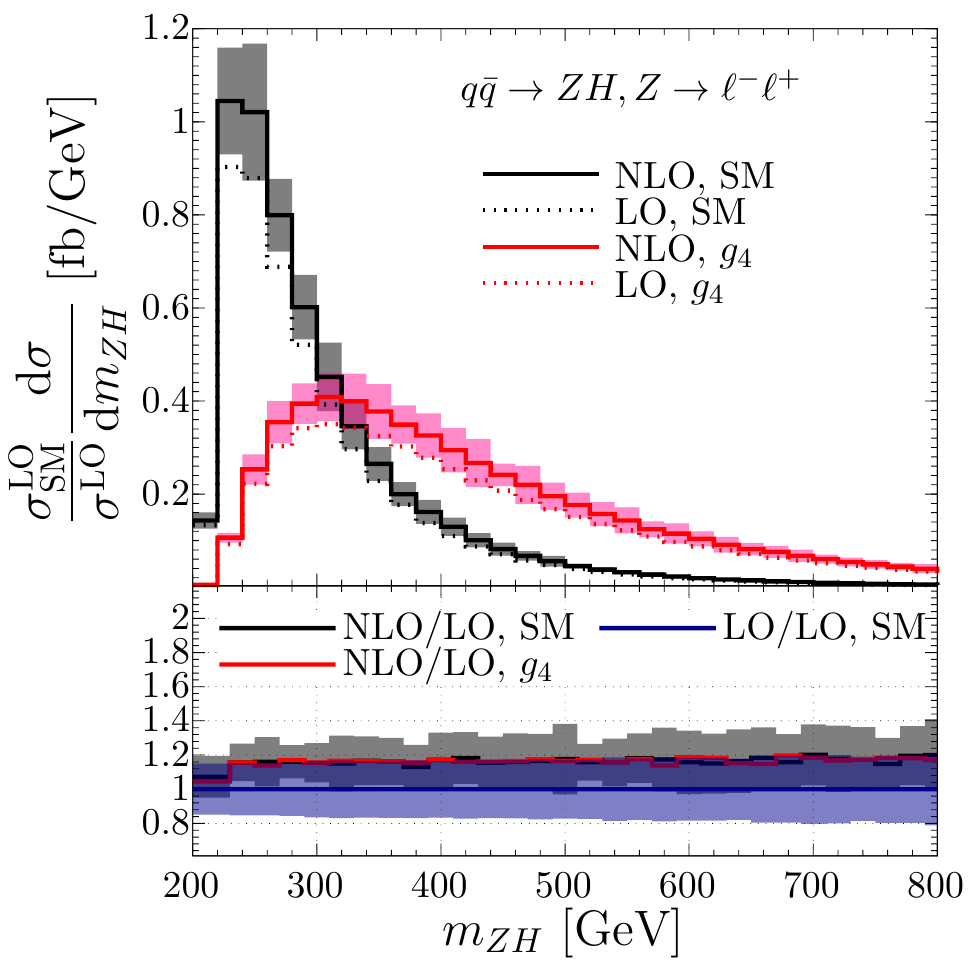,width=0.32\linewidth} \hspace{10mm}
\epsfig{figure=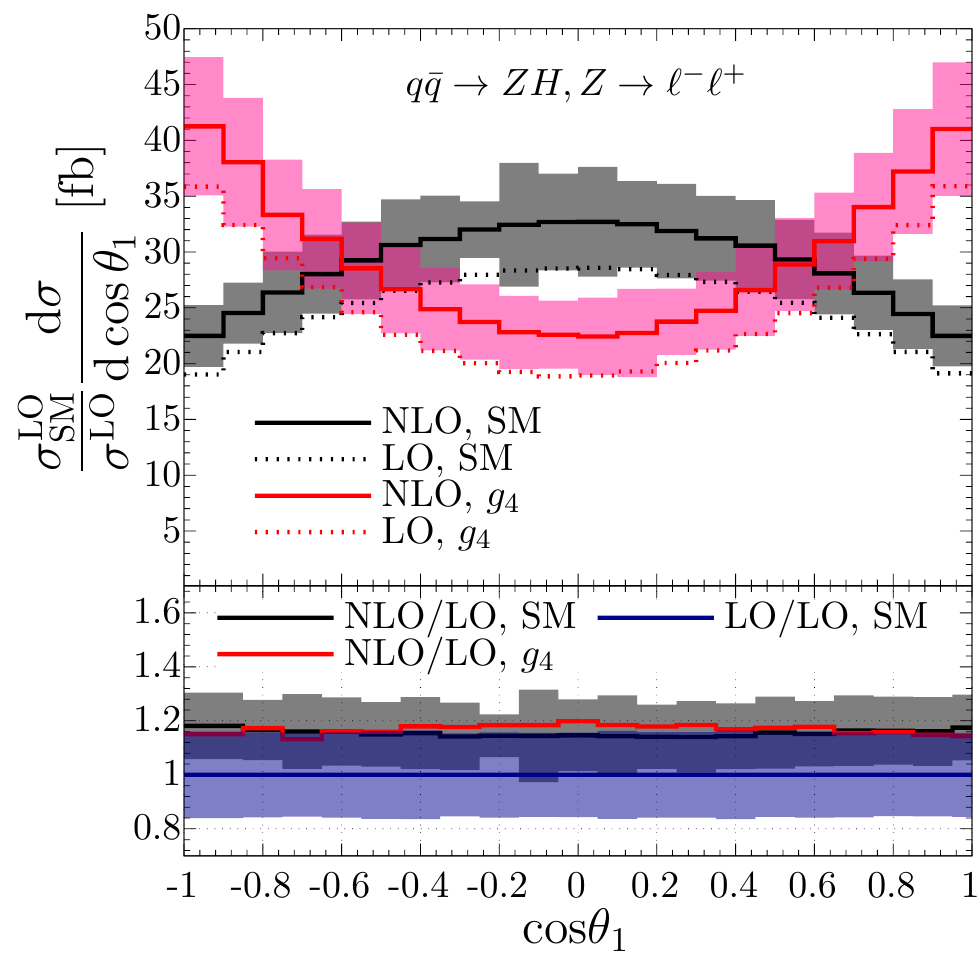,width=0.32\linewidth}
\\
\epsfig{figure=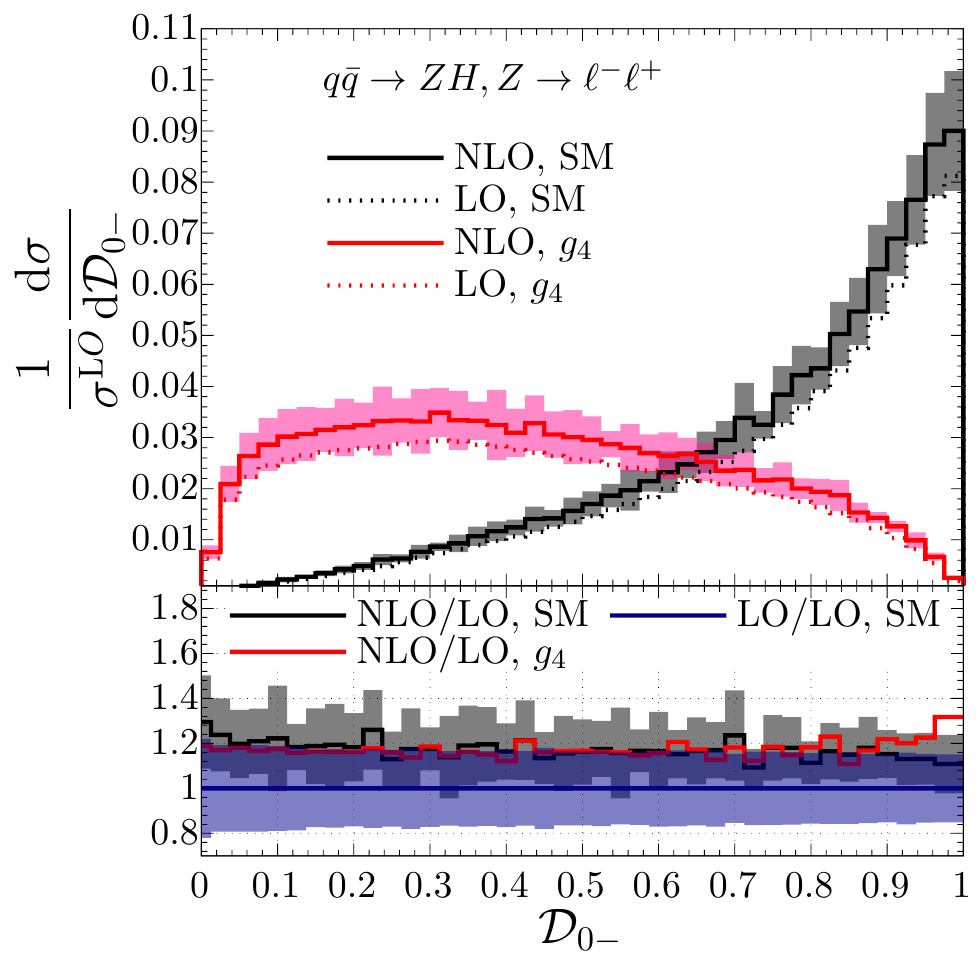,width=0.33\linewidth} \hspace{10mm}
\epsfig{figure=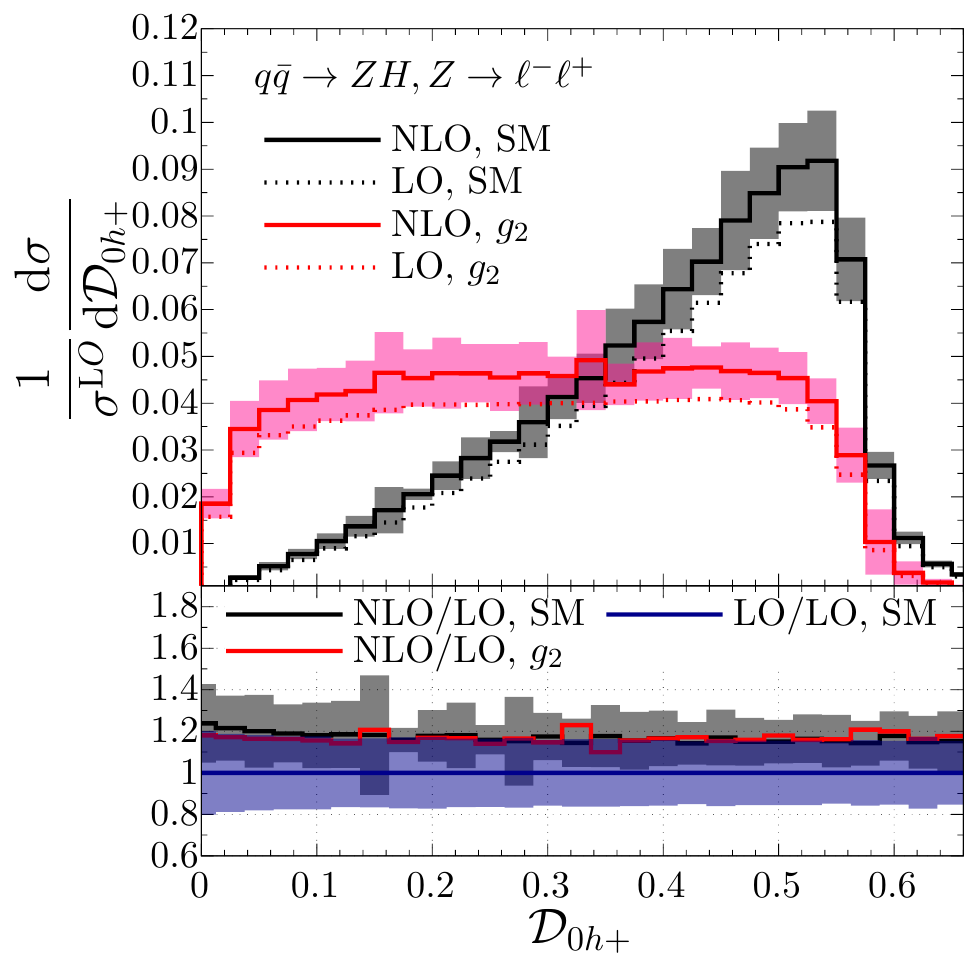,width=0.33\linewidth}
\captionsetup{justification=centerlast}
\caption{Selected kinematic distributions for simulated $q \bar q \to ZH \to \ell^-\ell^+ b \bar b$ events at 14 TeV
shown at LO (dotted) and NLO (solid) in QCD. The SM (black) and anomalous coupling model (red) are shown.
The anomalous coupling model shown is the pseudoscalar model $f_{g4}=1$ in all plots except for the ${\cal D}_{0h+}$ discriminant distribution,
where the $f_{g2}=1$ model is shown instead. The bottom panels show the $k$-factor ratios. 
}
\label{fig:VH_NLO}
\end{figure}

The above mentioned special features of the loop-induced $gg$ process motivate separating it from the $q\bar{q}$ production mode. 
While it is not feasible to isolate one process from the other on an event-by-event basis, it is possible to enhance (or decrease) 
the relative fraction of the two processes. 
We use the MELA approach with the ${\cal{D}_{\rm ggZH}}$ discriminant calculated according to Eq.~(\ref{eq:melaAlt})
with ${\rm sig}=gg \to ZH$ and ${\rm alt}=q\bar q \to ZH$. The distributions of ${\cal{D}_{\rm ggZH}}$ for the
 $gg$- and $q\bar q$-initiated processes are shown in Fig.~\ref{diag:mZHSM}. 
 As an example, in a restricted range of the ${\cal{D}_{\rm ggZH}}$ observable which keeps 80\% of the $gg$-initiated process,
 the fraction of this process is enhanced from 7\% to 22\%. As described in Section~\ref{sect:exp_kinematics},
this approach is superior to selection based on individual kinematic observables, such as
selecting higher values of $m_{ZH}$ or $p_\mathrm{T}^H$~\cite{Hespel-2015-gg}.

In the following, we present anomalous coupling results for the $gg$ and $q\bar{q}$ processes separately. 
This serves to illustrate the particular anomalous coupling features of our framework.
A full experimental result needs to include both processes together. 
In Fig.~\ref{fig:gg_ZH}, we show the effects of several combinations of anomalous $Ht\bar{t}$ couplings in the $gg \to ZH$ process and compare
the shape changes to the SM prediction. 
We use the general interaction structure from Eq.~(\ref{eq:ampl-spin0-qq}) with CP-even ($\kappa$) and CP-odd ($\tilde\kappa$) 
Yukawa-type couplings and consider the scenarios with a \textit{wrong sign} Yukawa coupling, pure CP-odd couplings, and mixtures of CP-even 
and CP-odd couplings.  
In Fig.~\ref{fig:gg_ZH}, we show the $ZH$ invariant mass and the two angles $\theta_1$ and $\theta_2$.
The angles are defined for the $VH$ process in Fig.~\ref{fig:kinematics} and Ref.~\cite{Anderson:2013afp}, but we note that
it is possible to define the sign of $\cos\theta_1$ when $\theta_1$ is the angle between the 
$Z$ boson and the longitudinal direction of the overall boost of the $V^* \to ZH$ system, defined in the $V^*$ rest frame. 
In the case where the $VH$ system has finite transverse momentum, 
we first boost the system in the transverse direction to set the transverse motion to zero. 
For a practical application of this approach at the LHC, see Ref.~\citep{Chatrchyan:2011ya}.

In the upper row of Fig.~\ref{fig:VH_NLO} we present the $q\bar{q} \to ZH $ process and compare the leading order with the next-to-leading QCD prediction.
The black and red curves correspond to the SM $HZZ$ coupling and a pure CP-odd coupling from $g_4^{ZZ}$ in Eq.~(\ref{eq:HVV}), respectively. 
Shape changes due to the different coupling structure are significant, even on a logarithmic scale. 
Hence, this process offers strong discrimination power even for small admixtures of $g_4^{ZZ}$ into the SM-like $g_1^{ZZ}$ coupling structure. 
The higher order corrections, shown in the differences between the solid and dotted curves and in the lower panes, are positive, fairly constant, and $\mathcal{O}(+10\%)$. 
In the lower row of Fig.~\ref{fig:VH_NLO}	we show the matrix element discriminants 
${\cal D}_{0h+}$ and ${\cal D}_{0-}$, defined in Eq.~(\ref{eq:melaAlt}), for 
the alternative hypotheses  $g_2^{ZZ}=1$ and $g_4^{ZZ}=1$, respectively. 
Again, the NLO QCD corrections are fairly constant over a wide range. 
The plots show strong discrimination power between each anomalous hypothesis and the SM. 
The results allow for a more accurate estimate of systematic uncertainties in future analyses,
and the flat correction reinforces previous leading order studies.

\section{Application to new resonance production}
\label{sect:exp_newres}

The techniques developed for the study of the $H(125)$ boson would apply to a search for or a study of a new 
resonance $X(m_X)$ which may arise in the extensions of the SM,
such as any Singlet model or Two Higgs Doublet model. 
For example, if any enhancement or modification of the di-boson spectrum or kinematics in the off-shell region is observed,
one would have to determine the source of this effect.
For example, it might come from a modification of the \Hboson couplings in the off-shell region, including anomalous tensor
structures; a modification of the continuum production, possibly from anomalous self-interactions;
or yet another resonance $X$ with a larger mass. This latter scenario is necessary 
to consider in order to complete the experimental studies. 

If a new state $X$ is observed, one would need to determine its spin and parity quantum 
numbers in all accessible final states. The techniques discussed in Section~\ref{sect:exp_onshell}
would be directly relevant. If the width of the resonance is sizable, interference with background, 
as discussed in Section~\ref{sect:exp_offshell}, will become relevant. 
Moreover, interference with the off-shell $H(125)$ boson tail would become important as well. 
All these effects are included in the coherent framework of the JHU generator 
with the modified MCFM matrix element library, and are available in the MELA package 
for MC re-weighting and optimal discriminant calculations. 
They have been employed in analyses of Run-II of LHC data~\cite{Sirunyan:2018qlb,Sirunyan:2019pqw}.

\begin{figure}[ht]
\centering
	\includegraphics[width=0.48\textwidth]{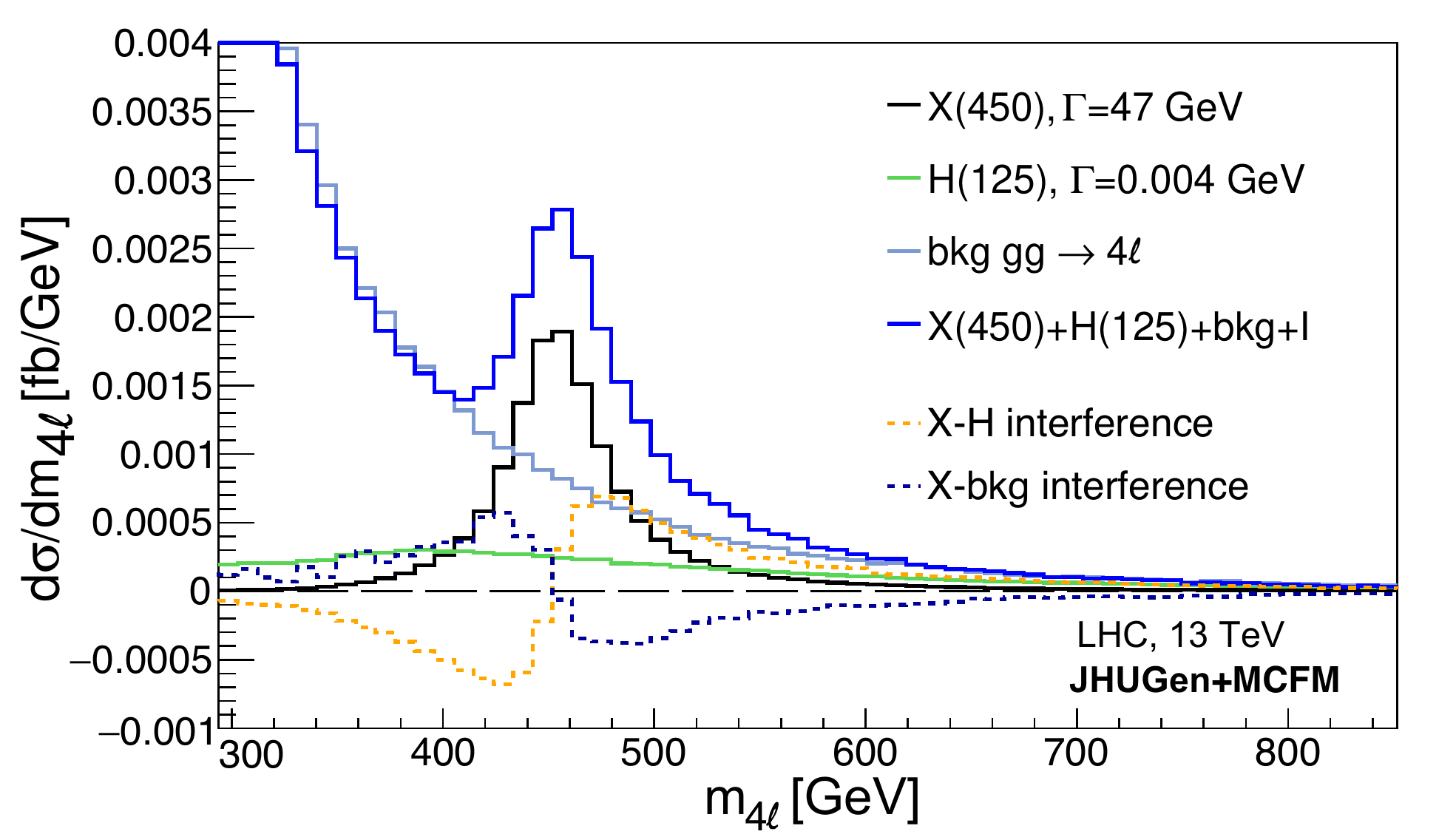}
\captionsetup{justification=centerlast}
\caption{
Differential cross section of the gluon fusion process 
$gg \to ZZ/Z\gamma^*/\gamma^*\gamma^*\to 4\ell$ as a function of invariant mass $m_{4\ell}$ generated with 
JHUGen+MCFM at LO in QCD. 
The distribution is shown in the presence of a hypothetical scalar $X(450)$ resonance with SM-like couplings,
$m_X=450$\,GeV, and  $\Gamma_X=47$\,GeV.
Several components are either isolated or combined as indicated in the legend.
Interference (I) of all contributing amplitudes is included. 
}
\label{fig:MCFM_BSM_GenLevel}
\end{figure}

\begin{figure}[ht]
\centering
	\includegraphics[width=0.48\textwidth]{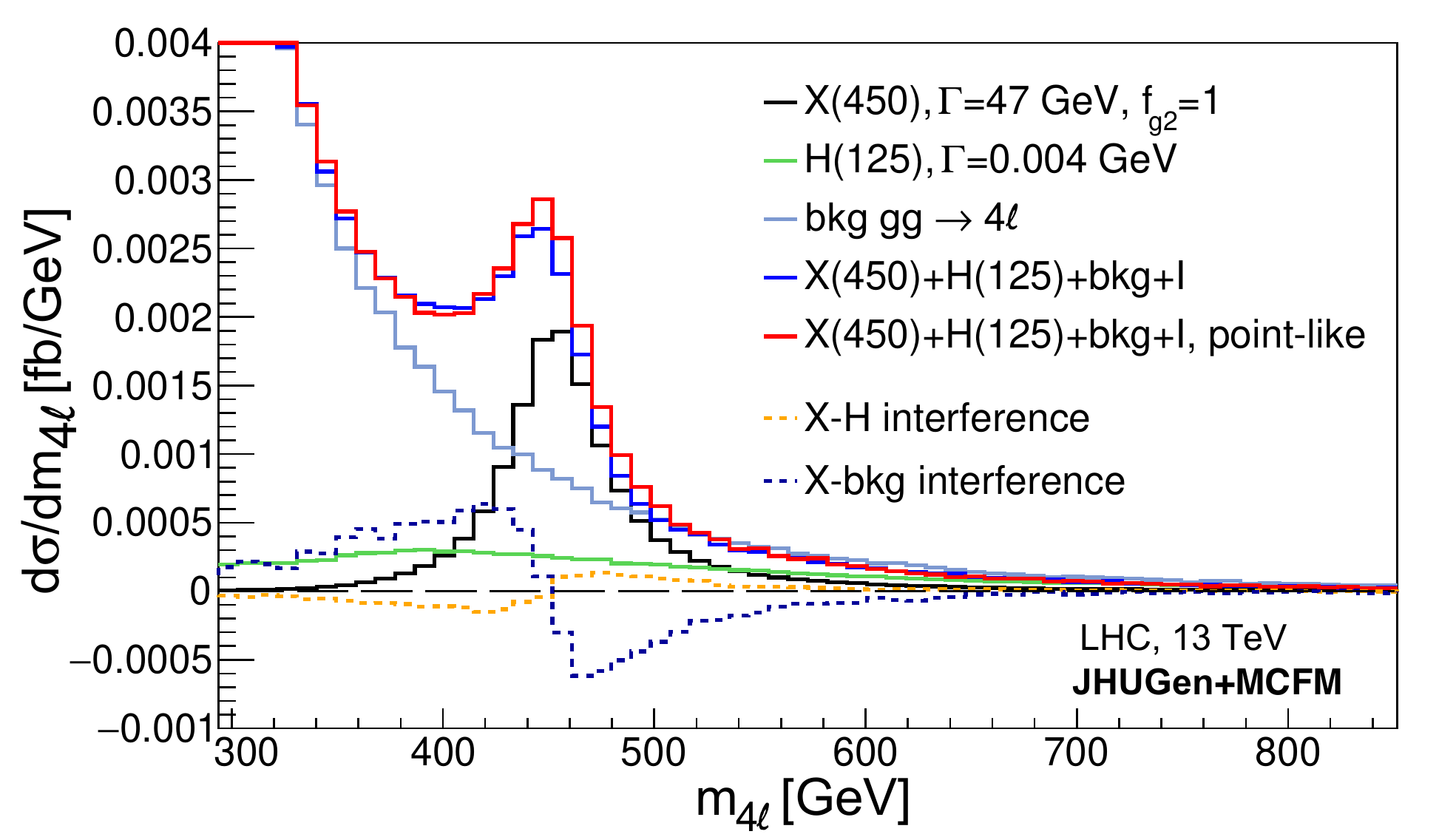}
	\includegraphics[width=0.48\textwidth]{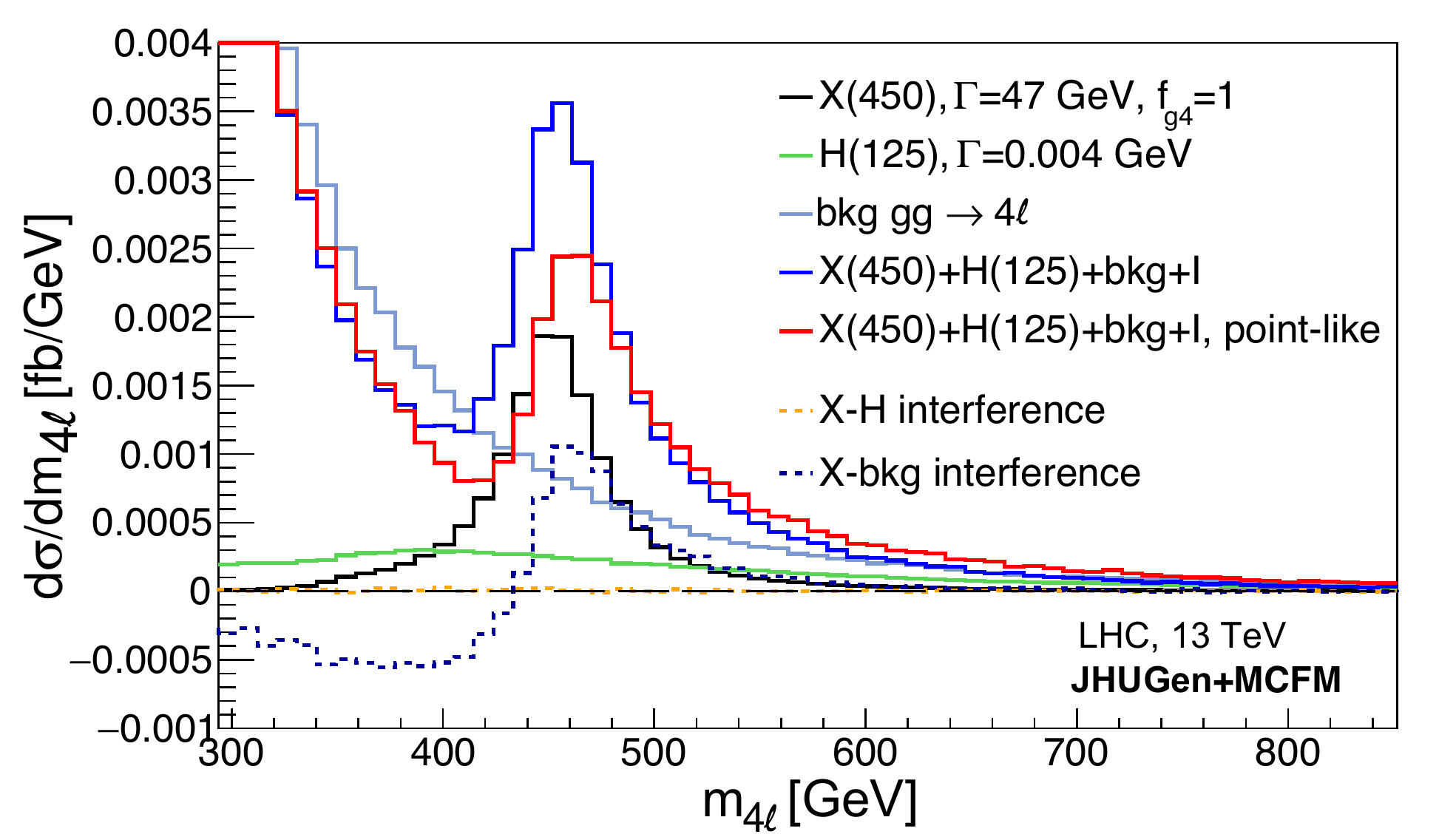}
\captionsetup{justification=centerlast}
\caption{
Same as Fig.~\ref{fig:MCFM_BSM_GenLevel}, but for the anomalous couplings of a new resonance $X(450)$.
A scalar resonance with $f_{g2}=1$ (left) and a pseudoscalar resonance with $f_{g4}=1$ (right) are considered. 
Both a top loop and a point-like interaction are considered in the gluon fusion production. 
}
\label{fig:MCFM_BSM_GenLevel_2}
\end{figure}

\begin{figure}[ht]
\centering
	\includegraphics[width=0.48\textwidth]{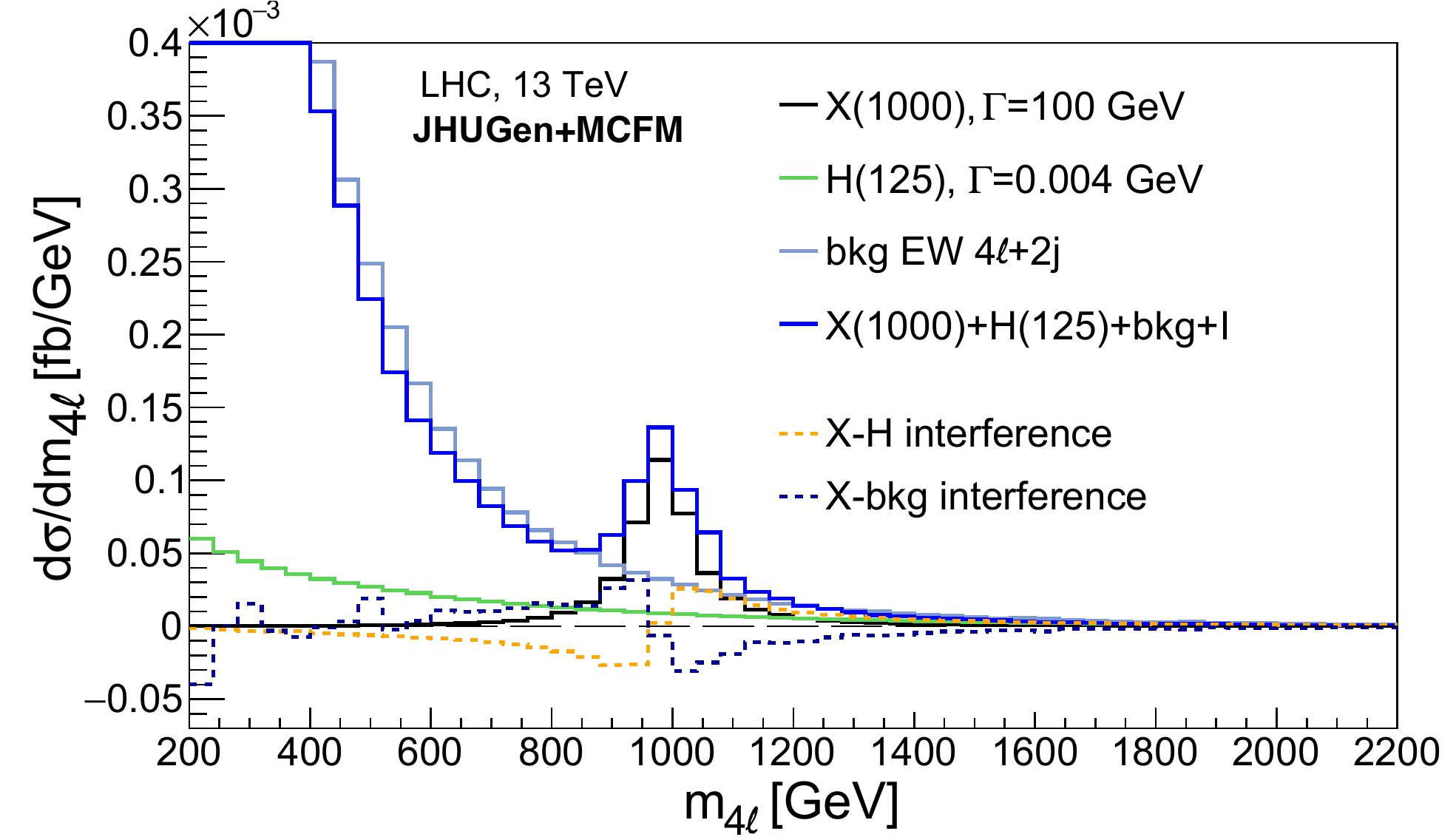}
	\includegraphics[width=0.48\textwidth]{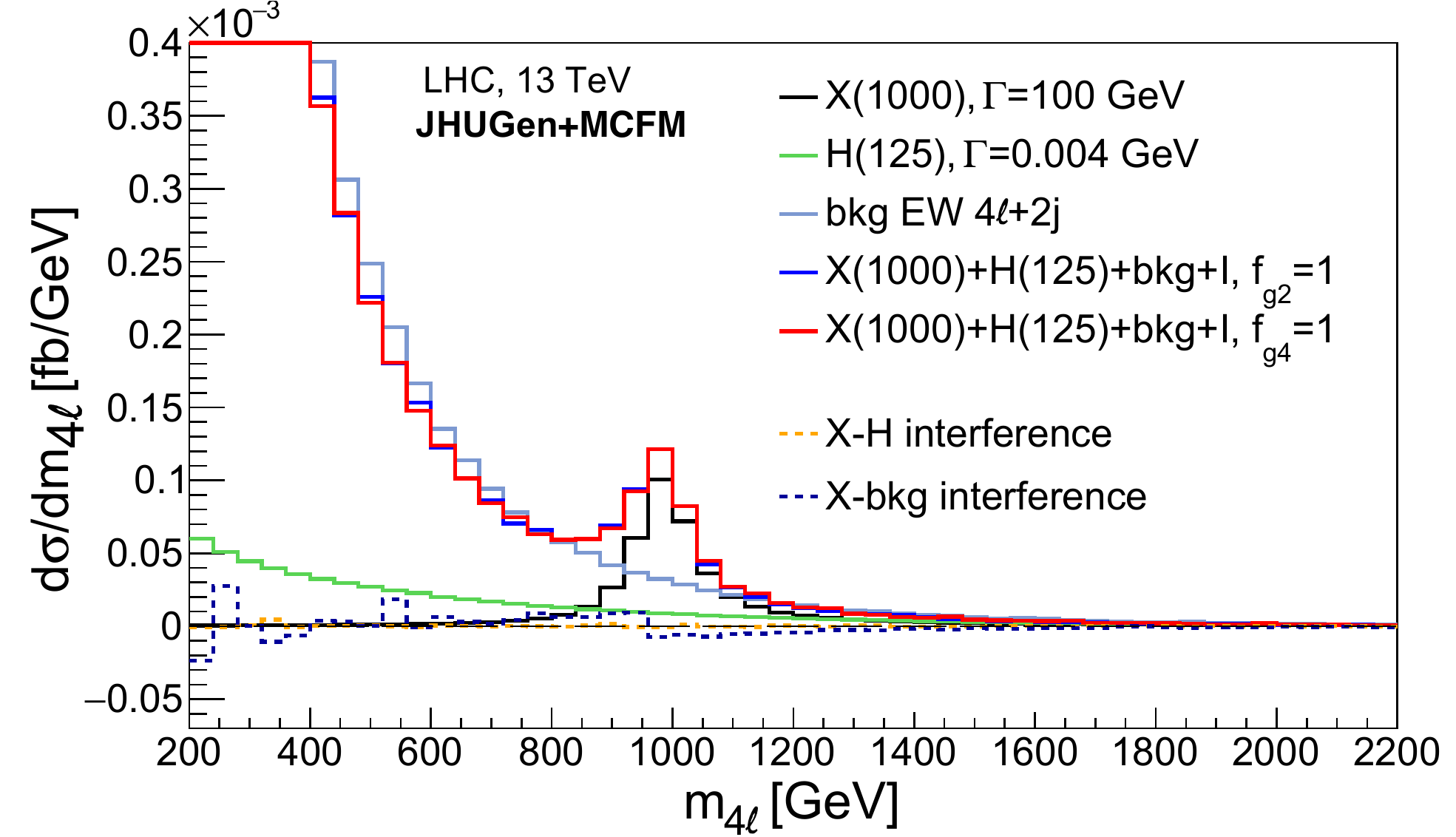}
\captionsetup{justification=centerlast}
\caption{
Differential cross section of the electroweak production process 
$qq \to qq(ZZ/Z\gamma^*/\gamma^*\gamma^*\to 4\ell)$
as a function of invariant mass $m_{4\ell}$ generated with JHUGen+MCFM. 
The distribution is shown in the presence of a hypothetical $X(1000)$ resonance with SM-like couplings (left)
and anomalous couplings ($f_{g2}=1$ and $f_{g4}=1$, right), $m_X=1000$\,GeV, and  $\Gamma_X=100$\,GeV.
Several components are either isolated or combined as indicated in the legend.
Interference (I) of all contributing amplitudes is included. 
}
\label{fig:VBF_BSM_GenLevel}
\end{figure}

Applications of off-shell $H(125)$ simulation with an additional broad $X(m_X)$ resonance
are shown in Figs.~\ref{fig:MCFM_BSM_GenLevel}, \ref{fig:MCFM_BSM_GenLevel_2}, and~\ref{fig:VBF_BSM_GenLevel}.
The cross section of the generated resonance $X$ corresponds to the limit obtained by the recent CMS search~\cite{Sirunyan:2018qlb},
which includes all interference effects of a broad resonance. 
The most general $XVV$ and $Xgg$ couplings discussed in application to $HVV$ and $Hgg$ 
in Sections~\ref{sect:exp_onshell} and~\ref{sect:exp_offshell} are possible. 
It is interesting to observe that in the scalar case, the interference of $X(m_X)$ with the $H(125)$ off-shell tail and
its interference with the background have opposite signs and partially cancel each other, but the net effect
still remains and alters the distributions. In the cases of anomalous $XVV$ couplings, 
the size of the interference changes, and in some particular cases, such as the $g_4$ coupling, 
even the sign of the interference flips.  
The point-like $Xgg$ couplings are also tested and shown in Fig.~\ref{fig:MCFM_BSM_GenLevel_2},
which models the scenario when new heavy states in the gluon fusion loop are responsible for 
production of the new state $X$. 

\section{Summary}
\label{sect:cp_summary}

In this paper, we have investigated the Higgs boson interactions at invariant masses both at and well beyond the mass peak.
We considered the weak vector boson fusion process $pp \to 4f +jj$, gluon fusion $pp \to 4f$, 
and associated production $pp \to VH$.
All three processes contribute to both the on-shell and off-shell signal regions.
NLO QCD effects, including the $gg$ initial state, are investigated in $pp \to VH$ production.
Through these processes, we study $HVV$ interactions in the regime of large momentum transfer, which exposes the unitarization 
feature in the Standard Model and is sensitive to the mechanism of electroweak symmetry breaking at high energies.
Our framework allows a general coupling parameterization for the 125 GeV Higgs boson and for 
a possible second spin-zero resonance. Modifications of the triple and quartic gauge boson couplings are also considered. 
Deviations from the SM expectation can be parameterized in terms of anomalous couplings, 
effective field theory operators, and pseudo observables. 
The framework of the JHUGen event generator and MELA library for the matrix element analysis enable simulation, 
optimal discrimination, reweighting techniques, and analysis with the most general  anomalous couplings of 
a bosonic resonance and the triple and quartic gauge boson interactions. The capabilities of the framework have been illustrated
with projections for measuring the EFT operators with the expected full data samples of the LHC and the High-Luminosity LHC. 

\bigskip

\noindent
{\bf Acknowledgments}:
We acknowledge the contributions of CMS collaboration colleagues to the MELA project development
and help with integration and validation of the JHU event generator. 
We thank Tianran Chen for updating the Hom4PS program for this analysis and providing support,
we thank Amitabh Basu for suggesting the cutting planes algorithm in application to our analysis,
we thank Michael Spira and Margarete M\"uhlleitner for help with the HDECAY and C2HDM\_HDECAY programs, 
we thank Andrew Gilbert for help with features of the Root program, 
we thank Giovanni Petrucciani for discussion of the Higgs cross section studies, 
we thank Jared Feingold for machine learning studies, 
we thank Savvas Kyriacou for help with the fit implementation,
and we thank Jeffrey Davis for help in cross section calculations and interface to different coupling conventions in the generator. 
This research is partially supported by the U.S. NSF under grants PHY-1404302 and PHY-1707887, 
by the Fundamental Research Funds for the Central Universities (China),
and by the U.S. DOE under grant DE-SC0011702.
Calculations reported in this paper were performed on the Maryland Advanced Research Computing Center (MARCC). 

\vspace{0.5cm}

Note added: A new application of this framework to LHC data appeared in Ref.~\cite{CMS-HIG-19-009}.
We would also like to point that the relative sign of the CP-even and CP-odd couplings in Eq.~(\ref{eq:g2gg})
is consistent with Refs.~\cite{Chetyrkin:1996ke,Chetyrkin:1998mw,Bernreuther:2015fts}, 
while the sign was reversed between Refs.~\cite{Artoisenet:2013puc,Demartin:2014fia}. 
We adopt the sign convention of the antisymmetric tensor $\varepsilon_{0123}=+1$
consistent with Refs.~\cite{Chetyrkin:1996ke,Chetyrkin:1998mw,Bernreuther:2015fts}
and thank Werner Bernreuther for pointing out the sign ambiguity. 

\appendix 

\section{Coupling relation to the Warsaw basis}
\label{sect:appendix_Warsaw}

Translations of the EFT operators between the Higgs basis and the Warsaw basis, which 
is defined in Ref.~\cite{Grzadkowski:2010es}, can be performed 
with tools such as Rosetta~\cite{Falkowski:2015wza}. 
For example, under the assumption that $\delta c_z = \delta c_w$ in Section~\ref{sect:cp_couplings},
 we have five independent CP-even and three CP-odd electroweak $HVV$ operators, as well as one CP-even and 
one CP-odd $Hgg$ operator in the Higgs basis. The same number
of independent \Hboson operators exists in the Warsaw basis. The relationship between the six
CP-even operators is quoted explicitly in Eq.~(14) of Ref.~\cite{Falkowski:2015wza}. 
Eliminating the assumption $\delta c_z = \delta c_w$ yields one additional degree of freedom:
$\delta m$ in the Higgs basis;
$\Delta M_W$ in our anomalous coupling approach in Eq.~(\ref{eq:deltaMW});
and a linear combination of three coefficients, called $\delta v$ in Ref.~\cite{Falkowski:2015wza}, in the Warsaw basis.

We extend the above equivalence to include the CP-odd operators and derive the translation 
of the four operators between the Higgs basis and the Warsaw basis as
\begin{eqnarray}
\label{eq:WarsawCPodd}
   g_4^{ZZ} &=& -2 \frac{v^2}{\Lambda^2} \left( s_w^2 w_{\phi \tilde{B}} + c_w^2 w_{\phi \tilde{W}} + s_w c_w w_{\phi B \tilde{W}} \right)\,,
   \nonumber \\
   g_4^{\gamma\gamma} &=& -2 \frac{v^2}{\Lambda^2} \left( c_w^2 w_{\phi \tilde{B}} + s_w^2 w_{\phi \tilde{W}} - s_w c_w w_{\phi B \tilde{W}} \right)\,,
   \nonumber \\
   g_4^{Z\gamma} &=& -2 \frac{v^2}{\Lambda^2} \left( s_w c_w ( w_{\phi \tilde{W}}-w_{\phi \tilde{B}})  + \frac12 (s_w^2-c_w^2) w_{\phi B \tilde{W}} \right)\,,   
   \nonumber \\
   g_4^{gg} &=& -2 \frac{v^2}{\Lambda^2} w_{\phi \tilde{G}}\,.
\end{eqnarray}
Another set of coefficients is sometimes used and is related to the Warsaw basis through
\begin{eqnarray}
  C_{\varphi \tilde{B}} &=& -\frac{c_w^2}{\Lambda^2 e^2} w_{\phi \tilde{B}}\,,
  \nonumber \\
  C_{\varphi \tilde{W}} &=& -\frac{s_w^2}{\Lambda^2 e^2} w_{\phi \tilde{W}}\,,
  \nonumber \\
  C_{\varphi B \tilde{W}} &=& -\frac{s_w c_w}{\Lambda^2 e^2} w_{\phi B \tilde{W}}\,,
  \nonumber \\
  C_{\varphi \tilde{G}} &=& -\frac{w_{\phi \tilde{G}}}{\Lambda^2 g_s^2} \,.  
\end{eqnarray}

We would like to note that it is a question of convenience which operators in the Higgs basis are chosen as independent ones
under the SU(2)$\times$U(1) symmetry. For studies performed in this paper, we find it convenient to pick  
$\delta c_z, c_{zz}, c_{z \Box}, \tilde c_{zz}, c_{z \gamma}, \tilde c_{z \gamma}, c_{\gamma \gamma}, \tilde c_{\gamma \gamma}, c_{gg},$ and~$\tilde c_{gg}$
as an independent set of $HVV$ and $Hgg$ couplings and use Eqs.~(\ref{eq:deltaMW}--\ref{eq:kappa2Zgamma}) 
to express the other couplings listed in Eq.~(\ref{eq:EFT_ci}).
In other analyses, another convention may be more convenient.
For instance, when performing measurements with the $WW$ final state, one could pick 
$\delta c_w, c_{ww}, c_{w \Box},$ and~$\tilde c_{ww}$ in place of $\delta c_z, c_{zz}, c_{z \Box},$ and~$\tilde c_{zz}$
to perform the measurements, and express the latter using Eqs.~(\ref{eq:deltaMW})--(\ref{eq:kappa1WW}).
In the end, all couplings can be translated into a common convention. 
In order to simplify translation between different coupling conventions and operator bases, including the Higgs and Warsaw bases,
within the JHU generator framework, we provide the JHUGenLexicon program,
which includes an interface to the generator and matrix element library
and can also be used for standalone or other applications. 

\input{paper_jhugen.bbl}

\end{document}

%% file: paper_jhugen.bbl
\providecommand{\href}[2]{#2}\begingroup\raggedright\endgroup

%% file: paper_jhugen.bbl
\begin{thebibliography}{100}%
\makeatletter
\providecommand{\hrefCMSnoop }[0]{\@secondoftwo}%
\makeatother
\providecommand{\doi}{\texttt{doi:}\begingroup \urlstyle{tt}\Url}

\bibitem{Gao:2010qx}
Y.~Gao\hrefCMSnoop {}{ {et~al.}, ``{Spin determination of single-produced
  resonances at hadron colliders}'',} \textit{ Phys. Rev. D} \textbf{ 81}
  (2010) 075022,
  \href{http://dx.doi.org/10.1103/PhysRevD.81.075022}{\doi{10.1103/PhysRevD.81.075022}},
\href{http://www.arXiv.org/abs/1001.3396}{\texttt{ arXiv:1001.3396}}.

\bibitem{Bolognesi:2012mm}
S.~Bolognesi\hrefCMSnoop {}{ {et~al.}, ``{Spin and parity of a single-produced
  resonance at the LHC}'',} \textit{ Phys. Rev. D} \textbf{ 86} (2012) 095031,
  \href{http://dx.doi.org/10.1103/PhysRevD.86.095031}{\doi{10.1103/PhysRevD.86.095031}},
\href{http://www.arXiv.org/abs/1208.4018}{\texttt{ arXiv:1208.4018}}.

\bibitem{Anderson:2013afp}
I.~Anderson\hrefCMSnoop {}{ {et~al.}, ``{Constraining anomalous HVV
  interactions at proton and lepton colliders}'',} \textit{ Phys. Rev. D}
  \textbf{ 89} (2014) 035007,
  \href{http://dx.doi.org/10.1103/PhysRevD.89.035007}{\doi{10.1103/PhysRevD.89.035007}},
\href{http://www.arXiv.org/abs/1309.4819}{\texttt{ arXiv:1309.4819}}.

\bibitem{Gritsan:2016hjl}
\hrefCMSnoop {}{A.~V. Gritsan, R.~R{\"o}ntsch, M.~Schulze, and M.~Xiao,
  ``{Constraining anomalous Higgs boson couplings to the heavy flavor fermions
  using matrix element techniques}'',} \textit{ Phys. Rev. D} \textbf{ 94}
  (2016) 055023,
  \href{http://dx.doi.org/10.1103/PhysRevD.94.055023}{\doi{10.1103/PhysRevD.94.055023}},
\href{http://www.arXiv.org/abs/1606.03107}{\texttt{ arXiv:1606.03107}}.

\bibitem{Campbell:2010ff}
\hrefCMSnoop {}{J.~M. Campbell and R.~K. Ellis, ``{MCFM for the Tevatron and
  the LHC}'',} \textit{ Nucl. Phys. Proc. Suppl.} \textbf{ 205-206} (2010)
  10--15,
  \href{http://dx.doi.org/10.1016/j.nuclphysbps.2010.08.011}{\doi{10.1016/j.nuclphysbps.2010.08.011}},
\href{http://www.arXiv.org/abs/1007.3492}{\texttt{ arXiv:1007.3492}}.

\bibitem{Campbell:2011bn}
\hrefCMSnoop {}{J.~M. Campbell, R.~K. Ellis, and C.~Williams, ``{Vector boson
  pair production at the LHC}'',} \textit{ JHEP} \textbf{ 07} (2011) 018,
  \href{http://dx.doi.org/10.1007/JHEP07(2011)018}{\doi{10.1007/JHEP07(2011)018}},
\href{http://www.arXiv.org/abs/1105.0020}{\texttt{ arXiv:1105.0020}}.

\bibitem{Campbell:2013una}
\hrefCMSnoop {}{J.~M. Campbell, R.~K. Ellis, and C.~Williams, ``{Bounding the
  Higgs width at the LHC using full analytic results for $gg \to e^- e^+ \mu^-
  \mu^+$}'',} \textit{ JHEP} \textbf{ 04} (2014) 060,
  \href{http://dx.doi.org/10.1007/JHEP04(2014)060}{\doi{10.1007/JHEP04(2014)060}},
\href{http://www.arXiv.org/abs/1311.3589}{\texttt{ arXiv:1311.3589}}.

\bibitem{Campbell:2015vwa}
\hrefCMSnoop {}{J.~M. Campbell and R.~K. Ellis, ``{Higgs constraints from
  vector boson fusion and scattering}'',} \textit{ JHEP} \textbf{ 04} (2015)
  030,
  \href{http://dx.doi.org/10.1007/JHEP04(2015)030}{\doi{10.1007/JHEP04(2015)030}},
\href{http://www.arXiv.org/abs/1502.02990}{\texttt{ arXiv:1502.02990}}.

\bibitem{Campbell:2015qma}
\hrefCMSnoop {}{J.~M. Campbell, R.~K. Ellis, and W.~T. Giele, ``{A
  Multi-Threaded Version of MCFM}'',} \textit{ Eur. Phys. J. C} \textbf{ 75}
  (2015) 246,
  \href{http://dx.doi.org/10.1140/epjc/s10052-015-3461-2}{\doi{10.1140/epjc/s10052-015-3461-2}},
\href{http://www.arXiv.org/abs/1503.06182}{\texttt{ arXiv:1503.06182}}.

\bibitem{Nelson:1986ki}
\hrefCMSnoop {}{C.~A. Nelson, ``{Correlation between decay planes in
  Higgs-boson decays into a W Pair (into a Z Pair)}'',} \textit{ Phys. Rev. D}
  \textbf{ 37} (1988) 1220,
\href{http://dx.doi.org/10.1103/PhysRevD.37.1220}{\doi{10.1103/PhysRevD.37.1220}}.

\bibitem{Soni:1993jc}
\hrefCMSnoop {}{A.~Soni and R.~M. Xu, ``{Probing CP violation via Higgs decays
  to four leptons}'',} \textit{ Phys. Rev. D} \textbf{ 48} (1993) 5259,
  \href{http://dx.doi.org/10.1103/PhysRevD.48.5259}{\doi{10.1103/PhysRevD.48.5259}},
\href{http://www.arXiv.org/abs/hep-ph/9301225}{\texttt{ arXiv:hep-ph/9301225}}.

\bibitem{Plehn:2001nj}
\hrefCMSnoop {}{T.~Plehn, D.~L. Rainwater, and D.~Zeppenfeld, ``{Determining
  the structure of Higgs couplings at the LHC}'',} \textit{ Phys. Rev. Lett.}
  \textbf{ 88} (2002) 051801,
  \href{http://dx.doi.org/10.1103/PhysRevLett.88.051801}{\doi{10.1103/PhysRevLett.88.051801}},
\href{http://www.arXiv.org/abs/hep-ph/0105325}{\texttt{ arXiv:hep-ph/0105325}}.

\bibitem{Choi:2002jk}
\hrefCMSnoop {}{S.~Y. Choi, D.~J. Miller, M.~M. M{\" u}hlleitner, and P.~M.
  Zerwas, ``{Identifying the Higgs spin and parity in decays to Z pairs}'',}
  \textit{ Phys. Lett. B} \textbf{ 553} (2003) 61,
  \href{http://dx.doi.org/10.1016/S0370-2693(02)03191-X}{\doi{10.1016/S0370-2693(02)03191-X}},
\href{http://www.arXiv.org/abs/hep-ph/0210077}{\texttt{ arXiv:hep-ph/0210077}}.

\bibitem{Buszello:2002uu}
\hrefCMSnoop {}{C.~P. Buszello, I.~Fleck, P.~Marquard, and J.~J. van~der Bij,
  ``{Prospective analysis of spin- and CP-sensitive variables in $H \to ZZ \to
  \ell_1^+ \ell_1^- \ell_2^+ \ell_2^-$ at the LHC}'',} \textit{ Eur. Phys. J.
  C} \textbf{ 32} (2004) 209,
  \href{http://dx.doi.org/10.1140/epjc/s2003-01392-0}{\doi{10.1140/epjc/s2003-01392-0}},
\href{http://www.arXiv.org/abs/hep-ph/0212396}{\texttt{ arXiv:hep-ph/0212396}}.

\bibitem{Hankele:2006ma}
\hrefCMSnoop {}{V.~Hankele, G.~Klamke, D.~Zeppenfeld, and T.~Figy, ``{Anomalous
  Higgs boson couplings in vector boson fusion at the CERN LHC}'',} \textit{
  Phys. Rev. D} \textbf{ 74} (2006) 095001,
  \href{http://dx.doi.org/10.1103/PhysRevD.74.095001}{\doi{10.1103/PhysRevD.74.095001}},
\href{http://www.arXiv.org/abs/hep-ph/0609075}{\texttt{ arXiv:hep-ph/0609075}}.

\bibitem{Accomando:2006ga}
\hrefCMSnoop {}{E.~Accomando {et~al.}, ``{Workshop on CP studies and
  non-standard Higgs physics}'',} (2006).
\href{http://www.arXiv.org/abs/hep-ph/0608079}{\texttt{ arXiv:hep-ph/0608079}}.

\bibitem{Godbole:2007cn}
\hrefCMSnoop {}{R.~M. Godbole, D.~J. Miller, and M.~M. M{\" u}hlleitner,
  ``{Aspects of CP violation in the HZZ coupling at the LHC}'',} \textit{ JHEP}
  \textbf{ 12} (2007) 031,
  \href{http://dx.doi.org/10.1088/1126-6708/2007/12/031}{\doi{10.1088/1126-6708/2007/12/031}},
\href{http://www.arXiv.org/abs/0708.0458}{\texttt{ arXiv:0708.0458}}.

\bibitem{Hagiwara:2009wt}
\hrefCMSnoop {}{K.~Hagiwara, Q.~Li, and K.~Mawatari, ``{Jet angular correlation
  in vector-boson fusion processes at hadron colliders}'',} \textit{ JHEP}
  \textbf{ 07} (2009) 101,
  \href{http://dx.doi.org/10.1088/1126-6708/2009/07/101}{\doi{10.1088/1126-6708/2009/07/101}},
\href{http://www.arXiv.org/abs/0905.4314}{\texttt{ arXiv:0905.4314}}.

\bibitem{DeRujula:2010ys}
A.~De~R{\' u}jula\hrefCMSnoop {}{ {et~al.}, ``{Higgs look-alikes at the
  LHC}'',} \textit{ Phys. Rev. D} \textbf{ 82} (2010) 013003,
  \href{http://dx.doi.org/10.1103/PhysRevD.82.013003}{\doi{10.1103/PhysRevD.82.013003}},
\href{http://www.arXiv.org/abs/1001.5300}{\texttt{ arXiv:1001.5300}}.

\bibitem{Christensen:2010pf}
\hrefCMSnoop {}{N.~D. Christensen, T.~Han, and Y.~Li, ``{Testing CP Violation
  in ZZH Interactions at the LHC}'',} \textit{ Phys. Lett. B} \textbf{ 693}
  (2010) 28,
  \href{http://dx.doi.org/10.1016/j.physletb.2010.08.008}{\doi{10.1016/j.physletb.2010.08.008}},
\href{http://www.arXiv.org/abs/1005.5393}{\texttt{ arXiv:1005.5393}}.

\bibitem{Ellis:2012xd}
\hrefCMSnoop {}{J.~Ellis, D.~S. Hwang, V.~Sanz, and T.~You, ``{A fast track
  towards the `Higgs' spin and parity}'',} \textit{ JHEP} \textbf{ 11} (2012)
  134,
  \href{http://dx.doi.org/10.1007/JHEP11(2012)134}{\doi{10.1007/JHEP11(2012)134}},
\href{http://www.arXiv.org/abs/1208.6002}{\texttt{ arXiv:1208.6002}}.

\bibitem{Chen:2012jy}
\hrefCMSnoop {}{Y.~Chen, N.~Tran, and R.~Vega-Morales, ``{Scrutinizing the
  Higgs signal and background in the $2e2\mu$ golden channel}'',} \textit{
  JHEP} \textbf{ 01} (2013) 182,
  \href{http://dx.doi.org/10.1007/JHEP01(2013)182}{\doi{10.1007/JHEP01(2013)182}},
\href{http://www.arXiv.org/abs/1211.1959}{\texttt{ arXiv:1211.1959}}.

\bibitem{Artoisenet:2013puc}
P.~Artoisenet\hrefCMSnoop {}{ {et~al.}, ``{A framework for Higgs
  characterisation}'',} \textit{ JHEP} \textbf{ 11} (2013) 043,
  \href{http://dx.doi.org/10.1007/JHEP11(2013)043}{\doi{10.1007/JHEP11(2013)043}},
\href{http://www.arXiv.org/abs/1306.6464}{\texttt{ arXiv:1306.6464}}.

\bibitem{Chen:2013waa}
M.~Chen\hrefCMSnoop {}{ {et~al.}, ``{Role of interference in unraveling the ZZ
  couplings of the newly discovered boson at the LHC}'',} \textit{ Phys. Rev.
  D} \textbf{ 89} (2014) 034002,
  \href{http://dx.doi.org/10.1103/PhysRevD.89.034002}{\doi{10.1103/PhysRevD.89.034002}},
\href{http://www.arXiv.org/abs/1310.1397}{\texttt{ arXiv:1310.1397}}.

\bibitem{Maltoni:2013sma}
\hrefCMSnoop {}{F.~Maltoni, K.~Mawatari, and M.~Zaro, ``{Higgs characterisation
  via vector-boson fusion and associated production: NLO and parton-shower
  effects}'',} \textit{ Eur. Phys. J. C} \textbf{ 74} (2014) 2710,
  \href{http://dx.doi.org/10.1140/epjc/s10052-013-2710-5}{\doi{10.1140/epjc/s10052-013-2710-5}},
\href{http://www.arXiv.org/abs/1311.1829}{\texttt{ arXiv:1311.1829}}.

\bibitem{Azatov:2014jga}
\hrefCMSnoop {}{A.~Azatov, C.~Grojean, A.~Paul, and E.~Salvioni, ``{Taming the
  off-shell Higgs boson}'',} \textit{ Zh. Eksp. Teor. Fiz.} \textbf{ 147}
  (2015) 410--425, \href{http://dx.doi.org/10.1134/S1063776115030140,
  10.7868/S0044451015030039}{\doi{10.1134/S1063776115030140,
  10.7868/S0044451015030039}},
  \href{http://www.arXiv.org/abs/1406.6338}{\texttt{ arXiv:1406.6338}}.
[J. Exp. Theor. Phys.120,354(2015)].

\bibitem{Cacciapaglia:2014rla}
\hrefCMSnoop {}{G.~Cacciapaglia, A.~Deandrea, G.~Drieu La~Rochelle, and J.-B.
  Flament, ``{Higgs couplings: disentangling New Physics with off-shell
  measurements}'',} \textit{ Phys. Rev. Lett.} \textbf{ 113} (2014) 201802,
  \href{http://dx.doi.org/10.1103/PhysRevLett.113.201802}{\doi{10.1103/PhysRevLett.113.201802}},
\href{http://www.arXiv.org/abs/1406.1757}{\texttt{ arXiv:1406.1757}}.

\bibitem{Denner:2014cla}
\hrefCMSnoop {}{A.~Denner, S.~Dittmaier, S.~Kallweit, and A.~M{\"u}ck, ``{HAWK
  2.0: A Monte Carlo program for Higgs production in vector-boson fusion and
  Higgs strahlung at hadron colliders}'',} \textit{ Comput. Phys. Commun.}
  \textbf{ 195} (2015) 161--171,
  \href{http://dx.doi.org/10.1016/j.cpc.2015.04.021}{\doi{10.1016/j.cpc.2015.04.021}},
\href{http://www.arXiv.org/abs/1412.5390}{\texttt{ arXiv:1412.5390}}.

\bibitem{Dolan:2014upa}
\hrefCMSnoop {}{M.~J. Dolan, P.~Harris, M.~Jankowiak, and M.~Spannowsky,
  ``{Constraining $CP$-violating Higgs sectors at the LHC using gluon
  fusion}'',} \textit{ Phys. Rev. D} \textbf{ 90} (2014) 073008,
  \href{http://dx.doi.org/10.1103/PhysRevD.90.073008}{\doi{10.1103/PhysRevD.90.073008}},
\href{http://www.arXiv.org/abs/1406.3322}{\texttt{ arXiv:1406.3322}}.

\bibitem{Englert:2014ffa}
\hrefCMSnoop {}{C.~Englert, Y.~Soreq, and M.~Spannowsky, ``{Off-Shell Higgs
  Coupling Measurements in BSM scenarios}'',} \textit{ JHEP} \textbf{ 05}
  (2015) 145,
  \href{http://dx.doi.org/10.1007/JHEP05(2015)145}{\doi{10.1007/JHEP05(2015)145}},
\href{http://www.arXiv.org/abs/1410.5440}{\texttt{ arXiv:1410.5440}}.

\bibitem{Gonzalez-Alonso:2014eva}
\hrefCMSnoop {}{M.~Gonzalez-Alonso, A.~Greljo, G.~Isidori, and D.~Marzocca,
  ``{Pseudo-observables in Higgs decays}'',} \textit{ Eur. Phys. J. C} \textbf{
  75} (2015) 128,
  \href{http://dx.doi.org/10.1140/epjc/s10052-015-3345-5}{\doi{10.1140/epjc/s10052-015-3345-5}},
\href{http://www.arXiv.org/abs/1412.6038}{\texttt{ arXiv:1412.6038}}.

\bibitem{Ballestrero:2015jca}
\hrefCMSnoop {}{A.~Ballestrero and E.~Maina, ``{Interference Effects in Higgs
  production through Vector Boson Fusion in the Standard Model and its Singlet
  Extension}'',} \textit{ JHEP} \textbf{ 01} (2016) 045,
  \href{http://dx.doi.org/10.1007/JHEP01(2016)045}{\doi{10.1007/JHEP01(2016)045}},
\href{http://www.arXiv.org/abs/1506.02257}{\texttt{ arXiv:1506.02257}}.

\bibitem{Greljo:2015sla}
\hrefCMSnoop {}{A.~Greljo, G.~Isidori, J.~M. Lindert, and D.~Marzocca,
  ``{Pseudo-observables in electroweak Higgs production}'',} \textit{ Eur.
  Phys. J. C} \textbf{ 76} (2016) 158,
  \href{http://dx.doi.org/10.1140/epjc/s10052-016-4000-5}{\doi{10.1140/epjc/s10052-016-4000-5}},
\href{http://www.arXiv.org/abs/1512.06135}{\texttt{ arXiv:1512.06135}}.

\bibitem{Hespel:2015zea}
\hrefCMSnoop {}{B.~Hespel, F.~Maltoni, and E.~Vryonidou, ``{Higgs and Z boson
  associated production via gluon fusion in the SM and the 2HDM}'',} \textit{
  JHEP} \textbf{ 06} (2015) 065,
  \href{http://dx.doi.org/10.1007/JHEP06(2015)065}{\doi{10.1007/JHEP06(2015)065}},
\href{http://www.arXiv.org/abs/1503.01656}{\texttt{ arXiv:1503.01656}}.

\bibitem{Kauer:2015dma}
\hrefCMSnoop {}{N.~Kauer, C.~O'Brien, and E.~Vryonidou, ``{Interference effects
  for $ H\to W\;W\to \ell \nu q{\overline{q}}^{\prime } $ and $ H\to ZZ\to \ell
  \overline{\ell}q\overline{q} $ searches in gluon fusion at the LHC}'',}
  \textit{ JHEP} \textbf{ 10} (2015) 074,
  \href{http://dx.doi.org/10.1007/JHEP10(2015)074}{\doi{10.1007/JHEP10(2015)074}},
\href{http://www.arXiv.org/abs/1506.01694}{\texttt{ arXiv:1506.01694}}.

\bibitem{Kauer:2015hia}
\hrefCMSnoop {}{N.~Kauer and C.~O'Brien, ``{Heavy Higgs signal–background
  interference in $gg\rightarrow VV$ in the Standard Model plus real
  singlet}'',} \textit{ Eur. Phys. J. C} \textbf{ 75} (2015) 374,
  \href{http://dx.doi.org/10.1140/epjc/s10052-015-3586-3}{\doi{10.1140/epjc/s10052-015-3586-3}},
\href{http://www.arXiv.org/abs/1502.04113}{\texttt{ arXiv:1502.04113}}.

\bibitem{Kilian:2015opv}
\hrefCMSnoop {}{W.~Kilian, T.~Ohl, J.~Reuter, and M.~Sekulla, ``{Resonances at
  the LHC beyond the Higgs boson: The scalar/tensor case}'',} \textit{ Phys.
  Rev. D} \textbf{ 93} (2016) 036004,
  \href{http://dx.doi.org/10.1103/PhysRevD.93.036004}{\doi{10.1103/PhysRevD.93.036004}},
\href{http://www.arXiv.org/abs/1511.00022}{\texttt{ arXiv:1511.00022}}.

\bibitem{Mimasu:2015nqa}
\hrefCMSnoop {}{K.~Mimasu, V.~Sanz, and C.~Williams, ``{Higher Order QCD
  predictions for Associated Higgs production with anomalous couplings to gauge
  bosons}'',} \textit{ JHEP} \textbf{ 08} (2016) 039,
  \href{http://dx.doi.org/10.1007/JHEP08(2016)039}{\doi{10.1007/JHEP08(2016)039}},
\href{http://www.arXiv.org/abs/1512.02572}{\texttt{ arXiv:1512.02572}}.

\bibitem{Degrande:2016dqg}
C.~Degrande\hrefCMSnoop {}{ {et~al.}, ``{Electroweak Higgs boson production in
  the standard model effective field theory beyond leading order in QCD}'',}
  \textit{ Eur. Phys. J. C} \textbf{ 77} (2017) 262,
  \href{http://dx.doi.org/10.1140/epjc/s10052-017-4793-x}{\doi{10.1140/epjc/s10052-017-4793-x}},
\href{http://www.arXiv.org/abs/1609.04833}{\texttt{ arXiv:1609.04833}}.

\bibitem{Dwivedi:2016xwm}
\hrefCMSnoop {}{S.~Dwivedi, D.~K. Ghosh, B.~Mukhopadhyaya, and A.~Shivaji,
  ``{Distinguishing $CP$-odd couplings of the Higgs boson to weak boson
  pairs}'',} \textit{ Phys. Rev. D} \textbf{ 93} (2016) 115039,
  \href{http://dx.doi.org/10.1103/PhysRevD.93.115039}{\doi{10.1103/PhysRevD.93.115039}},
\href{http://www.arXiv.org/abs/1603.06195}{\texttt{ arXiv:1603.06195}}.

\bibitem{deFlorian:2016spz}
\hrefCMSnoop {}{{LHC Higgs Cross Section Working Group} Collaboration,
  ``{Handbook of LHC Higgs Cross Sections: 4. Deciphering the Nature of the
  Higgs Sector}'',}
  \href{http://dx.doi.org/10.23731/CYRM-2017-002}{\doi{10.23731/CYRM-2017-002}},
\href{http://www.arXiv.org/abs/1610.07922}{\texttt{ arXiv:1610.07922}}.

\bibitem{Azatov:2016xik}
\hrefCMSnoop {}{A.~Azatov, C.~Grojean, A.~Paul, and E.~Salvioni, ``{Resolving
  gluon fusion loops at current and future hadron colliders}'',} \textit{ JHEP}
  \textbf{ 09} (2016) 123,
  \href{http://dx.doi.org/10.1007/JHEP09(2016)123}{\doi{10.1007/JHEP09(2016)123}},
\href{http://www.arXiv.org/abs/1608.00977}{\texttt{ arXiv:1608.00977}}.

\bibitem{Denner:2017vms}
\hrefCMSnoop {}{A.~Denner, J.-N. Lang, and S.~Uccirati, ``{NLO electroweak
  corrections in extended Higgs Sectors with RECOLA2}'',} \textit{ JHEP}
  \textbf{ 07} (2017) 087,
  \href{http://dx.doi.org/10.1007/JHEP07(2017)087}{\doi{10.1007/JHEP07(2017)087}},
\href{http://www.arXiv.org/abs/1705.06053}{\texttt{ arXiv:1705.06053}}.

\bibitem{Deutschmann:2017qum}
\hrefCMSnoop {}{N.~Deutschmann, C.~Duhr, F.~Maltoni, and E.~Vryonidou,
  ``{Gluon-fusion Higgs production in the Standard Model Effective Field
  Theory}'',} \textit{ JHEP} \textbf{ 12} (2017) 063,
  \href{http://dx.doi.org/10.1007/JHEP12(2017)063,
  10.1007/JHEP02(2018)159}{\doi{10.1007/JHEP12(2017)063,
  10.1007/JHEP02(2018)159}},
  \href{http://www.arXiv.org/abs/1708.00460}{\texttt{ arXiv:1708.00460}}.
[Erratum: JHEP02,159(2018)].

\bibitem{Greljo:2017spw}
A.~Greljo\hrefCMSnoop {}{ {et~al.}, ``{Electroweak Higgs production with
  HiggsPO at NLO QCD}'',} \textit{ Eur. Phys. J. C} \textbf{ 77} (2017) 838,
  \href{http://dx.doi.org/10.1140/epjc/s10052-017-5422-4}{\doi{10.1140/epjc/s10052-017-5422-4}},
\href{http://www.arXiv.org/abs/1710.04143}{\texttt{ arXiv:1710.04143}}.

\bibitem{Goncalves:2017gzy}
\hrefCMSnoop {}{D.~Goncalves, T.~Plehn, and J.~M. Thompson, ``{Weak boson
  fusion at 100 TeV}'',} \textit{ Phys. Rev. D} \textbf{ 95} (2017) 095011,
  \href{http://dx.doi.org/10.1103/PhysRevD.95.095011}{\doi{10.1103/PhysRevD.95.095011}},
\href{http://www.arXiv.org/abs/1702.05098}{\texttt{ arXiv:1702.05098}}.

\bibitem{Jager:2017owh}
\hrefCMSnoop {}{B.~J{\"a}ger, L.~Salfelder, M.~Worek, and D.~Zeppenfeld,
  ``{Physics opportunities for vector-boson scattering at a future 100 TeV
  hadron collider}'',} \textit{ Phys. Rev. D} \textbf{ 96} (2017) 073008,
  \href{http://dx.doi.org/10.1103/PhysRevD.96.073008}{\doi{10.1103/PhysRevD.96.073008}},
\href{http://www.arXiv.org/abs/1704.04911}{\texttt{ arXiv:1704.04911}}.

\bibitem{Brass:2018hfw}
S.~Brass\hrefCMSnoop {}{ {et~al.}, ``{Transversal Modes and Higgs Bosons in
  Electroweak Vector-Boson Scattering at the LHC}'',} \textit{ Eur. Phys. J. C}
  \textbf{ 78} (2018) 931,
  \href{http://dx.doi.org/10.1140/epjc/s10052-018-6398-4}{\doi{10.1140/epjc/s10052-018-6398-4}},
\href{http://www.arXiv.org/abs/1807.02512}{\texttt{ arXiv:1807.02512}}.

\bibitem{Gomez-Ambrosio:2018pnl}
\hrefCMSnoop {}{R.~Gomez-Ambrosio, ``{Studies of Dimension-Six EFT effects in
  Vector Boson Scattering}'',} \textit{ Eur. Phys. J. C} \textbf{ 79} (2019)
  389,
  \href{http://dx.doi.org/10.1140/epjc/s10052-019-6893-2}{\doi{10.1140/epjc/s10052-019-6893-2}},
\href{http://www.arXiv.org/abs/1809.04189}{\texttt{ arXiv:1809.04189}}.

\bibitem{Goncalves:2018pkt}
\hrefCMSnoop {}{D.~Gonçalves, T.~Han, and S.~Mukhopadhyay, ``{Higgs Couplings
  at High Scales}'',} \textit{ Phys. Rev. D} \textbf{ 98} (2018) 015023,
  \href{http://dx.doi.org/10.1103/PhysRevD.98.015023}{\doi{10.1103/PhysRevD.98.015023}},
\href{http://www.arXiv.org/abs/1803.09751}{\texttt{ arXiv:1803.09751}}.

\bibitem{Harlander:2018yns}
\hrefCMSnoop {}{R.~V. Harlander, J.~Klappert, C.~Pandini, and
  A.~Papaefstathiou, ``{Exploiting the WH/ZH symmetry in the search for New
  Physics}'',} \textit{ Eur. Phys. J. C} \textbf{ 78} (2018) 760,
  \href{http://dx.doi.org/10.1140/epjc/s10052-018-6234-x}{\doi{10.1140/epjc/s10052-018-6234-x}},
\href{http://www.arXiv.org/abs/1804.02299}{\texttt{ arXiv:1804.02299}}.

\bibitem{Harlander:2018yio}
\hrefCMSnoop {}{R.~V. Harlander, J.~Klappert, S.~Liebler, and L.~Simon,
  ``{vh@nnlo-v2: New physics in Higgs Strahlung}'',} \textit{ JHEP} \textbf{
  05} (2018) 089,
  \href{http://dx.doi.org/10.1007/JHEP05(2018)089}{\doi{10.1007/JHEP05(2018)089}},
\href{http://www.arXiv.org/abs/1802.04817}{\texttt{ arXiv:1802.04817}}.

\bibitem{Lee:2018fxj}
\hrefCMSnoop {}{S.~J. Lee, M.~Park, and Z.~Qian, ``{Probing unitarity violation
  in the tail of the off-shell Higgs boson in $V_LV_L$ mode}'',} \textit{ Phys.
  Rev. D} \textbf{ 100} (2019) 011702,
  \href{http://dx.doi.org/10.1103/PhysRevD.100.011702}{\doi{10.1103/PhysRevD.100.011702}},
\href{http://www.arXiv.org/abs/1812.02679}{\texttt{ arXiv:1812.02679}}.

\bibitem{Kalinowski:2018oxd}
J.~Kalinowski\hrefCMSnoop {}{ {et~al.}, ``{Same-sign WW scattering at the LHC:
  can we discover BSM effects before discovering new states?}'',} \textit{ Eur.
  Phys. J. C} \textbf{ 78} (2018) 403,
  \href{http://dx.doi.org/10.1140/epjc/s10052-018-5885-y}{\doi{10.1140/epjc/s10052-018-5885-y}},
\href{http://www.arXiv.org/abs/1802.02366}{\texttt{ arXiv:1802.02366}}.

\bibitem{Perez:2018kav}
\hrefCMSnoop {}{G.~Perez, M.~Sekulla, and D.~Zeppenfeld, ``{Anomalous quartic
  gauge couplings and unitarization for the vector boson scattering process
  $pp\rightarrow W^+W^+jjX\rightarrow \ell ^+\nu _\ell \ell ^+\nu _\ell
  jjX$}'',} \textit{ Eur. Phys. J. C} \textbf{ 78} (2018) 759,
  \href{http://dx.doi.org/10.1140/epjc/s10052-018-6230-1}{\doi{10.1140/epjc/s10052-018-6230-1}},
\href{http://www.arXiv.org/abs/1807.02707}{\texttt{ arXiv:1807.02707}}.

\bibitem{Jaquier:2019bfs}
\hrefCMSnoop {}{M.~Jaquier and R.~R{\"o}ntsch, ``{Mixed scalar-pseudoscalar Higgs
  boson production through next-to-next-to-leading order at the LHC}'',}
  \textit{ JHEP} \textbf{ 06} (2020) 005,
  \href{http://dx.doi.org/10.1007/JHEP06(2020)005}{\doi{10.1007/JHEP06(2020)005}},
\href{http://www.arXiv.org/abs/1911.10631}{\texttt{ arXiv:1911.10631}}.

\bibitem{Denner:2019fcr}
\hrefCMSnoop {}{A.~Denner, S.~Dittmaier, and A.~M{\"u}ck, ``{PROPHECY4F 3.0: A
  Monte Carlo program for Higgs-boson decays into four-fermion final states in
  and beyond the Standard Model}'',} \textit{ Comput. Phys. Commun.} \textbf{
  254} (2020) 107336,
  \href{http://dx.doi.org/10.1016/j.cpc.2020.107336}{\doi{10.1016/j.cpc.2020.107336}},
\href{http://www.arXiv.org/abs/1912.02010}{\texttt{ arXiv:1912.02010}}.

\bibitem{Banerjee:2019twi}
S.~Banerjee\hrefCMSnoop {}{ {et~al.}, ``{Towards the ultimate differential
  SMEFT analysis}'',}
\href{http://www.arXiv.org/abs/1912.07628}{\texttt{ arXiv:1912.07628}}.

\bibitem{Chatrchyan:2012xdj}
\hrefCMSnoop {}{{CMS} Collaboration, ``{Observation of a new boson at a mass of
  125 GeV with the CMS experiment at the LHC}'',} \textit{ Phys. Lett. B}
  \textbf{ 716} (2012) 30--61,
  \href{http://dx.doi.org/10.1016/j.physletb.2012.08.021}{\doi{10.1016/j.physletb.2012.08.021}},
\href{http://www.arXiv.org/abs/1207.7235}{\texttt{ arXiv:1207.7235}}.

\bibitem{Chatrchyan:2012jja}
\hrefCMSnoop {}{{CMS} Collaboration, ``{Study of the Mass and Spin-Parity of
  the Higgs Boson Candidate Via Its Decays to Z Boson Pairs}'',} \textit{ Phys.
  Rev. Lett.} \textbf{ 110} (2013) 081803,
  \href{http://dx.doi.org/10.1103/PhysRevLett.110.081803}{\doi{10.1103/PhysRevLett.110.081803}},
\href{http://www.arXiv.org/abs/1212.6639}{\texttt{ arXiv:1212.6639}}.

\bibitem{Chatrchyan:2013mxa}
\hrefCMSnoop {}{{CMS} Collaboration, ``{Measurement of the properties of a
  Higgs boson in the four-lepton final state}'',} \textit{ Phys. Rev. D}
  \textbf{ 89} (2014) 092007,
  \href{http://dx.doi.org/10.1103/PhysRevD.89.092007}{\doi{10.1103/PhysRevD.89.092007}},
\href{http://www.arXiv.org/abs/1312.5353}{\texttt{ arXiv:1312.5353}}.

\bibitem{Chatrchyan:2013iaa}
\hrefCMSnoop {}{{CMS} Collaboration, ``{Measurement of Higgs boson production
  and properties in the WW decay channel with leptonic final states}'',}
  \textit{ JHEP} \textbf{ 01} (2014) 096,
  \href{http://dx.doi.org/10.1007/JHEP01(2014)096}{\doi{10.1007/JHEP01(2014)096}},
\href{http://www.arXiv.org/abs/1312.1129}{\texttt{ arXiv:1312.1129}}.

\bibitem{Aad:2013xqa}
\hrefCMSnoop {}{{ATLAS} Collaboration, ``{Evidence for the spin-0 nature of the
  Higgs boson using ATLAS data}'',} \textit{ Phys. Lett. B} \textbf{ 726}
  (2013) 120--144,
  \href{http://dx.doi.org/10.1016/j.physletb.2013.08.026}{\doi{10.1016/j.physletb.2013.08.026}},
\href{http://www.arXiv.org/abs/1307.1432}{\texttt{ arXiv:1307.1432}}.

\bibitem{Khachatryan:2014iha}
\hrefCMSnoop {}{{CMS} Collaboration, ``{Constraints on the Higgs boson width
  from off-shell production and decay to Z-boson pairs}'',} \textit{ Phys.
  Lett. B} \textbf{ 736} (2014) 64,
  \href{http://dx.doi.org/10.1016/j.physletb.2014.06.077}{\doi{10.1016/j.physletb.2014.06.077}},
\href{http://www.arXiv.org/abs/1405.3455}{\texttt{ arXiv:1405.3455}}.

\bibitem{Khachatryan:2014ira}
\hrefCMSnoop {}{{CMS} Collaboration, ``{Observation of the diphoton decay of
  the Higgs boson and measurement of its properties}'',} \textit{ Eur. Phys. J.
  C} \textbf{ 74} (2014) 3076,
  \href{http://dx.doi.org/10.1140/epjc/s10052-014-3076-z}{\doi{10.1140/epjc/s10052-014-3076-z}},
\href{http://www.arXiv.org/abs/1407.0558}{\texttt{ arXiv:1407.0558}}.

\bibitem{Khachatryan:2014kca}
\hrefCMSnoop {}{{CMS} Collaboration, ``{Constraints on the spin-parity and
  anomalous HVV couplings of the Higgs boson in proton collisions at 7 and 8
  TeV}'',} \textit{ Phys. Rev. D} \textbf{ 92} (2015) 012004,
  \href{http://dx.doi.org/10.1103/PhysRevD.92.012004}{\doi{10.1103/PhysRevD.92.012004}},
\href{http://www.arXiv.org/abs/1411.3441}{\texttt{ arXiv:1411.3441}}.

\bibitem{Khachatryan:2015mma}
\hrefCMSnoop {}{{CMS} Collaboration, ``{Limits on the Higgs boson lifetime and
  width from its decay to four charged leptons}'',} \textit{ Phys. Rev. D}
  \textbf{ 92} (2015) 072010,
  \href{http://dx.doi.org/10.1103/PhysRevD.92.072010}{\doi{10.1103/PhysRevD.92.072010}},
\href{http://www.arXiv.org/abs/1507.06656}{\texttt{ arXiv:1507.06656}}.

\bibitem{Khachatryan:2015cwa}
\hrefCMSnoop {}{{CMS} Collaboration, ``{Search for a Higgs boson in the mass
  range from 145 to 1000 GeV decaying to a pair of W or Z bosons}'',} \textit{
  JHEP} \textbf{ 10} (2015) 144,
  \href{http://dx.doi.org/10.1007/JHEP10(2015)144}{\doi{10.1007/JHEP10(2015)144}},
\href{http://www.arXiv.org/abs/1504.00936}{\texttt{ arXiv:1504.00936}}.

\bibitem{Aad:2015mxa}
\hrefCMSnoop {}{{ATLAS} Collaboration, ``{Study of the spin and parity of the
  Higgs boson in diboson decays with the ATLAS detector}'',} \textit{ Eur.
  Phys. J. C} \textbf{ 75} (2015) 476,
  \href{http://dx.doi.org/10.1140/epjc/s10052-015-3685-1}{\doi{10.1140/epjc/s10052-015-3685-1}},
\href{http://www.arXiv.org/abs/1506.05669}{\texttt{ arXiv:1506.05669}}.

\bibitem{Khachatryan:2016tnr}
\hrefCMSnoop {}{{CMS} Collaboration, ``{Combined search for anomalous
  pseudoscalar HVV couplings in VH(H $\to \text{b}\bar{\text{b}}$) production
  and H $\to$ VV decay}'',} \textit{ Phys. Lett. B} \textbf{ 759} (2016) 672,
  \href{http://dx.doi.org/10.1016/j.physletb.2016.06.004}{\doi{10.1016/j.physletb.2016.06.004}},
\href{http://www.arXiv.org/abs/1602.04305}{\texttt{ arXiv:1602.04305}}.

\bibitem{Sirunyan:2017tqd}
\hrefCMSnoop {}{{CMS} Collaboration, ``{Constraints on anomalous Higgs boson
  couplings using production and decay information in the four-lepton final
  state}'',} \textit{ Phys. Lett. B} \textbf{ 775} (2017) 1,
  \href{http://dx.doi.org/10.1016/j.physletb.2017.10.021}{\doi{10.1016/j.physletb.2017.10.021}},
\href{http://www.arXiv.org/abs/1707.00541}{\texttt{ arXiv:1707.00541}}.

\bibitem{Sirunyan:2019nbs}
\hrefCMSnoop {}{{CMS} Collaboration, ``{Constraints on anomalous $HVV$
  couplings from the production of Higgs bosons decaying to $\tau$ lepton
  pairs}'',} \textit{ Phys. Rev. D} \textbf{ 100} (2019) 112002,
  \href{http://dx.doi.org/10.1103/PhysRevD.100.112002}{\doi{10.1103/PhysRevD.100.112002}},
\href{http://www.arXiv.org/abs/1903.06973}{\texttt{ arXiv:1903.06973}}.

\bibitem{Sirunyan:2019twz}
\hrefCMSnoop {}{{CMS} Collaboration, ``{Measurements of the Higgs boson width
  and anomalous HVV couplings from on-shell and off-shell production in the
  four-lepton final state}'',} \textit{ Phys. Rev. D} \textbf{ 99} (2019)
  112003,
  \href{http://dx.doi.org/10.1103/PhysRevD.99.112003}{\doi{10.1103/PhysRevD.99.112003}},
\href{http://www.arXiv.org/abs/1901.00174}{\texttt{ arXiv:1901.00174}}.

\bibitem{Sirunyan:2020sum}
\hrefCMSnoop {}{{CMS} Collaboration, ``{Measurements of $\mathrm{t\bar{t}}$H
  production and the CP structure of the Yukawa interaction between the Higgs
  boson and top quark in the diphoton decay channel}'',} \textit{ Phys. Rev.
  Lett.} \textbf{ 125} (2020) 061801,
  \href{http://dx.doi.org/10.1103/PhysRevLett.125.061801}{\doi{10.1103/PhysRevLett.125.061801}},
\href{http://www.arXiv.org/abs/2003.10866}{\texttt{ arXiv:2003.10866}}.

\bibitem{Sirunyan:2018qlb}
\hrefCMSnoop {}{{CMS} Collaboration, ``{Search for a new scalar resonance
  decaying to a pair of Z bosons in proton-proton collisions at $\sqrt{s}=13$
  TeV}'',} \textit{ JHEP} \textbf{ 06} (2018) 127,
  \href{http://dx.doi.org/10.1007/JHEP06(2018)127,
  10.1007/JHEP03(2019)128}{\doi{10.1007/JHEP06(2018)127,
  10.1007/JHEP03(2019)128}},
  \href{http://www.arXiv.org/abs/1804.01939}{\texttt{ arXiv:1804.01939}}.
[Erratum: JHEP03,128(2019)].

\bibitem{Sirunyan:2019pqw}
\hrefCMSnoop {}{{CMS} Collaboration, ``{Search for a heavy Higgs boson decaying
  to a pair of W bosons in proton-proton collisions at $\sqrt{s} =$ 13 TeV}'',}
\href{http://www.arXiv.org/abs/1912.01594}{\texttt{ arXiv:1912.01594}}.

\bibitem{Cepeda:2019klc}
\hrefCMSnoop {}{M.~Cepeda {et~al.}, ``{Higgs Physics at the HL-LHC and
  HE-LHC}'',} \textit{ CERN Yellow Rep. Monogr.} \textbf{ 7} (2019) 221--584,
  \href{http://dx.doi.org/10.23731/CYRM-2019-007.221}{\doi{10.23731/CYRM-2019-007.221}},
\href{http://www.arXiv.org/abs/1902.00134}{\texttt{ arXiv:1902.00134}}.

\bibitem{Khachatryan:2016vau}
\hrefCMSnoop {}{{ATLAS, CMS} Collaboration, ``{Measurements of the Higgs boson
  production and decay rates and constraints on its couplings from a combined
  ATLAS and CMS analysis of the LHC pp collision data at $ \sqrt{s}=7 $ and 8
  TeV}'',} \textit{ JHEP} \textbf{ 08} (2016) 045,
  \href{http://dx.doi.org/10.1007/JHEP08(2016)045}{\doi{10.1007/JHEP08(2016)045}},
\href{http://www.arXiv.org/abs/1606.02266}{\texttt{ arXiv:1606.02266}}.

\bibitem{Sirunyan:2018koj}
\hrefCMSnoop {}{{CMS} Collaboration, ``{Combined measurements of Higgs boson
  couplings in proton-proton collisions at $\sqrt{s}=13\,\text {Te}\text {V}
  $}'',} \textit{ Eur. Phys. J. C} \textbf{ 79} (2019) 421,
  \href{http://dx.doi.org/10.1140/epjc/s10052-019-6909-y}{\doi{10.1140/epjc/s10052-019-6909-y}},
\href{http://www.arXiv.org/abs/1809.10733}{\texttt{ arXiv:1809.10733}}.

\bibitem{Aad:2019mbh}
\hrefCMSnoop {}{{ATLAS} Collaboration, ``{Combined measurements of Higgs boson
  production and decay using up to $80$ fb$^{-1}$ of proton-proton collision
  data at $\sqrt{s}=$ 13 TeV collected with the ATLAS experiment}'',} \textit{
  Phys. Rev. D} \textbf{ 101} (2020) 012002,
  \href{http://dx.doi.org/10.1103/PhysRevD.101.012002}{\doi{10.1103/PhysRevD.101.012002}},
\href{http://www.arXiv.org/abs/1909.02845}{\texttt{ arXiv:1909.02845}}.

\bibitem{Sirunyan:2018kst}
\hrefCMSnoop {}{{CMS} Collaboration, ``{Observation of Higgs boson decay to
  bottom quarks}'',} \textit{ Phys. Rev. Lett.} \textbf{ 121} (2018) 121801,
  \href{http://dx.doi.org/10.1103/PhysRevLett.121.121801}{\doi{10.1103/PhysRevLett.121.121801}},
\href{http://www.arXiv.org/abs/1808.08242}{\texttt{ arXiv:1808.08242}}.

\bibitem{Aaboud:2018zhk}
\hrefCMSnoop {}{{ATLAS} Collaboration, ``{Observation of $H \rightarrow
  b\bar{b}$ decays and $VH$ production with the ATLAS detector}'',} \textit{
  Phys. Lett. B} \textbf{ 786} (2018) 59--86,
  \href{http://dx.doi.org/10.1016/j.physletb.2018.09.013}{\doi{10.1016/j.physletb.2018.09.013}},
\href{http://www.arXiv.org/abs/1808.08238}{\texttt{ arXiv:1808.08238}}.

\bibitem{Sirunyan:2018hoz}
\hrefCMSnoop {}{{CMS} Collaboration, ``{Observation of
  $\mathrm{t\overline{t}}$H production}'',} \textit{ Phys. Rev. Lett.} \textbf{
  120} (2018) 231801,
  \href{http://dx.doi.org/10.1103/PhysRevLett.120.231801}{\doi{10.1103/PhysRevLett.120.231801}},
\href{http://www.arXiv.org/abs/1804.02610}{\texttt{ arXiv:1804.02610}}.

\bibitem{Aaboud:2018urx}
\hrefCMSnoop {}{{ATLAS} Collaboration, ``{Observation of Higgs boson production
  in association with a top quark pair at the LHC with the ATLAS detector}'',}
  \textit{ Phys. Lett. B} \textbf{ 784} (2018) 173--191,
  \href{http://dx.doi.org/10.1016/j.physletb.2018.07.035}{\doi{10.1016/j.physletb.2018.07.035}},
\href{http://www.arXiv.org/abs/1806.00425}{\texttt{ arXiv:1806.00425}}.

\bibitem{Sirunyan:2017khh}
\hrefCMSnoop {}{{CMS} Collaboration, ``{Observation of the Higgs boson decay to
  a pair of $\tau$ leptons with the CMS detector}'',} \textit{ Phys. Lett. B}
  \textbf{ 779} (2018) 283--316,
  \href{http://dx.doi.org/10.1016/j.physletb.2018.02.004}{\doi{10.1016/j.physletb.2018.02.004}},
\href{http://www.arXiv.org/abs/1708.00373}{\texttt{ arXiv:1708.00373}}.

\bibitem{Aaboud:2018pen}
\hrefCMSnoop {}{{ATLAS} Collaboration, ``{Cross-section measurements of the
  Higgs boson decaying into a pair of $\tau$-leptons in proton-proton
  collisions at $\sqrt{s}=13$ TeV with the ATLAS detector}'',} \textit{ Phys.
  Rev. D} \textbf{ 99} (2019) 072001,
  \href{http://dx.doi.org/10.1103/PhysRevD.99.072001}{\doi{10.1103/PhysRevD.99.072001}},
\href{http://www.arXiv.org/abs/1811.08856}{\texttt{ arXiv:1811.08856}}.

\bibitem{jhugen}
\hrefCMSnoop {}{``{JHU generator}'',}. \url{http://spin.pha.jhu.edu/}.

\bibitem{Grzadkowski:2010es}
\hrefCMSnoop {}{B.~Grzadkowski, M.~Iskrzynski, M.~Misiak, and J.~Rosiek,
  ``{Dimension-Six Terms in the Standard Model Lagrangian}'',} \textit{ JHEP}
  \textbf{ 10} (2010) 085,
  \href{http://dx.doi.org/10.1007/JHEP10(2010)085}{\doi{10.1007/JHEP10(2010)085}},
\href{http://www.arXiv.org/abs/1008.4884}{\texttt{ arXiv:1008.4884}}.

\bibitem{Falkowski:2001958}
\hrefCMSnoop {}{A.~Falkowski, ``{Higgs Basis: Proposal for an EFT basis choice
  for LHC HXSWG}'',} Technical Report LHCHXSWG-INT-2015-001, 2015.
\newblock \url{https://cds.cern.ch/record/2001958}.

\bibitem{Gonzalez-Alonso:2015bha}
\hrefCMSnoop {}{M.~Gonzalez-Alonso, A.~Greljo, G.~Isidori, and D.~Marzocca,
  ``{Electroweak bounds on Higgs pseudo-observables and $h \to 4 \ell$
  decays}'',} \textit{ Eur. Phys. J. C} \textbf{ 75} (2015) 341,
  \href{http://dx.doi.org/10.1140/epjc/s10052-015-3555-x}{\doi{10.1140/epjc/s10052-015-3555-x}},
\href{http://www.arXiv.org/abs/1504.04018}{\texttt{ arXiv:1504.04018}}.

\bibitem{Alboteanu:2008my}
\hrefCMSnoop {}{A.~Alboteanu, W.~Kilian, and J.~Reuter, ``{Resonances and
  Unitarity in Weak Boson Scattering at the LHC}'',} \textit{ JHEP} \textbf{
  11} (2008) 010,
  \href{http://dx.doi.org/10.1088/1126-6708/2008/11/010}{\doi{10.1088/1126-6708/2008/11/010}},
\href{http://www.arXiv.org/abs/0806.4145}{\texttt{ arXiv:0806.4145}}.

\bibitem{Kauer:2012hd}
\hrefCMSnoop {}{N.~Kauer and G.~Passarino, ``{Inadequacy of zero-width
  approximation for a light Higgs boson signal}'',} \textit{ JHEP} \textbf{ 08}
  (2012) 116,
  \href{http://dx.doi.org/10.1007/JHEP08(2012)116}{\doi{10.1007/JHEP08(2012)116}},
\href{http://www.arXiv.org/abs/1206.4803}{\texttt{ arXiv:1206.4803}}.

\bibitem{Caola:2013yja}
\hrefCMSnoop {}{F.~Caola and K.~Melnikov, ``{Constraining the Higgs boson width
  with ZZ production at the LHC}'',} \textit{ Phys. Rev. D} \textbf{ 88} (2013)
  054024,
  \href{http://dx.doi.org/10.1103/PhysRevD.88.054024}{\doi{10.1103/PhysRevD.88.054024}},
\href{http://www.arXiv.org/abs/1307.4935}{\texttt{ arXiv:1307.4935}}.

\bibitem{Djouadi:2018xqq}
\hrefCMSnoop {}{A.~Djouadi, J.~Kalinowski, M.~M{\"u}hlleitner, and M.~Spira,
  ``{HDECAY: Twenty$_{++}$ years after}'',} \textit{ Comput. Phys. Commun.}
  \textbf{ 238} (2019) 214--231,
  \href{http://dx.doi.org/10.1016/j.cpc.2018.12.010}{\doi{10.1016/j.cpc.2018.12.010}},
\href{http://www.arXiv.org/abs/1801.09506}{\texttt{ arXiv:1801.09506}}.

\bibitem{Fontes:2017zfn}
D.~Fontes\hrefCMSnoop {}{ {et~al.}, ``{The C2HDM revisited}'',} \textit{ JHEP}
  \textbf{ 02} (2018) 073,
  \href{http://dx.doi.org/10.1007/JHEP02(2018)073}{\doi{10.1007/JHEP02(2018)073}},
\href{http://www.arXiv.org/abs/1711.09419}{\texttt{ arXiv:1711.09419}}.

\bibitem{Brivio:2019myy}
\hrefCMSnoop {}{I.~Brivio, T.~Corbett, and M.~Trott, ``{The Higgs width in the
  SMEFT}'',} \textit{ JHEP} \textbf{ 10} (2019) 056,
  \href{http://dx.doi.org/10.1007/JHEP10(2019)056}{\doi{10.1007/JHEP10(2019)056}},
\href{http://www.arXiv.org/abs/1906.06949}{\texttt{ arXiv:1906.06949}}.

\bibitem{CMS-HIG-12-041}
\href
  {https://cms-physics.web.cern.ch/cms-physics/public/HIG-12-041-pas.pdf}{{CMS}
  Collaboration, ``{Updated results on the new boson discovered in the search
  for the standard model Higgs boson in the $H\to ZZ\to 4\ell$ channel in pp
  collisions at $\sqrt{s}$ = 7 and 8 TeV}'',} Technical Report CMS-HIG-12-041,
  CERN, Geneva, 2012.

\bibitem{Frixione:2007vw}
\hrefCMSnoop {}{S.~Frixione, P.~Nason, and C.~Oleari, ``{Matching NLO QCD
  computations with Parton Shower simulations: the POWHEG method}'',} \textit{
  JHEP} \textbf{ 11} (2007) 070,
  \href{http://dx.doi.org/10.1088/1126-6708/2007/11/070}{\doi{10.1088/1126-6708/2007/11/070}},
\href{http://www.arXiv.org/abs/0709.2092}{\texttt{ arXiv:0709.2092}}.

\bibitem{Heinemeyer:2013tqa}
\hrefCMSnoop {}{{LHC Higgs Cross Section Working Group} Collaboration,
  ``{Handbook of LHC Higgs Cross Sections: 3. Higgs Properties}'',}
  \href{http://dx.doi.org/10.5170/CERN-2013-004}{\doi{10.5170/CERN-2013-004}},
\href{http://www.arXiv.org/abs/1307.1347}{\texttt{ arXiv:1307.1347}}.

\bibitem{Agashe:2014kda}
\hrefCMSnoop {}{{Particle Data Group} Collaboration, ``{Review of Particle
  Physics}'',} \textit{ Chin. Phys.} \textbf{ C38} (2014) 090001,
\href{http://dx.doi.org/10.1088/1674-1137/38/9/090001}{\doi{10.1088/1674-1137/38/9/090001}}.

\bibitem{Denner:2016kdg}
\hrefCMSnoop {}{A.~Denner, S.~Dittmaier, and L.~Hofer, ``{Collier: a
  fortran-based Complex One-Loop LIbrary in Extended Regularizations}'',}
  \textit{ Comput. Phys. Commun.} \textbf{ 212} (2017) 220--238,
  \href{http://dx.doi.org/10.1016/j.cpc.2016.10.013}{\doi{10.1016/j.cpc.2016.10.013}},
\href{http://www.arXiv.org/abs/1604.06792}{\texttt{ arXiv:1604.06792}}.

\bibitem{Brehmer:2019bvj}
J.~Brehmer\hrefCMSnoop {}{ {et~al.}, ``{Effective LHC measurements with matrix
  elements and machine learning}'',} \textit{ J. Phys. Conf. Ser.} \textbf{
  1525} (2020) 012022,
  \href{http://dx.doi.org/10.1088/1742-6596/1525/1/012022}{\doi{10.1088/1742-6596/1525/1/012022}},
\href{http://www.arXiv.org/abs/1906.01578}{\texttt{ arXiv:1906.01578}}.

\bibitem{Berger:2019wnu}
\hrefCMSnoop {}{N.~Berger {et~al.}, ``{Simplified Template Cross Sections -
  Stage 1.1}'',}
\href{http://www.arXiv.org/abs/1906.02754}{\texttt{ arXiv:1906.02754}}.

\bibitem{Neyman:1933}
\hrefCMSnoop {}{J.~Neyman and E.~S. Pearson, ``{On the Problem of the Most
  Efficient Tests of Statistical Hypotheses}'',} \textit{ Philosophical
  Transactions of the Royal Society} \textbf{ A231} (1933) 289--337,
  \href{http://dx.doi.org/10.1098/rsta.1933.0009}{\doi{10.1098/rsta.1933.0009}}.

\bibitem{Atwood:1991ka}
\hrefCMSnoop {}{D.~Atwood and A.~Soni, ``{Analysis for magnetic moment and
  electric dipole moment form-factors of the top quark via $e^+ e^- \to t
  \bar{t}$ }'',} \textit{ Phys. Rev. D} \textbf{ 45} (1992) 2405--2413,
\href{http://dx.doi.org/10.1103/PhysRevD.45.2405}{\doi{10.1103/PhysRevD.45.2405}}.

\bibitem{Davier:1992nw}
\hrefCMSnoop {}{M.~Davier, L.~Duflot, F.~Le~Diberder, and A.~Rouge, ``{The
  Optimal method for the measurement of tau polarization}'',} \textit{ Phys.
  Lett. B} \textbf{ 306} (1993) 411--417,
\href{http://dx.doi.org/10.1016/0370-2693(93)90101-M}{\doi{10.1016/0370-2693(93)90101-M}}.

\bibitem{Diehl:1993br}
\hrefCMSnoop {}{M.~Diehl and O.~Nachtmann, ``{Optimal observables for the
  measurement of three gauge boson couplings in $e^+ e^- \to W^+ W^-$ }'',}
  \textit{ Z. Phys.} \textbf{ C62} (1994) 397--412,
\href{http://dx.doi.org/10.1007/BF01555899}{\doi{10.1007/BF01555899}}.

\bibitem{Hocker:2007ht}
\hrefCMSnoop {}{A.~Hocker {et~al.}, ``{TMVA - Toolkit for Multivariate Data
  Analysis}'',}
\href{http://www.arXiv.org/abs/physics/0703039}{\texttt{
  arXiv:physics/0703039}}.

\bibitem{Verkerke:2003ir}
\hrefCMSnoop {}{W.~Verkerke and D.~P. Kirkby, ``{The RooFit toolkit for data
  modeling}'',} in \textit{ 13$^\text{th}$ International Conference for
  Computing in High-Energy and Nuclear Physics (CHEP03)}.
\newblock 2003.
\newblock \href{http://www.arXiv.org/abs/physics/0306116}{\texttt{
  arXiv:physics/0306116}}.
\newblock
{CHEP-2003-MOLT007}.

\bibitem{Brun:1997pa}
\hrefCMSnoop {}{R.~Brun and F.~Rademakers, ``{ROOT: An object oriented data
  analysis framework}'',} \textit{ Nucl. Instrum. Meth. A} \textbf{ 389} (1997)
  81,
\href{http://dx.doi.org/10.1016/S0168-9002(97)00048-X}{\doi{10.1016/S0168-9002(97)00048-X}}.

\bibitem{powhegvv}
\hrefCMSnoop {}{T.~Melia, P.~Nason, R.~R{\"o}ntsch, and G.~Zanderighi, ``{$ZZ$,
  $W^{\pm}Z$ and $W^+W^-$ production, including $\gamma$/Z interference, singly
  resonant contributions and interference for identical leptons}'',} \textit{
  JHEP} \textbf{ 1111} (2011) 078.

\bibitem{CMS-HIG-19-001}
\href {https://cds.cern.ch/record/2668684}{{CMS} Collaboration, ``{Measurements
  of properties of the Higgs boson in the four-lepton final state in
  proton-proton collisions at $\sqrt{s}=13$ TeV}'',} Technical Report
  CMS-HIG-19-001, CERN, Geneva, 2019.

\bibitem{Sjostrand:2014zea}
T.~Sj{\"o}strand\hrefCMSnoop {}{ {et~al.}, ``An introduction to {PYTHIA}
  8.2'',} \textit{ Comput. Phys. Commun.} \textbf{ 191} (2015) 159,
  \href{http://dx.doi.org/10.1016/j.cpc.2015.01.024}{\doi{10.1016/j.cpc.2015.01.024}},
\href{http://www.arXiv.org/abs/1410.3012}{\texttt{ arXiv:1410.3012}}.

\bibitem{HomotopyContinuation}
\hrefCMSnoop {}{T.~Chen and T.-Y. Li, ``Homotopy continuation method for
  solving systems of nonlinear and polynomial equations'',} \textit{ Commun.
  Inf. Syst.} \textbf{ 15} (2015), no.~2, 119--307,
  \href{http://dx.doi.org/10.4310/CIS.2015.v15.n2.a1}{\doi{10.4310/CIS.2015.v15.n2.a1}}.

\bibitem{Hom4PS1}
\hrefCMSnoop {}{T.~Chen, T.-L. Lee, and T.-Y. Li, ``{Hom4PS-3}: A Parallel
  Numerical Solver for Systems of Polynomial Equations Based on Polyhedral
  Homotopy Continuation Methods'',} in \textit{ Mathematical Software---ICMS
  2014}, H.~Hong and C.~Yap, eds., number 8592 in Lecture Notes in Computer
  Science, pp.~183--190.
\newblock Springer Berlin Heidelberg, January, 2014.
\newblock
  \href{http://dx.doi.org/10.1007/978-3-662-44199-2_30}{\doi{10.1007/978-3-662-44199-2_30}}.

\bibitem{Hom4PS2}
\hrefCMSnoop {}{T.~Chen, T.-L. Lee, and T.-Y. Li, ``Mixed cell computation in
  {Hom4PS-3}'',} \textit{ Journal of Symbolic Computation} \textbf{ 79, Part 3}
  (March, 2017) 516--534,
  \href{http://dx.doi.org/10.1016/j.jsc.2016.07.017}{\doi{10.1016/j.jsc.2016.07.017}}.

\bibitem{cuttingplanes}
A.~S. Nemirovsky and D.~B. Yudin, ``Problem complexity and method efficiency in
  optimization.''.
\newblock Wiley, 1983.

\bibitem{gurobi}
\href {http://www.gurobi.com}{{Gurobi Optimization, LLC}, ``Gurobi Optimizer
  Reference Manual'',} 2018.

\bibitem{Catani:2007vq}
\hrefCMSnoop {}{S.~Catani and M.~Grazzini, ``{An NNLO subtraction formalism in
  hadron collisions and its application to Higgs boson production at the
  LHC}'',} \textit{ Phys. Rev. Lett.} \textbf{ 98} (2007) 222002,
  \href{http://dx.doi.org/10.1103/PhysRevLett.98.222002}{\doi{10.1103/PhysRevLett.98.222002}},
\href{http://www.arXiv.org/abs/hep-ph/0703012}{\texttt{ arXiv:hep-ph/0703012}}.

\bibitem{Grazzini:2008tf}
\hrefCMSnoop {}{M.~Grazzini, ``{NNLO predictions for the Higgs boson signal in
  the $H \to WW \to\ell\nu\ell\nu$ and $H \to ZZ \to 4\ell$ decay channels}'',}
  \textit{ JHEP} \textbf{ 02} (2008) 043,
  \href{http://dx.doi.org/10.1088/1126-6708/2008/02/043}{\doi{10.1088/1126-6708/2008/02/043}},
\href{http://www.arXiv.org/abs/0801.3232}{\texttt{ arXiv:0801.3232}}.

\bibitem{Grazzini:2013mca}
\hrefCMSnoop {}{M.~Grazzini and H.~Sargsyan, ``{Heavy-quark mass effects in
  Higgs boson production at the LHC}'',} \textit{ JHEP} \textbf{ 09} (2013)
  129,
  \href{http://dx.doi.org/10.1007/JHEP09(2013)129}{\doi{10.1007/JHEP09(2013)129}},
\href{http://www.arXiv.org/abs/1306.4581}{\texttt{ arXiv:1306.4581}}.

\bibitem{Melnikov:2015laa}
\hrefCMSnoop {}{K.~Melnikov and M.~Dowling, ``{Production of two Z-bosons in
  gluon fusion in the heavy top quark approximation}'',} \textit{ Phys. Lett.
  B} \textbf{ 744} (2015) 43--47,
  \href{http://dx.doi.org/10.1016/j.physletb.2015.03.030}{\doi{10.1016/j.physletb.2015.03.030}},
\href{http://www.arXiv.org/abs/1503.01274}{\texttt{ arXiv:1503.01274}}.

\bibitem{Caola:2015psa}
\hrefCMSnoop {}{F.~Caola, K.~Melnikov, R.~R{\"o}ntsch, and L.~Tancredi, ``{QCD
  corrections to ZZ production in gluon fusion at the LHC}'',} \textit{ Phys.
  Rev. D} \textbf{ 92} (2015) 094028,
  \href{http://dx.doi.org/10.1103/PhysRevD.92.094028}{\doi{10.1103/PhysRevD.92.094028}},
\href{http://www.arXiv.org/abs/1509.06734}{\texttt{ arXiv:1509.06734}}.

\bibitem{Grazzini:2018owa}
\hrefCMSnoop {}{M.~Grazzini, S.~Kallweit, M.~Wiesemann, and J.~Y. Yook, ``{$ZZ$
  production at the LHC: NLO QCD corrections to the loop-induced gluon fusion
  channel}'',} \textit{ JHEP} \textbf{ 03} (2019) 070,
  \href{http://dx.doi.org/10.1007/JHEP03(2019)070}{\doi{10.1007/JHEP03(2019)070}},
\href{http://www.arXiv.org/abs/1811.09593}{\texttt{ arXiv:1811.09593}}.

\bibitem{Caola:2016trd}
F.~Caola\hrefCMSnoop {}{ {et~al.}, ``{QCD corrections to vector boson pair
  production in gluon fusion including interference effects with off-shell
  Higgs at the LHC}'',} \textit{ JHEP} \textbf{ 07} (2016) 087,
  \href{http://dx.doi.org/10.1007/JHEP07(2016)087}{\doi{10.1007/JHEP07(2016)087}},
\href{http://www.arXiv.org/abs/1605.04610}{\texttt{ arXiv:1605.04610}}.

\bibitem{Bishara:2016jga}
\hrefCMSnoop {}{F.~Bishara, U.~Haisch, P.~F. Monni, and E.~Re, ``{Constraining
  Light-Quark Yukawa Couplings from Higgs Distributions}'',} \textit{ Phys.
  Rev. Lett.} \textbf{ 118} (2017) 121801,
  \href{http://dx.doi.org/10.1103/PhysRevLett.118.121801}{\doi{10.1103/PhysRevLett.118.121801}},
\href{http://www.arXiv.org/abs/1606.09253}{\texttt{ arXiv:1606.09253}}.

\bibitem{Bylund:2016phk}
O.~Bessidskaia~Bylund\hrefCMSnoop {}{ {et~al.}, ``{Probing top quark neutral
  couplings in the Standard Model Effective Field Theory at NLO in QCD}'',}
  \textit{ JHEP} \textbf{ 05} (2016) 052,
  \href{http://dx.doi.org/10.1007/JHEP05(2016)052}{\doi{10.1007/JHEP05(2016)052}},
\href{http://www.arXiv.org/abs/1601.08193}{\texttt{ arXiv:1601.08193}}.

\bibitem{Englert:2016hvy}
\hrefCMSnoop {}{C.~Englert, R.~Rosenfeld, M.~Spannowsky, and A.~Tonero, ``{New
  physics and signal-background interference in associated $pp\to HZ$
  production}'',} \textit{ EPL} \textbf{ 114} (2016) 31001,
  \href{http://dx.doi.org/10.1209/0295-5075/114/31001}{\doi{10.1209/0295-5075/114/31001}},
\href{http://www.arXiv.org/abs/1603.05304}{\texttt{ arXiv:1603.05304}}.

\bibitem{Hespel-2015-gg}
\hrefCMSnoop {}{B.~Hespel, F.~Maltoni, and E.~Vryonidou, ``{Higgs and Z boson
  associated production via gluon fusion in the SM and the 2HDM}'',} \textit{
  JHEP} \textbf{ 06} (2015) 065,
  \href{http://dx.doi.org/10.1007/JHEP06(2015)065}{\doi{10.1007/JHEP06(2015)065}},
\href{http://www.arXiv.org/abs/1503.01656}{\texttt{ arXiv:1503.01656}}.

\bibitem{Chatrchyan:2011ya}
\hrefCMSnoop {}{{CMS} Collaboration, ``{Measurement of the weak mixing angle
  with the Drell-Yan process in proton-proton collisions at the LHC}'',}
  \textit{ Phys. Rev. D} \textbf{ 84} (2011) 112002,
  \href{http://dx.doi.org/10.1103/PhysRevD.84.112002}{\doi{10.1103/PhysRevD.84.112002}},
\href{http://www.arXiv.org/abs/1110.2682}{\texttt{ arXiv:1110.2682}}.

\bibitem{CMS-HIG-19-009}
\href {https://cds.cern.ch/record/2725543}{{CMS} Collaboration, ``{Constraints
  on anomalous Higgs boson couplings to vector bosons and fermions in
  production and decay in the $H\to4\ell$ channel}'',} Technical Report
  CMS-HIG-19-009, CERN, Geneva, 2020.

\bibitem{Chetyrkin:1996ke}
\hrefCMSnoop {}{K.~G. Chetyrkin, B.~A. Kniehl, and M.~Steinhauser, ``{Three
  loop O$(\alpha_s^2 G_F M_t^2)$ corrections to hadronic Higgs decays}'',}
  \textit{ Nucl. Phys. B} \textbf{ 490} (1997) 19,
  \href{http://dx.doi.org/10.1016/S0550-3213(97)00051-5}{\doi{10.1016/S0550-3213(97)00051-5}},
\href{http://www.arXiv.org/abs/hep-ph/9701277}{\texttt{ arXiv:hep-ph/9701277}}.

\bibitem{Chetyrkin:1998mw}
\hrefCMSnoop {}{K.~G. Chetyrkin, B.~A. Kniehl, M.~Steinhauser, and W.~A.
  Bardeen, ``{Effective QCD interactions of CP odd Higgs bosons at three
  loops}'',} \textit{ Nucl. Phys. B} \textbf{ 535} (1998) 3,
  \href{http://dx.doi.org/10.1016/S0550-3213(98)00594-X}{\doi{10.1016/S0550-3213(98)00594-X}},
\href{http://www.arXiv.org/abs/hep-ph/9807241}{\texttt{ arXiv:hep-ph/9807241}}.

\bibitem{Bernreuther:2015fts}
W.~Bernreuther\hrefCMSnoop {}{ {et~al.}, ``{Production of heavy Higgs bosons
  and decay into top quarks at the LHC}'',} \textit{ Phys. Rev. D} \textbf{ 93}
  (2016), no.~3, 034032,
  \href{http://dx.doi.org/10.1103/PhysRevD.93.034032}{\doi{10.1103/PhysRevD.93.034032}},
\href{http://www.arXiv.org/abs/1511.05584}{\texttt{ arXiv:1511.05584}}.

\bibitem{Demartin:2014fia}
F.~Demartin\hrefCMSnoop {}{ {et~al.}, ``{Higgs characterisation at NLO in QCD:
  CP properties of the top-quark Yukawa interaction}'',} \textit{ Eur. Phys.
  J.} \textbf{ C74} (2014), no.~9, 3065,
  \href{http://dx.doi.org/10.1140/epjc/s10052-014-3065-2}{\doi{10.1140/epjc/s10052-014-3065-2}},
\href{http://www.arXiv.org/abs/1407.5089}{\texttt{ arXiv:1407.5089}}.

\bibitem{Falkowski:2015wza}
A.~Falkowski\hrefCMSnoop {}{ {et~al.}, ``{Rosetta: an operator basis translator
  for Standard Model effective field theory}'',} \textit{ Eur. Phys. J. C}
  \textbf{ 75} (2015) 583,
  \href{http://dx.doi.org/10.1140/epjc/s10052-015-3806-x}{\doi{10.1140/epjc/s10052-015-3806-x}},
\href{http://www.arXiv.org/abs/1508.05895}{\texttt{ arXiv:1508.05895}}.

\end{thebibliography}
